%% file: main.tex
\documentclass[a4paper,11pt]{article}
\pdfoutput=1 

\usepackage{jheppub} 
\usepackage[T1]{fontenc} 
\usepackage{array}
\usepackage{amsmath}
\usepackage{slashed}
\usepackage{amsmath,amsthm,amssymb}
\usepackage{caption}
\usepackage{bbold}
\usepackage{amssymb}
\usepackage{verbatim}
\usepackage{graphicx}
\usepackage{enumerate}
\usepackage{longtable}
\usepackage{caption}
\usepackage{subcaption}
\usepackage{xcolor}
\usepackage{hyperref}
\usepackage{multirow}
\usepackage{mathtools}
\usepackage{breqn}
\usepackage{cases}
\newcommand{\nn}{\nonumber \\}
\newtheorem{thm}{Theorem}

\newcommand{\unipd}{Dipartimento di Fisica ed Astronomia, Universit\`a di Padova, Via Marzolo 8, 35131 Padova, Italy}
\newcommand{\pdinfn}{INFN, Sezione di Padova, Via Marzolo 8, 35131 Padova, Italy}
\newcommand{\higgsct}{Higgs Centre for Theoretical Physics, School of Physics and Astronomy, The University of Edinburgh, Edinburgh EH9 3JZ, Scotland, UK}

\def\la{\langle}
\def\ra{\rangle}
\def\lb{[}
\def\rb{]}
\def\nn{\nonumber \\}

\title{\boldmath Adaptive Integrand Decomposition in parallel and orthogonal space
	}

\author[a,b]{Pierpaolo Mastrolia,}
\author[c]{Tiziano Peraro,}
\author[a,b]{Amedeo Primo}

\affiliation[a]{\unipd}
\affiliation[b]{\pdinfn}
\affiliation[c]{\higgsct}

\emailAdd{pierpaolo.mastrolia@cern.ch}
\emailAdd{tiziano.peraro@ed.ac.uk}
\emailAdd{amedeo.primo@pd.infn.it}

\preprint{Edinburgh 2016/08}
\abstract{We present the integrand decomposition of multiloop
  scattering amplitudes in parallel and orthogonal space-time
  dimensions, $d=d_\parallel+d_\perp$, being $d_\parallel$
  the dimension of the parallel space spanned by the legs of the diagrams. 
  When the number $n$ of  external legs is
  $n\le 4$, the corresponding 
  representation of the multiloop integrals exposes a subset of integration variables
  which can be easily integrated away by means of Gegenbauer
  polynomials orthogonality condition. 
  By decomposing the integration momenta along parallel and orthogonal
  directions, the polynomial division algorithm is drastically
  simplified. Moreover, the orthogonality conditions of Gegenbauer
  polynomials can be suitably applied to integrate the decomposed
  integrand, yielding the systematic annihilation of spurious terms.
  Consequently, multiloop amplitudes are expressed in terms of
  integrals corresponding to irreducible scalar products
  of loop momenta and external momenta. We revisit the one-loop decomposition, which turns out to be
  controlled by the maximum-cut theorem in different dimensions, and we discuss the integrand
  reduction of two-loop planar and non-planar integrals up to $n=8$
  legs, for arbitrary external and internal kinematics. The proposed
  algorithm extends to all orders in perturbation theory.}

\begin{document} 
\maketitle
\newpage
\input{Intro.tex}

\input{Transverse.tex}
\input{Adaptive.tex}
\input{Example.tex}
\input{Conclusions.tex}

\clearpage
\appendix
\input{Appendix.tex}
\bibliographystyle{JHEP}
\bibliography{references}
\end{document}

%% file: Intro.tex
\section{Introduction} 
The decomposition of multiloop scattering amplitudes in terms of independent
functions, together with the subsequent determination of the
latter, is a viable alternative - often the only accessible one - to
the direct integration which, for non-trivial processes,
may require the calculation of a prohibitively large 
 number of complicated Feynman integr{\it als}.

Understanding the properties of Feynman integr{\it ands} has led to the
development of novel algorithms aiming to the automated determination
of partonic cross sections for high-multiplicity processes which have been
successfully applied, in the last
decade, to one-loop amplitudes. More generally, the use of 
{\it  unitarity-based methods} and 
{\it integrand decomposition algorithms} 
has shown that exploiting the algebraic properties of the integrands may
lead to the discovery of novel properties of the amplitudes, hidden beneath the superficial look
of Feynman integrals' representation, which, if properly
engineered, may turn into drastic simplifications for their evaluation.

In this paper, we elaborate on a representation of dimensionally
regulated Feynman integrals where, for any given diagram, the number of space-time dimensions
$d \ ( = 4 - 2 \epsilon)$ is split into
{\it parallel} (or longitudinal) and {\it orthogonal} (or transverse)
dimensions, as $d=d_{\parallel}+ d_{\perp}$ \ 
\cite{Collins:105730,Kreimer:1991wj,Kreimer:1992zv,Czarnecki:1994td,Frink:1996ya,kreimer:1996qy}. 
Accordingly, the parallel space is spanned by the independent external momenta of the
loop, namely $d_{\parallel} = n - 1$, where $n$ is the number of legs, whereas
the transverse space is spanned by the complementary orthogonal directions.
For diagrams with a number of legs 
$n \ge 5$, the 
orthogonal space embeds the $-2 \epsilon$ regulating dimensions, $d_\perp = - 2 \epsilon$,
while, for diagrams with $n \le 4$,
the orthogonal space is larger and it embeds, beside the regulating dimensions, also  
 the four-dimensional complement of the parallel space, namely  
 $d_\perp = (5-n) - 2 \epsilon$. In this sense, the decomposition of
 the space-time dimensions in parallel and orthogonal directions can
 be considered as {\it adaptive}, since it depends on the number
 of legs of the individual loop.

Decomposing the loop momenta $q_{i}^{\alpha}$ in terms of parallel and
orthogonal vectors, $q_{i}^{\alpha}=q_{\parallel\,i}^{\alpha}+\lambda_{i}^{\alpha}$, has the
immediate advantage of exposing a subset of integration variables
which can be trivially integrated away, hence they can be 
eliminated from the calculation before applying any reduction
algorithm.
In fact, multidimensional polar coordinates can be suitably introduced in order to
parametrize the integral over the orthogonal space in terms of integrations
over radial variables $\lambda_{ii}(=\lambda_{i}\cdot\lambda_{i})$ and
a generalised solid angle. This change of coordinates makes manifest that numerators and denominators of Feynman integrands do not depend on the same set of integration
variables.
Indeed, the quadratic Feynman denominators 
depend {\it only} on the parallel directions, on the radial variables $\lambda_{ii}$ and the {\it relative orientations} $\lambda_{ij}$, $i< j$, of the transverse vectors
but they do not depend on their individual components, which can be mapped into a set of angular variables $\boldsymbol{\Theta}_{\perp}$.
Conversely, the numerators may depend on {\it all} variables. 
In the case of diagrams with $n \le 4$, 
the dependence of the integrand on transverse angles,
say $\theta_i$, is polynomial in $\sin \theta_i$ and $\cos \theta_i$ and, therefore,
the integration over $\boldsymbol{\Theta}_{\perp}$ can be trivially
performed. In this article, we show how it can be carried out by means of
the {\it orthogonality relation for Gegenbauer polynomials}, as an
alternative to the Passarino-Veltman tensor reduction used in
ref.~\cite{Kreimer:1991wj}.

{\it After integrating over the transverse angles} $\boldsymbol{\Theta}_{\perp}$, the
integrand will solely depend on $q_{\parallel\,i}$  and on the $\lambda_{ij}$ variables appearing in the denominators. These variables correspond to
(reducible and irreducible) scalar products between loop momenta and
external momenta.\\ 

The integration over orthogonal and parallel space has been used to evaluate multi-scale Feynman integrals, up to two- and
three-point
functions~\cite{Kreimer:1991wj,Kreimer:1992zv,Czarnecki:1994td,Frink:1996ya,kreimer:1996qy}. The
goal of this communication is instead discussing how the decomposition of space-time into parallel and orthogonal
subspaces simplifies the multiloop integrand reduction algorithm 
~\cite{Mastrolia:2011pr,Badger:2012dp,Zhang:2012ce,Mastrolia:2012an,Mastrolia:2012wf}.
Namely, our objective is not the evaluation of Feynamn integrals, rather
their decomposition in terms of independent integrals. We show that this
procedure can be applied to arbitrarily complicated diagrams. In
particular, we consider the decomposition up to two-loop eight-point planar
and non-planar integrals and we discuss how the same procedure can be
extended to higher orders. Previous studies of two-loop integrands in four-dimensions can be found in~\cite{Feng:2012bm,ArkaniHamed:2010kv,ArkaniHamed:2010gh}.

Feynman integrals are multivariate integrals of rational integrands
and they can be decomposed in terms of a set of irreducible integrals
(IRIs) by multivariate polynomial division~\cite{,Zhang:2012ce,Mastrolia:2012an}. In fact, the partial fractioning
of Feynman integr{\it ands} amounts to iterative divisions (modulo Gr\"{o}bner basis)
between the numerator and the denominators, once they are written as
polynomials in the components of the integration momenta along a given
basis of momentum space.
 The resulting integrand decomposition is a sum of integrands 
whose denominators are given by all the possible partitions of the initial set of
denominators and whose numerators correspond to the
{\it remainders} of the division w.r.t.\ the set of denominators they
sit on. The remainders of the division contain, by
definition, terms which cannot be expressed in terms of
denominators. In fact, since each component of a given integration momentum
corresponds to a scalar product of that momentum with an element
of the momentum basis, the remainder should contain only irreducible
scalar products (ISPs). On the contrary, reducible
scalar products (RSPs) can be decomposed in terms of
denominators.

 The integrand decomposition is effectively a 
unitarity-based decomposition of the integrand, since 
each remainder can be considered as the {\it residue of the cut} identified by the
simultaneous vanishing of the corresponding denominators.
It should be observed that integrand reduction can be applied as well to the case of integrals whose
denominators are raised to powers higher than one
\cite{Mastrolia:2013kca}.
Integrating the decomposed integrand over the loop momenta corresponds
to the decomposition of the original integral in terms of IRIs. In fact, upon integration, some of the
ISPs in the residues may correspond to vanishing integrals: these
terms are called  {\it spurious}, because although
present in the integrand decomposition, they do not contribute to the
amplitude. Instead, the non-spurious ISPs correspond to
the (numerators of) IRIs appearing in the amplitude decomposition.
Therefore, within the integrand decomposition algorithm, the reduction of any
scattering amplitude in terms of IRIs turns into the algebraic problem
of determining the coefficients of the monomials of the residues. 

The basic elements of the integrand decomposition algorithm are:
{\it i)} the space-time dimensions, namely the number of integration variables;
{\it ii)} the momentum basis used for the decomposition of the loop
momenta;
{\it iii)} the structure of the numerators and the variables they
depend on;
{\it iv)} the form of the denominators and the variables they
depend on;
{\it v)} the structure of the residues;
{\it vi)} the solutions of the cut equations.
The integrand reduction algorithm was originally proposed for one-loop
integrals in four dimensions \cite{Ossola:2006us,Ossola:2007bb} and later extended to $d=4 - 2 \epsilon$
dimensions \cite{Ellis:2007br,Ellis:2008ir,Ossola:2008xq,Mastrolia:2008jb}, to deal with dimensionally regulated
amplitudes (see \cite{Ellis:2011cr} for a review). 
In the one-loop case, the residues were built by using two driving principles:
on the one side, the knowledge of the set of IRIs which could appear in the decomposition
of generic one-loop integrals \cite{Passarino:1978jh} and, on the other side,
the Lorentz covariance of spurious terms which could additionally appear in
the numerators.

The integrand reduction algorithm for one-loop integrals has been implemented in several
public libraries, like {\sc Cutools} \cite{Ossola:2007ax},
 {\sc Samurai} \cite{Mastrolia:2010nb} and {\sc Ninja}
 \cite{Mastrolia:2012bu,Peraro:2014cba}, 
which played an important role in the
development of codes for the automatic evaluation of scattering
amplitudes for generic scattering processes at NLO accuracy, as recently reviewed in \cite{vanDeurzen:2015jmn}.
In particular, {\sc Ninja} implements an ameliorated integrand
decomposition algorithm \cite{Mastrolia:2012bu}, which introduced the idea of the (univariate) polynomial
division for the calculation of the residues.

In order to extend the integrand decomposition at higher orders \cite{Mastrolia:2011pr,Badger:2012dp}, the
same driving principles could not be
applied. The first reason for this is that the
basis of independent integrals is not known. Moreover, the interplay of more integration
momenta makes the classification of the spurious terms less obvious.
One additional difference w.r.t.\ the one-loop case, which was indeed to be expected, is the contribution of integrals
corresponding to non-spurious ISPs~
\cite{Mastrolia:2011pr}.
Nevertheless, the systematic determination of the residues at higher order was systematized by means of algebraic geometry methods \cite{Zhang:2012ce,Mastrolia:2012an},
namely the polynomial division modulo Gr\"obner basis. An implementation of such method is provided by the public package {\sc BasisDet} \cite{Zhang:2012ce}. Integrand decomposition beyond one-loop has been successfully applied
to a first case of non trivial two-loop five-point helicity amplitude in \cite{Badger:2013gxa,Badger:2015lda}.

One of the main outcomes of the multivariate polynomial division
algorithm is the so called {\it maximum-cut theorem}
\cite{Mastrolia:2012an}, which can be applied
whenever the on-shell conditions are sufficient in order to
constrain all integration
variables. In this case, the system of equations is
zero-dimensional and the remainder of the division (of a numerator
that depends on all variables constrained by the cut-conditions) can be cast as a univariate polynomial of degree $n_s-1$, being $n_s$ the number of
solutions of the system. 
This theorem extends to all loops and to all dimensions the beauty of the
four-dimensional quadruple-cut \cite{Britto:2004nc}, which is
known to have {\it two} solutions and whose residue is parametrized in
terms of {\it two} monomials \cite{Ossola:2006us}.
The number of integration variables depends on the dimensions of the loop
momenta, therefore the number of denominators to be put on-shell in order to
fix them depends on the space-time dimensions as well. Therefore, maximum-cuts are
realized by cutting diagrams with different number of external legs,
according to the dimensionality of the integration momenta. \\

The use of the $d=d_{\parallel}+d_{\perp}$ representation of Feynman integrals in tandem with the integrand reduction technique has several interesting
consequences.  These allow to organize the algorithm in three steps,
and we will refer to it as \emph{divide-integrate-divide}.

{\bf Divide}.  First, one can see that any split of the loop
components with separates the physical directions from the
$(-2\epsilon)$-dimensional ones yields simpler on-shell cut conditions, hence
the division procedure becomes significantly simpler.  In fact, the
Gr\"obner basis trivialize, as they are linear in the variables to be
reduced and quadratic in the irreducible variables which will appear
in the residues, up to a choice of monomial order. In this case, the
Gr\"obner basis are built from differences of denominators (basic
${\cal S}$-polynomials).  Moreover, the form of the cut conditions and
the Gr\"obner basis is further simplified in the
$d=d_{\parallel}+d_{\perp}$ representation, due to the dependence of
the denominators on a reduced set of variables, hence the
determination of the cut-residues becomes lighter.  We can properly talk
of {\it adaptive cutting}, since the dimensions of the parallel space,
\textit{i.e.} the number of variables constrained by the on-shell
conditions, depend on the number of legs.

{\bf Integrate.}
Second, {\it after} the integrand reduction, the integration over the orthogonal solid angle of
the decomposed integrand allows for the automatic detection and
annihilation of the spurious
integrals, which vanish because of the orthogonality condition
enforced by the Gegenbauer polynomial integration. Within the proposed
parametrization, the spherical symmetry of the transverse angular integrations offers an explicit  
geometric interpretation of the spurious integrals as being related to monomials which
are odd under rotation group transformations, as observed in \cite{Ita:2015tya}. Alternatively, if the integration over $\boldsymbol{\Theta}_{\perp}$ is
performed {\it before} the reduction, the corresponding residue will
not contain any spurious term, therefore the number of non-vanishing
coefficients to be determined through the reduction algorithm will be significantly
smaller.

{\bf Divide.}
Finally, we notice that the integration of the residues over
the transverse angles $\boldsymbol{\Theta}_{\perp}$ reintroduces, in general, a dependence on
the variables $\lambda_{ij}$. The denominators depend on these
variables and, therefore, the integrated
residues may be subject to a {\it second polynomial division}, which
further simplifies them.  In some cases, namely when the variables
$\lambda_{ij}$ form a Gram determinant, this additional division can be shown to be
equivalent to applying dimensional shifting recurrence relations \cite{Tarasov:1996br,Lee:2009dh} at the
integrand level (the dimensions of any Feynman integral are controlled by
the power of the Gram determinant, characteristic of each loop).\\

We now observe that
{\it after the integrand decomposition} outlined above, the
integrand will depend on a subset of the parallel space variables and on the transverse variables $\lambda_{ij}$, which correspond just to
irreducible scalar products (ISPs) between loop momenta and
external momenta. Therefore,
any scattering amplitude, at any loop order and with arbitrary kinematics, 
admits an algebraic decomposition in terms of a set of 
irreducible integrals (IRIs), corresponding to these ISPs.

It is important to stress that, although independent under polynomial division, the IRIs are
not a minimal set.
Indeed, they can be related through identities which belong to the
general class of integration-by-parts relations
(IBPs), hence their number can be further reduced. The amplitude, in this case,
would be finally expressed in terms of a minimal set of Master
Integrals (MIs).
IBPs relation for IRIs can be suitably built by algebraic geometry
methods through sygyzy 
equations \cite{Gluza:2010ws,Ita:2015tya,Larsen:2015ped}.
In particular, the outcome of the proposed integrand reduction
algorithm is suitable for an IBP-reduction in the
parallel space along the lines of \cite{Baikov:1996iu,Larsen:2015ped}.
Progress on the improvement of system solving strategies for IBP
equations are under intense development \cite{vonManteuffel:2014ixa,Kant:2013vta}.
Moreover, should the reduction to MIs not be available for the process under
consideration, the representation of the amplitudes in terms of
IRIs can be employed in tandem with the numerical integration of the
latter. Promising advances on the numerical integration of Feynman
integrals have recently been applied to a non-trivial two-loop case \cite{Borowka:2016ehy}.\\

The paper is organized as follows. 
In sec.~\ref{sec:1}, we discuss the $d=d_{\parallel}+d_{\perp}$
representation of multiloop Feynman integrals and the integration over
the transverse directions by means of the orthogonality relation for
Gegenbauer polynomials. Besides analysing the properties of the
transverse space for general topologies with $n\leq 4$ external legs,
we discuss further simplifications that can be obtained for factorized
and ladder topologies. In fact, we show that the integration of Gegenbauer
polynomials can be used in all cases where the numerator depends on
more variables than the denominators. 
As an example of the considerably simplified form of Feynman
integrands achieved by integrating out the transverse directions
prior to the application of any reduction algorithm, we discuss a
four-point contribution to the helicity amplitude
$A(g_1^+,g_2^-,g_3^+,g_4^-)$ at two loops. 
In sec.~\ref{sec:2}, we
present the {\it adaptive integrand decomposition} algorithm for
multiloop scattering amplitudes. We revisit the well-know results for
the one-loop integrand decomposition, by showing that, in
$d=d_{\parallel}+d_{\perp}$, all unitarity cuts are reduced to
zero-dimensional systems and by providing an alternative
representation of the residues, dictated by the maximum-cut theorem,
as complete polynomials in the transverse variables. 
The novel parametrization of the residues emerging in
$d=d_{\parallel}+d_{\perp}$ yields a different, yet equivalent,
decomposition of one-loop amplitudes w.r.t.\ to the known
decomposition in $d=4-2\epsilon$. 
At two loops we provide a classification of the residues appearing in
the integrand decomposition formula for planar and non-planar
topologies with arbitrary kinematics, by considering the top-down
reduction from the eight-point maximum-cut topologies, down to the
product of two one-point topologies. As a concrete example of the application of the {\it adaptive} division algorithm, we provide the explicit expression of the coefficients of the residue of the double-box contribution to $A(g_1^+,g_2^-,g_3^+,g_4^-)$.
In sec.~\ref{sec:3}, we give our summary and conclusions.

We have collected in the appendices the detailed discussion of most of the calculations leading to the results presented in this work. In appendix~\ref{Ap:1}, we propose a new derivation of the parametric expression of Feynman integrals in terms of parallel- and transverse- space variables and we discuss the change of coordinates to be performed in the transverse space in order to map, at any loop order, all integrations over the four-dimensional transverse directions into simple angular integrals.
In appendices~\ref{Ap:1l}-\ref{Ap:2l}, we collect some useful formulae for one- and two- loop integrals respectively, including a list of tensor integrals which can be reduced by integrating over the transverse angles. In appendix~\ref{Ap:2} we recall the main properties of Gegenbauer polynomials and, finally, in appendix~\ref{Ap:3}, we provide a representation in terms of spinor variables of the momentum-basis to which we refer throughout the text. The calculations presented in this paper have been performed with the help of {\sc Singular}~\cite{DGPS}.

%% file: Transverse.tex
		\section{Parallel and orthogonal space for multiloop Feynman integrals}
		\label{sec:1}
		In this section we consider generic $\ell$-loop Feynman integrals with $n$ external legs in a $d$-dimensional Euclidean space,
		\begin{align}
		I_{n}^{d\,(\ell)}[\mathcal{N}]=\int\left( \prod_{i=1}^{\ell}\frac{d^dq_{i}}{\pi^{d/2}}\right)\frac{\mathcal{N}(q_{i})}{\prod_{j}D_{j}(q_{i})},
		\label{eq:Ilk+1}
		\end{align}
		where $\mathcal{N}(q_{i})$ is an arbitrary tensor numerator and the denominators $D_j(q_i)$ are defined as
		\begin{align}
		&D_j=l_j^2+m_j^2, \quad\text{with}\quad l_j^{\alpha}=\sum_{i}\alpha_{ij}q_i^{\alpha}+\sum_{i}\beta_{ij}p_i^{\alpha},
		\label{eq:dens2l}
		\end{align}
		being $\{p_1,\,\dots,\,p_{n-1}\}$ the set of independent external momenta and $\alpha$ and $\beta$ incidence matrices taking values in $\{0,\pm 1\}$. We first recall the usual parametrization of $I_{n}^{d\,(\ell)}$ obtained by formally splitting the $d$-dimensional space into the four-dimensional physical one, where external momenta and polarizations lie, and the corresponding orthogonal subspace, whose dimension is conventionally set to $d-4=-2\epsilon$. Later we show that, when a Feynman integral has $n\leq 4$ external legs which do not span the full physical space, $I_{n}^{d\,(\ell)}$ is more conveniently expressed in terms of vectors living in the $d_{\parallel}=n-1$ dimensional space described by the external kinematics and a set of transverse variables belonging to its orthogonal complement with dimension $d_{\perp}=d-n-1$. This alternative parametric representation of Feynman integrals remarkably simplifies, at any loop order, the integration over the transverse components of the loop momenta.
		\subsection{Feynman integrals in $d=4-2\epsilon$}
		 When dealing with a regularization scheme where the external kinematics is kept in four dimensions, it is customary to  split the $d$-dimensional loop momenta into a four-dimensional part and a $(-2\epsilon)$-dimensional one, 
		\begin{align}
		q^{\alpha}_{i}=&q_{[4]\,i}^{\alpha}+\mu_i^{\alpha},
		\label{4mu}
		\end{align}
		so that, by defining $\mu_{ij}=\mu_{i}\cdot\mu_{j}$, we have  
		\begin{align}
		q_i\cdot q_{j}=q_{[4]\, i}\cdot q_{[4]\, j}+\mu_{ij}.
		\end{align}
		The vectors $\mu_i^{\alpha}$ lie in a subspace which is completely orthogonal to the four-dimensional one, $\mu_i\cdot p_j=0$, hence we can rewrite the denominators \eqref{eq:dens2l} as
		\begin{align}
		&D_i=l_{i[4]}^2+\sum_{j,k}\alpha_{ij}\alpha_{ik}\,\mu_{jk}+m_i^2,\quad\text{with}\quad l_{i[4]}^{\alpha}=\sum_{j}\alpha_{ij}q_{i[4]}^{\alpha}+\sum_{j}\beta_{ij}p_j^{\alpha}.
		\label{eq:densmu}
		\end{align}
		For the same reason, the numerator appearing in \eqref{eq:Ilk+1} can depend on  $q_{i[4]}^{\alpha}$ and  $\mu_{ij}$ only. This means that we can express the integrals over the $(-2\epsilon)$-dimensional subspace into spherical coordinates and integrate out all directions orthogonal to the relative orientations of the vectors $\mu_i^{\alpha}$, obtaining
		\begin{align}
		I_{n}^{d\,(\ell)}[\mathcal{N}]=\Omega^{(l)}_d\int\prod_{i=1}^{\ell} d^4q_{[4]\,i}\int \prod_{1\leq i\leq j\leq \ell} d\mu_{ij}\left[G(\mu_{ij})\right]^{\frac{d-5-\ell}{2}} \frac{\mathcal{N}(q_{[4]\,i},\mu_{ij})}{\prod_{m}D_{m}(q_{[4]\,i},\mu_{ij})},
		\label{eq:oldpar}
		\end{align}
		where $G(\mu_{ij})=\text{det}[(\mu_i\cdot\mu_j)]$  is the Gram determinant and
		\begin{align}
		\Omega^{(\ell)}_d=\prod_{i=1}^{\ell}\frac{\Omega_{(d-4-i)}}{2\pi^{\frac{d}{2}}},\qquad \Omega_n=\frac{2\pi^{\frac{n+1}{2}}}{\Gamma\left(\frac{n+1}{2}\right)}.
		\end{align}
		As it is explicitly shown by \eqref{eq:densmu}, in this parametrization the set of denominators which characterizes the integral depends, in general, on the same ${\ell(\ell+9)}/{2}$ variables as the numerator, corresponding to the $4\ell$ four-dimensional components of the loop momenta, which are decomposed into some basis of four-dimensional vectors $\{e_i^{\alpha}\}$,
		\begin{align}
		q_{[4]\, i}^{\alpha}=\sum_{j=1}^{4}x_{ji}e_j^{\alpha},
		\end{align}
		 and the ${\ell(\ell+1)}/{2}$ scalar products $\mu_{ij}$. It should be noted that, at multiloop level, the denominators of particular classes of Feynman integrals, such as ladder topologies and factorized integrals, might depend on a reduced number of variables $\mu_{ij}$, due to the absence of propagators involving both loop momenta $q_{i}^{\alpha}$ and $q_{j}^{\alpha}$. In the following, we first derive an integral parametrization alternative to \eqref{eq:oldpar}, valid for Feynman integrals with $n\leq 4$ external legs, by assuming that the denominators depend on the maximal number of loop variables and then we show how this parametrization can be used, in a further simplified form, for ladder and factorized integrals as well.
		 \subsection{Feynman integrals in $d=d_{\parallel}+d_{\perp}$} \label{sec:feyintdpardperp}
		 For a number of external legs $n\leq 4$, it is possible to reparametrize the integral \eqref{eq:Ilk+1} in such a way that the number of variables appearing in the denominators is reduced to ${\ell(\ell+2d_{\parallel}+1)}/{2}$ and the integration over the remaining $\ell(4-d_{\parallel})$ variables, upon which the numerator will generally show a polynomial dependence, can be performed through a straightforward expansion of the numerator in terms of orthogonal polynomials. 
		In fact, the choice of the four-dimensional basis $\{e_i^{\alpha}\}$ is completely arbitrary and, if $d_{\parallel}\leq 3$, one can choose $4-d_{\parallel}$  vectors of such basis to lie into the subspace orthogonal to the external kinematics, \textit{i.e.}
		 \begin{subequations}
		 	\begin{align}
		 	&e_i\cdot p_j=0,\:\qquad i> d_{\parallel},\quad \forall j,\\
		 	&e_i\cdot e_j =\delta_{ij},\quad \;i,j> d_{\parallel}.
		 	\label{eq:perp}
		 	\end{align}
		 	\label{eq:ep}
		 \end{subequations}
		 In this way, we can rewrite the $d$-dimensional loop momenta as
		 \begin{align}
		 q_{i}^{\alpha}=q_{\parallel\, i}^{\alpha}+\lambda^{\alpha}_{i},
		 \label{eq:newdeco}
		 \end{align}
		 where $q_{\parallel\, i}^{\alpha}$ is a vector of the $d_{\parallel}$-dimensional space spanned by the external momenta,
		 \begin{align}
		 q_{\parallel\, i}^{\alpha}=\sum_{j=1}^{d_{\parallel}}x_{ji}e_{j}^{\alpha},
		 \end{align}
		 and 
		 \begin{align}
		 &\lambda^{\alpha}_i=\sum_{j=d_{\parallel}+1}^4x_{ji}e_j^{\alpha}+\mu_i^{\alpha},\qquad \lambda_{ij}=\sum_{l=d_{\parallel}+1}^4x_{li}x_{lj}+\mu_{ij},
		 \end{align}
		 belongs the $d_{\perp}$-dimensional orthogonal subspace. In this parametrization, all denominators become independent of the transverse components of the loop momenta,
		 \begin{align}
		 &D_i=l_{\parallel\, i}^2+\sum_{j,l}\alpha_{ij}\alpha_{il}\,\lambda_{jl}+m_i^2,\quad\text{with}\quad
		 &l_{\parallel\, i}^{\alpha}=\sum_{j}\alpha_{ij}q_{\parallel\, i}^{\alpha}+\sum_{j}\beta_{ij}p_j^{\alpha},
		 \label{eq:dens2l2}
		 \end{align}
		 and they depend on a reduced set of ${\ell(\ell+2d_{\parallel}+1)}/{2}$ variables, corresponding to the $\ell d_{\parallel}$ components of $q_{\parallel\, i}^{\alpha}$ and the $\ell(\ell+1)/2$ scalar products $\lambda_{ij}$. Once the decomposition \eqref{eq:newdeco} has been introduced, it can be shown that all transverse components $x_{ji}$ ($j>d_{\parallel}$) as well as the relative orientations of the vectors $\lambda_{i}^{\alpha}$ can be  mapped into angular variables, defined through a suitable orthonormalization procedure described in appendix~\ref{Ap:1}. In particular, by introducing the angles
		 \begin{align}
		 \boldsymbol{\Theta}_{\Lambda}=&\{\theta_{ij}\},\qquad\quad 1\leq i<j\leq \ell,\nn
		 \boldsymbol{\Theta}_{\perp}=&\{\theta_{ij}\}, \qquad j\leq i\leq j+3-d_{\parallel},\:\;1\leq j\leq \ell, 
		 \end{align}
		we can define a polynomial transformation of the type
		  \begin{align}
		  \begin{cases}
		 \lambda_{ij}\to& P\left[\lambda_{kk},\sin[\boldsymbol{\Theta}_{\Lambda}],\cos[\boldsymbol{\Theta}_{\Lambda}]\right],\qquad\quad i\neq j, \\
		 x_{ji}\to& P\left[\lambda_{kk},\sin[\boldsymbol{\Theta}_{\perp,\,\Lambda}],\cos[\boldsymbol{\Theta}_{\perp,\,\Lambda}]\right],\quad  j>d_{\parallel},\quad k=1,\dots\ell \\
		 \end{cases}
		 \label{eq:poltr}
		 \end{align}
		such that the integral \eqref{eq:Ilk+1} can be rewritten as
		\begin{align}
		I_{n}^{d\,(\ell)}[\mathcal{N}]=\Omega^{(\ell)}_d\!\int\prod_{i=1}^{\ell} d^{n-1}q_{\parallel \, i}\int\! d^{\frac{\ell(\ell+1)}{2}} \boldsymbol{\Lambda}\int\! d^{(4-d_{\parallel})\ell}\boldsymbol{\Theta}_{\perp}\frac{\mathcal{N}(q_{i \,\parallel},\boldsymbol{\Lambda},\boldsymbol{\Theta}_{\perp})}{\prod_{j}D_{j}(q_{ \parallel\, i},\boldsymbol{\Lambda})},
		\label{eq:lambth}
		\end{align}
		where
		\begin{subequations}
		\begin{align}
		\label{eq:lambdaShout}
		\int d^{\frac{\ell(\ell+1)}{2}} \boldsymbol{\Lambda}
		=&\int \prod_{1\leq i\leq j}d\lambda_{ij}\left[G(\lambda_{ij})\right]^{\frac{d_{\perp}-1-\ell}{2}}\\
		=&\int_{0}^{\infty}\!\prod_{i=1}^{\ell}d\lambda_{ii}(\lambda_{ii})^{\frac{d_{\perp}-2}{2}}
		\int \!d^{\frac{\ell(\ell-1)}{2}}\boldsymbol{\Theta}_{\Lambda},
		\label{eq:Lambdaint}
		\end{align}
		\end{subequations}
		with
		\begin{align}
		\int \!d^{\frac{\ell(\ell-1)}{2}}\boldsymbol{\Theta}_{\Lambda}=&\int_{-1}^{1}\!\prod_{1\leq i<j\leq \ell}^\ell \!d\!\cos\theta_{ij}(\sin\theta_{ij})^{d_{\perp}-2-i},
		\end{align}
		defines the integral over the variables $\boldsymbol{\Lambda}=\{\lambda_{ii},\boldsymbol{\Theta}_\Lambda\}$, corresponding the norm of the transverse vectors $\lambda_{i}^{\alpha}$ and their relative orientations, and
		\begin{align}
		\int d^{(4-d_{\parallel})\ell}\boldsymbol{\Theta}_{\perp}=\int_{-1}^{1}\prod_{i=1}^{4-d_{\parallel}}\prod_{j=1}^{\ell}\!d\!\cos\theta_{i+j-1\;j}(\sin\theta_{i+j-1\;j})^{d_{\perp}-i-j-1}
		\end{align}
		parametrizes the integral over the components of $\lambda_{i}^{\alpha}$ lying in the four-dimensional space. We observe that, since the choice of the four-dimensional basis $\{e_i^{\alpha}\}$ and the consequent definition of the transverse space variables $\boldsymbol{\Lambda}$ and $\boldsymbol{\Theta}_{\perp}$ are determined by the external kinematics and do not depend on the specific set of denominators which characterizes the integral, the parametrization \eqref{eq:lambth} can be used for both planar and non-planar topologies. Moreover it should be noted that, in the case of two-point integrals with vanishing external momentum $p^2=0$, the r.h.s. of \eqref{eq:lambth} holds for $d_{\parallel}=2$, since we can define only two directions orthogonal to a massless vector.
	
	    An important observation regarding eq.~\eqref{eq:newdeco} is that it allows to express a subset of components of $q_{\parallel i}^\alpha$ and $\lambda_{ij}$ as combinations of loop denominators by solving linear relations. One can indeed always build differences of denominators which are linear in the loop momenta and independent of $\lambda_{ij}$, while the relation between $\lambda_{ij}$ and the denominators is always linear by definition, as apparent from eq.~\eqref{eq:dens2l2}.
	    
	      More explicitly, at one-loop all the loop denominators can be taken to have the form
	    \begin{equation}
	      D_j = \big(q_1+\sum_i \beta_{ij} p_i\big)^2 + m_j^2, \qquad j=1,\ldots,r
	    \end{equation}
	    where $r$ is the total number of loop denominators.  Hence one can choose any denominator $D_{\bar j}$ and consider $r-1$ differences of the form $D_j-D_{\bar j}$.  These differences have no quadratic terms in the loop momenta and can thus be used to express $r-1$ of the variables $\{x_{ji} , j\leq d_\parallel\}$ as linear combinations of denominators.  By applying one more independent equation, given by the definition of any of the denominators, the variable $\lambda_{11}$ is written as a linear combination of the variables $\{x_{ji} , j\leq d_\parallel\}$, as one can see from eq.~\eqref{eq:dens2l2}. 
	    
	     At higher loops one can split the $r$ loop denominators into partitions identified by the subset of loop momenta each denominator depends on, and similarly consider differences of denominators belonging to the same partition which will again generate a set of linear relations between physical loop components and denominators.  By solving these relations, one can express a subset of the variables $\{x_{ji} , j\leq d_\parallel\}$ as linear combinations of denominators.  Finally one can, again, consider eq.~\eqref{eq:dens2l2} for a representative of each partition of denominators, completing the set of linear relations which can thus be solved for a subset of the variables $\lambda_{ij}$.  It is straightforward to see that the complete set of relations is equivalent to the definition of the loop denominators themselves. 
	     
	      As an explicit example, at two loops one can have at most three partitions $P_1,P_2,P_3$, which respectively correspond to denominators having the following forms
	    \begin{align}
	      D_j ={}& \big(q_1+\sum_i \beta_{ij} p_i\big)^2 + m_j^2, \qquad j\in P_1, \nonumber \\
	      D_j ={}& \big(q_2+\sum_i \beta_{ij} p_i\big)^2 + m_j^2, \qquad j\in P_2,  \nonumber \\
	      D_j ={}& \big(q_1+q_2+\sum_i \beta_{ij} p_i\big)^2 + m_j^2, \qquad j\in P_3.
	    \end{align}
	    Therefore one can choose a representative for each partition, say $D_{\bar j_i}\in P_i$ for $i=1,2,3$, and observe that for any $j\in P_i$ the difference $D_j-D_{\bar j_i}$ is linear in the loop momenta.  This allows to write $r-3$ linear equations which can be solved for a subset of the variables $\{x_{ji} , j\leq d_\parallel\}$ in terms of the other physical directions and denominators.  One can thus complete this set of relations with 3 more equations (or possibly less, if any of the partitions is empty) which are defined by eq.~\eqref{eq:dens2l2} applied to one denominator for each partition $P_i$.  In the case when none of the partitions is empty, these three equations can be solved for the variables $\lambda_{11},\lambda_{12},\lambda_{22}$ which are thus written as a combination of denominators and irreducible components of $q_{\parallel i}^\alpha$ by solving linear relations.  If the denominators are independent of $\lambda_{12}$, this variable cannot obviously be written in terms of denominators but it can be integrated out by means of the techniques presented later on in this paper.  As we shall see in sec.~\ref{sec:factorizedandladder}, this is true at any number of loops, whenever the loop momenta are independent of one of these variables. 
	    
	    The observations made in this paragraph imply that solving the $d$-dimensional cut constraints for integrand reduction methods is never more complex than solving a linear system of equations.  It is worth noticing, for completeness, that a similar procedure can also be applied to the decomposition of eq.~\eqref{4mu}, the main difference being that the resulting relations for $\mu_{ij}$ will not only depend on the components of the loop momenta along the physical directions, but also on the orthogonal directions.
	
		\subsection{Angular integration over the transverse space}
		As we have explicitly indicated in \eqref{eq:lambth}, the denominators of Feynman integrals, being completely independent of the transverse components of the four-dimensional loop momenta, do not depend on any of variables $\boldsymbol{\Theta}_{\perp}$, which are in one-by-one correspondence with $\{x_{ji}\}$, $j>d_{\parallel}$. In addition, since the dependence of the numerator on the transverse variables is a polynomial one and \eqref{eq:poltr} is a polynomial transformation, after the change of variables the integrand is mapped into a polynomial in (\textit{sine} and \textit{cosine} of) $\boldsymbol{\Theta}_{\perp}$, with rational coefficients depending on $\boldsymbol{\Lambda}$ and on the physical directions $\{x_{ji}\}$, $j\leq d_{\parallel}$. Finally, we observe that all the integrals over $\boldsymbol{\Theta}_{\perp}$ can be performed \textit{independently} from each other and that they are all of the type
		\begin{align}
		\int_{-1}^{1}\!d\!\cos\theta_{ij} (\sin\theta_{ij})^{\alpha}(\cos\theta_{ij})^{\beta}.
		\label{eq:angularint}
		\end{align}
		The values of the exponents $\alpha$ and $\beta$ appearing in eq.\eqref{eq:angularint} depend both on the angle $\theta_{ij}$ under consideration and on the specific expression of the numerator. Nevertheless, these  integrals can be computed once and for all up to the desired rank and then re-used in every concrete calculation.  One algorithmic way to perform these integrals consists first in expanding the numerator in terms of \textit{Gegenbauer polynomials} $C^{(\alpha)}_{n}(\cos\theta)$, a particular class of orthogonal polynomials over the interval $[-1,1]$ (see appendix~\ref{Ap:2}), and then repeatedly make use of the orthogonality relation
			\begin{align}
			\int_{-1}^{1}\!d\!\cos\theta (\sin\theta)^{2\alpha-1}C^{(\alpha)}_{n}(\cos\theta)C^{(\alpha)}_{m}(\cos\theta)=\delta_{mn}\frac{2^{1-2\alpha}\pi\Gamma(n+2\alpha)}{n!(n+\alpha)\Gamma^2(\alpha)}.
			\label{eq:ortcos}
			\end{align}
		In this way, all integrations over $\boldsymbol{\Theta}_{\perp}$, \textit{i.e.} over all components of the loop momenta orthogonal to the external kinematics, are brought back to a unique integral formula which is able to automatically set to zero all \textit{spurious} contributions to the Feynman integral. Equivalently, one can show that this angular integration is in fact analogous to a \textit{tensor decomposition} of the subspace orthogonal to the external legs of the diagram~\cite{Kreimer:1991wj}. Consider a topology with $n\leq 4$ external legs.  A generic term contributing to a $\ell$-loop integral of such topology has the form
		\begin{equation} \label{eq:angularpvfactorized}
		\int \left(\prod_{i=1}^{\ell}\frac{d^dq_{i}}{\pi^{d/2}}\right) \frac{\mathcal{N}(q_{\parallel\, i},\lambda_{ij}) }{\prod_{j}D_{j}(q_{i})} \Big( \prod_{r=d_{\parallel}+1}^4 \prod_{t=1}^\ell (e_r \cdot q_t)^{\alpha_{r,t}} \Big).
		\end{equation}
		In the first factor on the right of the integration measure, we collected the dependence on the variables $\lambda_{ij}$ and on the components of the loop momenta along the directions of the external momenta, while the second one depends on the transverse components which can be integrated out. Because of the obvious relation
		\begin{equation}
		(q_i\cdot e_j) = (\lambda_i\cdot e_j), \quad \textrm{if $j > d_{\parallel}$},
		\end{equation}
		the angular integration can also be performed via a tensor decomposition restricted to the $d_{\perp}$-dimensional orthogonal subspace.  In particular, this decomposition only depends on the $d_{\perp}$-dimensional projection of the metric tensor and it is independent of the external legs of the diagram, which makes it significantly simpler than a full $d$-dimensional tensorial reduction. This implies that we can easily perform the transverse integration by considering the decomposition
		\begin{align}
		\int \left(\prod_{i=1}^{\ell}\frac{d^dq_{i}}{\pi^{d/2}}\right) \frac{\mathcal{N}(q_{\parallel\, i},\lambda_{ij}) }{\prod_{j}D_{j}(q_{i})} & \Big( \lambda_1^{\nu_{11}}\cdots \lambda_1^{\nu_{1 \alpha_1}} \cdots \lambda_l^{\nu_{l 1}}\cdots \lambda_l^{\nu_{l \alpha_l}}  \Big) \nn &  \qquad = \sum_{\sigma \in S} a_\sigma\; g_{[d_{\perp}]}^{\nu_{\sigma (11)} \nu_{\sigma (12)}} \cdots g_{[d_{\perp}]}^{{\mu_{\sigma(l) \sigma(\alpha_l-1)}} {\mu_{\sigma(l) \sigma(\alpha_l)}}},
		\end{align}
		where $\alpha_i=\sum_t \alpha_{i,t}$ (cfr.\ with eq.~\eqref{eq:angularpvfactorized}) and $S$ is the set of non-equivalent permutations of the Lorentz indexes $\nu_i$ appearing on the l.h.s..  One can thus solve for the coefficients $a_\sigma$ in the traditional way, \textit{i.e.}\ by contracting both sides of the equation with each term on the r.h.s.\ side and using the identities
		\begin{equation}
		g^{[d_{\perp}]}_{\mu \nu} \lambda_i^\mu \lambda_j ^\nu = \lambda_{ij}, \qquad (g_{[d_{\perp}]}^{\mu \nu})^2 = d_{\perp},
		\end{equation}
		which allow to replace the second factor in the product of \eqref{eq:angularpvfactorized} with a combination of variables $\lambda_{ij}$.  Notice that this combination only depends on the number $n$ of external legs and on the powers of loop momenta appearing in the product of the transverse component, while it is completely independent of the expression of the other factors appearing in the integrand. This implies that, similarly to the explicit angular integration discussed above, this decomposition can be performed for the occurring rank once and for all and it is independent of the internal details of the topology and the particular process under consideration.\\
		
		In the following we provide some concrete examples of the integral representation \eqref{eq:lambth} and of the integration procedure in the case four-point integrals up to three loops. We refer the reader to appendix~\ref{Ap:1} for the derivation of eq.~\eqref{eq:lambth} as well as of the explicit expression of the change of variables \eqref{eq:poltr}.	General results for one- and two- loop integrals in all kinematic configurations, including a list of integrals over the transverse directions, are collected in Appendices~\ref{Ap:1l}-\ref{Ap:2l}.
		\subsection{Four-point examples}
		\label{Sec:example4pt}
\begin{figure*}[ht!]
			\centering
			\begin{subfigure}[t]{0.25\textwidth}
				\centering
				\includegraphics[height=1.2in]{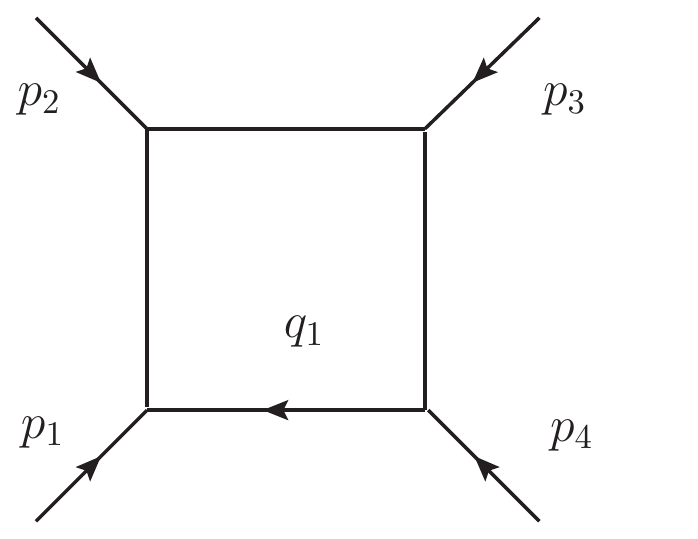}
				\caption{$\ell=1,\,d_{\parallel}=3$}
				\label{fig:1lbox}
			\end{subfigure}%
			\begin{subfigure}[t]{0.33\textwidth}
				\centering
				\includegraphics[height=1.2in]{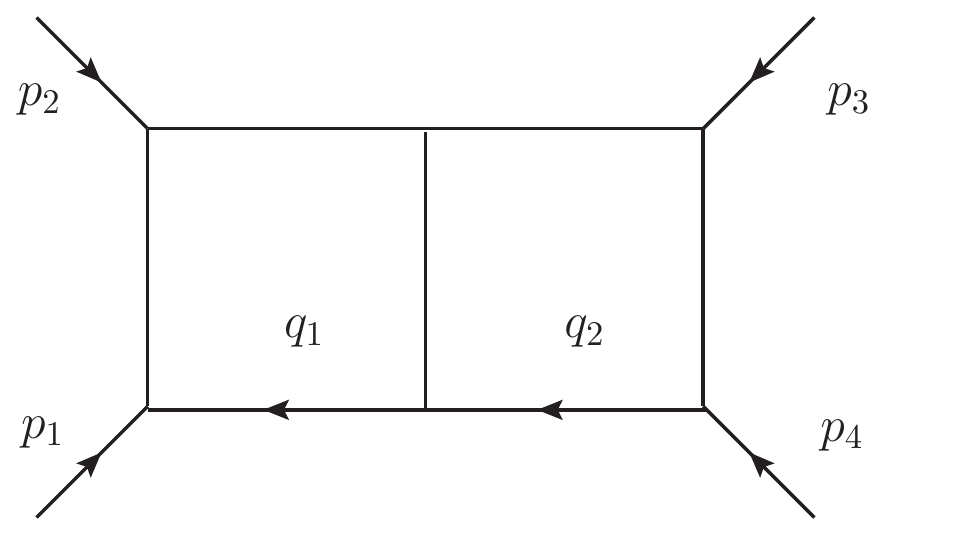}
				\caption{$\ell=2,\,d_{\parallel}=3$}
				\label{fig:2lbox} 
			\end{subfigure}
			\begin{subfigure}[t]{0.33\textwidth}
				\centering
				\includegraphics[height=1.2in]{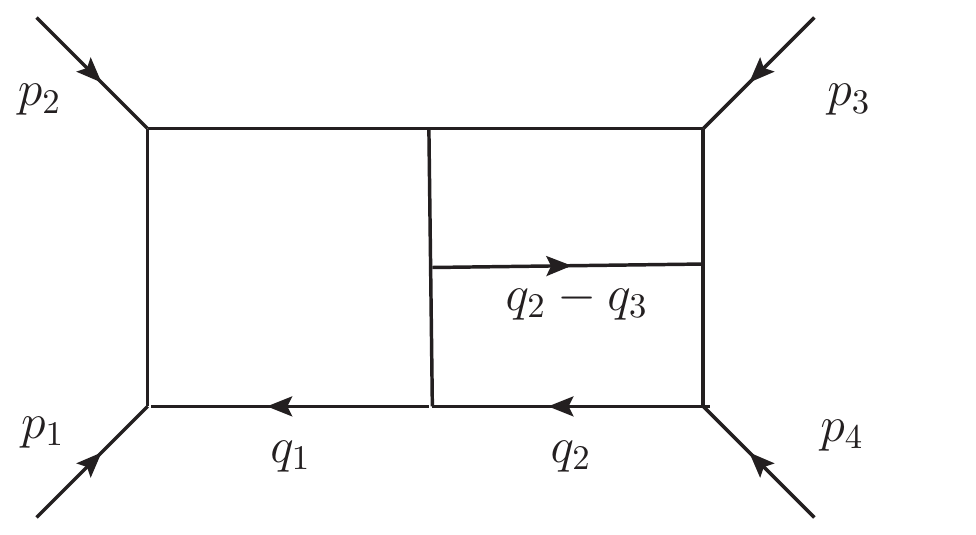}
				\caption{$\ell=3,\,d_{\parallel}=3$}
				\label{fig:3lcourt}
			\end{subfigure}
			\caption{Four-point diagrams}
			\label{fig:123boxes}
		\end{figure*}

		As an example, we consider the four-point topologies depicted in fig.~\ref{fig:123boxes}. Due to momentum conservation, the external momenta $\{p_1,p_2,p_3,p_4\}$ span a subspace with dimension $d_{\parallel}=3$ and, as a consequence, we can build a four-dimensional basis $\{e_i^{\alpha}\}$ containing one single transverse direction $e_{4}^{\alpha}$,
		\begin{align}
		&p_i\cdot e_{4}=0 \qquad \forall i=1,2,3.
		\label{eq:vperp4}
		\end{align}
		Thus, in all the three cases, we can decompose the $d$-dimensional loop momenta according to \eqref{eq:newdeco}, where $q_{\parallel\, i}^{\alpha}\equiv q_{[3]\,i}^{\alpha}$ are three-dimensional vectors defined as
		\begin{align}
		q_{[3]\,i}^{\alpha}=&\sum_{j=1}^{3}x_{ji}e_{j}^{\alpha}, \quad i=1,\,\dots,\, \ell
		\label{eq:q4pt}
		\end{align}
		and $\lambda_i^{\alpha}$ are vectors in the $d_{\perp}=d-3$ dimensional orthogonal space,
		\begin{align}
		\lambda_i^{\alpha}=x_{4i}e_4^{\alpha}+\mu_i^{\alpha},\quad i=1,\,\dots,\, \ell.
		\label{eq:lambda4pt}
		\end{align}
		In this way, all denominators become independent of the component $x_{4i}$ of each loop momentum. The $d_{\parallel}+d_{\perp}$ parametrization of the integrals can now be read directly from \eqref{eq:lambth} with $d_{\parallel}=3$, by changing $\ell=1,2,3$ according to the case. The particular form of the change of variables \eqref{eq:poltr}, which is needed in order to reduce the integrals over the transverse directions to the orthogonality relation \eqref{eq:ortcos}, are derived in appendix~\ref{Ap:1}.
		\begin{enumerate}[(a)] 
			\item
			 For the one loop integral of fig.~\ref{fig:1lbox} we define $\boldsymbol{\Lambda}=\{\lambda_{11}\}$ and $\boldsymbol{\Theta}_\perp=\{\theta_{11}\}$  and we write 
				\begin{align}
				I_4^{d\,(1)}[\,\mathcal{N}\,]=& \frac{1}{\pi^2\Gamma\big(\frac{d-4}{2}\big)}\int\! d^3\!q_{[3]\,1}\int_{0}^{\infty}\!d\lambda_{11} (\lambda_{11})^{\frac{d-5}{2}}\int_{-1}^{1}\!d\!\cos\theta_{11}(\sin\theta_{11})^{d-6}\times\nn
				&\frac{\mathcal{N}(q_{[3]\,1},\lambda_{11},\cos\theta_1)}{\prod_{m=0}^3D_m(q_{[3]\,1},\lambda_{11})}.
				\label{eq:1Lbox}
				\end{align}
				In this case, the set of transformations \eqref{eq:poltr} is reduced to 
			\begin{align}
			x_{41}=\sqrt{\lambda_{11}}\cos\theta_{11},
			\label{eq:x4pol}
			\end{align}
			which expresses the transverse component $x_{41}$ of the loop momentum in terms of the single angular variable $\theta_{11}$. As we have already discussed, the numerator of a general Feynman integral corresponding to the box topology can have at most a polynomial dependence on $x_{41}$ (and hence on $\cos\theta_{11}$), so that the angular integration can always be reduced to the orthogonality relation \eqref{eq:ortcos}. In particular, for the case of a scalar integral we obtain
			\begin{align}
			I_4^{d\,(1)}[1]= \frac{1}{\pi^{3/2}\Gamma\big(\frac{d-3}{2}\big)}\int \!d^3q_{[3]\,1 }\int_{0}^{\infty}\!d\lambda_{11} (\lambda_{11})^{\frac{d-5}{2}}\frac{1}{\prod_{m=0}^3D_m(q_{[3]\,1},\lambda_{11})}.
			\label{eq:4ptsc}
			\end{align}
			Moreover, as we recall in appendix~\ref{Ap:2}, odd powers of $x_{41}$ can be expressed in terms of (products of) Gegenbauer polynomials with different indices and vanish by orthogonality, so that only even powers of the transverse variable produce non-zero contributions. As an example, which will be later become useful, let us consider the integrals
			\begin{subequations}
				\begin{align}
				\label{sub4t2}
				I_4^{d\,(1)}[\,x_{41}^2\,]
				=&\frac{1}{\pi^2\Gamma\big(\frac{d-4}{2}\big)}\int \!d^3q_{[3]\,1}\int_{0}^{\infty}\!d\lambda_{11} (\lambda_{11})^{\frac{d-3}{2}}\int_{-1}^{1}\!d\!\cos\theta_{11}\frac{(\sin\theta_{11})^{d-6}(\cos\theta_{11})^2}{\prod_{m=0}^3D_m(q_{[3]\,1},\lambda_{11})},\\
				I_4^{d\,(1)}[\,x_{41}^4\,]=
				&\frac{1}{\pi^2\Gamma\big(\frac{d-4}{2}\big)}\int\! d^3q_{[3]\,1}\int_{0}^{\infty}\!d\lambda_{11} (\lambda_{11})^{\frac{d-1}{2}}\int_{-1}^{1}\!d\!\cos\theta_{11}\frac{(\sin\theta_{11})^{d-6}(\cos\theta_{11})^4}{\prod_{m=0}^3D_m(q_{[3]\,1},\lambda_{11})}.
				\label{eq:sub4t4}
				\end{align}
			\end{subequations}
			Once powers of $\cos\theta_{11}$ are expressed terms of Gegenbauer polynomials,
			\begin{subequations}
				\begin{align}
				(\cos\theta_{11})^2=&\frac{1}{(d-5)^2}\big[C_{1}^{(\frac{d-5}{2})}(\cos\theta_{11})\big]^2,\\
				(\cos\theta_{11})^4=&\frac{1}{(d-3)^2}\bigg[C_{0}^{(\frac{d-5}{2})}(\cos\theta_{11})+\frac{4}{(d-5)^2}C_{2}^{(\frac{d-5}{2})}(\cos\theta_{11})\bigg]^2,
				\end{align}
			\end{subequations}
			we can evaluate the angular integrals by means of the orthogonality relations \eqref{eq:ortcos} and obtain
			\begin{subequations}
				\begin{align}
				I_4^{d\,(1)}[\,x_{41}^2\,]=&
				\frac{1}{d-3}I_4^{d\,(1)}[\,\lambda_{11}\,]=\frac{1}{2}I_4^{d+2\,(1)}[1],\\
				I_4^{d\,(1)}[\,x_{41}^4\,]=&
				\frac{3}{(d-3)(d-1)}I_4^{d\,(1)}[\,\lambda_{11}^2\,]=\frac{3}{4}I_4^{d+4\,(1)}[1].
				\end{align}
				\label{sub4t2r}
			\end{subequations}
			In the second equality, we have identified additional powers of  $\lambda_{11}$ in the numerator, produced by the integration over the transverse component, with higher-dimensional scalar integrals, as it can be easily checked from the explicit expression of the $d$-dimensional integral \eqref{eq:4ptsc}. Results for higher rank numerators can be found in appendix~\ref{Ap:1l}. 
			\item 
			At two loops the transverse space of the topology shown in fig.~\ref{fig:2lbox} is described by the variables $\boldsymbol{\Lambda}=\{\lambda_{11},\lambda_{22},\theta_{12}\}$ and $\boldsymbol{\Theta}_{\perp}=\{\theta_{11},\theta_{22}\}$ and we have
			\begin{align}
			I_4^{d\,(2)}[\,\mathcal{N}\,]=&\frac{2^{d-6}}{\pi^5\Gamma(d-5)}\int\! d^3q_{[3]\,1}d^3q_{[3]\, 2}
			\int_{0}^{\infty}\!d\lambda_{11}d\lambda_{22} (\lambda_{11})^{\frac{d-5}{2}}(\lambda_{22})^{\frac{d-5}{2}}\times\nn
			&\int_{-1}^{1}\!d\!\cos\theta_{12}d\!\cos\theta_{22}d\!\cos\theta_{11}\left(\sin\theta_{12}\right)^{d-6}(\sin\theta_{11})^{d-6}
			(\sin\theta_{22})^{d-7}\times\nn
			&\frac{\mathcal{N}(q_{[3]\,i},\lambda_{ii},\cos\theta_{ij},\sin\theta_{ij})}{\prod_{m=0}^{7}D_m(q_{[3]\,i},\lambda_{ii},\cos\theta_{12})}.
			\end{align}
			In this case, \eqref{eq:poltr} reads
			\begin{align}
			\begin{cases}
			\lambda_{12}=&\sqrt{\lambda_{11}\lambda_{22}}\cos\theta_{12}\\
			x_{41}=&\sqrt{\lambda_{11}}\cos\theta_{11}\\
			x_{42}=&\sqrt{\lambda_{22}}\big(\cos\theta_{11}\cos\theta_{12}+\sin\theta_{11}\sin\theta_{12}\cos\theta_{22}\big),
			\end{cases}
			\end{align}
			so that, after the change of variables, any term in the numerator depending on $x_{41}$ and $x_{42}$ is mapped into a polynomial in (\textit{sine} and \textit{cosine} of) $\boldsymbol{\Theta}_\perp$, with coefficients depending on $\boldsymbol{\Lambda}$, which can be easily integrated through the expansion in terms of Gegenbauer polynomials. In this way we find, for the scalar integral,
			\begin{align}
			I_4^{d\,(2)}[\,1\,]=\frac{2^{d-5}}{\pi^4\Gamma(d-4)}&\int d^3q_{[3] \,1}d^3q_{[3]\,2}
			\int_{0}^{\infty}\!d\lambda_{11}d\lambda_{22}(\lambda_{11})^{\frac{d-5}{2}}(\lambda_{22})^{\frac{d-5}{2}}\times\nn
			&\int_{-1}^{1}\!d\!\cos\theta_{12}\left(\sin\theta_{12}\right)^{d-6}\frac{1}{\prod_{m=0}^{6}D_m(q_{[3]\,i},\lambda_{ii},\cos\theta_{12})},
			\end{align}
			whereas the first non-spurious monomials in $x_{41}$ and $x_{42}$ produce
				\begin{align}
				\label{eq:2L4pttensor1}
				&I_4^{d\,(1)}[x_{4i}x_{4_j}]=\frac{1}{d-3}I_4^{d\,(1)}[\lambda_{ij}].
				\end{align}
			Results for higher rank numerators can be found in appendix~\ref{Ap:2l}.
			\item The transverse space of the three-loop topology shown in Fig~\ref{fig:3lcourt} is parametrized in terms of $\boldsymbol{\Lambda}=\{\lambda_{11},\lambda_{22},\lambda_{33},\theta_{12},\theta_{13},\theta_{23}\}$ and $\boldsymbol{\Theta}_{\perp}=\{\theta_{11},\theta_{22},\theta_{33}\}$,
			\begin{align}
			I_4^{d\,(3)}[\,\mathcal{N}\,]=&\frac{2^{d-7}}{\pi^8\Gamma(d-6)\Gamma\left(\frac{d-4}{2}\right)}\int\prod_{i=1}^3\! d^3q_{[3]\, i}
			\int_{0}^{\infty}\prod_{i=1}^3d\lambda_{ii}(\lambda_{ii})^{\frac{d-5}{2}}\times\nn
			&\int_{-1}^{1}\prod_{1\leq i\leq j\leq 3}\!d\!\cos\theta_{ij}(\sin\theta_{ij})^{d-5-i}\frac{\mathcal{N}(q_{[3]\,i},\lambda_{ii},\cos\theta_{ij},\sin\theta_{ij})}{\prod_{m=0}^{9}D_m(q_{[3],i},\lambda_{ii},\cos\theta_{12},\cos\theta_{13},\cos\theta_{23})}.
			\label{court}
			\end{align}
			The change of variables
			\begin{align}
			\begin{cases}
			\lambda_{12}=&\sqrt{\lambda_{11}\lambda_{22}}\cos\theta_{12}\\
			\lambda_{23}=&\sqrt{\lambda_{22}\lambda_{33}}\cos\theta_{13}\\
			\lambda_{13}=&\sqrt{\lambda_{11}\lambda_{33}}(\cos\theta_{12}\cos\theta_{13}+\sin\theta_{12}\sin\theta_{13}\cos\theta_{23})\\
			x_{41}=&\sqrt{\lambda_{11}}\cos\theta_{11}\\
			x_{42}=&\sqrt{\lambda_{22}}(\cos\theta_{11}\cos\theta_{12}+\sin\theta_{11}\sin\theta_{12}\cos\theta_{22})\\
			x_{43}=&\sqrt{\lambda_{33}} (\cos\theta_{11} \cos\theta_{12} \cos\theta_{13}+\sin\theta_{11} \sin\theta_{12} \cos\theta_{22} \cos\theta_{13}\\
			&-\sin\theta_{11} \sin \theta_{13} \cos\theta_{12}
			\cos\theta_{22} \cos\theta_{23}+\sin\theta_{12} \sin\theta_{13} \cos\theta_{11} \cos\theta_{23}\\
			&+\sin\theta_{11} \sin\theta_{13} \sin\theta_{22} \sin\theta_{23}\cos \theta_{33})
			\end{cases}
			\label{eq:3lvariables}
			\end{align}
			allows us to express the transverse components $x_{4i}$ in terms of the angular variables and then integrate over $\boldsymbol{\Theta}_{\perp}$ with the help of \eqref{eq:ortcos}. For the scalar integral we obtain
			\begin{align}
				I_4^{d\,(3)}[\,1\,]=&\frac{2^{d-5}}{\pi^{13/2}\Gamma(d-4)\Gamma\left(\frac{d-5}{2}\right)}\int\prod_{i=1}^3\! d^3q_{[3]\, i}
				\int_{0}^{\infty}\prod_{i=1}^3d\lambda_{ii}(\lambda_{ii})^{\frac{d-5}{2}}\times\nn
				&\int_{-1}^{1}\prod_{1\leq i< j\leq 3}\!d\!\cos\theta_{ij}(\sin\theta_{ij})^{d-5-i}\frac{1}{\prod_{m=0}^{9}D_m(q_{[3]\,i},\lambda_{ii},\cos\theta_{12},\cos\theta_{13},\cos\theta_{23})},
				\end{align}
				and, similarly to the previous case, it can be verified that
		 \begin{align}
		   &I_4^{d\,(3)}[x_{4i}x_{4j}]=\frac{1}{d-3}I_4^{d\,(3)}[\lambda_{ij}],\qquad \forall i,j=1,2,3.
		 \end{align}
		\end{enumerate}
		    \subsection{Factorized integrals and ladder topologies}\label{sec:factorizedandladder}
		    \begin{figure}[ht!]
		    	\centering
		    	\begin{subfigure}[t]{0.33\textwidth}
		    		\centering
		    		\includegraphics[height=1.in]{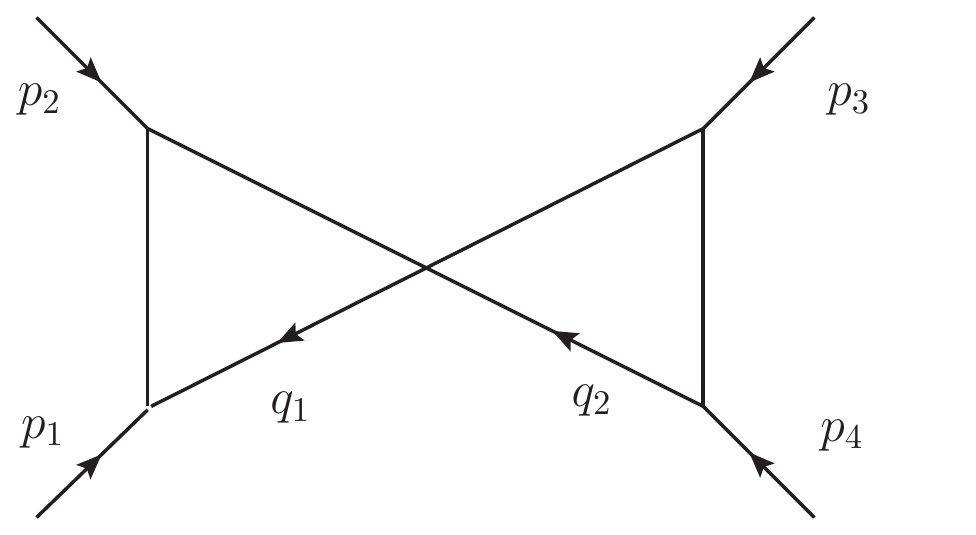}
		    		\caption{$\ell=2\,d_{\parallel}=3$ bowtie}
		    		\label{fig:bowties} 
		    		\end{subfigure}
		    		\qquad
		    	\begin{subfigure}[t]{0.33\textwidth}
		    		\centering
		    		\includegraphics[height=1.in]{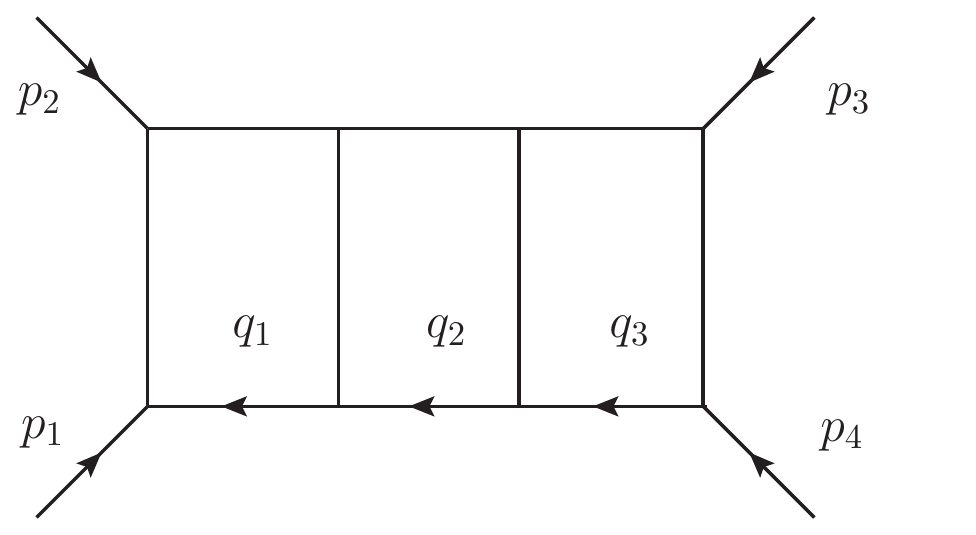}
		    		\caption{$\ell=3\,d_{\parallel}=3$ ladder}
		    		\label{fig:3lbox}
		    	\end{subfigure}%
		    	\caption{Bow-tie topology~\ref{fig:bowties} and three-loop ladder~\ref{fig:3lbox}.}
		        \end{figure}
		     The $d=d_{\parallel}+d_{\perp}$ parametrization \eqref{eq:lambth} applies to all Feynman integrals with $n\leq 4$ but, as we have already mentioned, there are special classes of multiloop integrals, associated to factorized and ladder topologies, which allow further simplifications. These integrals are characterized by a set of denominators which are independent of a certain number of  transverse orientations $\lambda_{ij}$, \textit{i.e.} on a subset of the angular variables $\boldsymbol{\Theta}_{\Lambda}$. This implies that, as it can be immediately understood from the properties of the change of variables \eqref{eq:poltr} and the integration measure \eqref{eq:Lambdaint}, the integration via expansion in Gegenbauer polynomials can be applied, besides to all $\boldsymbol{\Theta}_{\perp}$ angles, also the angles $\boldsymbol{\Theta}_{\Lambda}$ which do not appear in the denominators. In the following, in order to better emphasise the different strategies to be adopted for factorized and ladder integrals, we discuss the $d=d_{\parallel}+d_{\perp}$ parametrization in two concrete examples.
		      \subsubsection{Factorized integrals}
		      \label{sec:fact}
		      When the loop corresponding to $q_{i}^{\alpha}$ is \textit{factorized}, no denominator depends on $q_{i}\cdot q_{j}$, with $j\neq i$. In general, whether a factorized integral originates from Feynman diagrams or from the pinching of all propagators connecting one loop to the rest of the graph, the integrand is not completely factorized, since the numerator can still depend on the $(d-4)$-dimensional part of $q_{i}\cdot q_{j}$, corresponding to $\mu_{ij}$. Nevertheless, it can be shown that, after using the orthogonality relation \eqref{eq:ortcos} to integrate out $\mu_{ij}$, the $d=d_{\parallel}+d_{\perp}$ parametrization of a factorized integral is given by the product of the $d=d_{\parallel}+d_{\perp}$ parametrizations of the integrals corresponding to the subtopologies, whose transverse space can have different dimensions.\\
		      
		      As an example, let us consider a bow tie integral of the type shown in fig.~\ref{fig:bowties}, for which the
		      $d=4-2\epsilon$ parametrization \eqref{eq:oldpar} reads
		       \begin{align}
		       I_{4,\text{fact}}^{d\,(2)}[\,\mathcal{N}]=&\frac{2^{d-6}}{\pi^5\Gamma(d-5)}\int{d^4q_{[4]\,1}}d^4q_{[4]\,2}\int_{0}^{\infty}\!d\mu_{11}\int_{0}^{\infty}\!d\mu_{22}\times\nn
		       &\int_{-\sqrt{\mu_{11}\mu_{22}}}^{\sqrt{\mu_{11}\mu_{22}}}\!d\mu_{12}(\mu_{11}\mu_{22}-\mu_{12}^2)^{\frac{d-7}{2}}\frac{\mathcal{N}}{\prod_{i=0}^{2}D_i(q_{[4]\,1},\mu_{11})\prod_{j=3}^5D_j(q_{[4]\,2},\mu_{22})}.
		       \end{align} 
		       Any tensor numerator can always be split into terms of the form 
		       \begin{align}
		      \mathcal{N}(q_{[4],1},q_{[4]\,2},\mu_{ij})=(\mu_{12})^\alpha\mathcal{N}_1(q_{[4],1},\mu_{11})\mathcal{N}_2(q_{[4],2},\mu_{22}),\quad \alpha \in \mathbb{N}.
		       \label{eq:numbowtie}
		    \end{align}
		       so that, if we introduce the change of variable $\cos \phi \equiv\mu_{12}/\sqrt{\mu_{11}\mu_{22}}$, the integral over $\mu_{12}$ can be reduced to an integral of the type \eqref{eq:angularint}, which can be evaluated through the usual orthogonality relation \eqref{eq:ortcos},
		       \begin{align}
		       \int_{-1}^{1}\!d\!\cos\phi(\sin\phi)^{d-7}(\cos\phi)^{\alpha}=
		       \begin{cases}
		       0\qquad &\text{for}\quad \alpha=2n+1\\
		       \frac{\Gamma\left(\frac{\alpha+1}{2}\right)\Gamma\left(\frac{d-5}{2}\right)}{\Gamma\left(\frac{d+\alpha-4}{2}\right)}\qquad &\text{for}\quad \alpha=2n.
		       \end{cases}
		       \label{eq:factint}
		       \end{align}
		        After inserting this result in \eqref{eq:numbowtie}, the integral over each loop momentum is completely factorized and, by comparison with the $d=4-2\epsilon$ parametrization of one-loop integrals, we can identify, for the non-trivial case $\alpha=2n$,
		       \begin{align}
		       I_{4\,\text{fact}}^{d\,(2)}[\,(\mu_{12})^{\alpha}\mathcal{N}_1\mathcal{N}_2]=&\frac{2^{5-d-\alpha}B\left(\frac{1+\alpha}{2},\frac{d-4}{2}\right)}{B\left(\frac{d-4+\alpha}{2},\frac{d-4+\alpha}{2}\right)}\left(\int\frac{d^dq_{1}}{\pi^{d/2}}\frac{(\mu_{11})^{\frac{\alpha}{2}}\mathcal{N}_1}{\prod_{i=0}^{2}D_i(q_{1})}\right)\!\left(\int\frac{d^dq_{2}}{\pi^{d/2}}\frac{(\mu_{22})^{\frac{\alpha}{2}}\mathcal{N}_2}{\prod_{j=3}^{5}D_j(q_{2})}\right).
		       \end{align}
		       For each of the term in brackets we can now introduce the $d=d_{\parallel}+d_{\perp}$ parametrization $\eqref{eq:lambth}$ by working with two completely different basis, each one containing two vectors orthogonal to the external legs connected to the corresponding loop. We remark again that, in more general cases, the transverse space associated two factorized subdiagrams might have different dimensions.
		         \subsubsection{Ladder integrals}
		    Starting from $\ell\geq 3$, \textit{ladder} topologies corresponds to integrals whose denominators depend on a limited number variables $\lambda_{ij}$. In these cases, the $d=d_{\parallel}+d_{\perp}$ parametrization \eqref{eq:lambth} reads exactly as in the general case \eqref{eq:lambth} but the integration in terms of Gegenbauer polynomials can be extended to the subsets of angles $\boldsymbol{\Theta}_{\Lambda}$ corresponding to the $\lambda_{ij}$ which do not appear in the denominators. As an example, we consider the three-loop ladder box shown in Fig~\ref{fig:3lbox}, for which we introduce the same set of transverse variables as for the three-loop diagram of fig.~\ref{fig:3lcourt},
		    	\begin{align}
		    	\boldsymbol{\Lambda}=&\{\lambda_{11},\lambda_{22},\lambda_{33},\theta_{12},\theta_{13},\theta_{23}\},\nn \boldsymbol{\Theta}_{\perp}=&\{\theta_{11},\theta_{22},\theta_{33}\}
		    	\end{align}
		    	and parametrize the integral exactly as in \eqref{court}.
		    	This integral has no propagator depending on both $q_{1}^{\alpha}$ and $q_{3}^{\alpha}$, \textit{i.e.} the denominators are independent of $\lambda_{13}$ and hence of $\theta_{23}$, as it can be seen from \eqref{eq:3lvariables}. Therefore, the integral over $\theta_{23}$ is reduced to the form \eqref{eq:angularint}, 
		    	and it can be evaluated in the usual way
		    	 \begin{align}
		    	 \int_{-1}^{1}\!d\!\cos\theta_{23}(\sin\theta_{23})^{d-7-\beta}(\cos\theta_{23})^{\alpha}=
		    	 \begin{cases}
		    	 0\qquad &\text{for}\quad \alpha=2n+1\\
		    	 \frac{\Gamma\left(\frac{\alpha+1}{2}\right)\Gamma\left(\frac{d-5+\beta}{2}\right)}{\Gamma\left(\frac{d+\alpha+\beta-4}{2}\right)}\qquad &\text{for}\quad \alpha=2n.
		    	 \end{cases}
		    	 \label{eq:ladtint}
		    	 \end{align}
		    	 In \eqref{eq:ladtint} the indices $\alpha$ and $\beta$ are determined by the specific form of the numerator. In the scalar case ($\alpha=\beta=0$), this additional integration returns
		    	\begin{align}
		    	I_{4\,\text{ladder}}^{d\,(3)}[\,1\,]&=\frac{2^{d-5}}{\pi^{6}\Gamma(d-4)\Gamma\left(\frac{d-4}{2}\right)}\int\prod_{i=1}^3\! d^3q_{[3]\, i}
		    	\int_{0}^{\infty}\prod_{i=1}^3d\lambda_{ii}(\lambda_{ii})^{\frac{d-5}{2}}\times\nn
		    	&\int_{-1}^{1}\!d\!\cos\theta_{12}d\!\cos\theta_{13}(\sin\theta_{12})^{d-6}(\sin\theta_{13})^{d-6}\!\frac{1}{\prod_{m=0}^{9}D_m(q_{[3]\,i},\lambda_{ii},\cos\theta_{12},\cos\theta_{13})}.
		    	\end{align}  
		\subsection{Simplified integrand form}
		The $d=d_{\parallel}+d_{\perp}$ parametrization of Feynman integrals and the angular integration over transverse directions can be used in order to decompose scattering amplitudes in terms of a  reduced number of scalar integrals without explicitly performing any tensor reduction.
		In fact, transverse integration can be used \textit{ab initio} in order to obtain a simplified form of the integrand free of \textit{spurious} contributions, which can be more easily reduced, by means of traditional methods such as integration by parts, in terms of a minimal set of master integrals. In particular, as we show in the following example, this procedure is suited for application to helicity amplitudes which, in general, may enjoy better properties than the form factors defined in the usual tensor decomposition.
		Alternatively, transverse integration can be applied in tandem with algebraic methods, such as integrand decomposition, in order to achieve a step-by-step simplification of the reduction algorithm. The interplay of transverse integration and integrand decomposition will be the object of the next section. 
		\subsubsection{Example: the four-point integrand for $A^{2-\text{loop}}(g_1^+,g_2^-,g_3^+,g_4^-)$}
		\label{sec:esempioAggg}
		\begin{figure}[ht!]
			\centering
			\begin{subfigure}[t]{0.33\textwidth}
				\centering
				\includegraphics[height=1.in]{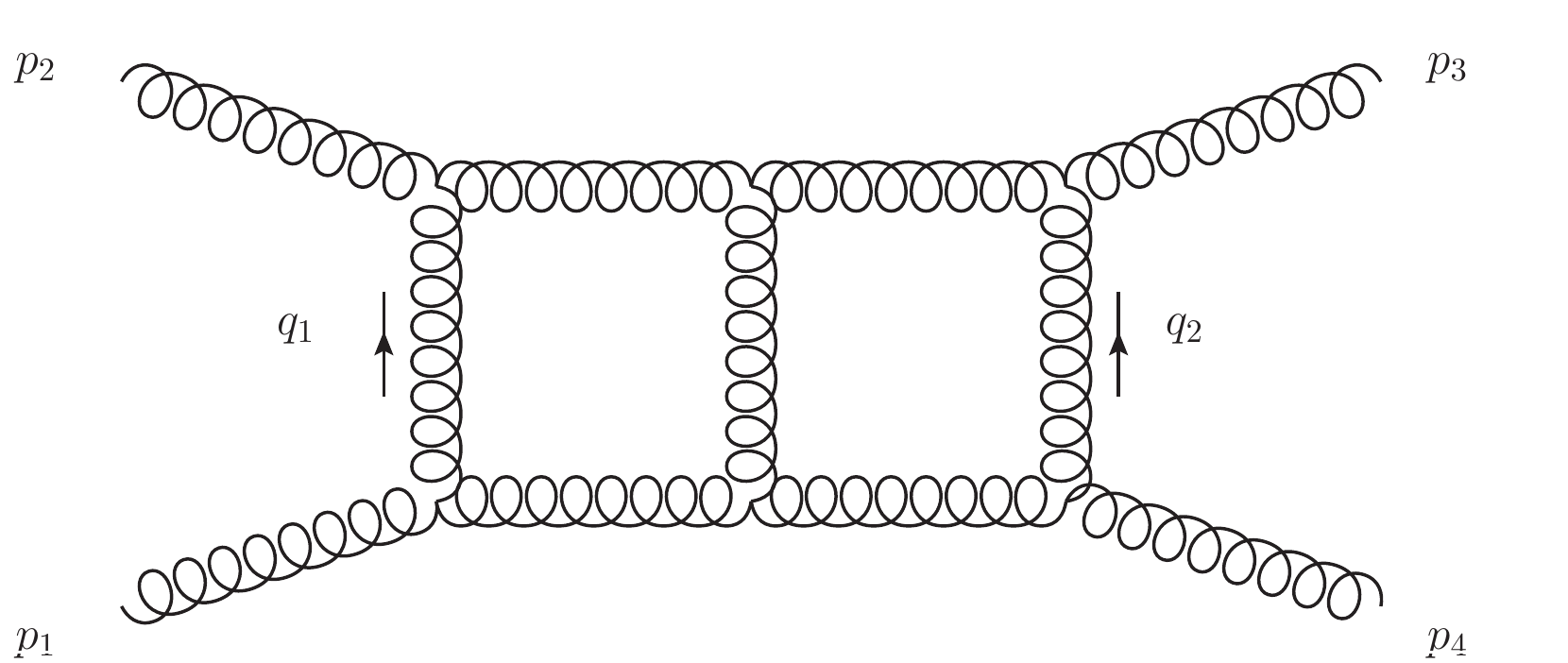}
				\caption{}
				\label{fig:ex1} 
			\end{subfigure}
			\qquad\qquad
			\begin{subfigure}[t]{0.33\textwidth}
				\centering
				\includegraphics[height=1.in]{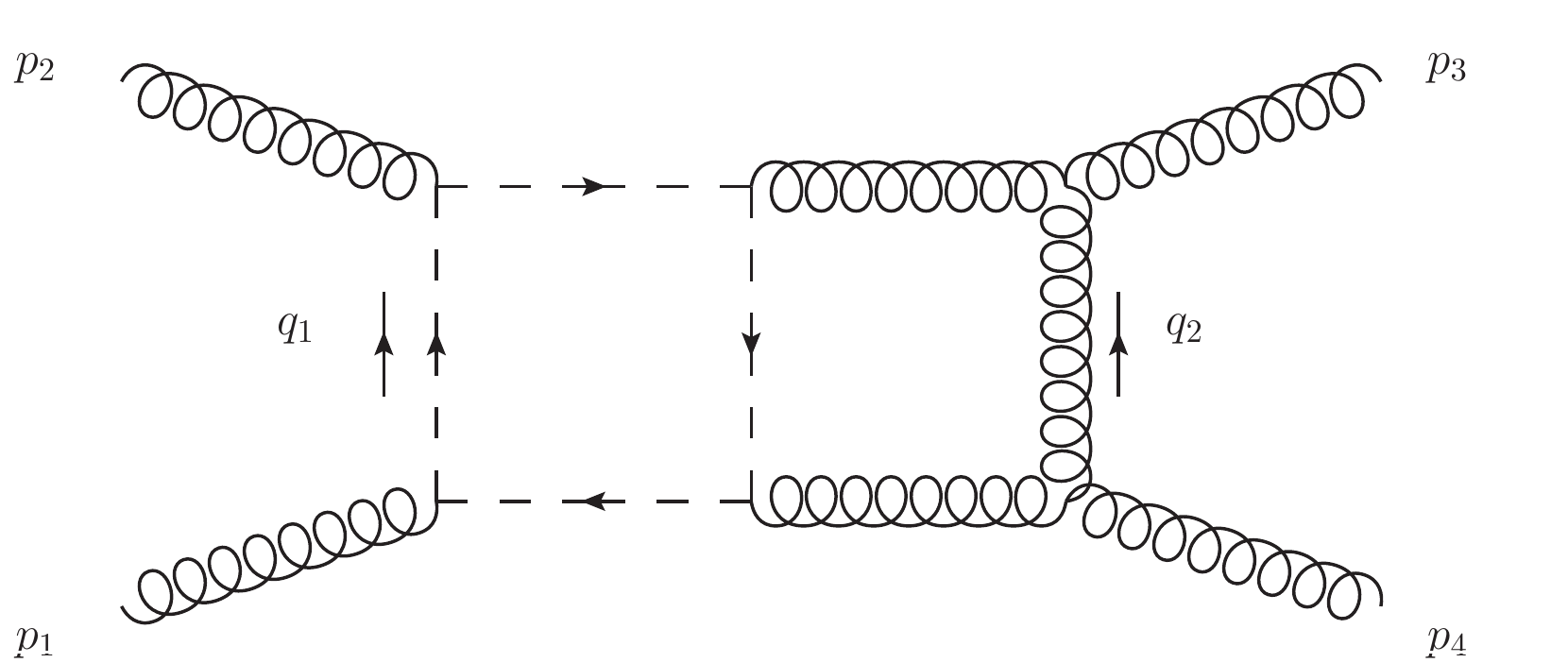}
				\caption{}
				\label{fig:ex2}
			\end{subfigure}
			\\
			\begin{subfigure}[t]{0.33\textwidth}
				\centering
				\includegraphics[height=1.in]{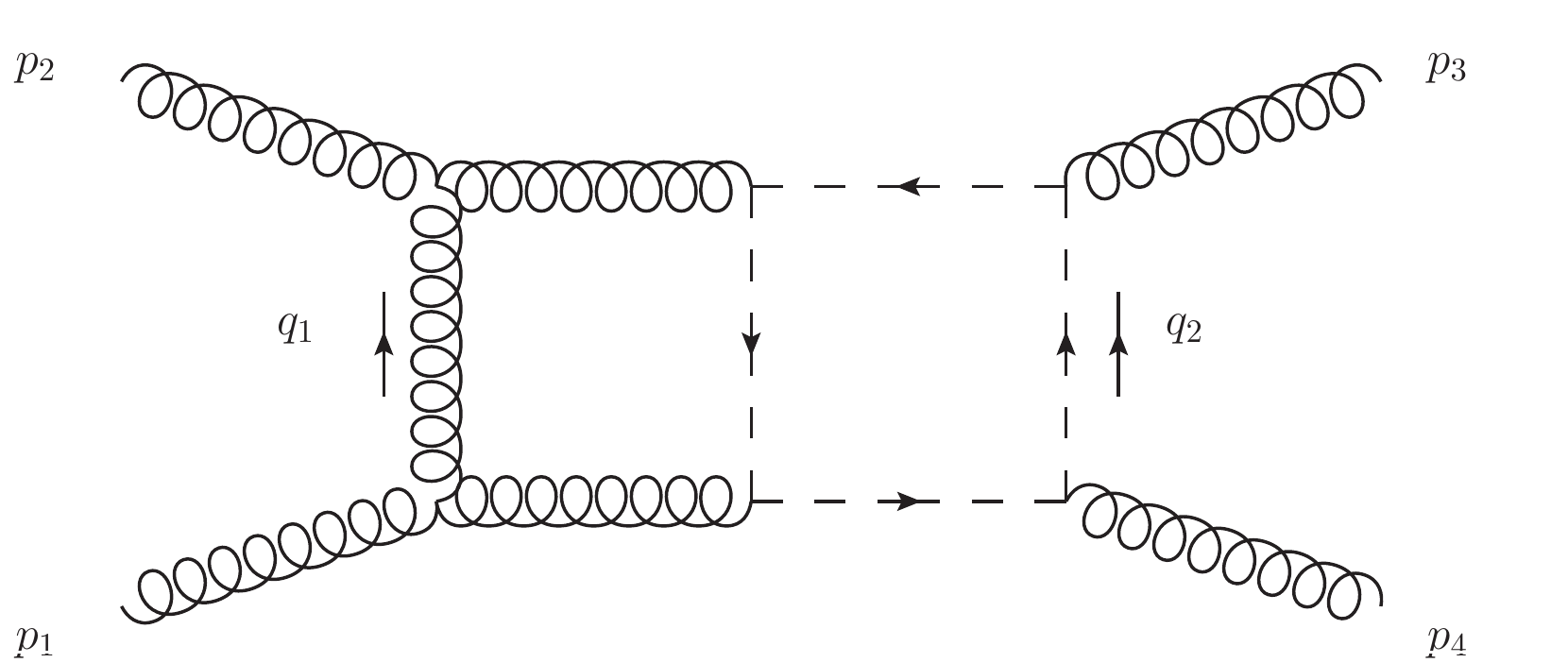}
				\caption{}
				\label{fig:ex3} 
			\end{subfigure}
			\qquad\qquad
			\begin{subfigure}[t]{0.33\textwidth}
				\centering
				\includegraphics[height=1.in]{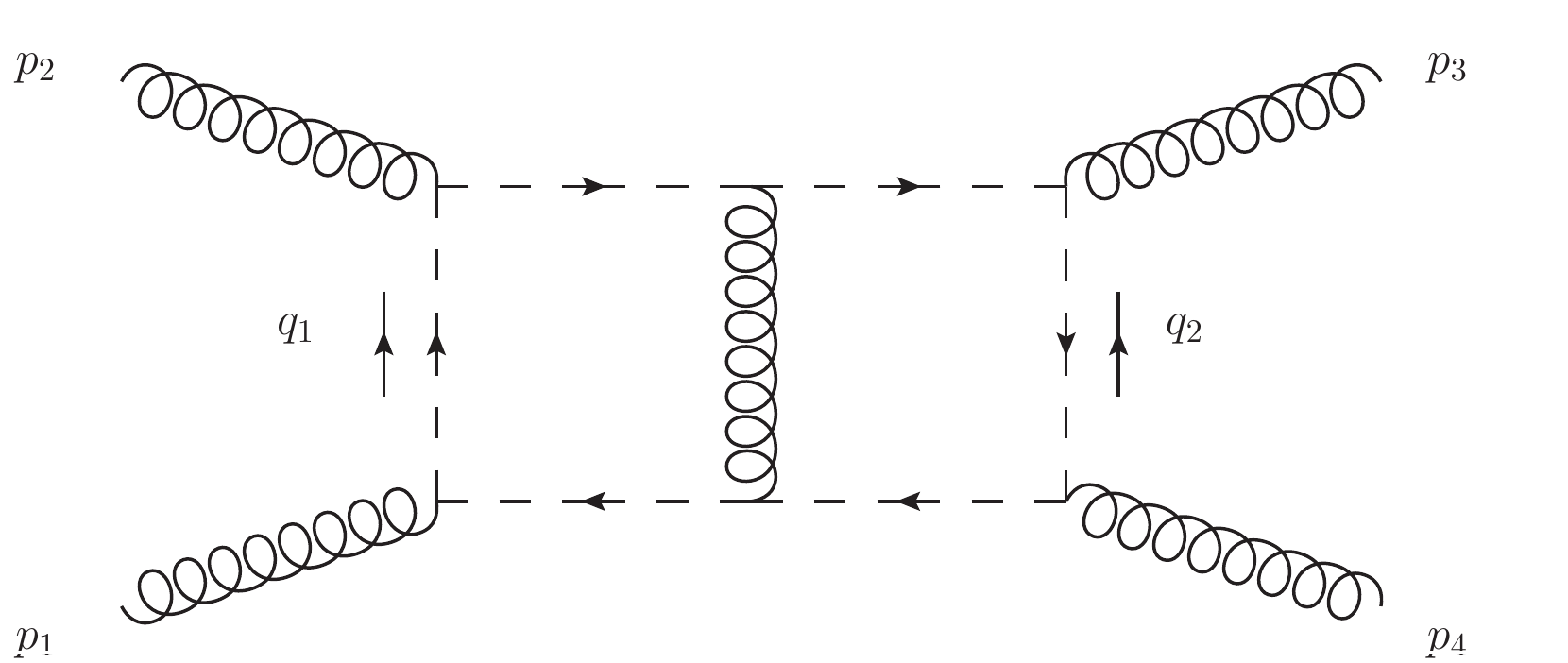}
				\caption{}
				\label{fig:ex4}
			\end{subfigure}
			\caption{Double-box contributions to $A^{2-\text{loop}}(g_1^+,g_2^-,g_3^+,g_4^-)$.}
		\end{figure}
		As an example, we consider the double-box contribution to a four-gluon color-ordered helicity amplitude and we show that the integration over the transverse variables can lead, prior to the application of any reduction algorithm, to a simplified representation of the integrand. 
		The topology, in $d=4-2\, \epsilon$ dimensions, is defined by the seven
		denominators
		\begin{align}
				D_1 = & {} (q_1+p_1)^2,  &\qquad 		D_2 = & {} q_1^2, \nn
			    D_3 = & {} (q_1-p_2)^2,  &\qquad     	D_4 = & {} (q_2-p_3)^2, \nn
				D_5 = & {} q_2^2,        &\qquad      	D_6 = & {} (q_2+p_4)^2, \nn
				D_7 = & {} (q_1+q_2+p_1+p_4)^2 
		\label{eq:densex}		
		\end{align}
		and the four irreducible scalar products
		\begin{align} \label{eq:2ldbisps}
			(q_1 \cdot p_4)\, \qquad (q_2\cdot p_1), \qquad (q_1 \cdot v_\perp)\, \qquad (q_2\cdot v_\perp),
		\end{align}
		where $v_\perp$ is orthogonal to the external momenta and it can be chosen as
		\begin{equation}
				v_\perp^\mu = -4\, i\, \epsilon^{\mu}{}_{\nu \rho \sigma} p_{1}^\nu\, p_{2}^\rho\, p_{3}^\sigma = \textrm{tr}_5(\mu\, p_1 \, p_2\, p_3).
		\end{equation}
		Notice that, conversely to the transverse vector $e_4$ introduced in the general discussion of sec.~\ref{Sec:example4pt}, with this definition $v_\perp$ is \emph{not}
		normalized 
		since, when dealing with realistic processes, it is convenient to use a representation which can be easily expressed in
		terms of spinor variables without introducing spurious square roots. It is worth observing that, while the first two scalar products in
		eq.~\eqref{eq:2ldbisps} live in the physical space defined by the
		external momenta, the last two lie along the orthogonal
		direction and will be integrated out using the technique previously discussed. Finally, as it is implied
		by the definition of the denominators \eqref{eq:densex}, all the external momenta $p_i$
		are taken as outgoing.
		\\
		
		We consider the helicity amplitude
		$A^{2-\text{loop}}(p_1^+,p_2^-,p_3^+,p_4^-)$. The double-box contribution to the amplitude is given, in a pure Yang-Mills theory, by the sum of the four
		diagrams shown in Figs.~\ref{fig:ex1}-\ref{fig:ex4}, namely a diagram involving only gluons and three diagrams
		with ghosts circulating in the loop.  The calculation can be easily
		carried out \textit{e.g.}\ in Feynman gauge and with an explicit choice of
		polarization vectors in terms of spinor variables such as
		\begin{align}
			\varepsilon^+_1 = \frac{\la 2\, \gamma^\mu 1\, \rb}{\sqrt{2} \, \la 2\, 1\ra}, \qquad   \varepsilon^-_2 = \frac{\la 2\, \gamma^\mu 1\, \rb}{\sqrt{2} \lb 2\, 1\rb}, \qquad   \varepsilon^+_3 = \frac{\la 2\, \gamma^\mu 3\, \rb}{\sqrt{2}  \, \la 2\, 3\ra}, \qquad   \varepsilon^-_4 = \frac{\la 4\, \gamma^\mu 1\, \rb}{\sqrt{2} \lb 4\, 1\rb}.
		\end{align}
		We remark, however, that the final result  for the on-shell residue, which we will discuss in sec.~\ref{sec:example}, is \emph{gauge
			invariant} and thus independent of the previous choices. \\
		
		After inserting the Feynman rules for each diagram and decomposing the loop
		momenta as
		\begin{align} \label{eq:2ldbqdec}
			q_1^\alpha ={}& x_{11}\, p_1^\alpha + x_{21}\, p_2^\alpha + x_{31}\, p_4^\alpha + x_{41}\, v_\perp^\alpha + \mu_i^\alpha,\nn
			q_2^\alpha ={}& x_{12}\, p_1^\alpha + x_{22}\, p_2^\alpha + x_{32}\, p_4^\alpha + x_{42}\, v_\perp^\alpha + \mu_i^\alpha
		\end{align}
		the numerator becomes a function of the coordinates $x_{ij}$ appearing
		in eq.~\eqref{eq:2ldbqdec} and the $(-2\epsilon)$-dimensional scalar
		products $\mu_{ij}$. According to eq.~\eqref{eq:lambda4pt}, the transverse vectors $\lambda_{i}$ can be identified with 
		\begin{align}
		\lambda_{1}^{\alpha}=& x_{41}\, v_\perp^\alpha + \mu_i^\alpha,\nn
		\lambda_{2}^{\alpha}=& x_{42}\, v_\perp^\alpha + \mu_i^\alpha
		\end{align}
		and the $d=d_{\parallel}+d_{\perp}$ parametrization of the integrand is simply obtained by applying on the integrand the shift
		\begin{equation}
			\mu_{ij} \to \lambda_{ij} - x_{4i}\, x_{4j}\, v_\perp^2,
		\end{equation}
		which, as we have already observed, makes the denominators independent of
		the transverse components $x_{41}(v_\perp\cdot q_1)/v_\perp^2$ and $x_{42}=(v_\perp\cdot q_2)/v_\perp^2$ of the loop momenta. \\
		
		After this change of variables, the numerators is given by a sum of 2025 distinct terms in the loop variables
		\begin{align}
		\mathbf{z}=\{x_{11},x_{21},x_{31},x_{41},x_{12},x_{22},x_{32},x_{42},\lambda_{11},\lambda_{22},\lambda_{12}\}.
		\end{align}
		The terms proportional to the transverse variables
		 $x_{4i}$ can be integrated out using the
		 results listed in appendix~\ref{Ap:2l}. Nevertheless, we want to remark that, when using a non-trivial normalization of $v_\perp$, the right hand sides of the formulas for
		$I_{4}^{d\,(2)}[(q\cdot v_\perp)^\alpha\, (q\cdot v_\perp)^\beta]$ must be
		multiplied by a factor $(v_\perp^2)^{(\alpha+\beta)/2}$. After integrating out transverse directions, the numerator is reduced to a sum of 773 terms in
		the variables 
		\begin{align}
		\boldsymbol{\tau}=\{x_{11},x_{21},x_{31},x_{12},x_{22},x_{32},\lambda_{11},\lambda_{22},\lambda_{12}\},
		\end{align}
		which, as we will explain in more detailed in the next sections, can be easily expressed in terms of denominators
		and physical scalar products. This procedure yields to a linear
		combination of integrals of this topology as well as subtopologies,
		which only depend on loop propagators and on the two irreducible scalar products
		of the loop momenta with the external legs.

%% file: Adaptive.tex
	\section{Adaptive Integrand Decomposition}
	\label{sec:2}
	\subsection{Integrand recurrence relation}\label{sec:integrrecrel}
	In the framework of the integrand reduction method \cite{Ossola:2006us,Ellis:2007br, Mastrolia:2011pr, Badger:2012dp, Zhang:2012ce, Mastrolia:2013kca}, the computation of dimensionally regulated $\ell$-loop integrals
	\begin{align}
	I^{d\,(\ell)}_{i_1\dots i_r}=\int \prod_{1=j}^{\ell}\frac{d^dq_{j}}{\pi^{d/2}}\,\frac{\mathcal{N}_{i_1\dots i_r}(q_j)}{D_{i_1}(q_j)\cdots D_{i_r}(q_j)}
	\end{align}
	 is rephrased in terms of the reconstruction of the integrand function as a sum of \textit{residues}, \textit{i.e.} irreducible numerators which cannot be expressed in terms of denominators $D_{i_k}$, sitting over all possible subsets of denominators,
	 \begin{align}
	 \mathcal{I}_{i_1\dots i_r}(q_j)\equiv\frac{\mathcal{N}_{i_1\dots i_r}(q_j)}{D_{i_1}(q_j)\cdots D_{i_r}(q_j)}=\sum_{k=0}^{r}\sum_{\{i_1\cdots i_k\}}\frac{\Delta_{j_1\cdots j_k}(q_j)}{D_{j_1}\cdots D_{j_{k}}(q_j)}.
	 \label{eq:intdec}
	 \end{align}
	For an integral with an arbitrary number $n$ of external legs, the integrand decomposition formula \eqref{eq:intdec} can be obtained by observing that both numerator and denominators are polynomials in the components of the loop momenta with respect to some basis, which we collectively label as $\mathbf{z}=\{z_1,\dots,z_{\frac{\ell(\ell+9)}{2}}\}$. Thus, we can fix a monomial ordering and  build a Gr\"{o}ebner basis $\mathcal{G}_{i_1\cdots i_r}(\mathbf{z})$ of the ideal $\mathcal{J}_{i_1\cdots i_r}$ generated by the set of denominators,
	\begin{align}
	\mathcal{J}_{i_1\cdot\cdot \cdot i_r}\equiv \bigg\{\sum_{k=1}^{r}h_k(\mathbf{z})D_{i_k}(\mathbf{z})\,:\,h_k(\mathbf{z})\in P[\mathbf{z}]\bigg\},
	\end{align}
	being $P[\mathbf{z}]$ the ring of polynomials in $\mathbf{z}$. By performing the polynomial division of $\mathcal{N}_{i_1\cdots i_r}(\mathbf{z})$ modulo $\mathcal{G}_{i_1\cdot\cdot\cdot i_r}(\mathbf{z})$,
	\begin{align}
	\mathcal{N}_{i_1\cdots i_r}(\mathbf{z})=\sum_{k=1}^{r}\mathcal{N}_{i_1\cdots i_{k-1}i_{k+1}\cdots i_{r}}(\mathbf{z})D_{i_k}(\mathbf{z})+\Delta_{i_1\cdots i_r}(\mathbf{z})
	\label{eq:q+r}
	\end{align}
	we obtain the recurrence relation
	\begin{align}
	\mathcal{I}_{i_1\cdots i_r}=\sum_{k=1}^{r}\mathcal{I}_{i_1\cdots i_{k-1}i_{k+1}\cdots i_{r}}+\frac{\Delta_{i_1\cdots i_r}(\mathbf{z})}{D_{i_1}(\mathbf{z})\,\cdots D_{i_n}(\mathbf{z})\,},
	\label{eq:rec}
	\end{align}
	whose iterative application to the integrands corresponding to subtopologies with fewer loop propagators yields to the complete decomposition \eqref{eq:intdec}.\\
	
	The properties of the ideal $\mathcal{J}_{i_1\cdot\cdot \cdot i_r}$ ~\cite{Cox:1997,Cox:2005,Buchberger:1970,Sturmfels:2002} allow to derive an important result concerning the parametric form of the residues corresponding to \textit{maximum-cuts}. We define as a maximum-cut a zero-dimensional system of equations
	\begin{align}
	D_{i_1}(\mathbf{z})=\dots=D_{i_{r}}(\mathbf{z})=0,
	\end{align}
	which completely constraints the loop variables $\mathbf{z}$. If a maximum-cut admits a finite number of solutions $n_s$, each with multiplicity one, it satisfies  the following~\cite{Mastrolia:2013kca}
	\begin{thm}\textbf{(Maximum cut)} The residue of a maximum-cut is a polynomial parametrized by $n_s$ coefficients, which admits an univariate representation of degree $(n_s-1)$.\\
	\end{thm}
    
   Depending on the choice of variables $\mathbf{z}$ and the monomial order, the picture presented in this section can significantly simplify.  A particular convenient choice of variables turns out to be the one presented in sec.~\ref{sec:feyintdpardperp}.  Indeed, as we observed at the end of that section, we can always express a subset of the components of $q_{\parallel i}^\alpha$ and $\lambda_{ij}$ as a combination of denominators by solving linear relations.  This set of relations is in turn equivalent to the definition of the denominators themselves.  This implies that if we choose the lexicographic monomial order with $\lambda_{ij}\prec x_{kl}$ for $k\leq d_{\parallel}$, the polynomials in the Gr\"obner bases are linear in the $\lambda_{ij}$ and the reducible components of $q_{\parallel i}^\alpha$.  The polynomial division can thus equivalently be performed by applying the aforementioned set of linear relations without explicitly computing the corresponding Gr\"obner basis.

	\subsection{Divide, integrate and divide}
	\label{sec:divetinetdiv}
	As we have seen in sec.~\ref{sec:1}, when dealing with an integral with $n\leq 4$ external legs, we can introduce the $d=d_{\parallel}+d_{\perp}$ parametrization which removes the dependence of the denominators on the transverse components of the loop momenta. Thus, if we indicate with $\mathbf{z}$ the full set of $\ell(\ell+9)/2$ variables
	\begin{align}
	\mathbf{z}=&\{\mathbf x_{\parallel\,i},\mathbf x_{\perp\,i},\lambda_{ij}\},\quad i,j=1,\dots\ell,
	\end{align}
	where $\mathbf{x}_{\parallel\,i}$($\mathbf{x}_{\perp\,i}$) are the components of the loop momenta parallel(orthogonal) to the external kinematics, the denominators are reduced to polynomials in the subset of variables
	\begin{align}
	\boldsymbol{\tau}=&\{\mathbf{x}_{\parallel},\lambda_{ij}\},\quad \boldsymbol{\tau}\subset \mathbf{z},
	\end{align}
	so that the general $r$ denominators integrand has the form
	\begin{align}
	 \mathcal{I}_{i_1\dots i_r}(\boldsymbol{\tau},\mathbf{x}_{\perp})\equiv\frac{\mathcal{N}_{i_1\dots i_r}(\boldsymbol{\tau},\mathbf{x}_{\perp})}{D_{i_1}(\boldsymbol{\tau})\cdots D_{i_r} (\boldsymbol{\tau})}.
	\end{align}
	This observation suggests an \textit{adaptive} version of the integrand decomposition algorithm, where the polynomial division is simplified by working on the reduced set of variables $\boldsymbol{\tau}$ and the expansion of the residues in terms of Gegenbauer polynomials allows the systematic identification of spurious terms. The algorithm is organized in three steps:
	\begin{enumerate}[1)]
		\item \textbf{Divide:} we adopt lexicographic ordering $\lambda_{ij}\prec \mathbf x_{\parallel}$ for the $\boldsymbol{\tau}$  variables and we divide the numerator $\mathcal{N}_{i_1\dots i_r}(\boldsymbol{\tau},\mathbf{x}_{\perp})$ modulo the Gr\"{o}ebner basis $\mathcal{G}_{i_1\cdots i_r}(\boldsymbol{\tau})$ of the ideal $\mathcal{J}_{i_1\cdots i_r}(\boldsymbol{\tau})$ generated by the denominators,
		\begin{align}
		\mathcal{N}_{i_1\dots i_r}(\boldsymbol{\tau},\mathbf{x}_{\perp})=\sum_{k=1}^{r}\mathcal{N}_{i_1\dots i_{k-1}i_{k+1}\dots i_r}(\boldsymbol{\tau},\mathbf{x}_{\perp})D_{i_k}(\boldsymbol{\tau})+\Delta_{i_1\dots i_{r}}(\mathbf x_{\parallel},\mathbf x_{\perp}).
		\end{align}
		As a consequence of the specific monomial ordering, the residue $\Delta_{i_1\dots i_{r}}$ can depend on the transverse components $\mathbf x_{\perp\,i}$, which are left untouched by the polynomial division, as well as on $\mathbf x_{\parallel\,i}$ but not on $\lambda_{ij}$ that are expressed in terms of denominators and irreducible physical scalar products. Conversely, the quotient, from which the numerators corresponding to the subdiagrams to be further divided are obtained, still depends on the full set of loop variables.  As we explained at the end of sec.~\ref{sec:integrrecrel}, the Gr\"obner basis does not need to be explicitly computed, since, with the choice of variables and the ordering described here, the division is equivalent to applying the set of linear relations described at the end of sec.~\ref{sec:feyintdpardperp}.
		\item\textbf{Integrate:} we write the contribution of the residue $\Delta_{i_1\dots i_{r}}$ to the integral in the $d=d_{\parallel}+d_{\perp}$ parametrization which, according to \eqref{eq:poltr}, maps
		\begin{align}
		\mathbf x_{\perp i}\to P[\boldsymbol{\tau},\sin[\boldsymbol{\Theta}_{\perp}],\cos[\boldsymbol{\Theta}_{\perp}]].
		\end{align} 
		In this way, we can integrate over transverse directions through the expansion of $\Delta_{i_1\dots i_{r}}$ in terms of Gegenbauer polynomials, which sets to zero spurious terms and reduce all non-vanishing contributions to monomials in $\lambda_{ij}$,
		\begin{equation}
		\resizebox{.9\hsize}{!}
		{$
		\begin{split}
		\int \prod_{1=j}^{\ell}\frac{d^dq_{j}}{\pi^{d/2}}\frac{\Delta_{i_1\dots i_{r}}(\mathbf x_{\parallel},\mathbf x_{\perp})}{D_{i_1}(\boldsymbol{\tau})\dots D_{i_r}(\boldsymbol{\tau})}=& 
		\Omega^{(\ell)}_d\!\int\prod_{i=1}^{\ell} d^{n-1}q_{\parallel \, i}\int\! d^{\frac{\ell(\ell+1)}{2}} \boldsymbol{\Lambda}\int\! d^{(4-d_{\parallel})\ell}\boldsymbol{\Theta}_{\perp}\frac{\Delta_{i_1\dots i_{r}}(\boldsymbol{\tau},\boldsymbol{\Theta}_{\perp})}{D_{i_1}(\boldsymbol{\tau})\dots D_{i_r}(\boldsymbol{\tau})}\nn
		=&
		\Omega^{(\ell)}_d\!\int\prod_{i=1}^{\ell} d^{n-1}q_{\parallel \, i}\int\! d^{\frac{\ell(\ell+1)}{2}}\boldsymbol{\Lambda}\frac{\Delta^{\text{int}}_{i_1\dots i_{n}}(\boldsymbol{\tau})}{D_{i_1}(\boldsymbol{\tau})\dots D_{i_r}(\boldsymbol{\tau})}.
		\end{split}
	    $}
		\end{equation}
		It should be noted that, due to the angular prefactors produced by the integration of the transverse directions, the integrated residue
		\begin{align}
		\Delta^{\text{int}}_{i_1\dots i_{r}}(\boldsymbol{\tau})=\int\! d^{(4-d_{\parallel})\ell}\boldsymbol{\Theta}_{\perp}\Delta_{i_1\dots i_{r}}(\boldsymbol{\tau},\boldsymbol{\Theta}_{\perp})
		\end{align}
		is, in general, a polynomial in $\boldsymbol{\tau}$ whose coefficients depend explicitly on the space-time dimension $d$. The full set of $\Delta^{\text{int}}_{i_1\dots i_{r}}(\boldsymbol{\tau})$, obtained by iterating on each subdiagram numerator the polynomial division and the integration over the transverse space, provides already a spurious term-free representation of the integrand \eqref{eq:intdec}.
		\item\textbf{Divide:} however, since $\Delta^{\text{int}}_{i_1\dots i_{r}}(\boldsymbol{\tau})$ depends on the same variables as the denominators $D_{i_k}(\boldsymbol{\tau})$, we can perform a further division modulo the Gr\"{o}ebner basis $\mathcal{G}_{i_1\cdots i_r}(\boldsymbol{\tau})$ and get
		\begin{align}
		\Delta_{i_1\dots i_r}^{\text{int}}(\boldsymbol{\tau})=\sum_{k=1}^{r}\mathcal{N}^{\text{int}}_{i_1\dots i_{k-1}i_{k+1}\dots i_r}(\boldsymbol{\tau})D_{i_k}(\boldsymbol{\tau})+\Delta^{'}_{i_1\dots i_r}(\mathbf x_{\parallel}),
		\end{align}
		where, due to the choice of lexicographic ordering, the new residue $\Delta^{'}_{i_1\dots i_r}(\mathbf x_{\parallel})$ can only depend on $\mathbf x_{\parallel}$. Therefore, this additional polynomial division allows us to obtain an integrand decomposition formula \eqref{eq:intdec}, where all irreducible numerators are function of the components of the loop momenta parallel to the external kinematics.  As in the previous case, the division modulo Gr\"obner can equivalently be implemented via a set of linear relations.
	\end{enumerate}
	The interpretation of the monomials appearing in the residues $\Delta^{'}_{i_1\cdots i_r}(\mathbf{x}_{\parallel})$ in terms of a basis of tensor integrals can be additionally simplified by making use of the Gram determinant $G[\lambda_{ij}]$ (or $G[\mu_{ij}]$ for cases with more than four external legs, where $\mathbf{x}_\parallel\equiv \mathbf{x}$). In fact, as it can be easily understood from \eqref{eq:oldpar} and \eqref{eq:lambdaShout}, $G[\mu_{ij}]$ and  $G[\lambda_{ij}]$ can be interpreted as operators that, when acting on an arbitrary numerator of a $d$-dimensional integral, produce a dimensional shift $d\to d+2$. Therefore, Gram determinants can be used in order to trade some of the $d$-dimensional tensor integrals originating from the residues with lower rank integrals in higher dimensions.
	\subsection{Integrate and divide}
	In the three-step algorithm {\it divide-integrate-divide}, outlined in
	the previous section, the integration over the transverse angles is
	performed after the integrand reduction, namely after determining the
	residues. This first option follows the standard integrand reduction
	procedure, where the spurious monomials are present in the decomposed
	integrand, although they do not contribute to the integrated
	amplitude. Alternatively, if the dependence of the numerators on the
	loop momenta is known, then the integration over the orthogonal angles
	can be carried out before the reduction. Within this second option,
	which we can refer to as {\it integrate-divide}, after eliminating the
	orthogonal angles from the integrands, the residues resulting from the
	polynomial divisions contain only non-spurious monomials. In order to
	integrate before the reduction, the dependence of the numerator on the
	loop momenta should be either known analytically or reconstructed
	semi-analytically \cite{Heinrich:2010ax, Hirschi:2016mdz}. Such situation may indeed occur when
	the integrands to be reduced are built
	by automatic generators or they emerge as quotients of the subsequent
	divisions.
	\subsection{One-Loop adaptive integrand decomposition}
	We hereby apply the \textit{adaptive} integrand decomposition algorithm in order to determine an alternative parametrization of one-loop residues. As an exceptional property of one-loop integrands, we find that by working with $\boldsymbol{\tau}$ variables, all unitarity cuts are reduced to zero-dimensional systems. Moreover, we show that the last step of the algorithm, \textit{i.e.} the further polynomial division after angular integration over the transverse space, provides an implementation of the dimensional recurrence relations at the integrand level.  
	\subsubsection*{One-loop residues in $d=4-2\epsilon$}
	A general one-loop integral with $n$ external legs is characterized by a set of $n$ denominators,
	\begin{align}
	I_{i_1\cdots i_{n}}^{d\,(1)}=\int \frac{d^d q}{\pi^{d/2}}\frac{\mathcal{N}_{i_1\cdot\cdot\cdot i_{n}}(q)}{D_{i_1}(q)\dots D_{i_{n}}(q)}.
	\end{align}
	The integrand depends on five variables which, in the standard $d=4-2\epsilon$ parametrization, are identified with
	 \begin{align}
	 \mathbf{z}=\{z_1,z_2,z_3,z_4,z_5\}\equiv\{x_1,x_2,x_3,x_4,\mu^2\},
	 \end{align}
	 where $x_i$ are the components of the four-dimensional part of the loop momentum with respect to a basis $\{e_i^{\alpha}\}$ of massless vectors~\cite{Ossola:2006us,Mastrolia:2010nb,delAguila:2004nf,Peraro:2014cba} (one particular definition of such basis can be found in Appendix~\ref{Ap:31}. The denominators $D_{i_k}(\mathbf{z})$ are quadratic in $\mathbf{z}$ whereas, for any renormalizable theory, the most general numerator is a polynomial of the type
	\begin{align}
	&\mathcal{N}_{i_1\cdot\cdot\cdot i_{n}}(\mathbf{z})=\sum_{\vec{j}\in J_5(n)}\alpha_{\vec{j}}z_1^{j_1}z_2^{j_2}z_3^{j_3}z_4^{j_4}z_5^{j_5},
	\label{eq:numnpar}
	\end{align}
	with 
	\begin{align}
	J_5(n)=\{\vec j=(j_1,\,...,\,j_5)/j_1+j_2+j_3+j_4+2j_5\leq n\}.
	\label{eq:J5}
	\end{align}
	Higher rank numerators, such as the one appearing in effective theories can be treated in a similar way, along the lines of \cite{Mastrolia:2013kca}.
	The polynomial division of the numerators $\mathcal{N}_{i_1\cdot\cdot\cdot i_{n}}(\mathbf{z})$ modulo the Gro\"{o}bener basis  $\mathcal{G}_{i_1\dots i_n}(\mathbf{z})=\{g_1(\mathbf{z}),g_2(\mathbf{z}),\dots g_{n}(\mathbf{z})\} $ returns the universal parametrization of the residues~\cite{Ossola:2006us,Ellis:2007br,Mastrolia:2012an}
	\begin{align}
	\Delta_{ijklm}=&c_0\mu^2,\nn
	\Delta_{ijkl}=&c_0+c_1x_{4}+c_2\mu^2+c_3x_{4}\mu^2+c_4\mu^4,\nn
	\Delta_{ijk}=&c_0+c_1x_4+c_2x_4^2+c_3x_4^3+c_4x_3+c_5x_3^2+c_6x_3^3+c_7\mu^2+c_8x_4\mu^2+c_9x_3\mu^2,\nn
	\Delta_{ij}=&c_0+c_1x_1+c_2x_1^2+c_3x_4+c_4x_4^2+c_5x_3+c_6x_3^3+c_7x_1x_4+c_8x_1x_3+c_9\mu^2,\nn
	\Delta{i}=&c_0+c_1x_1+c_2x_2+c_3x_3+c_4x_4.
	\label{eq:oldresidue}
	\end{align}
	Most of the terms appearing in \eqref{eq:oldresidue} are \textit{spurious}, \textit{i.e.} they vanish upon integration. Therefore, we can write the decomposition of an arbitrary one-loop amplitude with $n$ external legs as a linear combinations of master integrals (IRIs), corresponding to the non-spurious terms of the integrand, 
	\begin{subequations}
	\begin{align}
		 \label {integraldeco}
		\mathcal{A}_{n}^{(1)}=&\sum_{i\ll l}^{n}\left[c^{(ijlm)}_0I^{d\,(1)}_{ijlm}[1]+c^{(ijlm)}_2I^{d\,(1)}_{ijlm}[\mu^2]+c^{(ijlm)}_4I^{d\,(1)}_{ijlm}[\mu^4]\right]\nn 
		&+\sum_{i\ll l}^{n}\left[c^{(ijl)}_{0}I_{ijl}^{d\,(1)}[1]+c^{(ijl)}_{7}I_{ijl}^{d\,(1)}[\mu^2]\right]\nn
		&+\sum_{i\ll j}^{n}\bigg[c^{(ij)}_{0}I_{ij}^{d\,(1)}[1]+c^{(ij)}_{1}I_{ij}^{d\,(1)}[(q+p_i)\cdot e_2]+c^{(ij)}_{2}I_{ij}^{d\,(1)}[((q+p_i)\cdot e_2)^2]\nn
		&+c^{(ij)}_{9}I_{ij}[\mu^2]\bigg]+\sum_{i}^{k}c^{(i)}_{0}I_{i}^{d\,(1)}[1]\\
		=&\sum_{i\ll l}^{n}\left[c^{(ijlm)}_0I^{d\,(1)}_{ijlm}[1]+ c^{(ijlm)}_2(-\epsilon)I^{d+2\,(1)}_{ijlm}[1]+c^{(ijlm)}_4(-\epsilon)(1-\epsilon)I^{d+4\,(1)}_{ijlm}[1]\right]\nn
		&+\sum_{i\ll l}^{n}\left[c^{(ijl)}_{0}I_{ijl}^{d\,(1)}[1]+c^{(ijl)}_{7}(-\epsilon)I_{ijl}^{d\,(1)}[1]\right]
		\nn&+\sum_{i\ll j}^{n}\bigg[c^{(ij)}_{0}I_{ij}^{d\,(1)}[1]+c^{(ij)}_{1}I_{ij}^{d\,(1)}[(q+p_i)\cdot e_2]+c^{(ij)}_{2}I_{ij}^{d\,(1)}[((q+p_i)\cdot e_2)^2]+\nn
		&c^{(ij)}_{9}(-\epsilon)I_{ij}[1]\bigg]+\sum_{i}^{k}c^{(i)}_{0}I_{i}^{d\,(1)}[1],
		\label{integraldecoshift}
	\end{align}
	\end{subequations}
	where, in the second equality, we have identified $\mu^2$ numerators with higher-dimensional integrals~\cite{Bern:1995db}.\\
	
	The particular simple form of the five-point residue appearing eq.\eqref{eq:oldresidue} is due to the fact that the quintuple cut $D_{i}(\mathbf{z})=\dots=D_{m}(\mathbf{z})=0$ is a maximum-cut which admits a unique solution ($n_s=1$). Hence, $\Delta_{ijklm}$ is parametrized by a single coefficient, which is conventionally chosen as the one corresponding to the $\mu^2$ numerator.
	\subsubsection*{One-loop residues in $d=d_{\parallel}+d_{\perp}$}
	In $d=4-2\epsilon$ dimensions the maximum-cut theorem can only fix the parametric form of the residue of the quintuple cut, since the cut conditions for all lower-point integrands form an underdetermined system for the variables $\mathbf{z}$. However, all these integrands have $n\leq 4$ external legs and we can introduce the $d=d_{\parallel}+d_{\perp}$ parametrization in terms of the variables
	\begin{align}
	\mathbf{z}=\{\mathbf{x}_{\parallel},\mathbf{x}_{\perp},\lambda^2\},
	\end{align}
	where $\mathbf{x}_{\parallel}$ and $\mathbf{x}_{\perp}$ are defined according to the four-dimensional basis given in Appendix~\ref{Ap:32}. In this way, the $n$ denominators depend on the subset of variables
	\begin{align}
	\boldsymbol{\tau}=\{\mathbf{x}_\parallel,\lambda^2\}
	\end{align}
	containing exactly $n$ elements.  Since, as explained at the end of sec.~\ref{sec:feyintdpardperp}, these variables can be written as combinations of denominators via linear relations, and because the cut $D_{i_1}(\boldsymbol{\tau})=\cdots=D_{i_n}(\boldsymbol{\tau})=0$ with $n\leq 4$ is a maximum-cut, the corresponding set of cut equations is equivalent to a determined linear system and therefore has a single solution ($n_s=1$), which constrains all $\boldsymbol{\tau}$ variables. This means that we are in the \textit{Shape Lemma} position and, as implied by the discussion at the end of sec.~\ref{sec:integrrecrel}, a Gr\"{o}bener basis $\mathcal{G}_{i_1\dots i_n}(\boldsymbol{\tau})=\{g_1(\boldsymbol{\tau}),g_2(\boldsymbol{\tau}),\dots g_{n}(\boldsymbol{\tau})\}$ of $\mathcal{J}_{i_1\dots i_n}$ is found in the \textit{linear} form
	\begin{align}
	g_i(\boldsymbol{\tau})=\alpha_{i}+\tau_{i},\quad i=1,\dots,n.
	\label{eq:lineargb}
	\end{align}
    Hence, according to the maximum-cut theorem, the residues of all cuts with $n\leq 4$ are constant in $\boldsymbol{\tau}$.

    Although it is independent of $\boldsymbol{\tau}$, $\Delta_{i_1\cdot\cdot\cdot i_{n}}$ can still depend on the $4-d_{\parallel}$ four-dimensional transverse variables, which are left unconstrained by the cut conditions. However, the parametrization of the residue is terms of $\mathbf{x}_{\perp}$ is completely fixed by the renormalizability condition,
	\begin{align}
	&\Delta_{i_1\cdot\cdot\cdot i_{n}}(\mathbf{x}_\perp)=\sum_{\text{\tiny{$\vec{j}\in J_{4-d_{\parallel}}(n)$}}}\alpha_{\vec{j}}x_{\perp\,1}^{j_1}x_{\perp\,2}^{j_2}\dots x_{\perp\,4-d_{\parallel}}^{j_{4-d_{\parallel}}},\quad n\leq 4,
	\label{eq:respar}
	\end{align}
	with $J_{4-d_{\parallel}}(n)=\{\vec j=(j_1,\,...,\,j_{4-d_{\parallel}})/j_1+j_2+\dots j_{4-d_{\parallel}}\leq n\}$. Accordingly, from the polynomial division of the numerators modulo  $\mathcal{G}_{i_1\dots i_n}(\boldsymbol{\tau})$, with lexicographic ordering $\lambda^2\prec \mathbf{x}_{\perp}$, we find a parametric expression of the residues alternative to \eqref{eq:oldresidue},
	\begin{align}
	\Delta_{ijklm}=&c_0\mu^2,\nn
	\Delta_{ijkl}=&c_0+c_1x_{4}+c_2x_4^2+c_3x_{4}^3+c_4x_4^4,\nn
	\Delta_{ijk}=&c_0+c_1x_3+c_2x_4+c_3x_3^2+c_4x_3x_4+c_5x_4^2+c_6x_3^3+c_7x_3^2x_4+c_8x_3x_4^2+c_9x_4^3,\nn
	\Delta_{ij}=&c_0+c_1x_2+c_2x_3+c_3x_4+c_4x_2^2+c_5x_2x_3+c_6x_2x_4+c_7x_3^2+c_8x_3x_4+c_9x_4^2,\nn
	\Delta_{ij}|_{p^2=0}&=c_0+c_1x_1+c_2x_3+c_3x_4+c_4x_1^2+c_5x_1x_3+c_6x_1x_4+c_7x_3^2+c_8x_3x_4+c_9x_4^2,\nn
	\Delta{i}=&c_0+c_1x_1+c_2x_2+c_3x_3+c_4x_4.
	\label{eq:newres}
	\end{align}
	We observe that the two-point integrand with massless external momentum $p^2=0$, whose residue is indicated as $\Delta_{ij}|_{p^2=0}$, is the only exception to \eqref{eq:respar}, since the residue depends on the components $x_{1}$ parallel to $p$. In fact, due to the reduced dimension of the transverse space, the denominators depend on three variables $\boldsymbol{\tau}=\{x_1,x_2,\lambda^2\}$ so that, in this degenerate kinematic configuration, the double cut is not maximum any more.\\
	
	The residues \eqref{eq:newres} can now be integrated over the transverse directions by means of the orthogonality relation \eqref{eq:ortcos} for Gegenbauer polynomials, which removes spurious terms and reduce non-vanishing contributions to powers of $\lambda^2$. Hence, by making use of the results collected in Appendix~\ref{Ap:1l}, we obtain the decomposition of a generic one-loop amplitude in terms of IRIs, 
	\begin{subequations}
	\begin{align}
	\label{newintegraldeco}
		\mathcal{A}_{n}^{(1)}=&\sum_{i\ll l}^{n}\left[c^{(ijlm)}_0I^{d\,(1)}_{ijlm}[1]+c^{(ijlm)}_2\frac{1}{1-2\epsilon}I^{d\,(1)}_{ijlm}[\lambda^2]+c^{(ijlm)}_4\frac{3}{(1-2\epsilon)(3-2\epsilon)}I^{d\,(1)}_{ijlm}[\lambda^4]\right]\nn
		&+\sum_{i\ll l}^{n}\left[c^{(ijl)}_{0}I_{ijl}^{d\,(1)}[1]+c^{(ijl)}_{7}\frac{1}{2-2\epsilon}I_{ijl}^{d\,(1)}[\lambda^2]\right]\nn
		&+\sum_{i\ll j}^{n}\bigg[c^{(ij)}_{0}I_{ij}^{d\,(1)}[1]+c^{(ij)}_{1}I_{ij}^{d\,(1)}[((q+p_i)\cdot e_2)]+c^{(ij)}_{2}I_{ij}^{d\,(1)}[((q+p_i)\cdot e_2)^2]+\nn
		&\qquad\quad c^{(ij)}_{9}\frac{1}{3-2\epsilon}I_{ij}^{d\,(1)}[\lambda^2]\bigg]+\sum_{i}^{k}c^{(i)}_{0}I_{i}^{d\,(1)}[1]\\
		=&\sum_{i\ll l}^{n}\left[c^{(ijlm)}_0I^{d\,(1)}_{ijlm}[1]+c^{(ijlm)}_2\frac{1}{2}I^{d+2\,(1)}_{ijlm}[1]+c^{(ijlm)}_4\frac{3}{4}I^{d+4\,(1)}_{ijlm}[1]\right]\nn
		&+\sum_{i\ll l}^{n}\left[c^{(ijl)}_{0}I_{ijl}^{d\,(1)}[1]+c^{(ijl)}_{7}\frac{1}{2}I_{ijl}^{d+2\,(1)}[1]\right]\nn
		&+\sum_{i\ll j}^{n}\bigg[c^{(ij)}_{0}I_{ij}^{d\,(1)}[1]+c^{(ij)}_{1}I_{ij}^{d\,(1)}[(q+p_i)\cdot e_2]+c^{(ij)}_{2}I_{ij}^{d\,(1)}[((q+p_i)\cdot e_2)^2]\nn
		&\quad\qquad+c^{(ij)}_{9}\frac{1}{2}I_{ij}^{d+2\,(1)}[1]\bigg]+\sum_{i}^{k}c^{(i)}_{0}I_{i}^{d\,(1)}[1],
	\label{newintegraldecohd}
	\end{align}
	\end{subequations}
	where, in the second equality, we have identified monomials in $\lambda^2$ in the numerator with higher-dimensional integrals. \\
	
	The number of IRIs in which the amplitude is decomposed can be further reduced by observing that, due to the choice of lexicographic ordering, $\lambda^2$ is reducible, \textit{i.e.} it can be rewritten, modulo a constant remainder, in terms of denominators. Therefore, all higher-dimensional integrals appearing in \eqref{newintegraldecohd} are reduced to a combination of the corresponding scalar integral in $d$-dimensions and integrals with fewer denominators. As a consequence, this additional polynomial division can be interpreted an implementation of dimensional recurrence relations at the integrand level. The final decomposition of the amplitude in terms of the minimal set IRIs reads 
	\begin{align}
		\mathcal{A}_{n}^{(1)}=&\sum_{i\ll l}^{n}c^{(ijlm)}_0(\epsilon)I^{d\,(1)}_{ijlm}[1]+\sum_{i\ll l}^{n}c^{(ijl)}_{0}(\epsilon)I_{ijl}^{d\,(1)}[1]
		\nn
		&+\sum_{i\ll j}^{n}\left[c^{(ij)}_{0}(\epsilon)I_{ij}^{d\,(1)}[1]+c^{(ij)}_{1}I_{ij}^{d\,(1)}[(q+p_i)\cdot e_2]+c^{(ij)}_{2}I_{ij}^{d\,(1)}[((q+p_i)\cdot e_2)^2]\right]\nn
		&+\sum_{i}^{n}c^{(i)}_{0}I_{i}^{d\,(1)}[1].
		\label{eq:Fintegraldeco}
	\end{align}
	 It should be remarked that, although we have used similar a labelling, the coefficients the master integrals appearing \eqref{eq:Fintegraldeco} are different from the ones in \eqref{newintegraldeco}. Moreover, in \eqref{eq:Fintegraldeco}, these coefficients can present an explicit dependence on the space-time dimension, due to the angular prefactors produced by the integration over the transverse variables. We give a summary of the results obtained from the application of the \textit{adaptive} integrand reduction algorithm at one loop in Table~\ref{tab:1l}.
	\begin{table}[h]
		\centering
		\renewcommand{\arraystretch}{1.2}
		\scalebox{0.77}{
			\begin{tabular}{|c c||c|c|c|c|}
				\hline
				\multicolumn{2}{|c||}{$\mathcal{I}_{i_1\,\cdots\,i_n}$}&$\boldsymbol{\tau}$&$\Delta_{i_1\,\cdots\,i_n}$ &$\Delta^{\text{int}}_{i_1\,\cdots\,i_n}$& $\Delta^{'}_{i_1\,\cdots\,i_n}$ \\
				\hline
				\hline
				\multirow{2}{0.9cm}{\centering $\mathcal{I}_{i_1i_2i_3i_4i_5}$}&\multirow{2}{2.cm}{\centering\includegraphics[height=0.37in]{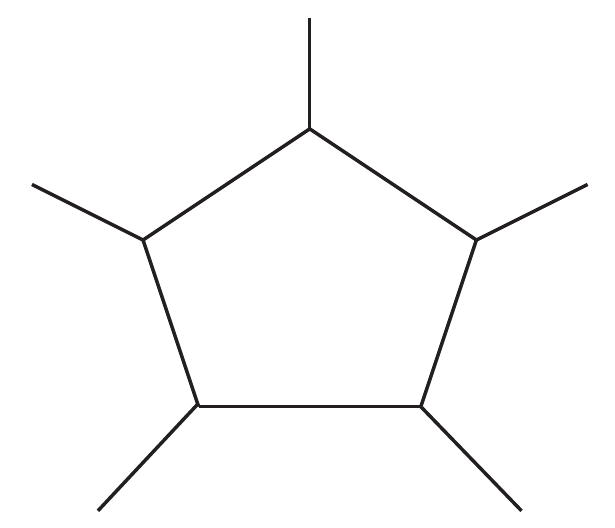}}&  &$1$&$-$&$-$\\
				&                &$\{x_1,x_2,x_3,x_4,\mu^2\}$ &$\{1\}$&$-$&$-$\\
				\hline
				\multirow{2}{0.9cm}{\centering $\mathcal{I}_{i_1i_2i_3i_4}$}&\multirow{2}{2.cm}{\centering\includegraphics[height=0.35in]{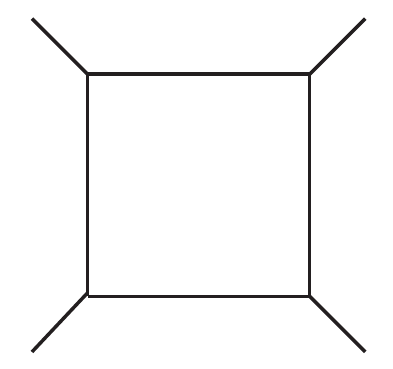}} & &$5$&$3$&$1$\\
				&     &$\{x_1,x_2,x_3,\lambda^2\}$ &$\{1,x_4,x_4^2,x_4^3,x_4^4\}$&$\{1,\lambda^{2},\lambda^4\}$&$\{1\}$\\
				\hline
				\multirow{2}{0.9cm}{\centering $\mathcal{I}_{i_1i_2i_3}$}&\multirow{2}{2.cm}{\centering\includegraphics[height=0.35in]{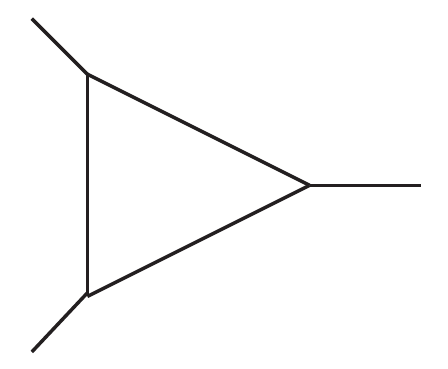}} &&$10$&$2$&$1$\\
				&    &  $\{x_1,x_2,\lambda^2\}$  &$\{1,x_3,x_4,x_3^2,x_3x_4,x_4^2,x_3^3,x_3^2x_4,x_3x_4^2,x_4^3\}$&$\{1,\lambda^{2}\}$&$\{1\}$\\
				\hline
				\multirow{2}{0.9cm}{\centering $\mathcal{I}_{i_1i_2}$}&\multirow{2}{2.cm}{\centering\includegraphics[height=0.35in]{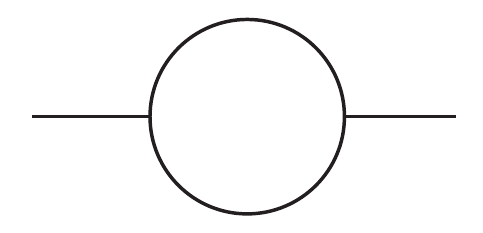}}&&$10$&$2$&$1$\\
				&        &$\{x_1,\lambda^2\}$&$\{1,x_2,x_3,x_4,x_2^2,x_2x_3,x_2x_4,x_3^2,x_3x_4,x_4^2\}$&$\{1,\lambda^2\}$&$\{1\}$\\
				\hline
				\multirow{2}{0.9cm}{\centering $\mathcal{I}_{i_1i_2}$}&\multirow{2}{2.cm}{\centering\includegraphics[height=0.35in]{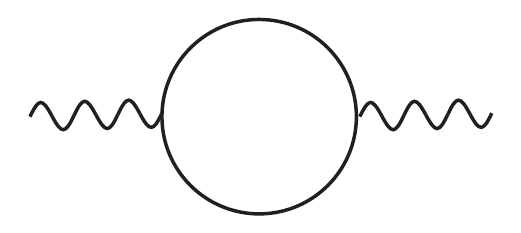}} &&$10$&$4$&$3$\\
				&     & $\{x_1,x_2,\lambda^2\}$  &$\{1,x_1,x_3,x_4,x_1^2,x_1x_3,x_1x_4,x_3^2,x_3x_4,x_4^2\}$&$\{1,x_1,x_1^2,\lambda^{2}\}$&$\{1,x_1,x_1^2\}$\\
				\hline
				\multirow{2}{0.9cm}{\centering $\mathcal{I}_{i_1}$}&\multirow{2}{2.cm}{\centering\includegraphics[height=0.35in]{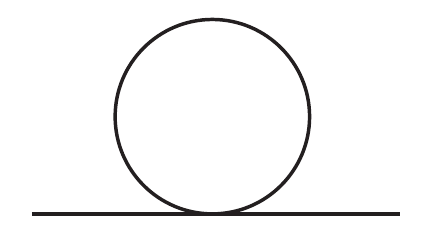}} &&$5$&$1$&$-$\\
				&    & $\{\lambda^2\}$   &$\{1,x_1,x_2,x_3,x_4\}$&$\{1\}$&$-$\\
				\hline
			\end{tabular}
		}
		\caption{\small{Residue parametrization for irreducible one-loop topologies. In the first column, $\boldsymbol{\tau}$ labels the variables the denominators depend on. $\Delta_{i_1\,\cdots\,i_n}$ indicates the residue obtained after the polynomial division of an arbitrary $n$-rank numerator and $\Delta_{i_1\,\cdots\,i_n}^{\text{int}}$ the result of its integral over transverse directions. $\Delta_{i_1\,\cdots\,i_n}^{'}$ corresponds to the minimal residue obtained from a further division of $\Delta_{i_1\,\cdots\,i_n}^{\text{int}}$. In the figures, wavy lines indicate massless particles, whereas solid ones stands for arbitrary masses.}}
		\label{tab:1l}
	\end{table}
	 \subsection{Two-loop adaptive integrand decomposition}
\begin{figure*}[ht!]
	 	\centering
	 	\begin{subfigure}[t]{0.33\textwidth}
	 		\centering
	 		\includegraphics[height=1.0in]{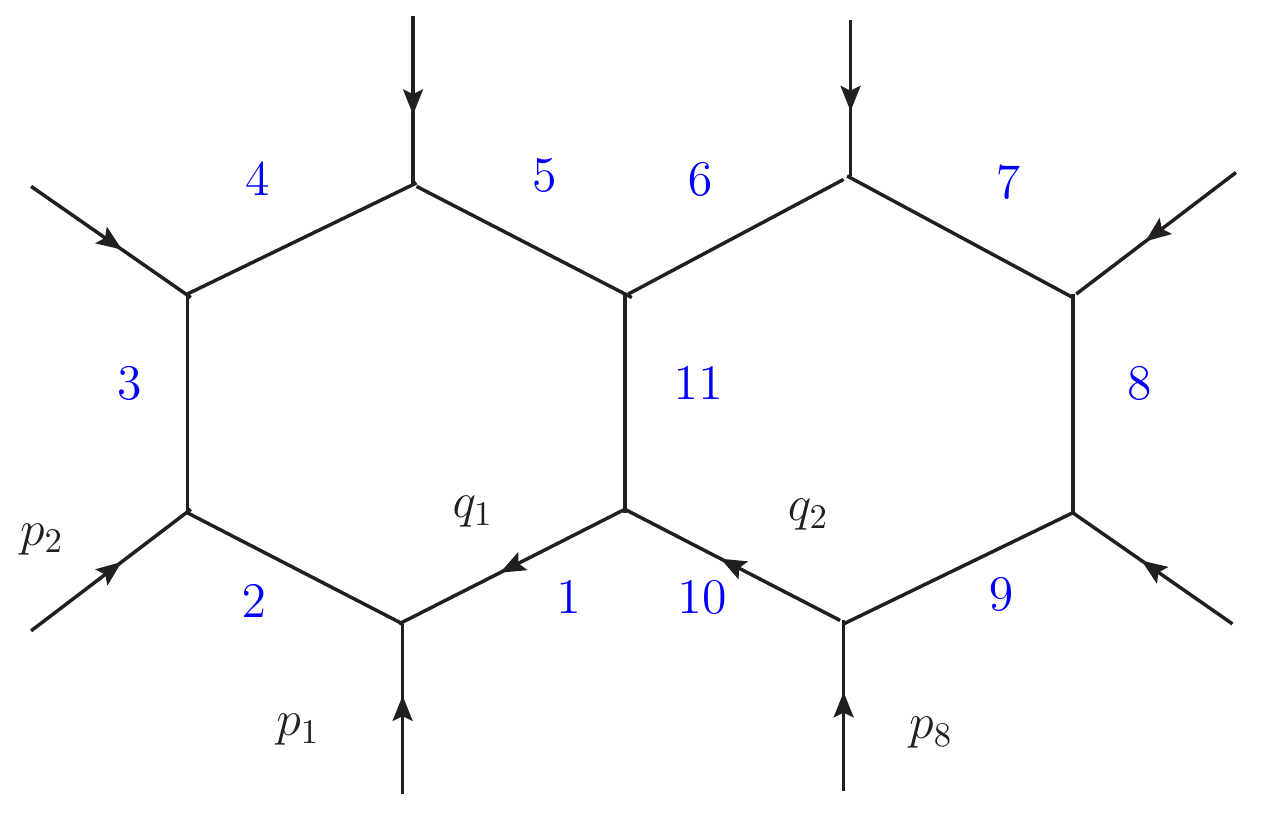}
	 		\caption{$\mathcal{I}_{12345678910\, 11}^{\text{P}}$}
	 		\label{fig:maxcutP}
	 	\end{subfigure}%
	 	\begin{subfigure}[t]{0.33\textwidth}
	 		\centering
	 		\includegraphics[height=1.1in]{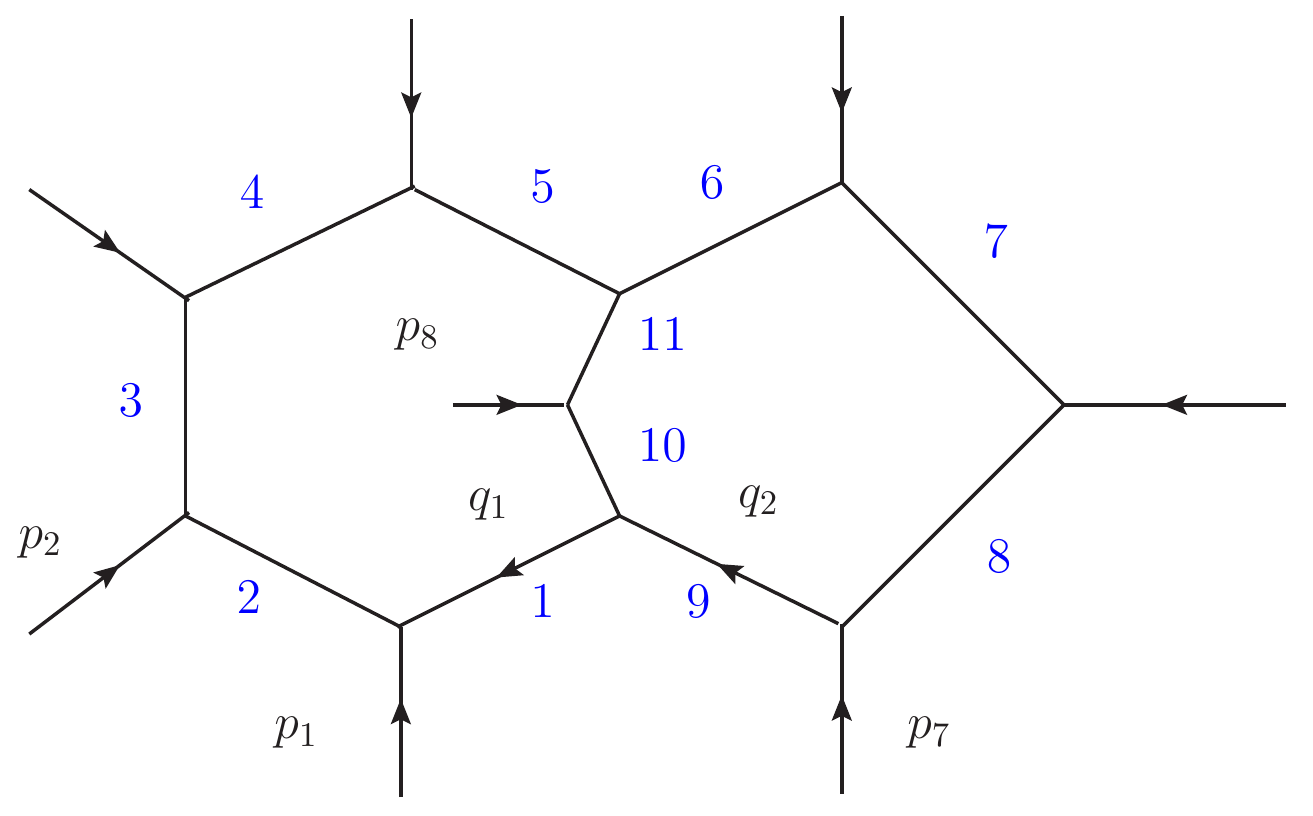}
	 		\caption{$\mathcal{I}_{12345678910\, 11}^{\text{NP}1}$}
	 		\label{fig:maxcutNP1} 
	 	\end{subfigure}
	 	\begin{subfigure}[t]{0.33\textwidth}
	 		\centering
	 		\includegraphics[height=1.1in]{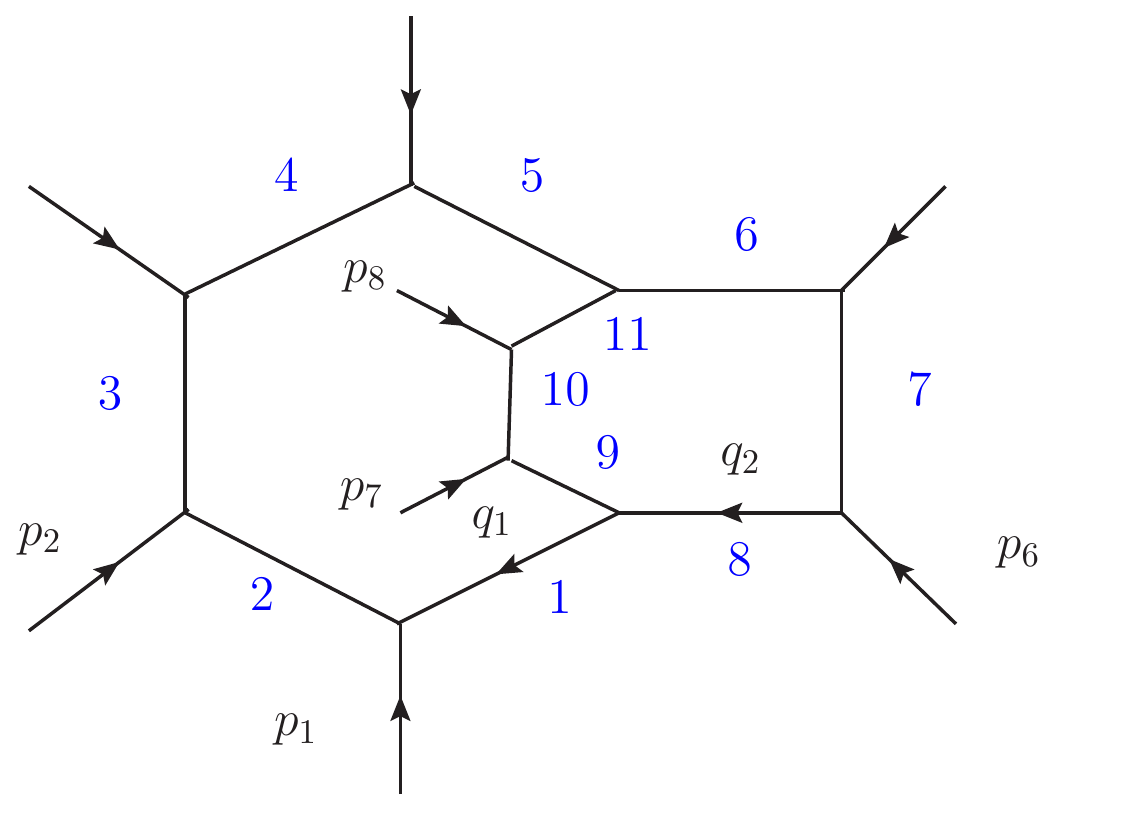}
	 		\caption{$\mathcal{I}_{12345678910\, 11}^{\text{NP2}}$}
	 		\label{fig:maxcutNP2}
	 	\end{subfigure}
	 	\caption{Maximum-cut topologies}
	 	\label{fig:maxcut}
	 \end{figure*}

	 In this section we  use the \textit{adaptive} integrand decomposition algorithm in order to determine the universal parametrization of the residues appearing in the integrand representation \eqref{eq:intdec} of the three eight-point topologies shown in fig.~\ref{fig:maxcutP}-\ref{fig:maxcutNP2}. The results hereby presented are valid for arbitrary (internal and external) kinematic configuration.\\
	 
	 At two-loops, we generally deal with $r$ denominators Feynman integrals of the type
	 \begin{align}
	 I_{i_1\cdots i_{r}}^{d\,(2)}=\int \frac{d^d q_1d^dq_2}{\pi^{d}}\frac{\mathcal{N}_{i_1\cdot\cdot\cdot i_{r}}(q_1,q_2)}{D_{i_1}\dots D_{i_{r}}},
	 \end{align}
	 with $r_1$ denominators depending on $q_{1}^{\alpha}$, $r_2$ denominators depending on $q_2^{\alpha}$ and $r_{12}=r-r_1-r_2$ depending on both loop momenta. The general numerator of a non-factorized integrand ($r_{12}\neq 0$) of a renormalizable theory is given by a polynomial of the type
	 \begin{align}
	 &\mathcal{N}_{i_1\cdot\cdot\cdot i_{r}}(\mathbf{z})=\sum_{\vec{j}\in J_{11}(s_1,s_2,s_{\text{tot}})}\alpha_{\vec{j}}z_1^{j_1}z_2^{j_2}\dots z_{11}^{j_{11}},
	 \label{eq:numnpar2L}
	 \end{align}
	 where $\mathbf{z}=\{z_1,\dots,z_{11}\}$ labels the full set of loop variables (defined according to the number of external legs) and $J_{11}(s_1,s_2,s_{\text{tot}})$ denotes the vectors of integers $(j_1,\dots, j_{11})$ satisfying
	\begin{subnumcases}{}
	 \label{eq:renr1}
	 \sum_{i=1}^4j_i+2j_9+j_{11}\leq s_1,\qquad\quad\quad\:\;\text{with}\quad s_1=r_1+r_{12}\\
	 \label{eq:renr2}
	 \sum_{i=5}^8j_i+2j_{10}+j_{11}\leq s_2,\qquad\qquad\,\text{with}\quad s_2=r_2+r_{12}\\
	 \label{eq:renrtot}
	 \sum_{i=1}^8j_i+2(j_{9}+j_{10}+j_{11})\leq s_{12},\quad\text{with}\quad s_{\text{tot}}=r_1+r_2+r_{12}-1.
	\end{subnumcases}
	As we have already observed in the one-loop case, the present discussion can be easily extended to the case of higher rank numerators.
	 Depending on the number of external legs, we determine the residue parametrization in two different ways:
	 \begin{itemize}
	 	\item 
	  for the parent topologies $\mathcal{I}_{1\cdots 11}^{\text{P}}$, $\mathcal{I}_{1\cdots 11}^{\text{NP}1}$ and $\mathcal{I}_{1\cdots 11}^{\text{NP}2}$ as well as for all subtopologies with $n>4$ four external legs we use the $d=4-2\epsilon$ parametrization and we define $\mathbf{z}$ as
	  \begin{align}
	  \mathbf{z}=\{x_{11},\dots x_{41},x_{12},\dots,x_{42},\mu_{11},\mu_{22},\mu_{12}\},
	  \end{align}
	  where $\{x_{1\,i},\dots,x_{4\,i}\}$ are the components of the four-dimensional vector $q_{[4]\,i}^{\alpha}$ with respect to the basis defined in Appendix~\ref{Ap:31}. In these cases, the denominators depend on the full set of variables $\mathbf{z}$ and the parametric form of the residues is determined through a single polynomial division of $\mathcal{N}_{i_1\cdot\cdot\cdot i_{r}}(\mathbf{z})$ modulo a Gr\"{o}ebner basis $\mathcal{G}_{i_1\dots i_r}(\mathbf{z})$ of the ideal generated by the denominators. The results obtained for the eight- seven- six- and five-point integrands are summarized in Tables~\ref{Tab:87pt}-\ref{Tab:65pt}. We observe that, according to the maximum-cut theorem, the residues of the master topologies $\mathcal{I}_{1\cdots 11}^{\text{P}}$, $\mathcal{I}_{1\cdots 11}^{\text{NP}1}$ and $\mathcal{I}_{1\cdots 11}^{\text{NP}2}$ contain one single coefficient, since the zero-dimensional systems $D_{1}(\mathbf{z})=\cdots=D_{11}(\mathbf{z})=0$ admit only one solution. 
	  \item For any subdiagram with $n\leq 4$ external legs we introduce the $d=d_{\parallel}+d_{\perp}$ parametrization and define $\mathbf{z}$ as
	 \begin{align}
	 \mathbf{z}=\{\mathbf x_{\parallel\,1},\mathbf x_{\perp\,1},\mathbf x_{\parallel\,2},\mathbf x_{\perp\,2},\lambda_{11},\lambda_{22},\lambda_{12}\},
	 \end{align}
	 being $\mathbf{x}_{\parallel\,i}=\{x_{1\,i},\dots,x_{d_{\parallel}\,i}\}$ the components of the vector $q_{\parallel\,i}^{\alpha}$ lying in the space spanned by the external momenta and $\mathbf{x}_{\perp\,i}=\{x_{d_{\parallel}+1\,i},\dots,x_{4\,i}\}$ the four-dimensional components of the transverse vector $\lambda_{i}^{\alpha}$ (see Appendix~\ref{Ap:32} for the explicit definition of the basis). In these cases, the denominators depend on the reduced set of variables
	 \begin{align}
	 \boldsymbol{\tau}=\{\mathbf x_{\parallel\,1},\mathbf x_{\parallel\,2},\lambda_{11},\lambda_{22},\lambda_{12}\},\qquad \boldsymbol{\tau}\subset \mathbf{z},
	 \end{align}
	 and we can go through the full \textit{adaptive} integrand decomposition algorithm described in sec.~\ref{sec:divetinetdiv}. We refer the reader to the Appendix~\ref{Ap:2l} for the most relevant formulae regarding the $d=d_{\parallel}+d_{\perp}$ parametrization of two-loop integrals and the integration over transverse variables. It should be noted that, conversely to the one-loop case, the $n$-ple cut $D_{i_1}(\boldsymbol{\tau})=\cdots=D_{i_r}(\boldsymbol{\tau})=0$ is, generally, non-maximum, since it does not constrain all variables $\boldsymbol{\tau}$. However, the choice of lexicographic ordering $\lambda_{ij}\prec \mathbf{x}_{\perp\,i}$ guarantees that the final residues $\Delta^{'}_{i_1\cdot\cdot\cdot i_{r}}(\mathbf{x}_{\parallel\, i})$ appearing in the integrand decomposition formula \eqref{eq:intdec} depend on the components of loop momenta parallel to the external kinematics only. All results are summarized in Tables~\ref{Tab:4pt}-\ref{Tab:1pt}.
	 \end{itemize}
	  Finally, in the case of an integrand factorized into two one-loop diagrams with $n_1$ and $n_2$ external legs, we can assume, as discussed in sec.~\ref{sec:fact}, the most general numerator to have the form 
	 \begin{align}
	 &\mathcal{N}_{i_1\cdot\cdot\cdot i_{r}}^{\text{fact}}(\mathbf{z}_1,\mathbf{z}_2)=\sum_{\substack{\text{$\vec{j_1}\in J_5(n_1)$}\\\text{$\vec{j_2}\in J_5(n_2)$}}}\alpha_{\vec{j_1},\,\vec{j_2}}z_{11}^{j_{11}}\dots z_{51}^{j_{51}}z_{12}^{j_{12}}\dots z_{52}^{j_{52}},
	 \label{eq:numnpar2Lfact}
	 \end{align}
	 where $\mathbf{z}_i=\{z_{1i},\dots,z_{5i}\}$ labels the set of variables parametrizing $q_{i}$ and $J_5(n_i)$ is defined by \eqref{eq:J5}. In this way we can introduce the $d=d_{\parallel}+d_{\perp}$ parametrization independently on the two loops and then proceed with the \textit{adaptive} integrand decomposition algorithm. As expected, the resulting residues, which are shown in Table~\ref{Tab:fact}, are simply given by the product of the corresponding one-loop residues collected in Table~\ref{tab:1l}.\\
	 
	 We would like to mention that the residues $\Delta_{i_1\cdots i_r}(\mathbf{x}_{\parallel})$ of non planar topologies, which are written in 
	 terms of a minimal set of physical components, produce an apparent violation of one the renormalizability conditions \eqref{eq:renr1}-\eqref{eq:renr2} satisfied by
	 the original numerators. This effect is due to the fact that, when the cut conditions are imposed, the presence of a number $r_{12}>1$ of denominators depending both on $q_{1}$ and $q_{2}$ implies the existence of linear relations between the physical components of the two loop momenta. This means that, up to subdiagrams contributions, the residues can always be rewritten in terms of a larger number of variables, in such a way to satisfy all renormalizabilty constraints \eqref{eq:renr1}-\eqref{eq:renrtot}.\\
		 \begin{table}[!ht]
		 	\centering
		 	\renewcommand{\arraystretch}{1.2}
		 	\scalebox{0.75}{
		 		\begin{tabular}{cc}%
		 	    	\begin{tabular}[t]{|c c||c|}
		 					\hline
		 						\multicolumn{2}{|c||}{$\mathcal{I}_{i_1\cdots i_n}$}&$\Delta_{i_1\cdots i_r}$ \\
		 						\hline
		 						\hline
		 						\multirow{2}{1.4cm}{\centering $\mathcal{I}_{12345678910\,11}^{\text{P}}$}&\multirow{2}{2.cm}{\centering\includegraphics[height=0.4in]{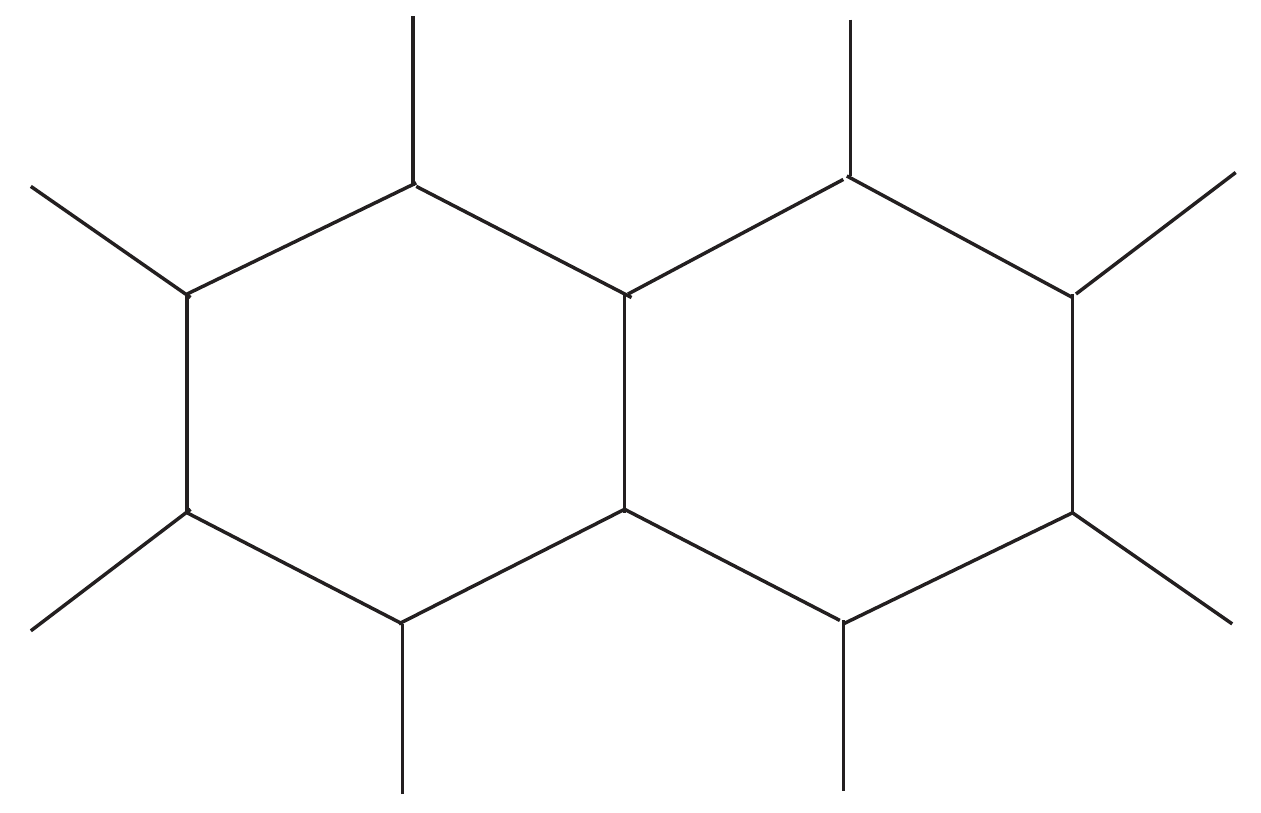}}&$1$\\
		 						&                        &$\{1\}$\\
		 						\hline
		 						\multirow{2}{1.4cm}{\centering $\mathcal{I}_{12345678910\,11}^{\text{NP}1}$}&\multirow{2}{2.cm}{\centering\includegraphics[height=0.35in]{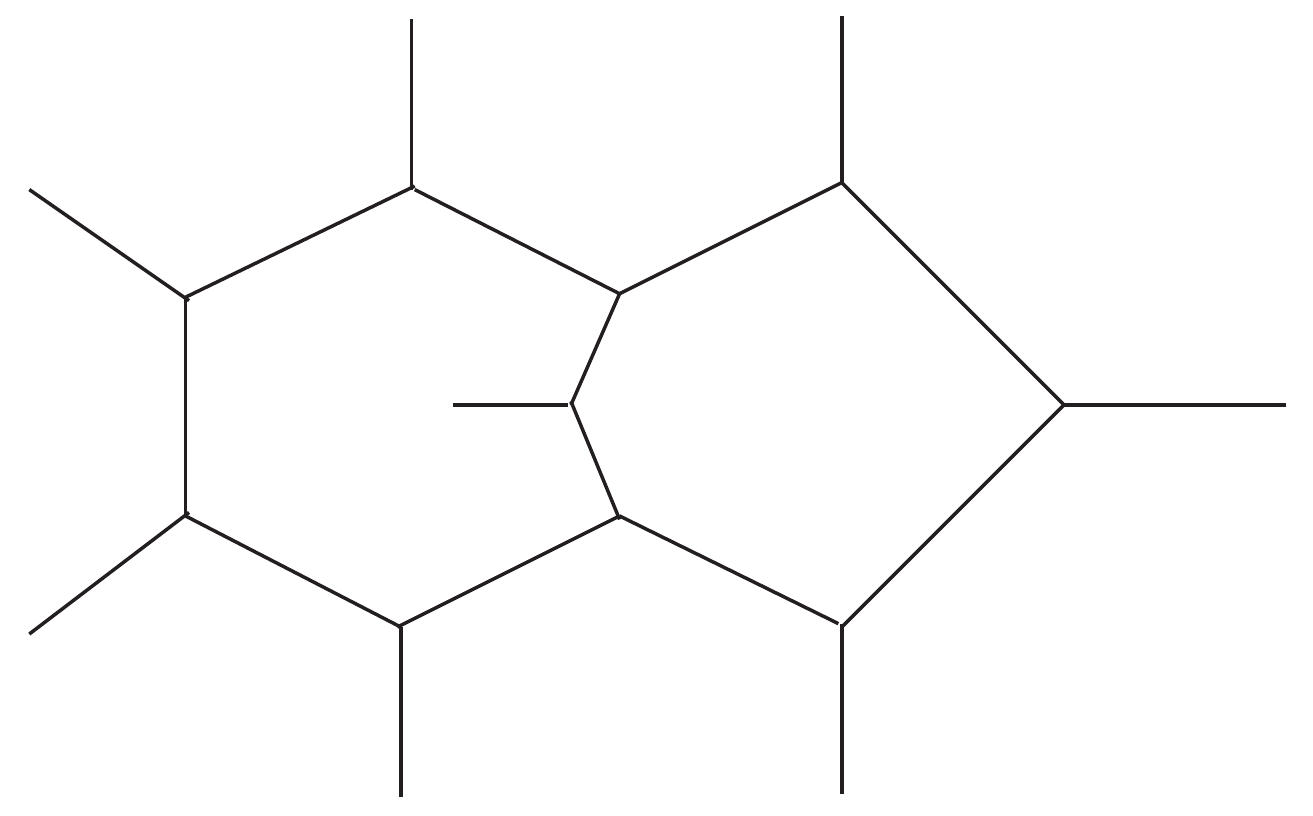}}&$1$\\
		 						&                        &$\{1\}$\\
		 						\hline
		 						\multirow{2}{1.4cm}{\centering $\mathcal{I}_{12345678910\,11}^{\text{NP}2}$}&\multirow{2}{2.cm}{\centering\includegraphics[height=0.40in]{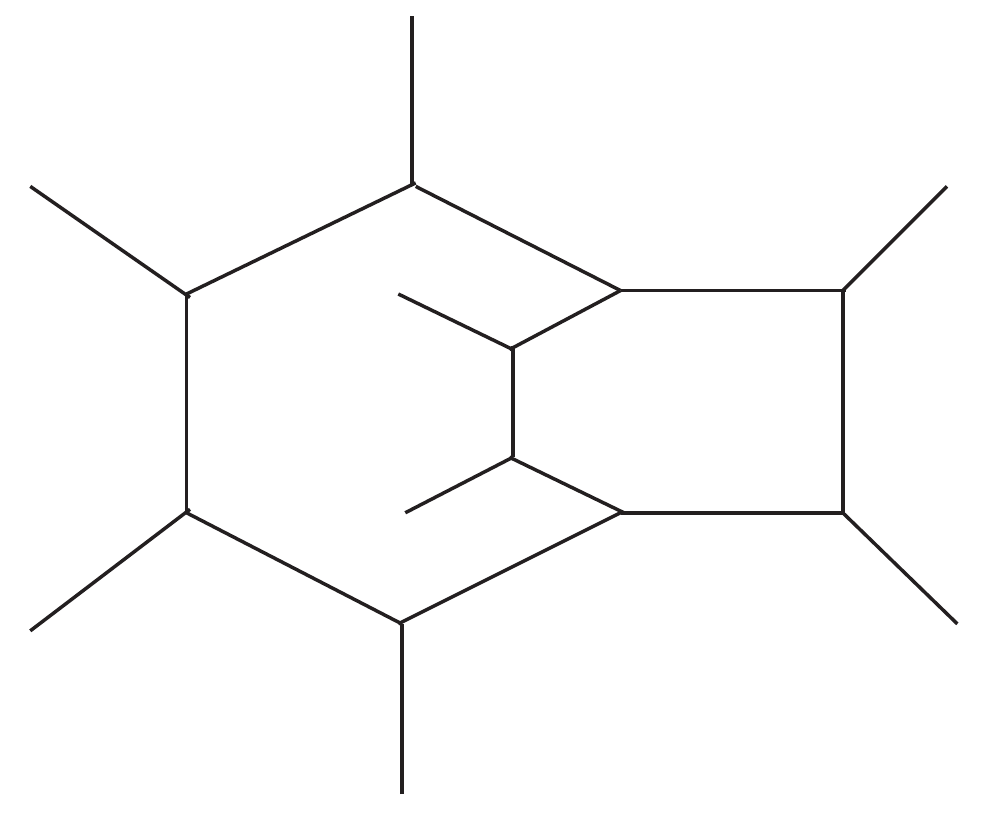}}&$1$\\
		 						&                        &$\{1\}$\\
		 						\hline
		 						\multirow{2}{1.4cm}{\centering $\mathcal{I}_{2345678910\,11}^{\text{P}}$}&\multirow{2}{2.cm}{\centering\includegraphics[height=0.35in]{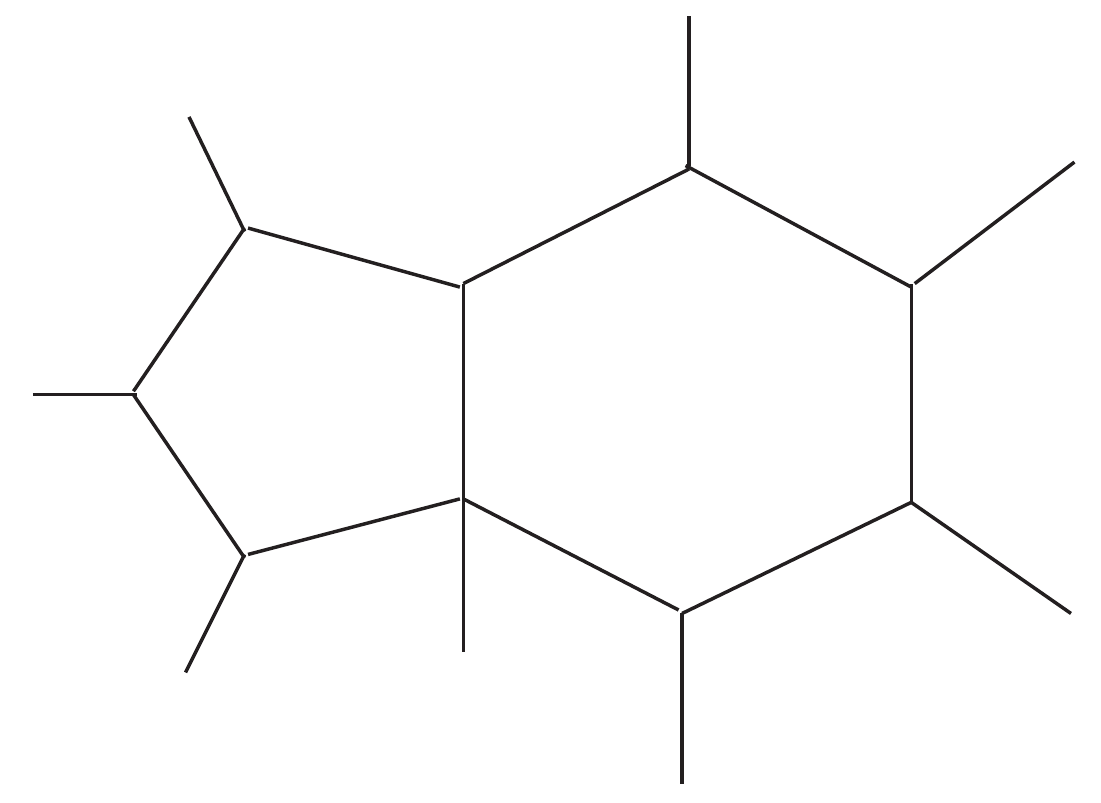}}&$6$\\
		 						&                        &$\{1,x_{41}\}$\\
		 						\hline
		 						\multirow{2}{1.4cm}{\centering $\mathcal{I}_{2345678910\,11}^{\text{NP}1}$}&\multirow{2}{2.cm}{\centering\includegraphics[height=0.35in]{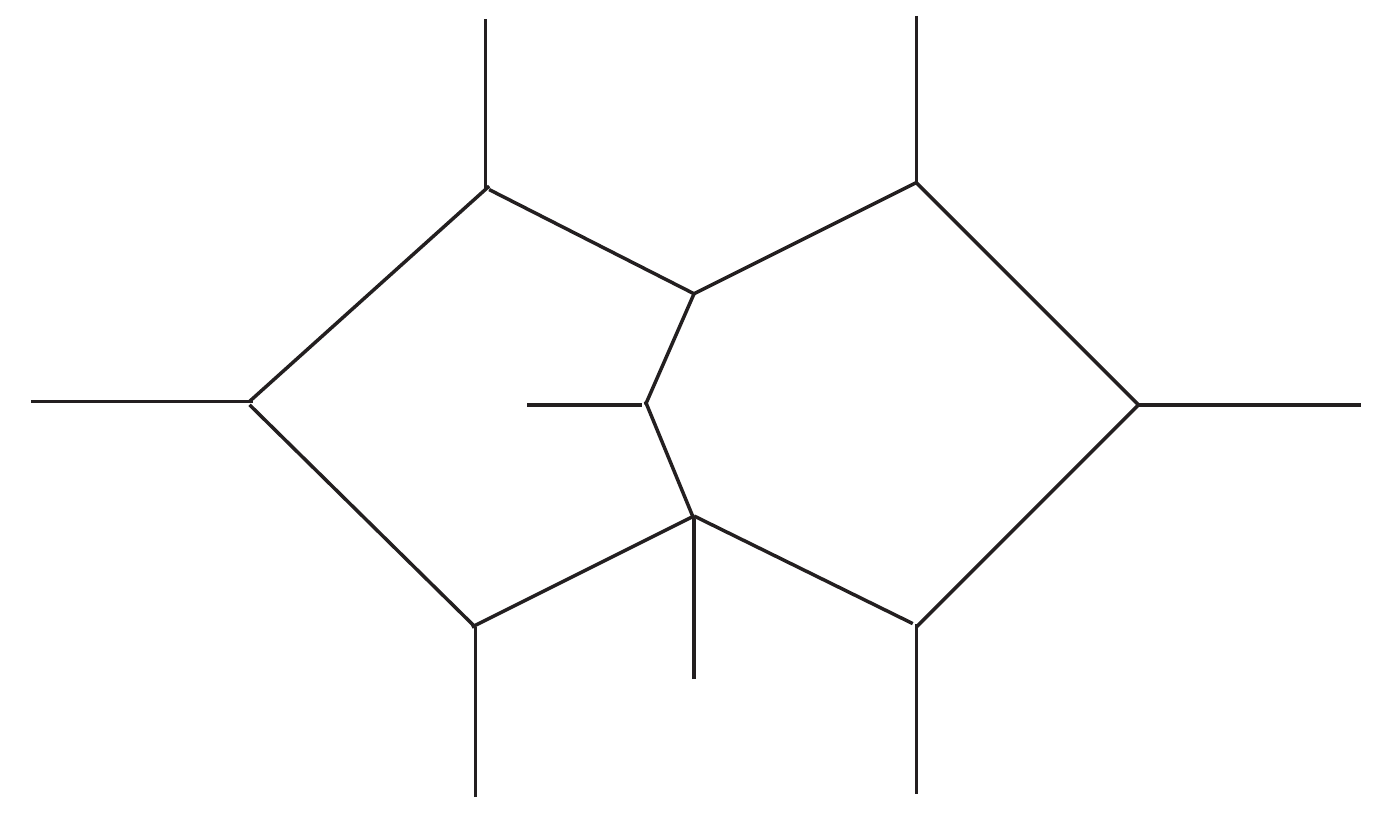}}&$10$\\
		 						&                        &$\{1,x_{42}\}$\\
		 						\hline
		 						\multirow{2}{1.4cm}{\centering $\mathcal{I}_{1234578910\,11}^{\text{NP2}}$}&\multirow{2}{2.cm}{\centering\includegraphics[height=0.35in]{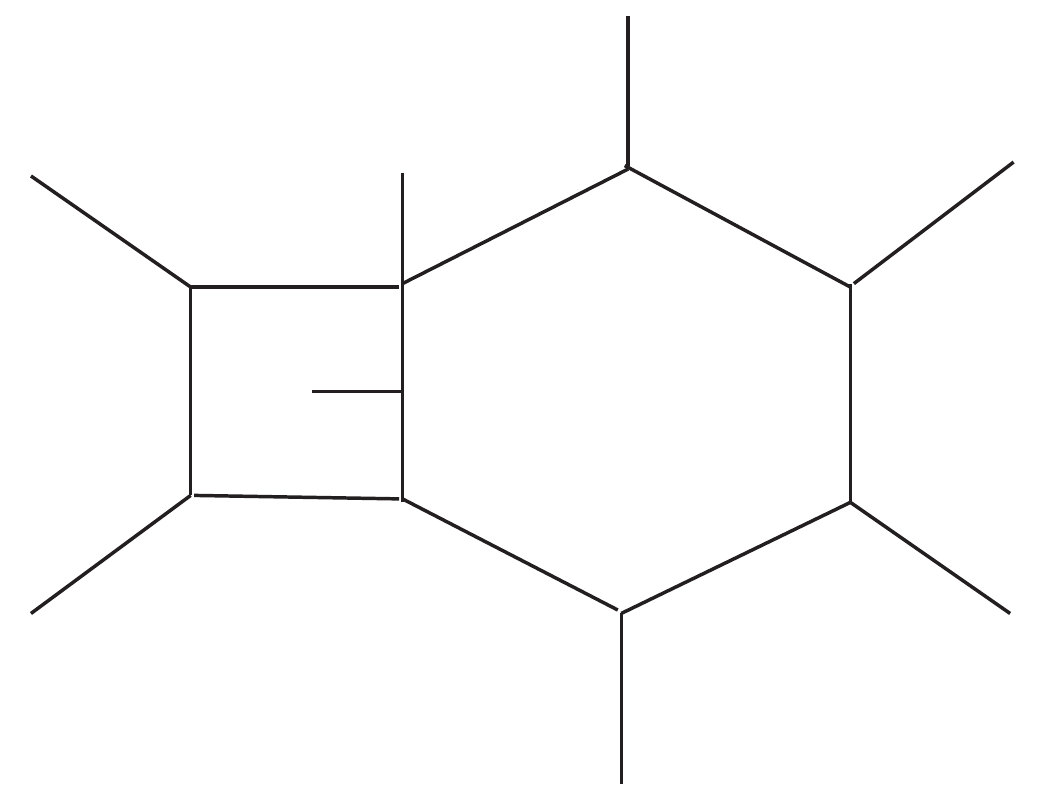}}&$6$\\
		 						&                        &$\{1,x_{42}\}$\\
		 						\hline
		 						\multirow{2}{1.4cm}{\centering $\mathcal{I}_{1234678910\,11}^{\text{NP}2}$}&\multirow{2}{2.cm}{\centering\includegraphics[height=0.35in]{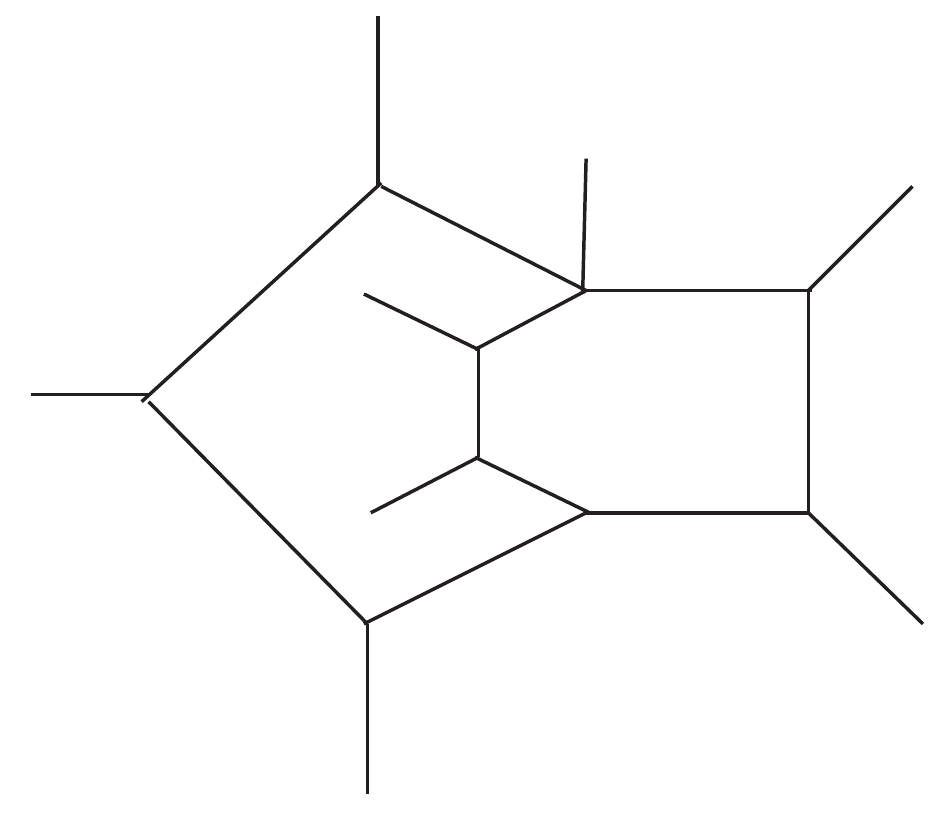}} &$10$\\
		 						&        &$\{1,x_{42}\}$\\
		 						\hline
		 						\multirow{2}{1.4cm}{\centering $\mathcal{I}_{234678910\,11}^{\text{P}}$}&\multirow{2}{2.cm}{\centering\includegraphics[height=0.35in]{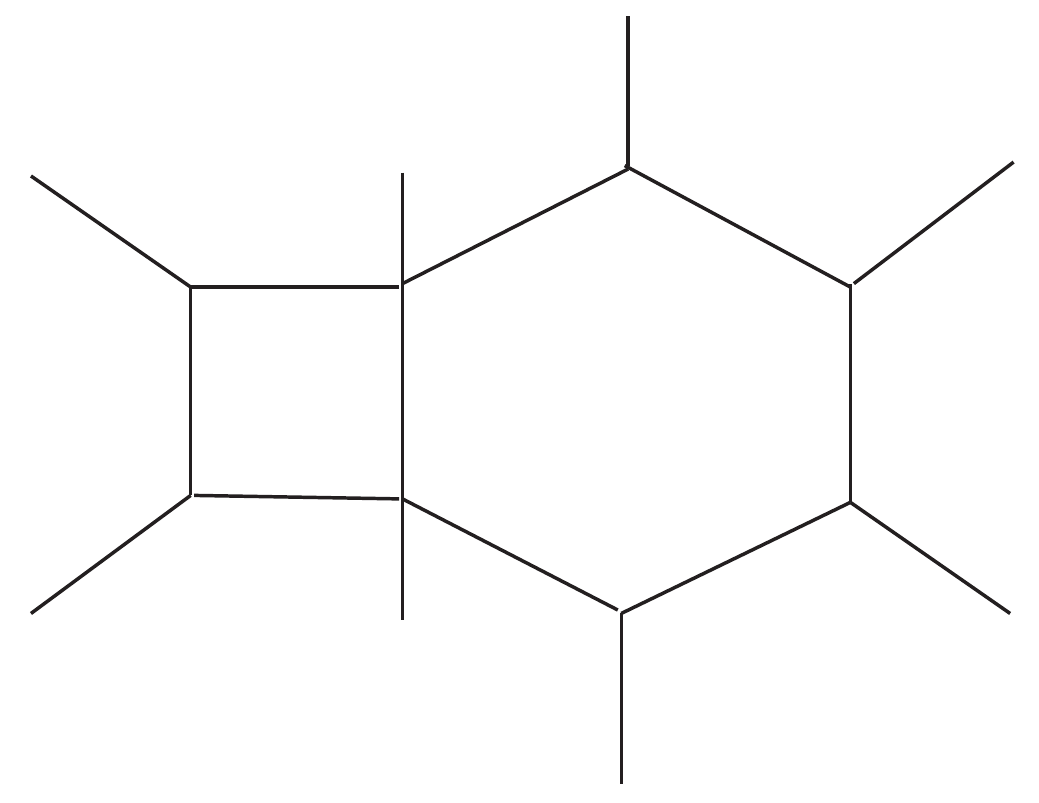}} &$15$\\
		 						&        &$\{1,x_{31},x_{41}\}$\\
		 						\hline
		 						\multirow{2}{1.4cm}{\centering $\mathcal{I}_{234578910\,11}^{\text{P}}$}&\multirow{2}{2.cm}{\centering\includegraphics[height=0.38in]{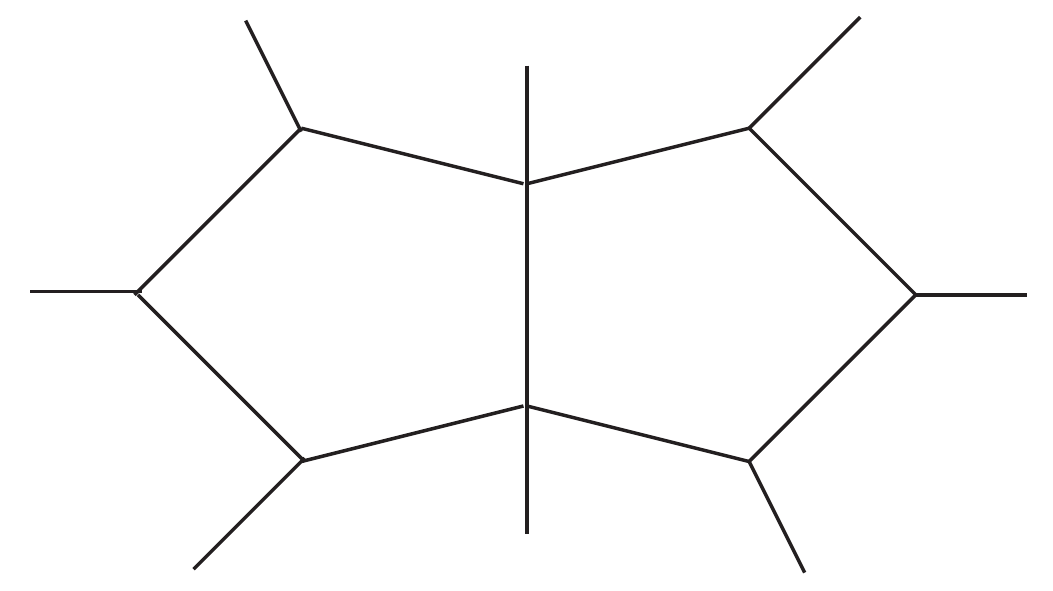}} 
		 						&$33$\\
		 						&        &$\{1,x_{41},x_{42}\}$\\
		 						\hline
		 						\multirow{2}{1.4cm}{\centering $\mathcal{I}_{234578910\,11}^{\text{NP}1}$}&\multirow{2}{2.cm}{\centering\includegraphics[height=0.40in]{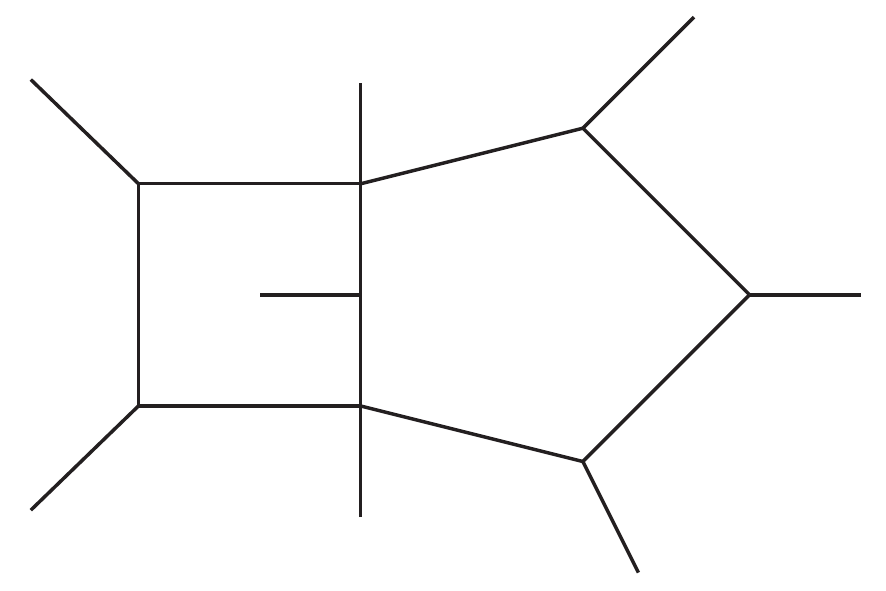}}
		 						&$39$\\
		 						&        &$\{1,x_{41},x_{42}\}$\\
		 						\hline
		 						\multirow{2}{1.4cm}{\centering $\mathcal{I}_{123456910\,11}^{\text{NP}1}$}&\multirow{2}{2.cm}{\centering\includegraphics[height=0.40in]{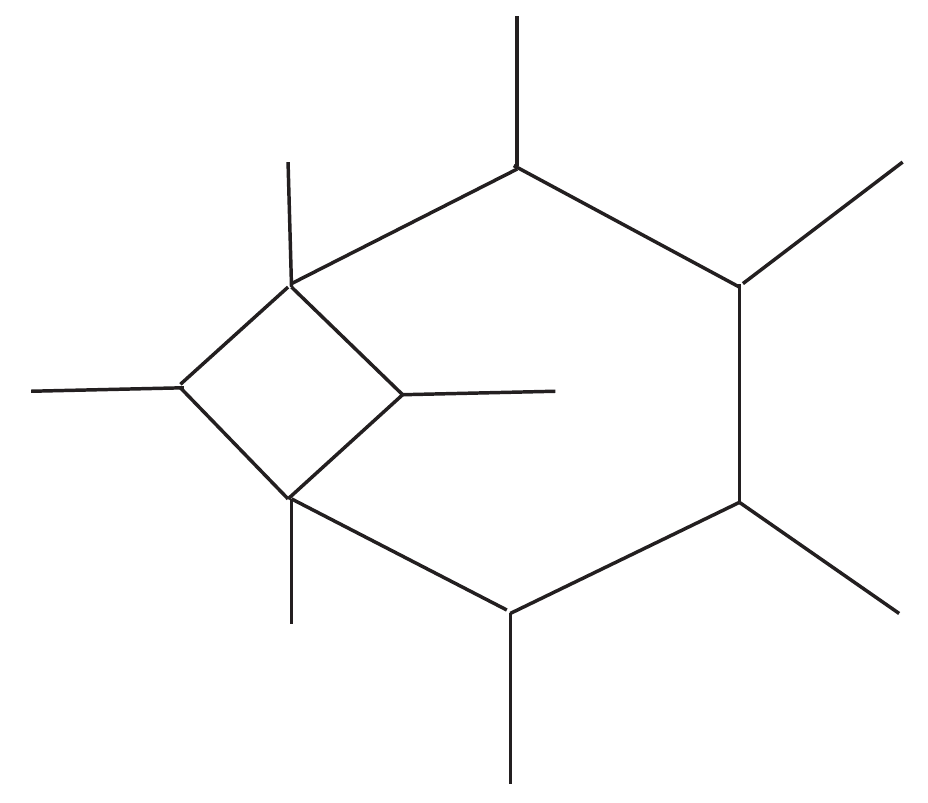}}
		 						&$15$\\
		 						&        &$\{1,x_{32},x_{42}\}$\\
		 						\hline
		 						\multirow{2}{1.4cm}{\centering $\mathcal{I}_{234678910\,11}^{\text{NP}2}$}&\multirow{2}{2.cm}{\centering\includegraphics[height=0.40in]{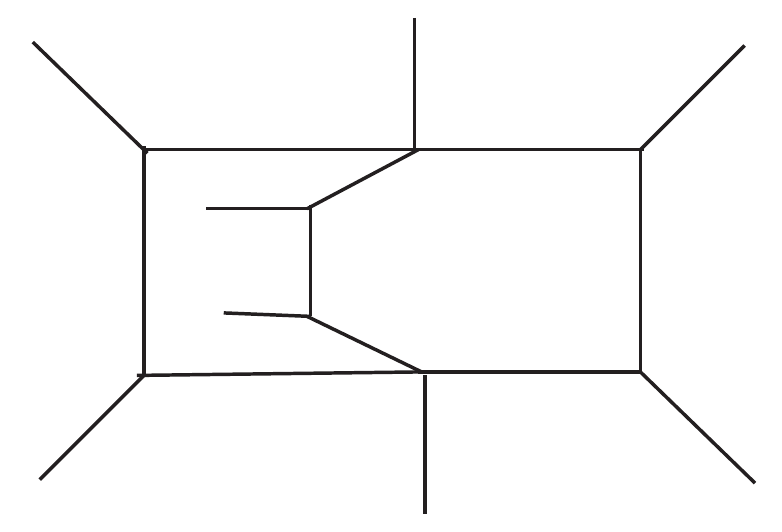}}
		 						&$45$\\
		 						&        &$\{1,x_{41},x_{42}\}$\\
		 						\hline
		 					\end{tabular}%
		 			&
		 				\begin{tabular}[t]{|c c||c|}
		 					\hline
		 					\multicolumn{2}{|c||}{$\mathcal{I}_{i_1\cdots i_r}$}&$\Delta_{i_1\cdots i_r}$ \\
		 					\hline
		 					\hline
		 					\multirow{2}{1.4cm}{\centering $\mathcal{I}_{1245678910\,11}^{\text{P}}$}&\multirow{2}{2.cm}{\centering\includegraphics[height=0.4in]{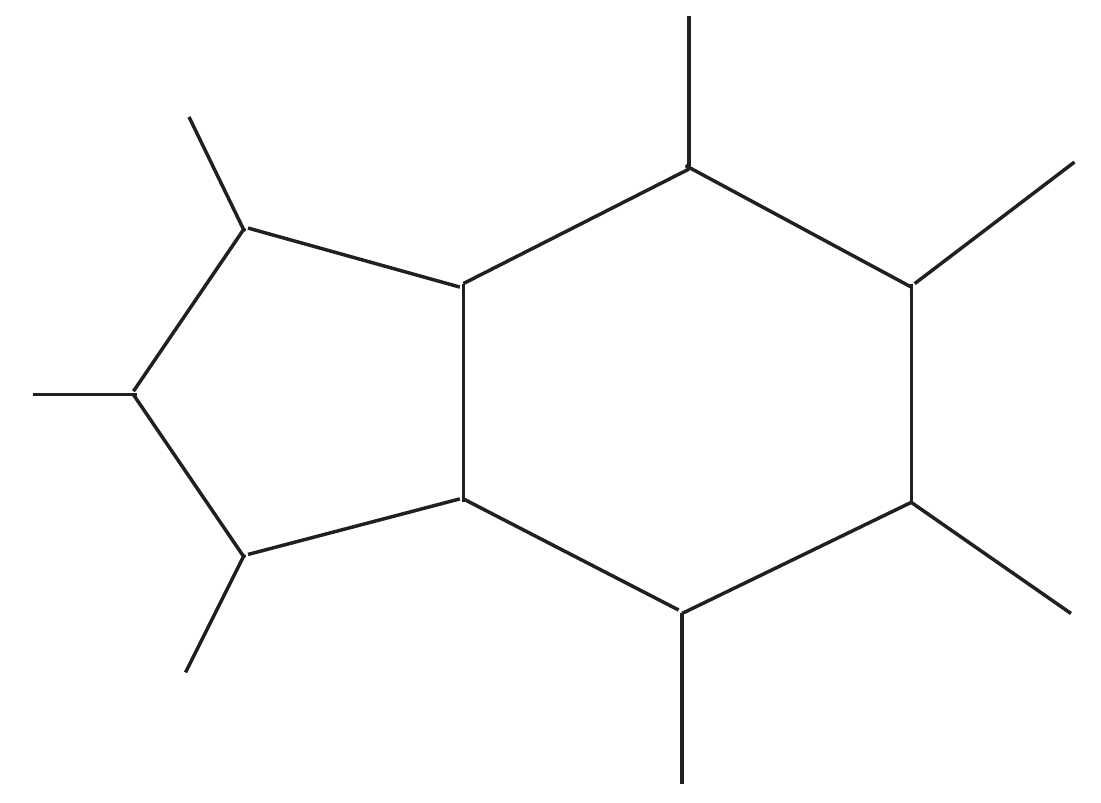}}&$6$\\
		 					&                        &$\{1,x_{41}\}$\\
		 					\hline
		 					\multirow{2}{1.4cm}{\centering $\mathcal{I}_{1245678910\,11}^{\text{NP}1}$}&\multirow{2}{2.cm}{\centering\includegraphics[height=0.35in]{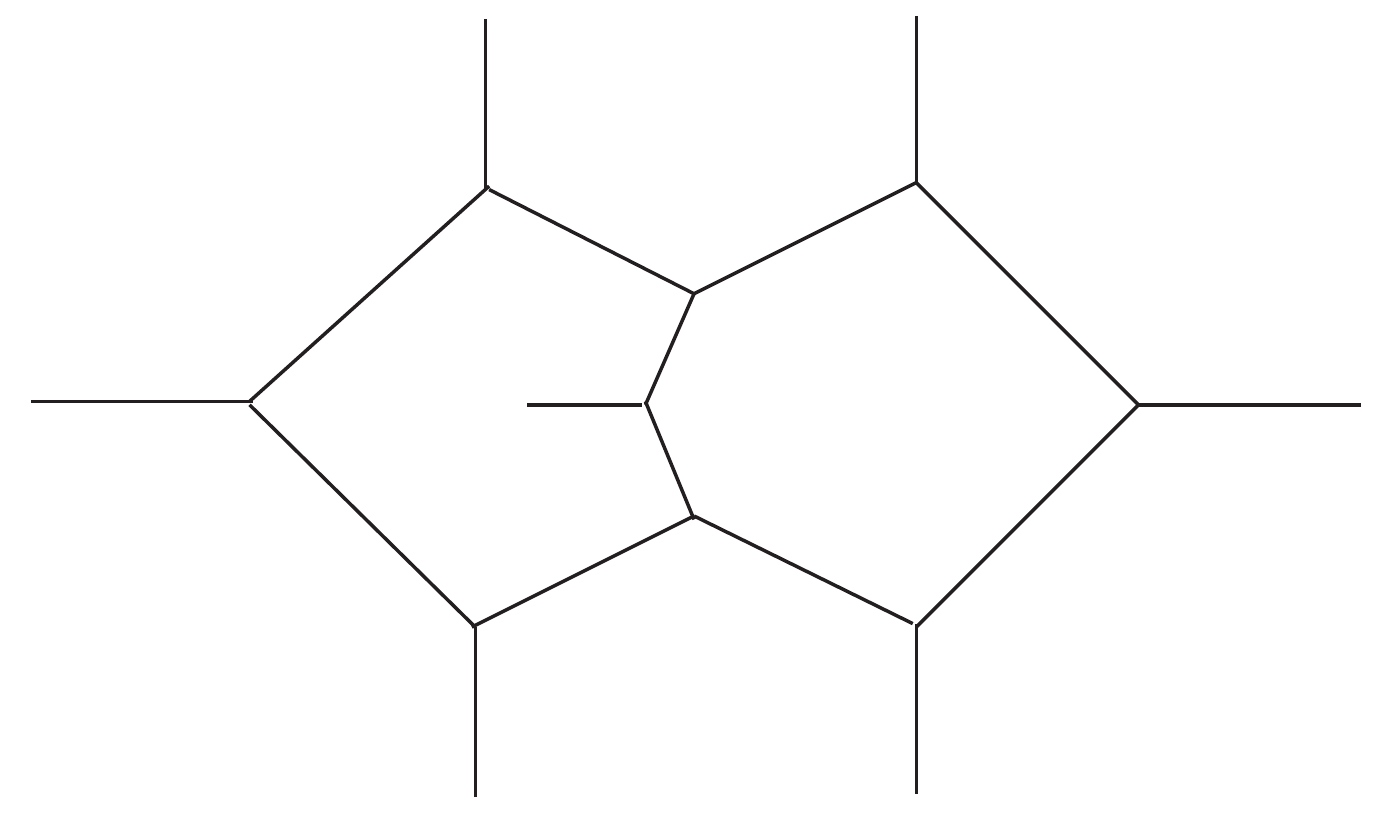}}&$10$\\
		 					&                        &$\{1,x_{42}\}$\\
		 					\hline
		 					\multirow{2}{1.4cm}{\centering $\mathcal{I}_{1234568910\,11}^{\text{NP}1}$}&\multirow{2}{2.cm}{\centering\includegraphics[height=0.40in]{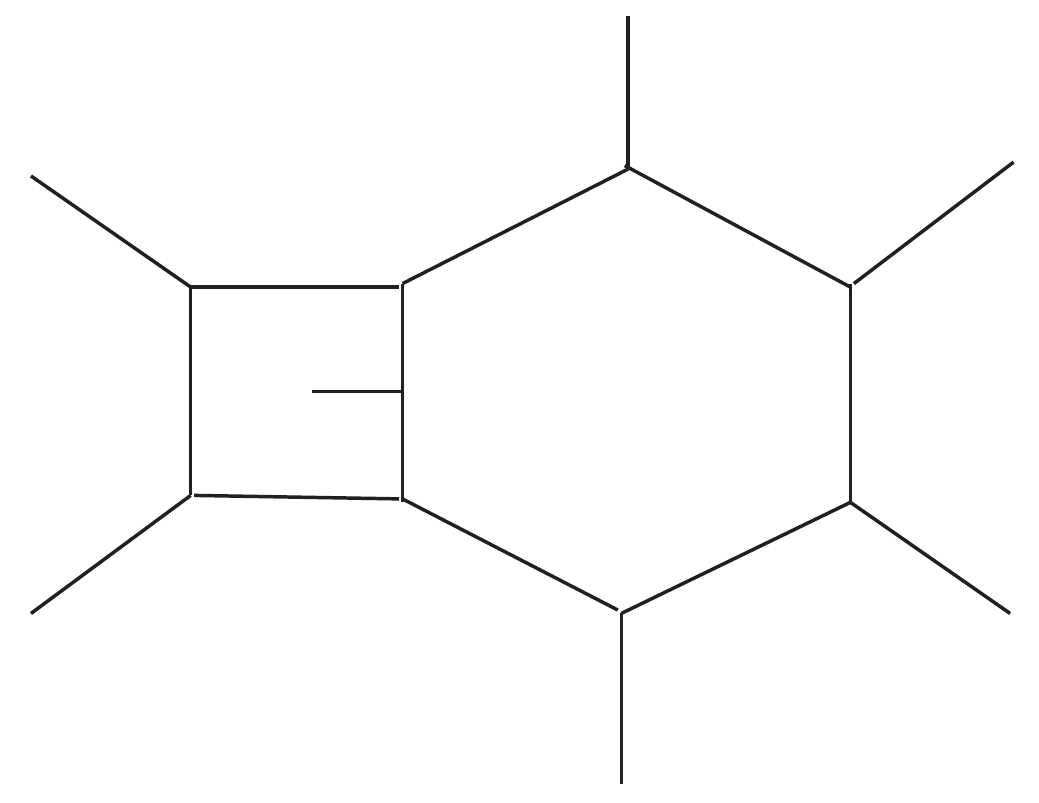}}&$6$\\
		 					&                        &$\{1,x_{42}\}$\\
		 					\hline
		 					\multirow{2}{1.4cm}{\centering $\mathcal{I}_{1245678910\,11}^{\text{NP}2}$}&\multirow{2}{2.cm}{\centering\includegraphics[height=0.35in]{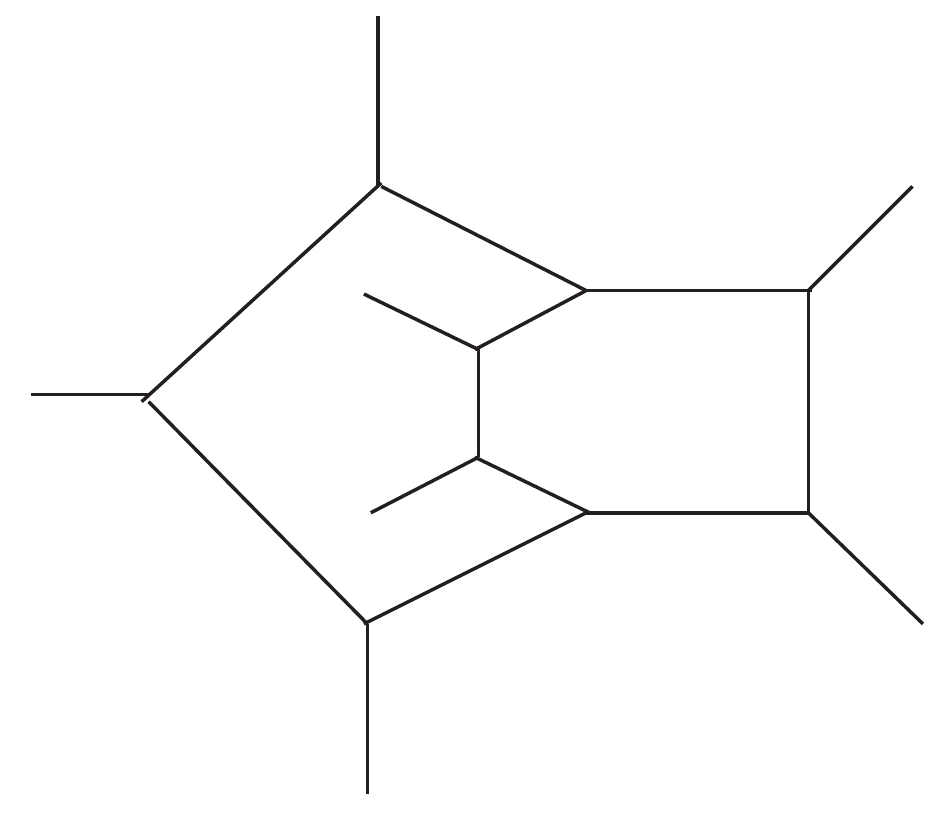}}&$10$\\
		 					&                        &$\{1,x_{42}\}$\\
		 					\hline
		 					\multirow{2}{1.4cm}{\centering $\mathcal{I}_{245678910\,11}^{\text{P}}$}&\multirow{2}{2.cm}{\centering\includegraphics[height=0.35in]{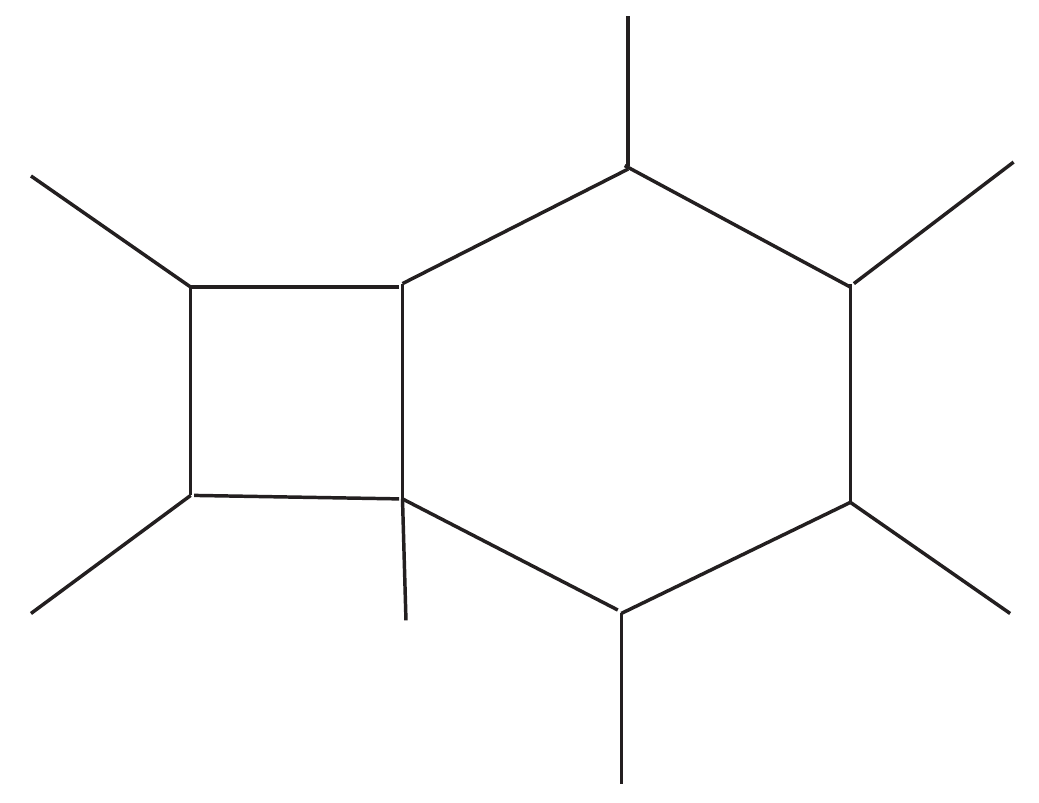}}&$15$\\
		 					&                        &$\{1,x_{31},x_{41}\}$\\
		 					\hline
		 					\multirow{2}{1.4cm}{\centering $\mathcal{I}_{123478910\,11}^{\text{P}}$}&\multirow{2}{2.cm}{\centering\includegraphics[height=0.35in]{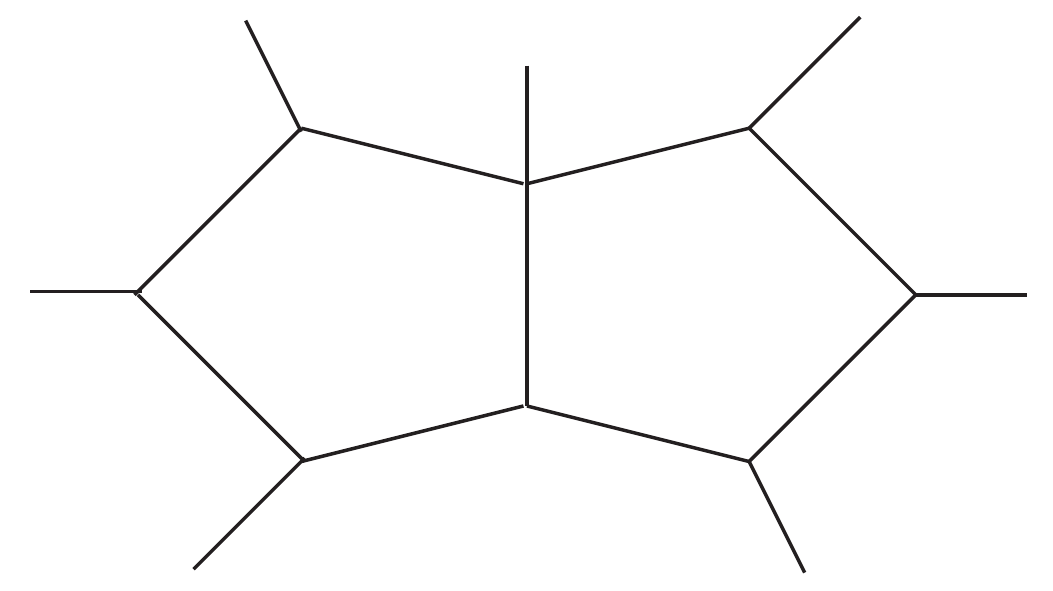}}&$33$\\
		 					&                        &$\{1,x_{41},x_{42}\}$\\
		 					\hline
		 					\multirow{2}{1.4cm}{\centering $\mathcal{I}_{124568910\,11}^{\text{NP}1}$}&\multirow{2}{2.cm}{\centering\includegraphics[height=0.35in]{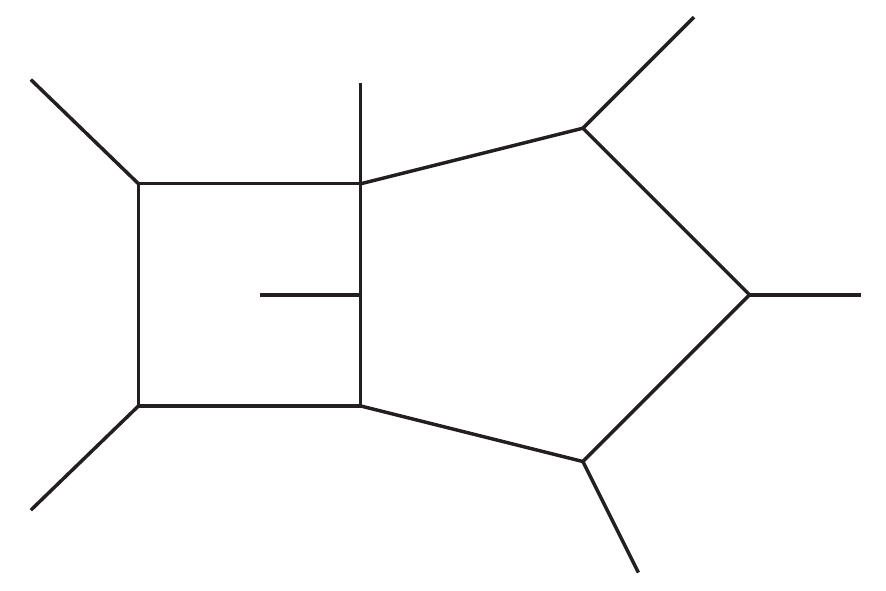}} &$39$\\
		 					&        &$\{1,x_{41},x_{42}\}$\\
		 					\hline
		 					\multirow{2}{1.4cm}{\centering $\mathcal{I}_{123456810\,11}^{\text{NP}1}$}&\multirow{2}{2.cm}{\centering\includegraphics[height=0.35in]{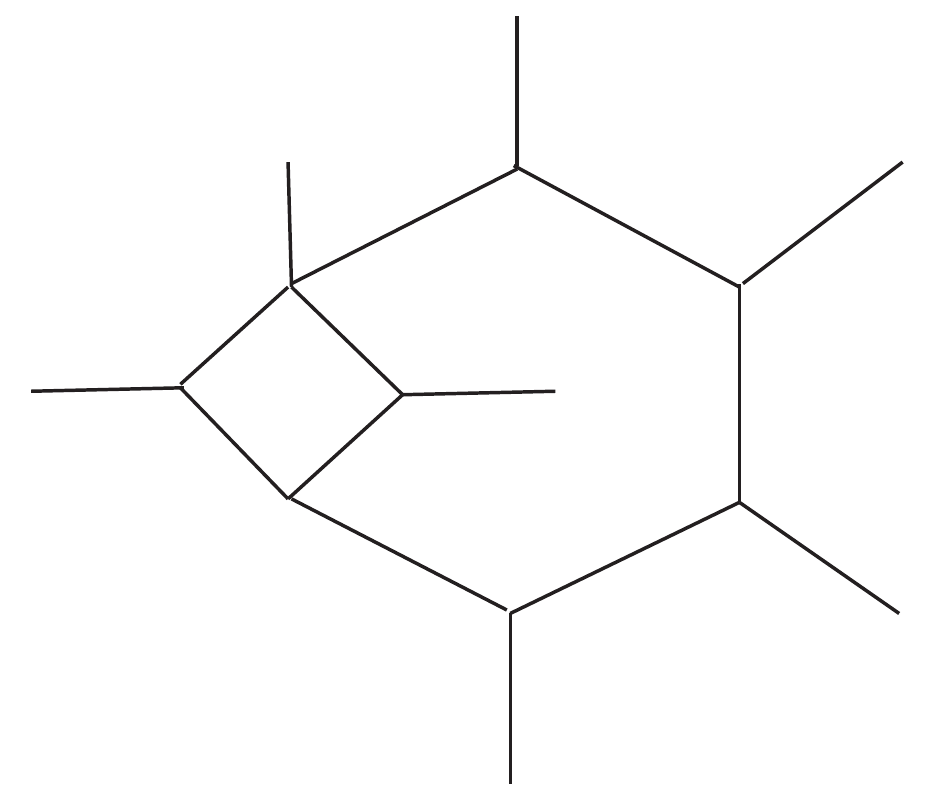}} &$15$\\
		 					&        &$\{1,x_{32},x_{42}\}$\\
		 					\hline
		 					\multirow{2}{1.4cm}{\centering $\mathcal{I}_{124678910\,11}^{\text{NP}2}$}&\multirow{2}{2.cm}{\centering\includegraphics[height=0.38in]{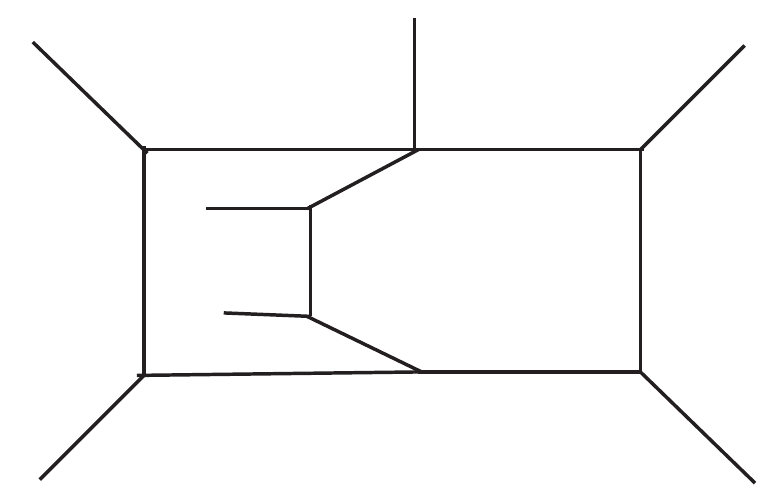}} 
		 					&$45$\\
		 					&        &$\{1,x_{41},x_{42}\}$\\
		 					\hline
		 					\multirow{2}{1.4cm}{\centering $\mathcal{I}_{2478910\,11}^{\text{NP}1}$}&\multirow{2}{2.cm}{\centering\includegraphics[height=0.40in]{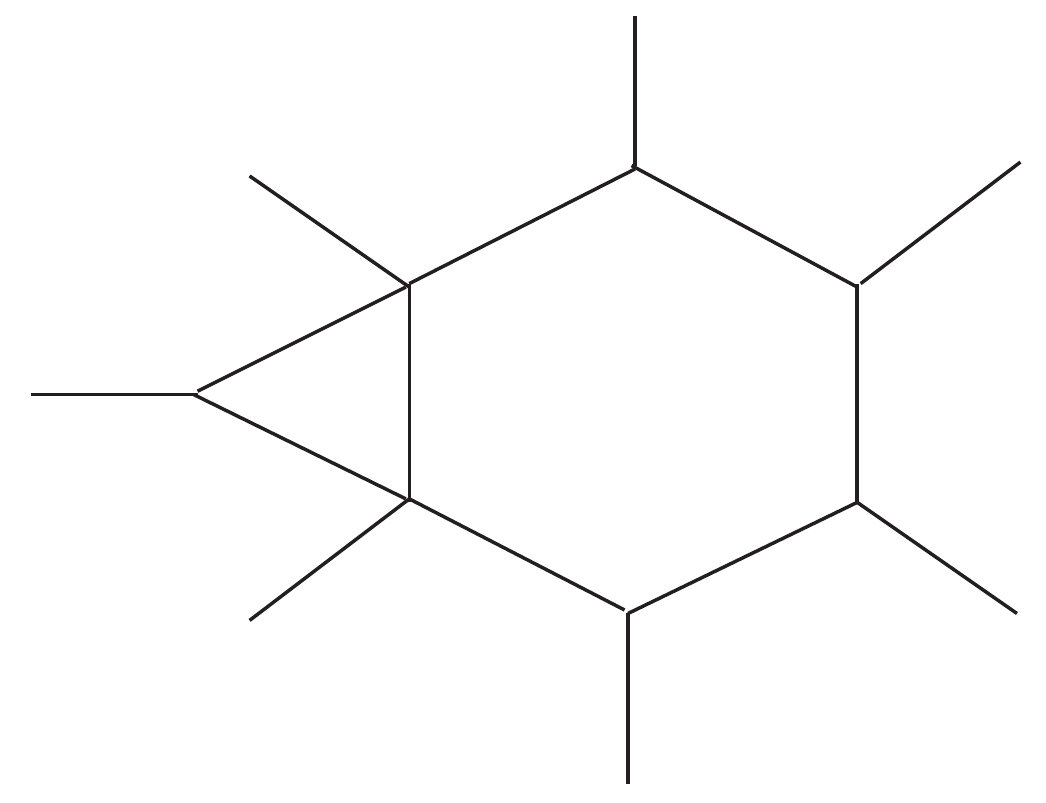}}
		 					&$20$\\
		 					&        &$\{1,x_{21},x_{31},x_{41}\}$\\
		 					\hline
		 					\multirow{2}{1.4cm}{\centering $\mathcal{I}_{23478910\,11}^{\text{NP}1}$}&\multirow{2}{2.cm}{\centering\includegraphics[height=0.40in]{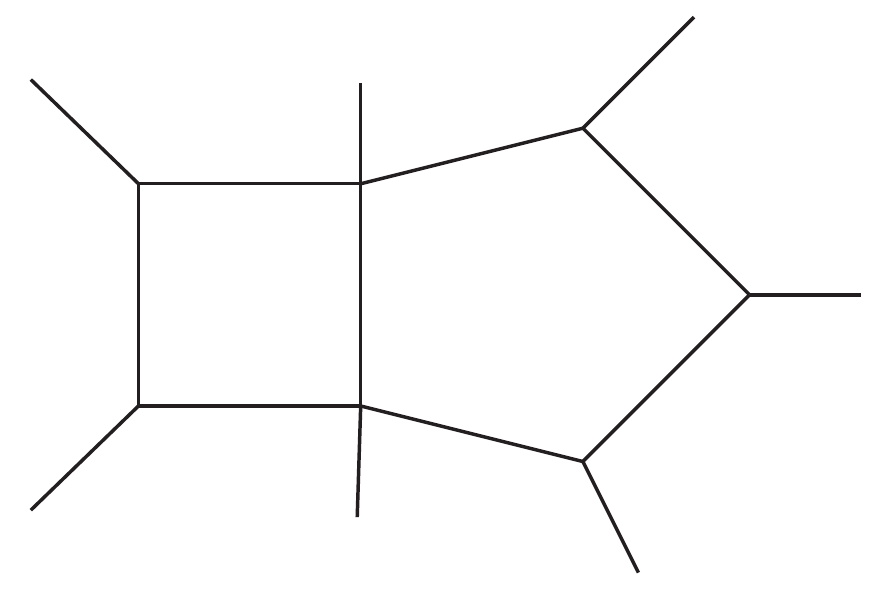}}
		 					&$76$\\
		 					&        &$\{1,x_{31},x_{41},x_{42}\}$\\
		 					\hline
		 					\multirow{2}{1.4cm}{\centering $\mathcal{I}_{24578910\,11}^{\text{NP}1}$}&\multirow{2}{2.cm}{\centering\includegraphics[height=0.40in]{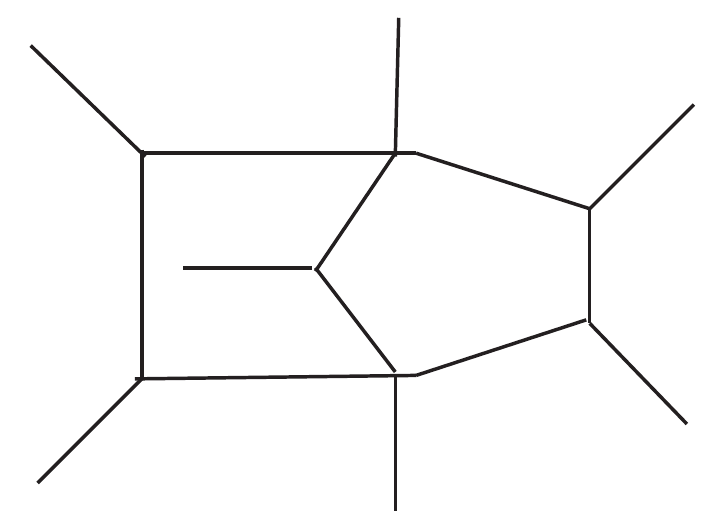}}
		 					&$116$\\
		 					&        &$\{1,x_{41},x_{32},x_{42}\}$\\
		 					\hline
		 					\multirow{2}{1.4cm}{\centering $\mathcal{I}_{12457810\,11}^{\text{NP}1}$}&\multirow{2}{2.cm}{\centering\includegraphics[height=0.40in]{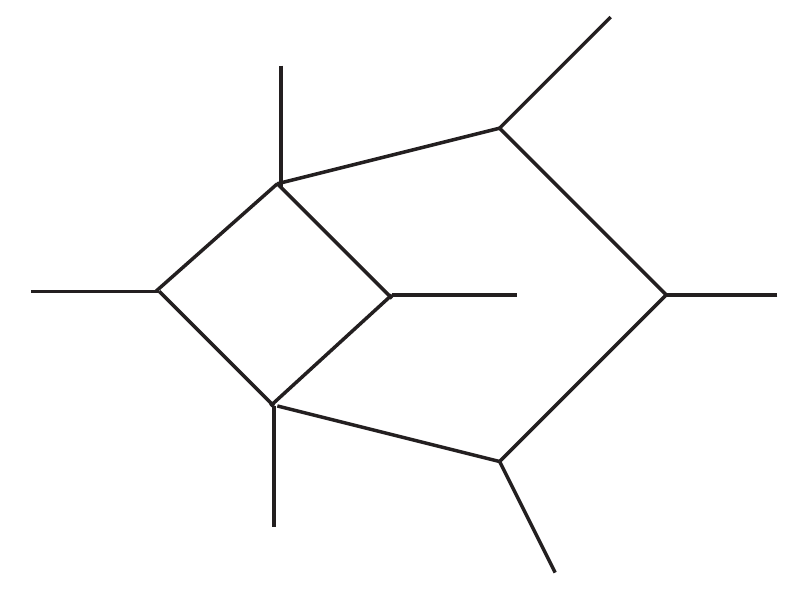}}
		 					&$80$\\
		 					&        &$\{1,x_{31},x_{41},x_{42}\}$\\
		 					\hline
		 				\end{tabular}
		 		\end{tabular}
		 	}
		  	\caption{\small{Residue parametrization for irreducible eight- and seven-point two-loop topologies. Denominators depend on the variables $\mathbf{z}=\{x_{11},x_{21},x_{31},x_{41},x_{12},x_{22},x_{32},x_{42},\mu_{11},\mu_{22},\mu_{12}\}$. In the second column we list the number of monomials of each residue and the set of variables appearing in it.}}
		  	\label{Tab:87pt}
		 \end{table}
		\begin{table}[!ht]
			\centering
			\renewcommand{\arraystretch}{1.2}
			\scalebox{0.75}{
		\begin{tabular}{cc}%
		  		\begin{tabular}[t]{|c c||c|}
		 			\hline
		 			\multicolumn{2}{|c||}{$\mathcal{I}_{i_1\cdots i_r}$}&$\Delta_{i_1\cdots i_r}$ \\
		 			\hline
		 			\hline
		 			\multirow{2}{1.4cm}{\centering $\mathcal{I}_{135678910\,11}^{\text{P}}$}&\multirow{2}{2.cm}{\centering\includegraphics[height=0.4in]{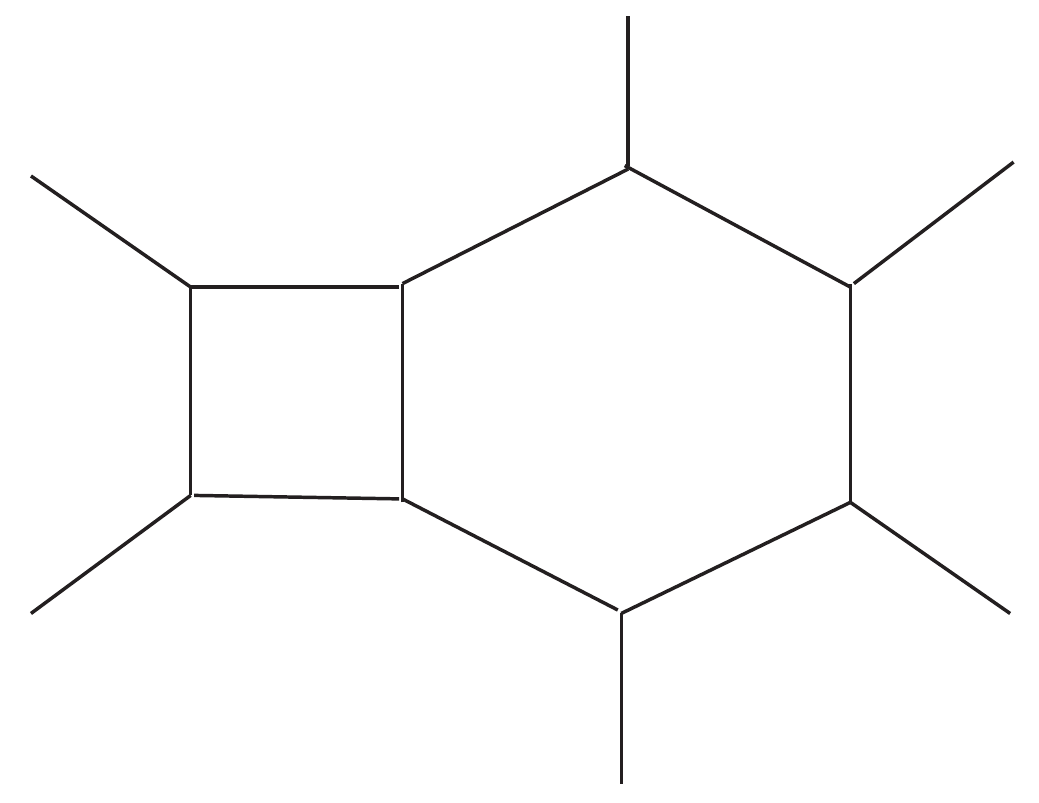}}&$15$\\
		 			&                        &$\{1,x_{31},x_{41}\}$\\
		 			\hline
		 			\multirow{2}{1.4cm}{\centering $\mathcal{I}_{124567910\,11}^{\text{P}}$}&\multirow{2}{2.cm}{\centering\includegraphics[height=0.35in]{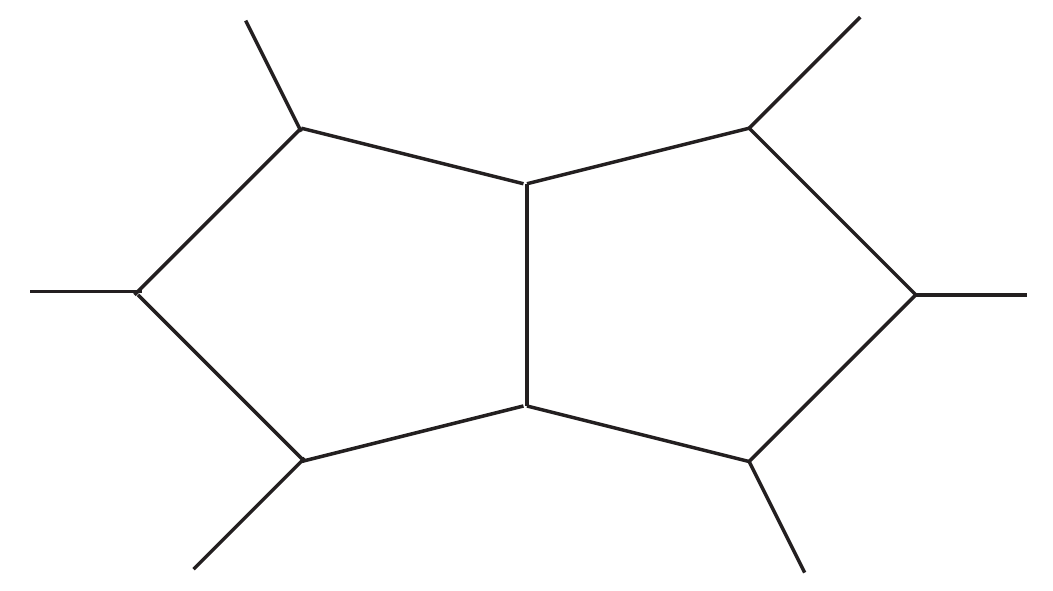}}&$62$\\
		 			&                        &$\{1,x_{41},x_{42}\}$\\
		 			\hline
		 			\multirow{2}{1.4cm}{\centering $\mathcal{I}_{23568910\,11}^{\text{NP}1}$}&\multirow{2}{2.cm}{\centering\includegraphics[height=0.35in]{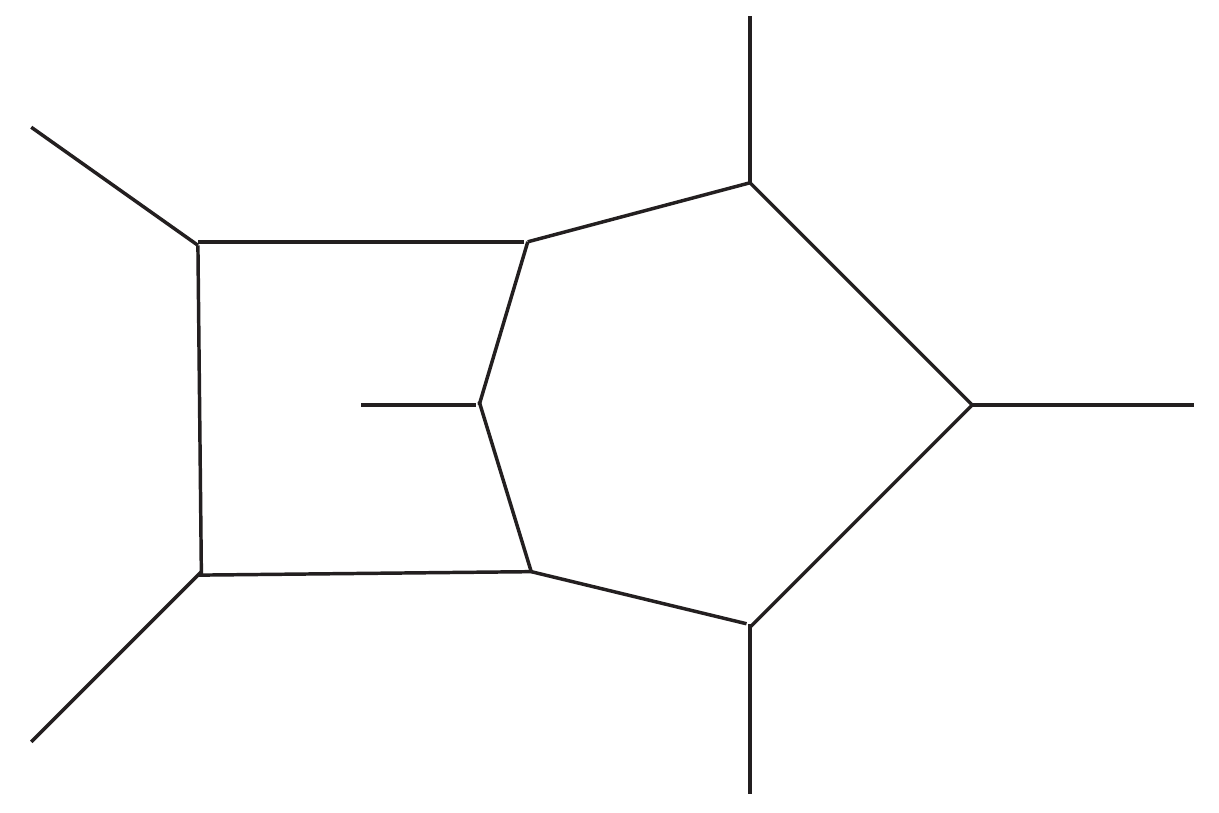}}&$39$\\
		 			&                        &$\{1,x_{41},x_{42}\}$\\
		 			\hline
		 			\multirow{2}{1.4cm}{\centering $\mathcal{I}_{123456910\,11}^{\text{NP}1}$}&\multirow{2}{2.cm}{\centering\includegraphics[height=0.35in]{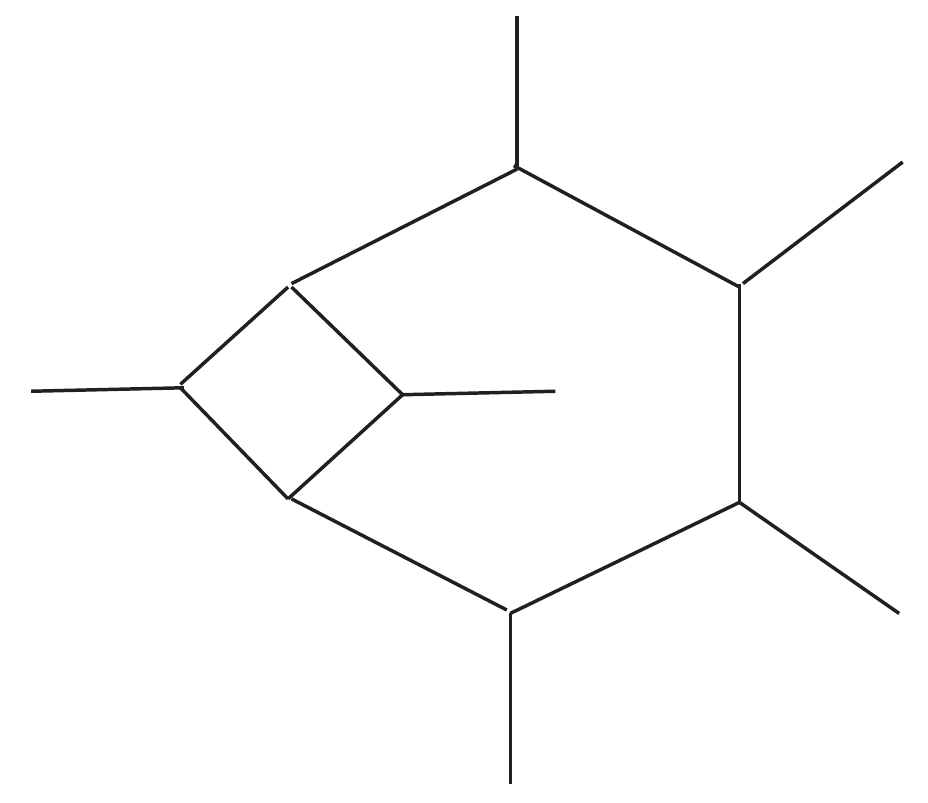}}&$15$\\
		 			&                        &$\{1,x_{32},x_{42}\}$\\
		 			\hline
		 			\multirow{2}{1.4cm}{\centering $\mathcal{I}_{135678910\,11}^{\text{NP}2}$}&\multirow{2}{2.cm}{\centering\includegraphics[height=0.35in]{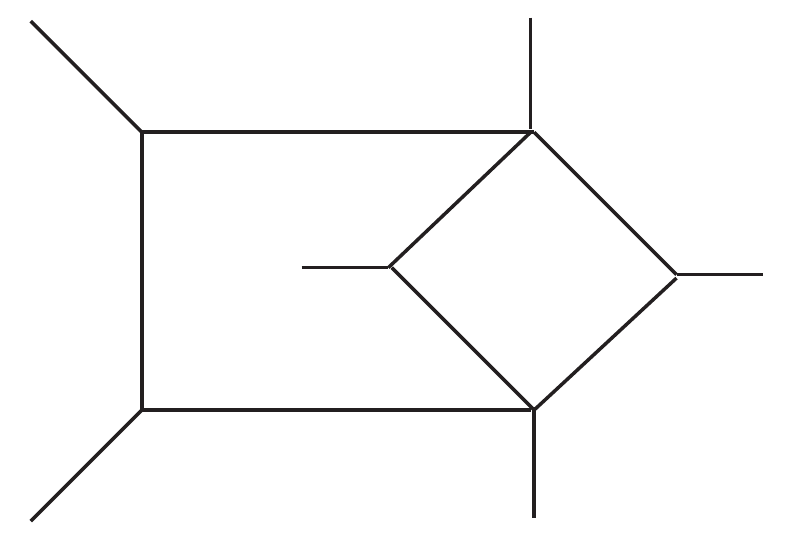}}&$45$\\
		 			&                        &$\{1,x_{41},x_{42}\}$\\
		 			\hline
		 			\multirow{2}{1.4cm}{\centering $\mathcal{I}_{25678910\,11}^{\text{P}}$}&\multirow{2}{2.cm}{\centering\includegraphics[height=0.40in]{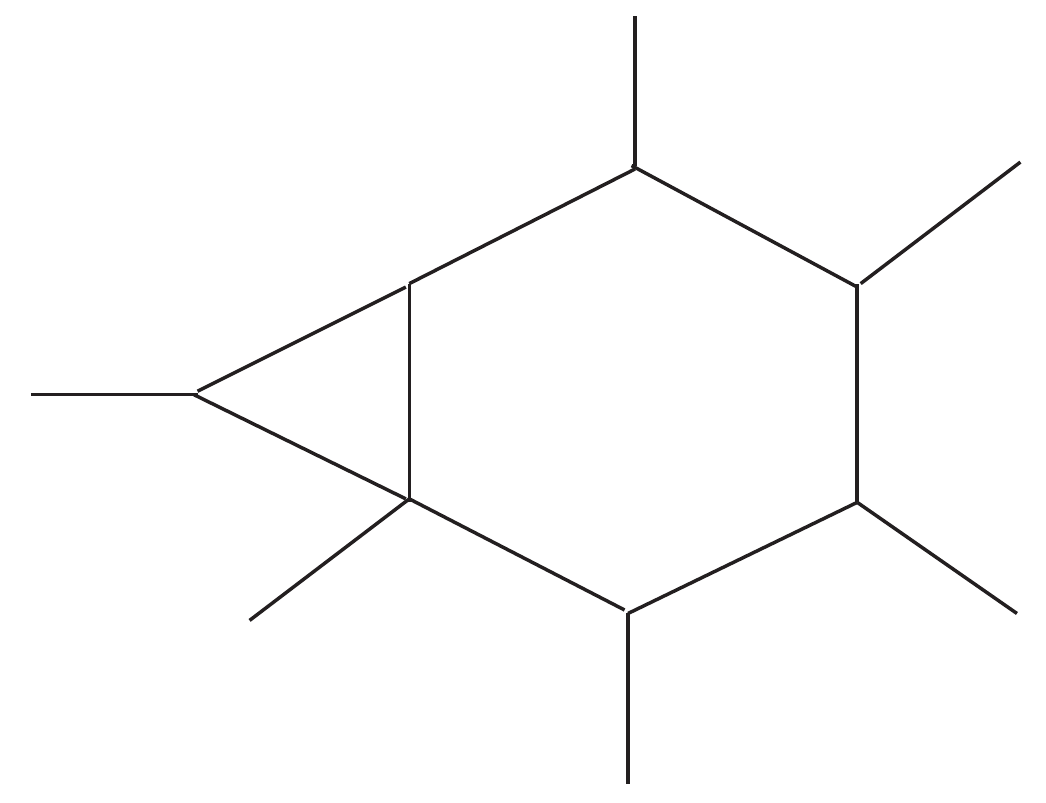}}&$20$\\
		 			&                        &$\{1,x_{21},x_{31},x_{41}\}$\\
		 			\hline
		 			\multirow{2}{1.4cm}{\centering $\mathcal{I}_{23568910\,11}^{\text{P}}$}&\multirow{2}{2.cm}{\centering\includegraphics[height=0.35in]{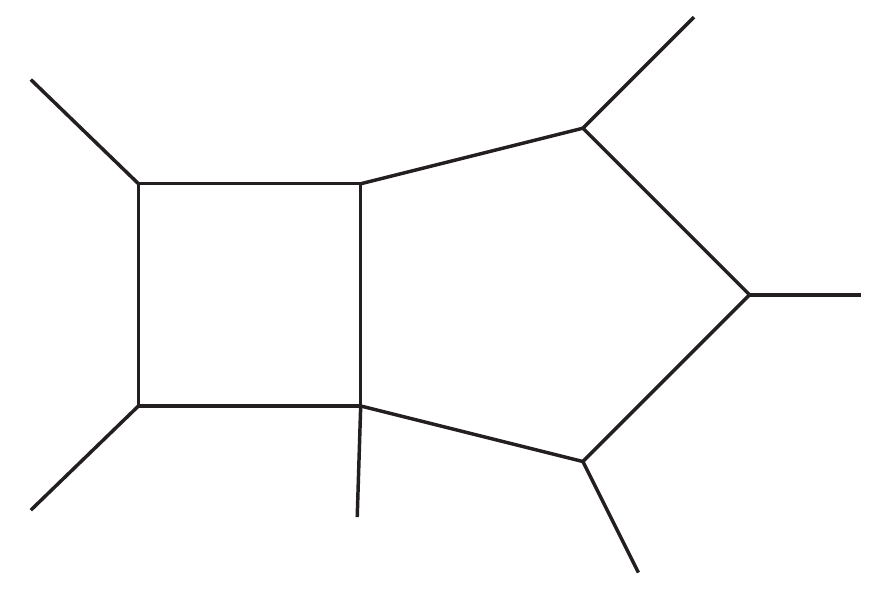}}&$76$\\
		 			&                        &$\{1,x_{31},x_{41},x_{42}\}$\\
		 			\hline
		 			\multirow{2}{1.4cm}{\centering $\mathcal{I}_{25678910\,11}^{\text{NP}1}$}&\multirow{2}{2.cm}{\centering\includegraphics[height=0.35in]{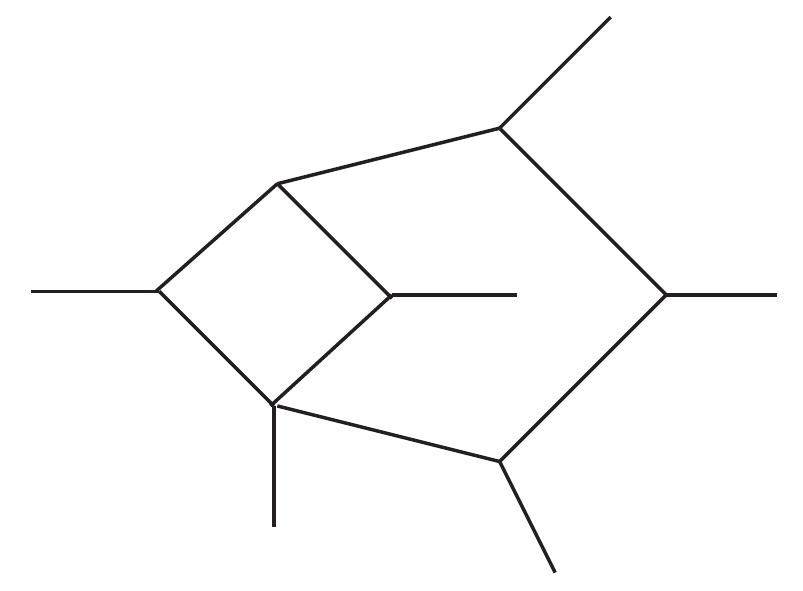}}&$80$\\
		 			&                        &$\{1,x_{31},x_{41},x_{42}\}$\\
		 			\hline
		 			\multirow{2}{1.4cm}{\centering $\mathcal{I}_{24568910\,11}^{\text{NP}1}$}&\multirow{2}{2.cm}{\centering\includegraphics[height=0.35in]{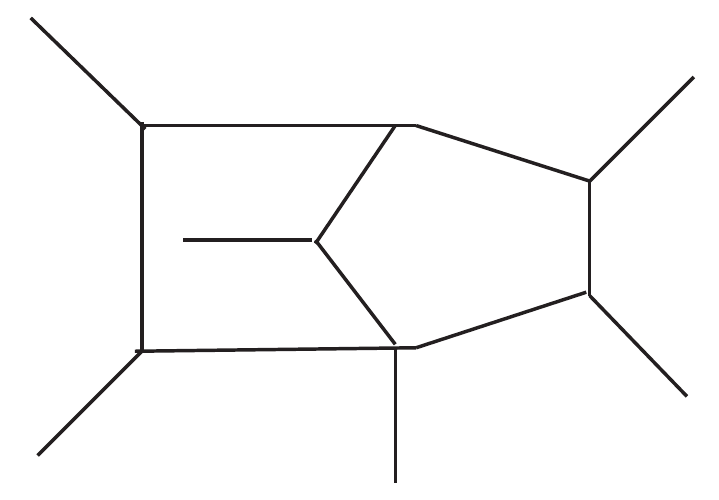}}&$116$\\
		 			&                        &$\{1,x_{41},x_{32},x_{42}\}$\\
		 			\hline
		 			\multirow{2}{1.4cm}{\centering $\mathcal{I}_{3678910\,11}^{\text{P}}$}&\multirow{2}{2.cm}{\centering\includegraphics[height=0.35in]{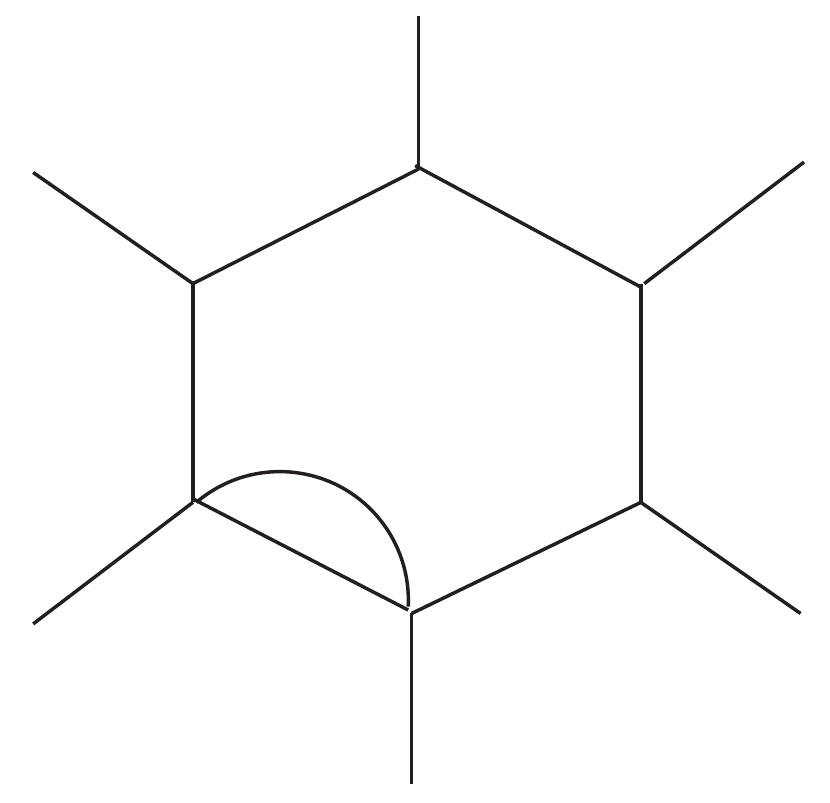}}&$15$\\
		 			&                        &$\{1,x_{11},x_{21},x_{31},x_{41}\}$\\
		 			\hline
		 			\multirow{2}{1.4cm}{\centering $\mathcal{I}_{2578910\,11}^{\text{P}}$}&\multirow{2}{2.cm}{\centering\includegraphics[height=0.35in]{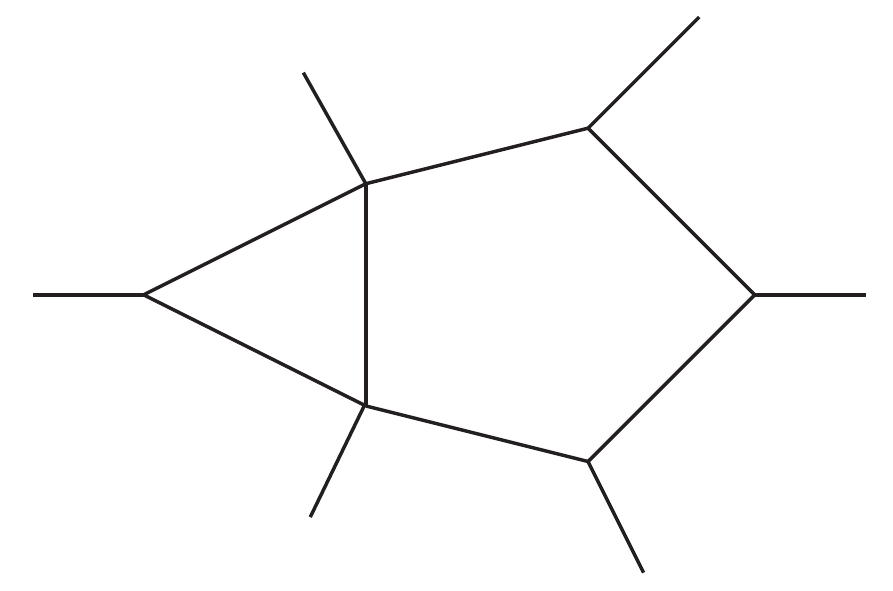}} &$94$\\
		 			&        &$\{1,x_{21},x_{31},x_{41},x_{42}\}$\\
		 			\hline
		 			\multirow{2}{1.4cm}{\centering $\mathcal{I}_{2357910\,11}^{\text{P}}$}&\multirow{2}{2.cm}{\centering\includegraphics[height=0.35in]{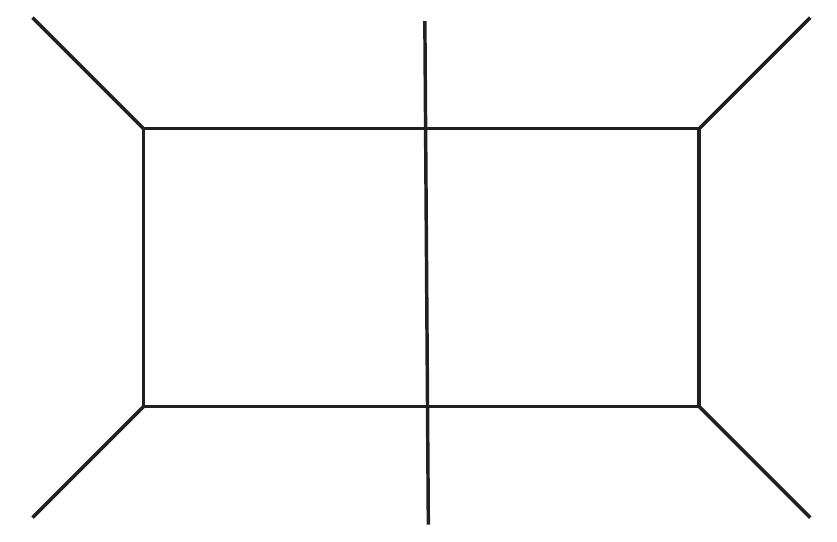}}&$160$\\
		 			&                        &$\{1,x_{31},x_{41},x_{32},x_{42}\}$\\
		 			\hline
		 			\multirow{2}{1.4cm}{\centering $\mathcal{I}_{2457910\,11}^{\text{NP}1}$}&\multirow{2}{2.cm}{\centering\includegraphics[height=0.35in]{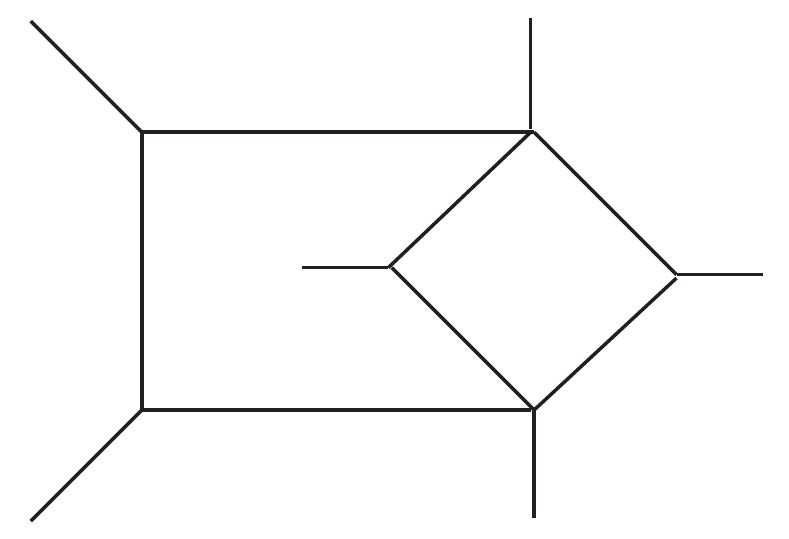}}&$185$\\
		 			&                        &$\{1,x_{31},x_{41},x_{32},x_{42}\}$\\
		 			\hline
		 	 	\end{tabular}%
		 	 	&
		 		\begin{tabular}[t]{|c c||c|}
		 				\hline
		 				\multicolumn{2}{|c||}{$\mathcal{I}_{i_1\cdots i_r}$}&$\Delta_{i_1\cdots i_r}$ \\
		 				\hline
		 				\hline
		 				\multirow{2}{1.4cm}{\centering $\mathcal{I}_{15678910\,11}^{\text{P}}$}&\multirow{2}{2.cm}{\centering\includegraphics[height=0.4in]{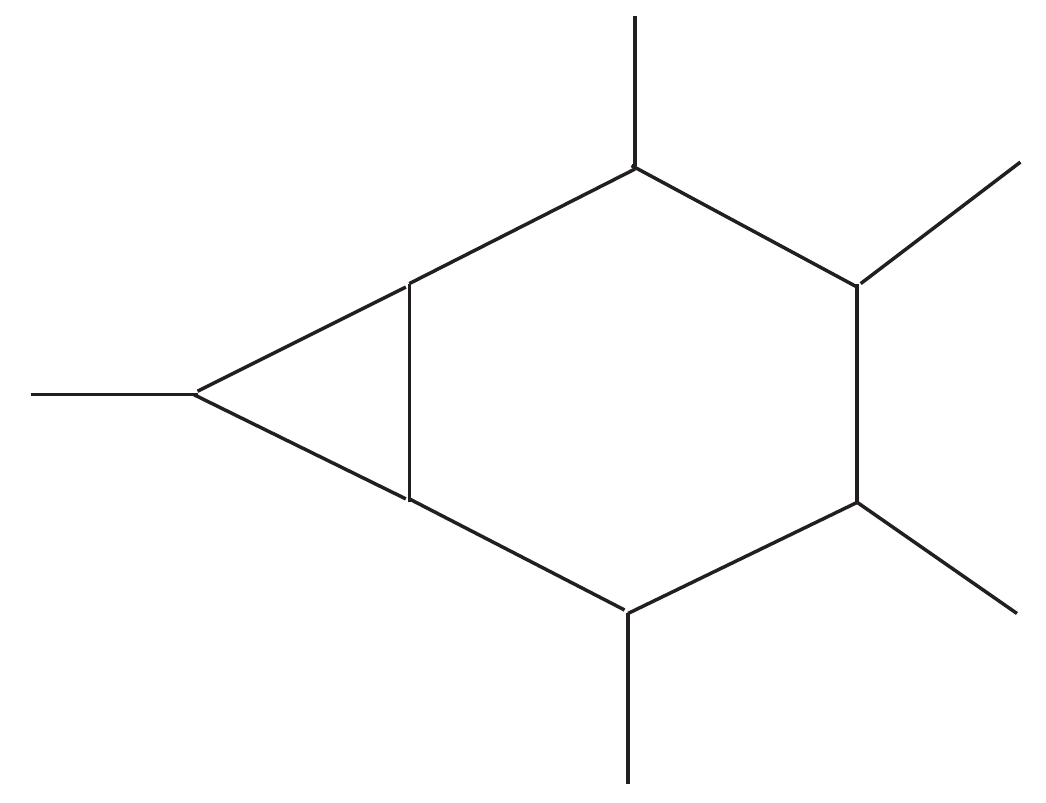}}&$20$\\
		 				&                        &$\{1,x_{21},x_{31},x_{41}\}$\\
		 				\hline
		 				\multirow{2}{1.4cm}{\centering $\mathcal{I}_{13567910\,11}^{\text{P}}$}&\multirow{2}{2.cm}{\centering\includegraphics[height=0.35in]{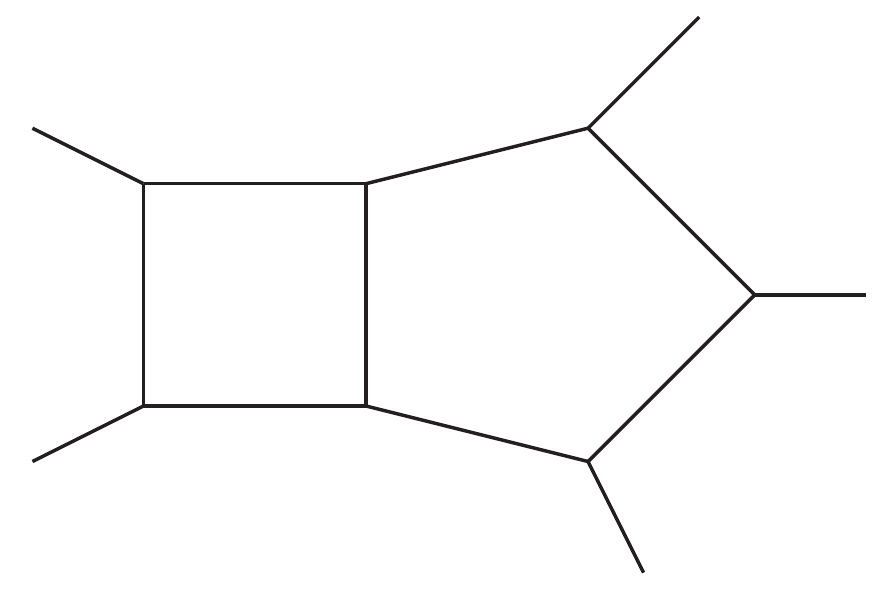}}&$76$\\
		 				&                        &$\{1,x_{31},x_{41},x_{42}\}$\\
		 				\hline
		 				\multirow{2}{1.4cm}{\centering $\mathcal{I}_{15678910\,11}^{\text{NP}1}$}&\multirow{2}{2.cm}{\centering\includegraphics[height=0.40in]{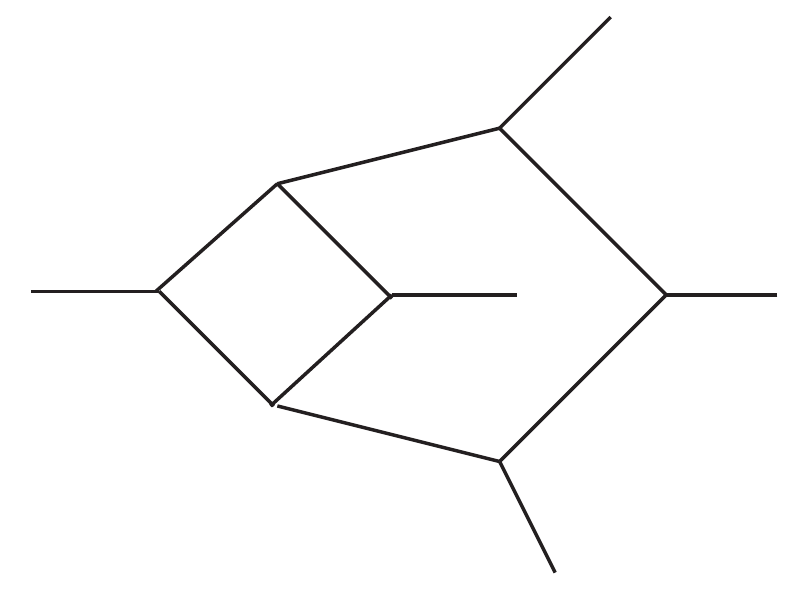}}&$80$\\
		 				&                        &$\{1,x_{31},x_{41},x_{42}\}$\\
		 				\hline
		 				\multirow{2}{1.4cm}{\centering $\mathcal{I}_{1678910\,11}^{\text{P}}$}&\multirow{2}{2.cm}{\centering\includegraphics[height=0.35in]{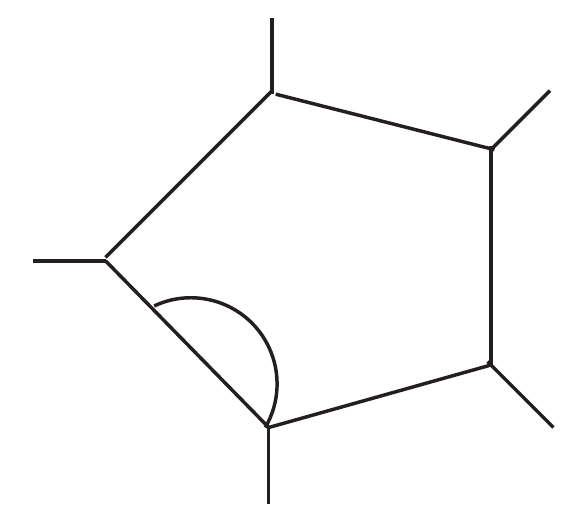}}&$15$\\
		 				&                        &$\{1,x_{11},x_{21},x_{31},x_{41}\}$\\
		 				\hline
		 				\multirow{2}{1.4cm}{\centering $\mathcal{I}_{13568910\,11}^{\text{NP}1}$}&\multirow{2}{2.cm}{\centering\includegraphics[height=0.35in]{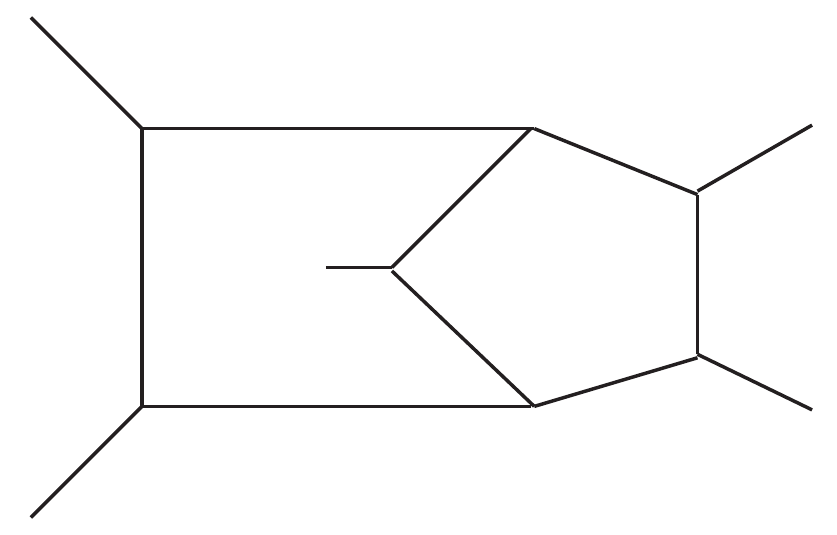}}&$116$\\
		 				&                        &$\{1,x_{31},x_{32},x_{42}\}$\\
		 				\hline
		 				\multirow{2}{1.4cm}{\centering $\mathcal{I}_{1467910\,11}^{\text{P}}$}&\multirow{2}{2.cm}{\centering\includegraphics[height=0.35in]{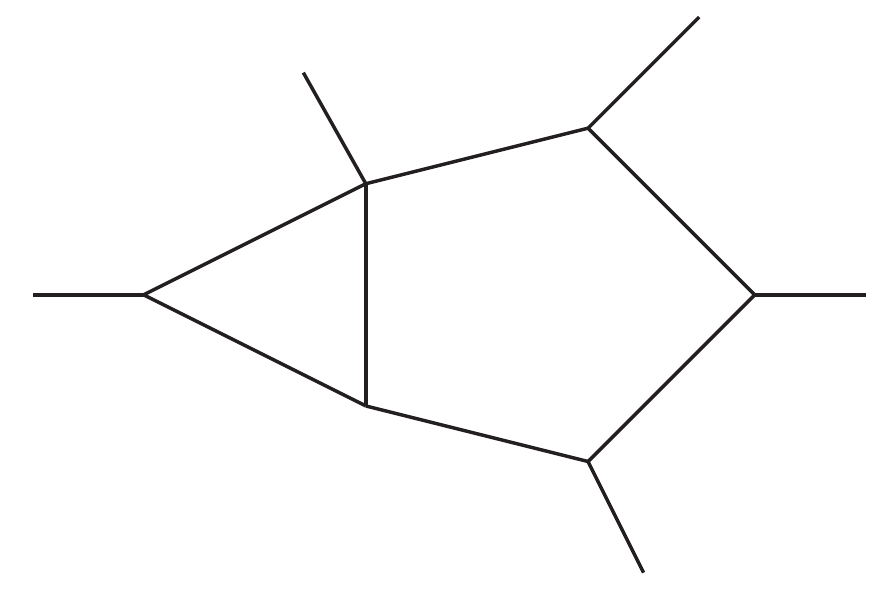}} &$94$\\
		 				&        &$\{1,x_{21},x_{31},x_{41},x_{42}\}$\\
		 				\hline
		 				\multirow{2}{1.4cm}{\centering $\mathcal{I}_{1678911}^{\text{P}}$}&\multirow{2}{2.cm}{\centering\includegraphics[height=0.35in]{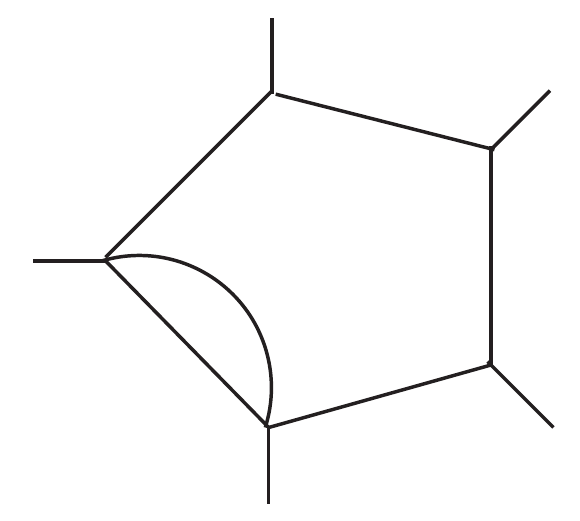}}&$66$\\
		 				&                        &$\{1,x_{11},x_{21},x_{31},x_{41},x_{42}\}$\\
		 				\hline
		 				\multirow{2}{1.4cm}{\centering $\mathcal{I}_{1256910\,11}^{\text{P}}$}&\multirow{2}{2.cm}{\centering\includegraphics[height=0.35in]{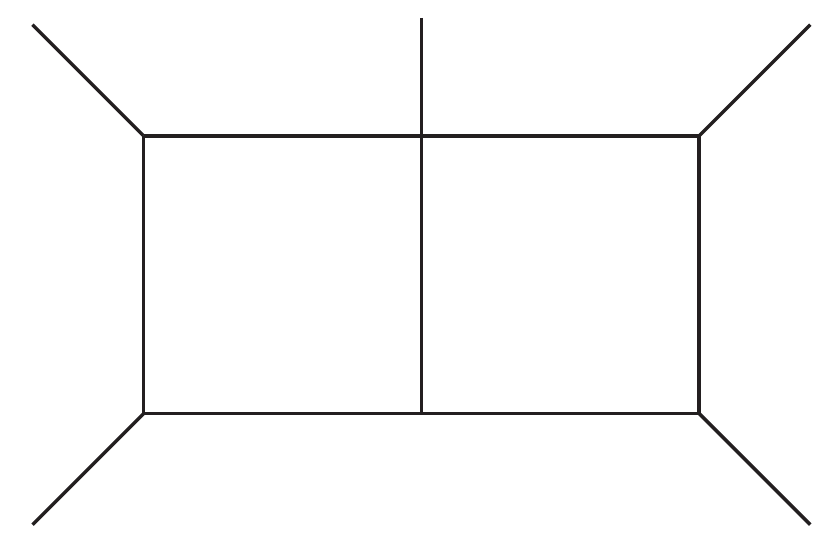}} &$160$\\
		 				&        &$\{1,x_{31},x_{41},y_{32},x_{42}\}$\\
		 				\hline
		 				\multirow{2}{1.4cm}{\centering $\mathcal{I}_{1357910\,11}^{\text{NP}1}$}&\multirow{2}{2.cm}{\centering\includegraphics[height=0.35in]{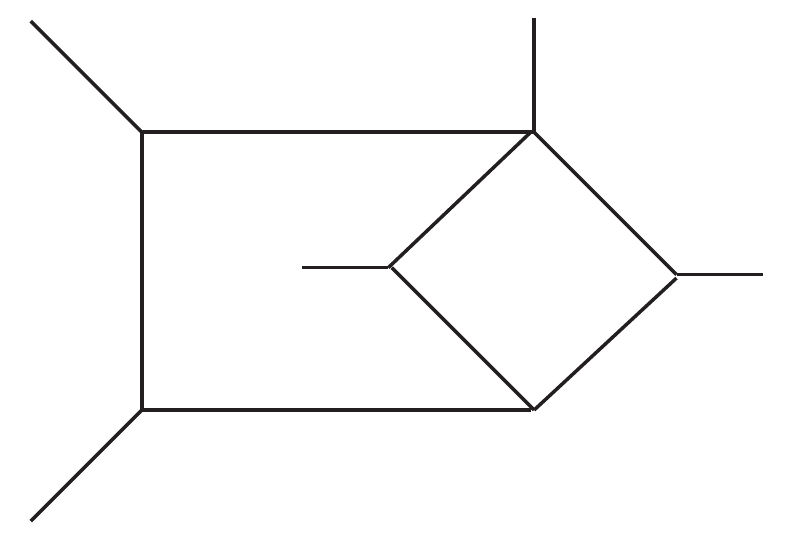}} &$185$\\
		 				&        &$\{1,x_{31},x_{41},x_{32},x_{42}\}$\\
		 				\hline
		 				\multirow{2}{1.4cm}{\centering $\mathcal{I}_{1256911}^{\text{P}}$}&\multirow{2}{2.cm}{\centering\includegraphics[height=0.38in]{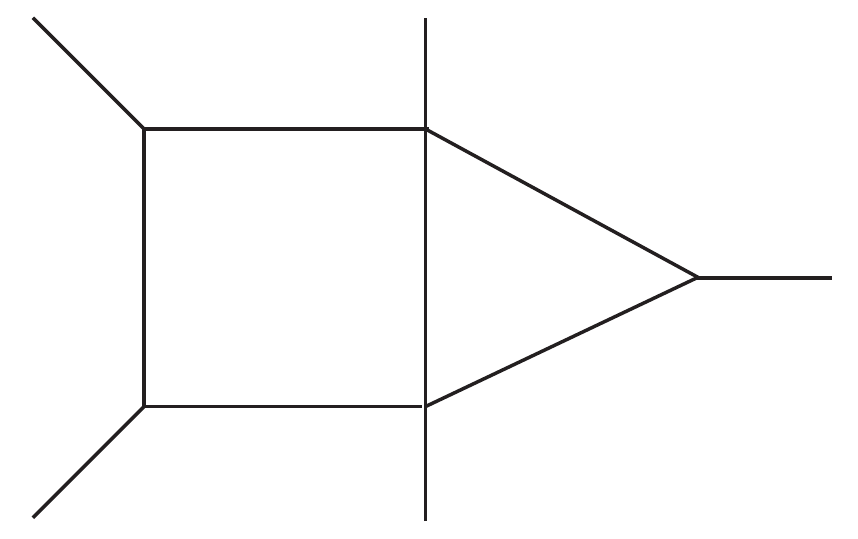}} 
		 				&$180$\\
		 				&        &$\{1,x_{11},x_{31},x_{41},x_{32},x_{42}\}$\\
		 				\hline
		 				\multirow{2}{1.4cm}{\centering $\mathcal{I}_{246910\,11}^{\text{NP}1}$}&\multirow{2}{2.cm}{\centering\includegraphics[height=0.40in]{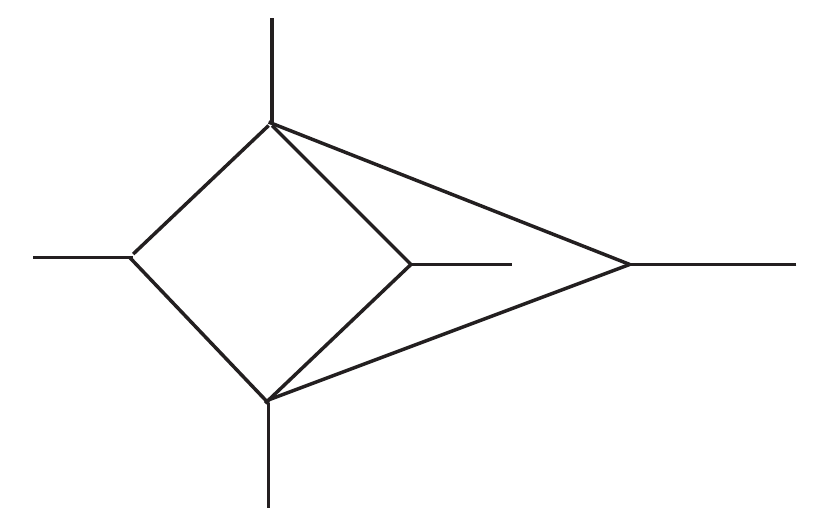}}
		 				&$246$\\
		 				&        &$\{1,x_{31},x_{41},x_{22},x_{32},x_{42}\}$\\
		 				\hline
		 			 	\end{tabular}
		 \end{tabular}
		}
		\caption{\small{Residue parametrization for irreducible six- and five-point two-loop topologies. Denominators depend on the variables $\mathbf{z}=\{x_{11},x_{21},x_{31},x_{41},x_{12},x_{22},x_{32},x_{42},\mu_{11},\mu_{22},\mu_{12}\}$. In the second column we list the number of monomials of each residue and the set of variables appearing in it.}}
		\label{Tab:65pt}
		 \end{table}
	\begin{table}[!ht]
		\centering
		\renewcommand{\arraystretch}{1.2}
		\scalebox{0.75}{
			\begin{tabular}{|c c||c|c|c|}
				\hline
				\multicolumn{2}{|c||}{$\mathcal{I}_{i_1\cdots i_r}$}&$\Delta_{i_1\cdots i_r}$ &$\Delta^{\text{int}}_{i_1\cdots i_r}$& $\Delta'_{i_1\cdots i_r}$ \\
				\hline
				\hline
				\multirow{2}{0.9cm}{\centering $\mathcal{I}_{1567910\,11}^{\text{P}}$}&\multirow{2}{2.cm}{\centering\includegraphics[height=0.35in]{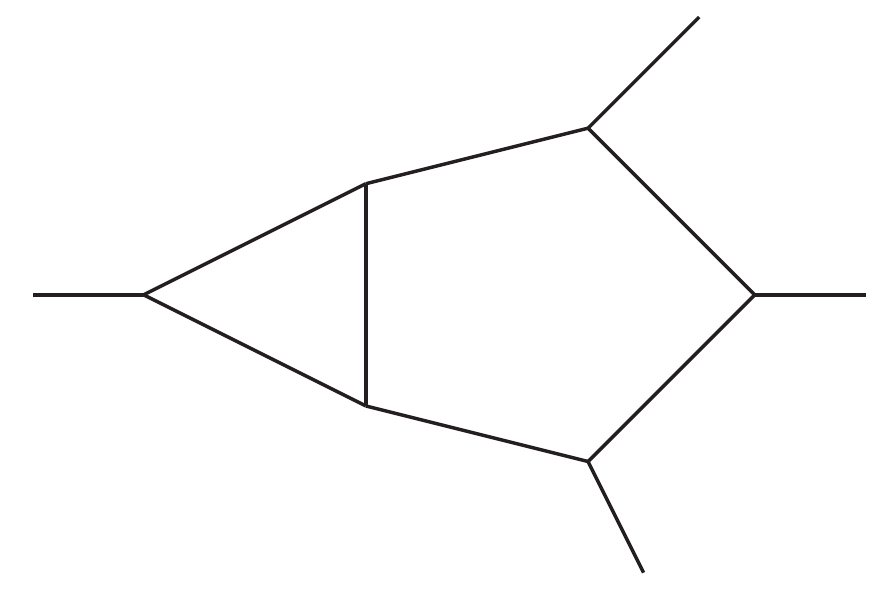}} &$94$&$53$&$10$\\ 
				&        &$\{1,x_{21},x_{31},x_{41},x_{42}\}$&$\{1,x_{21},x_{31},\lambda_{11},\lambda_{22},\lambda_{12}\}$&$\{1,x_{21},x_{31}\}$\\
				\hline
				\multirow{2}{0.85cm}{\centering $\mathcal{I}_{12256910\,11}^{\text{P}}$}&\multirow{2}{1.cm}{\centering\includegraphics[height=0.3in]{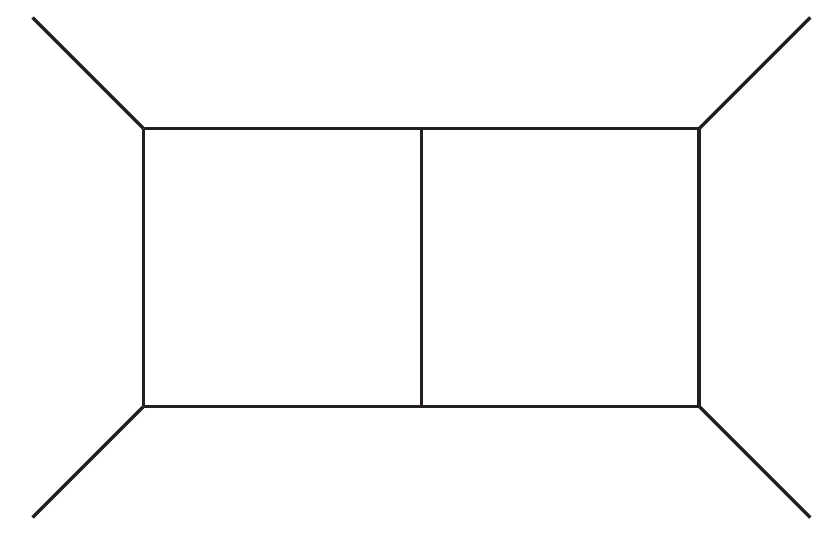}}&$160$&$93$&$22$\\
				&                        &$\{1,x_{31},x_{41},x_{32},x_{42}\}$&$\{1,x_{31},x_{32},\lambda_{11},\lambda_{22},\lambda_{12}\}$&$\{1,x_{31},x_{32}\}$\\
				\hline
				\multirow{2}{0.9cm}{\centering $\mathcal{I}_{1356910\,11}^{\text{NP}1}$}&\multirow{2}{2.cm}{\centering\includegraphics[height=0.35in]{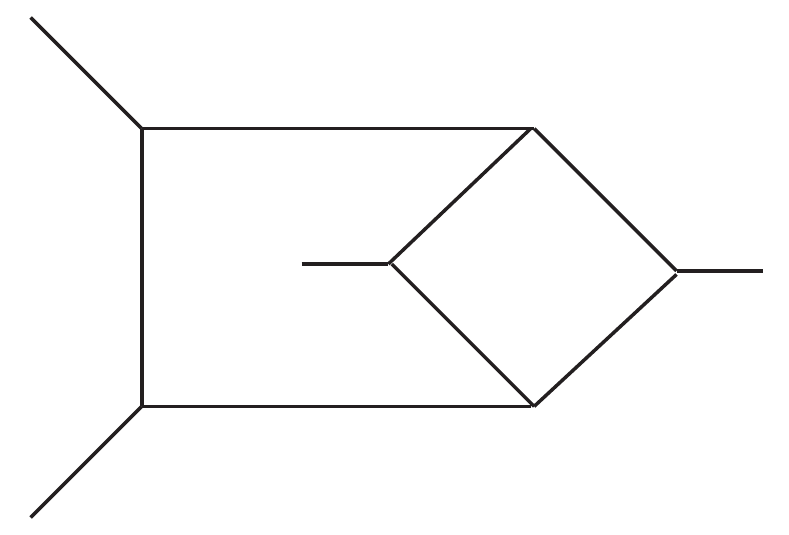}}&$184$&$105$&$25$\\
				&                        &$\{1,x_{31},x_{42},x_{32},x_{42}\}$&$\{1,x_{31},x_{32},\lambda_{11},\lambda_{22},\lambda_{12}\}$&$\{1,x_{31},x_{32}\}$\\
				\hline
				\multirow{2}{0.95cm}{\centering $\mathcal{I}_{1356811}^{\text{P}}$}&\multirow{2}{1.6cm}{\centering\includegraphics[height=0.35in]{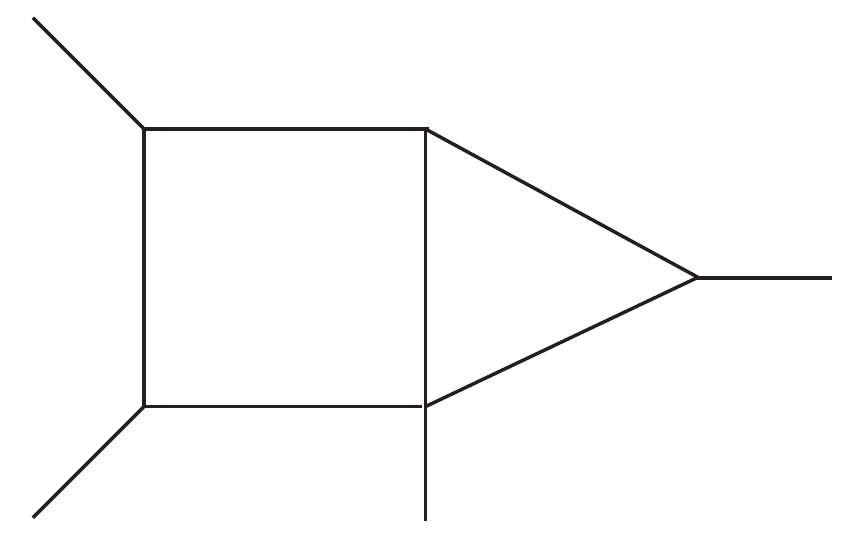}} &$180$&$101$&$39$\\
				&        &$\{1,x_{31},x_{41},x_{22},x_{32},x_{42}\}$&$\{1,x_{31},x_{22},x_{32},\lambda_{11},\lambda_{22},\lambda_{12}\}$&$\{1,x_{31},x_{22},y_{32}\}$\\
				\hline
				\multirow{2}{0.9cm}{\centering $\mathcal{I}_{168910\,11}^{\text{P}}$}&\multirow{2}{2.cm}{\centering\includegraphics[height=0.38in]{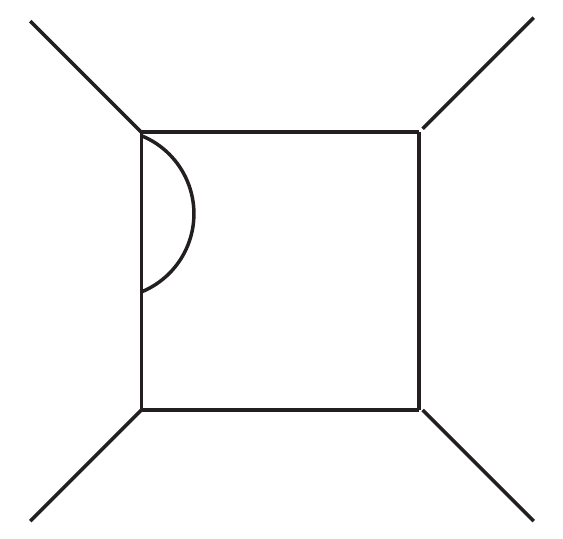}} &$66$&$35$&$10$\\
				&        &$\{1,x_{11},x_{21},x_{31},x_{41},x_{42}\}$&$\{1,x_{11},x_{21},x_{31},\lambda_{11},\lambda_{22},\lambda_{12}\}$&$\{1,x_{11},x_{21},x_{31}\}$\\
				\hline
				\multirow{2}{0.9cm}{\centering $\mathcal{I}_{246910\,11}^{\text{NP}1}$}&\multirow{2}{2.cm}{\centering\includegraphics[height=0.35in]{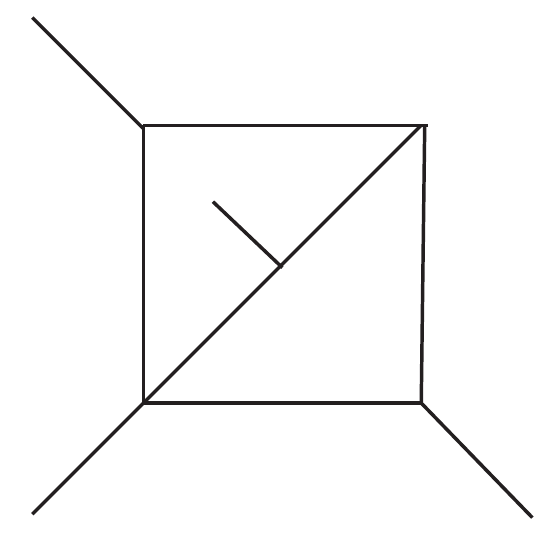}} &$245$&$137$&$55$\\
				&        &$\{1,x_{31},x_{41},x_{21},x_{32},x_{42}\}$&$\{1,x_{31},x_{22},x_{32},\lambda_{11},\lambda_{22},\lambda_{12}\}$&$\{1,x_{31},x_{22},y_{32}\}$\\
				\hline
				\multirow{2}{0.85cm}{\centering $\mathcal{I}_{36810\,11}^{\text{P}}$}&\multirow{2}{1.6cm}{\centering\includegraphics[height=0.35in]{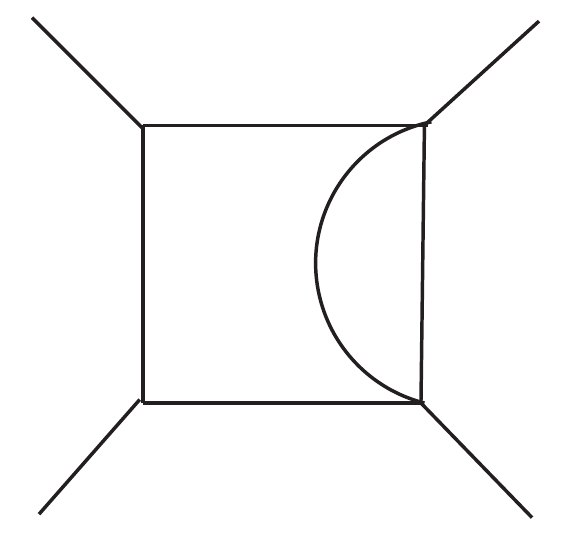}} &$115$&$66$&$35$\\
				&        &$\{1,x_{31},x_{41},x_{12},x_{22},x_{32},x_{42}\}$&$\{1,x_{31},x_{12},x_{22},x_{32},\lambda_{11},\lambda_{22},\lambda_{12}\}$&$\{1,x_{31},x_{12},x_{22},x_{32}\}$\\
				\hline
				\multirow{2}{0.85cm}{\centering $\mathcal{I}_{136811}^{\text{P}}$}&\multirow{2}{1.6cm}{\centering\includegraphics[height=0.35in]{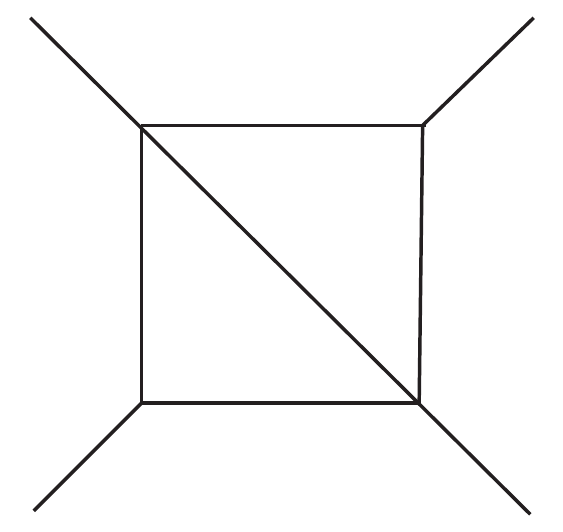}} &$180$&$103$&$60$\\
				&        &$\{1,x_{11},x_{31},x_{41},x_{22},x_{32},x_{42}\}$&$\{1,x_{11},x_{31},x_{22},x_{32},\lambda_{11},\lambda_{22},\lambda_{12}\}$&$\{1,x_{11},x_{31},x_{22},x_{32}\}$\\
				\hline
			\end{tabular}
		}
		\caption{\small{Residue parametrization for irreducible four-point two-loop topologies. Denominators depend on the variables $\boldsymbol{\tau}=\{x_{11},x_{21},x_{31},x_{12},x_{22},x_{32},\lambda_{11},\lambda_{22},\lambda_{12}\}$. For every step of the reduction algorithm, we list the number of monomials of each residues and the set of variables appearing in it.
				}}
		\label{Tab:4pt}
	\end{table}
	
	\begin{table}[!ht]
		\centering
		\renewcommand{\arraystretch}{1.2}
		\scalebox{0.75}{
			\begin{tabular}{|c c||c|c|c|}
				\hline
				\multicolumn{2}{|c||}{$\mathcal{I}_{i_1\cdots i_r}$}&$\Delta_{i_1\cdots i_r}$ &$\Delta^{\text{int}}_{i_1\cdots i_r}$& $\Delta'_{i_1\cdots i_r}$ \\
				\hline
				\hline 
				\multirow{2}{0.85cm}{\centering $\mathcal{I}_{1356911}^{\text{P}}$}&\multirow{2}{1.6cm}{\centering\includegraphics[height=0.4in]{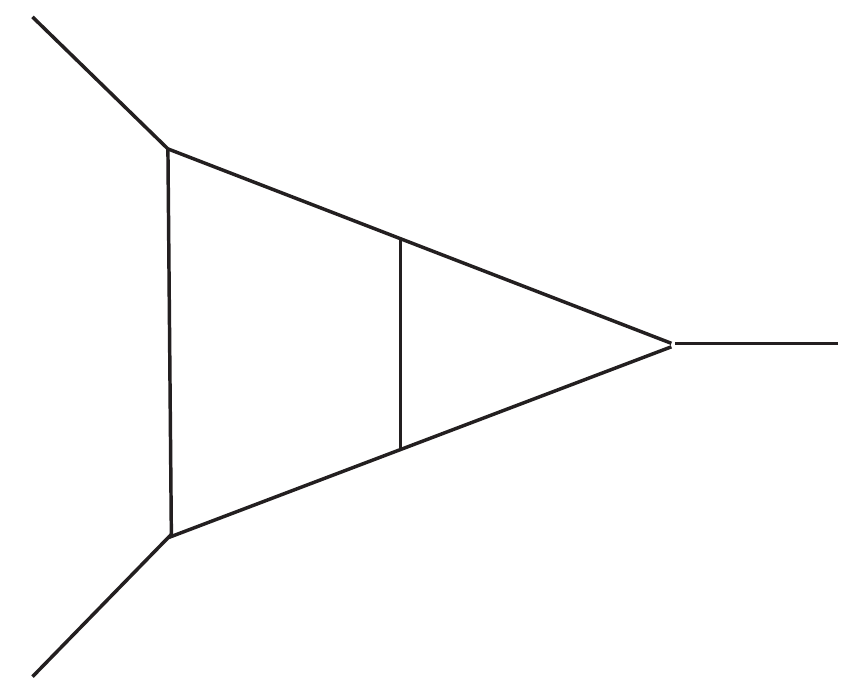}} &$180$&$22$&$4$\\
				&        &$\{1,x_{31},x_{41},x_{22},x_{32},x_{42}\}$&$\{1,x_{22},\lambda_{11},\lambda_{22},\lambda_{12}\}$&$\{1,x_{22}\}$\\
			    \hline
				\multirow{2}{0.9cm}{\centering $\mathcal{I}_{156910\,11}^{\text{NP}1}$}&\multirow{2}{2.cm}{\centering\includegraphics[height=0.25in]{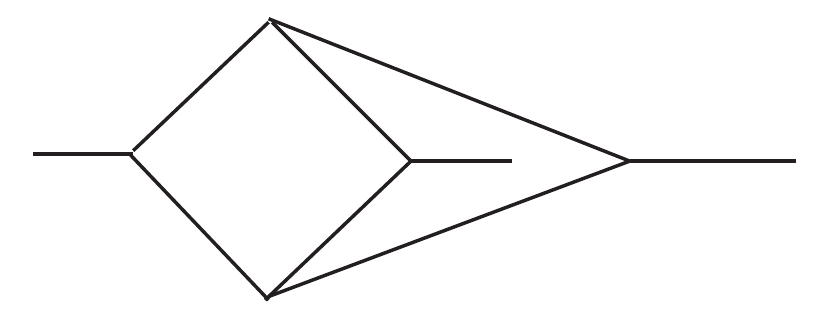}} &$240$&$30$&$6$\\
				&        &$\{1,x_{31},x_{41},x_{22},x_{32},x_{42}\}$&$\{1,x_{22},\lambda_{11},\lambda_{22},\lambda_{12}\}$&$\{1,x_{22}\}$\\
				\hline
				\multirow{2}{0.85cm}{\centering $\mathcal{I}_{15710\,11}^{\text{P}}$}&\multirow{2}{1.6cm}{\centering\includegraphics[height=0.4in]{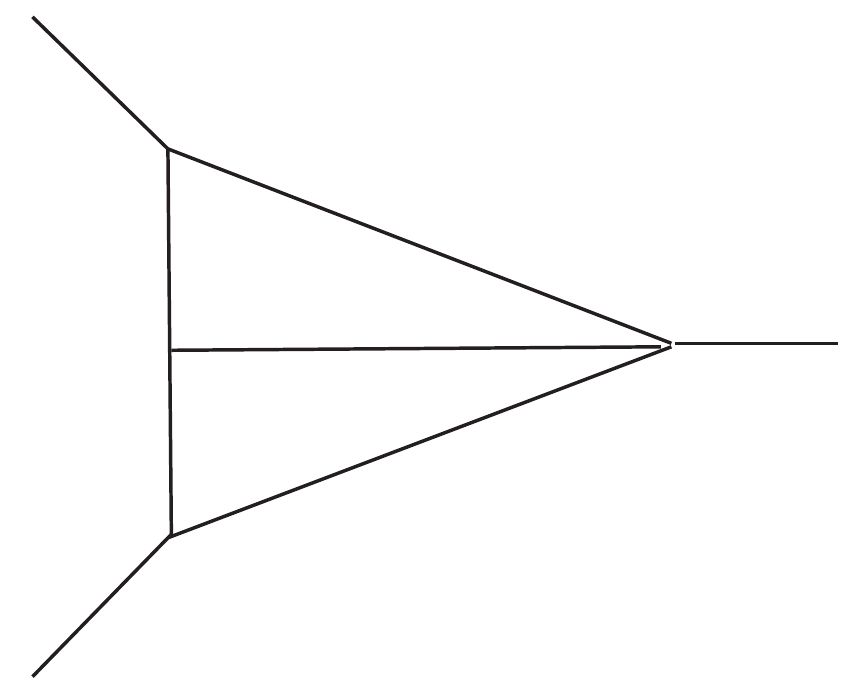}} &$180$&$33$&$13$\\
				&        &$\{1,x_{21},x_{31},x_{41},x_{12},x_{32},x_{42}\}$&$\{1,x_{21},x_{12},\lambda_{11},\lambda_{22},\lambda_{12}\}$&$\{1,x_
			{21},x_{12}\}$\\
				\hline
				\multirow{2}{0.85cm}{\centering $\mathcal{I}_{16910\,11}^{\text{P}}$}&\multirow{2}{1.6cm}{\centering\includegraphics[height=0.4in]{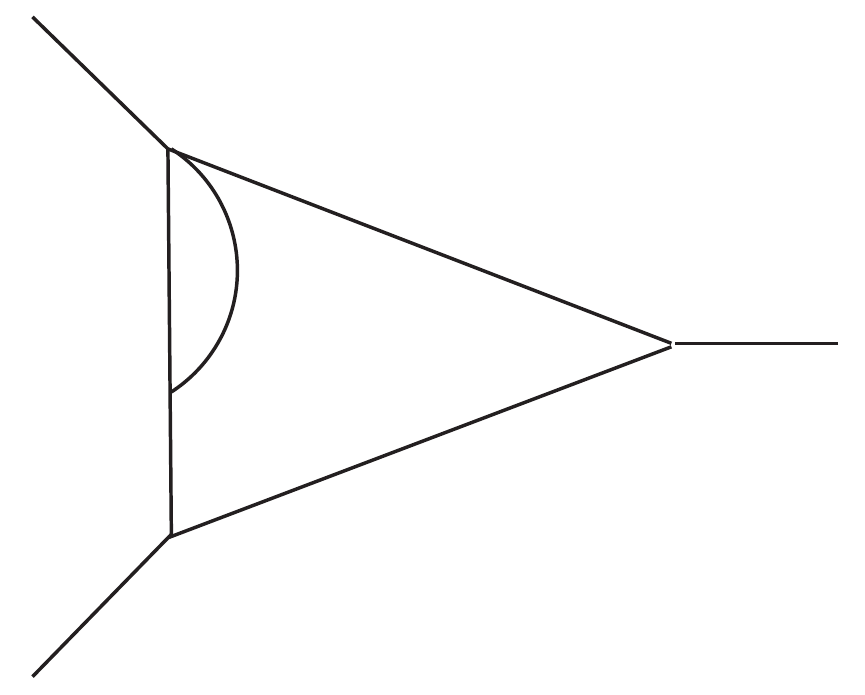}} &$115$&$20$&$6$\\
				&        &$\{1,x_{31},x_{41},x_{12},x_{22},x_{32},x_{42}\}$&$\{1,x_{11},x_{22}\lambda_{11},\lambda_{22},\lambda_{12}\}$&$\{1,x_{12},x_{22}\}$\\
				\hline
				\multirow{2}{0.85cm}{\centering $\mathcal{I}_{3610\,11}^{\text{P}}$}&\multirow{2}{1.6cm}{\centering\includegraphics[height=0.4in]{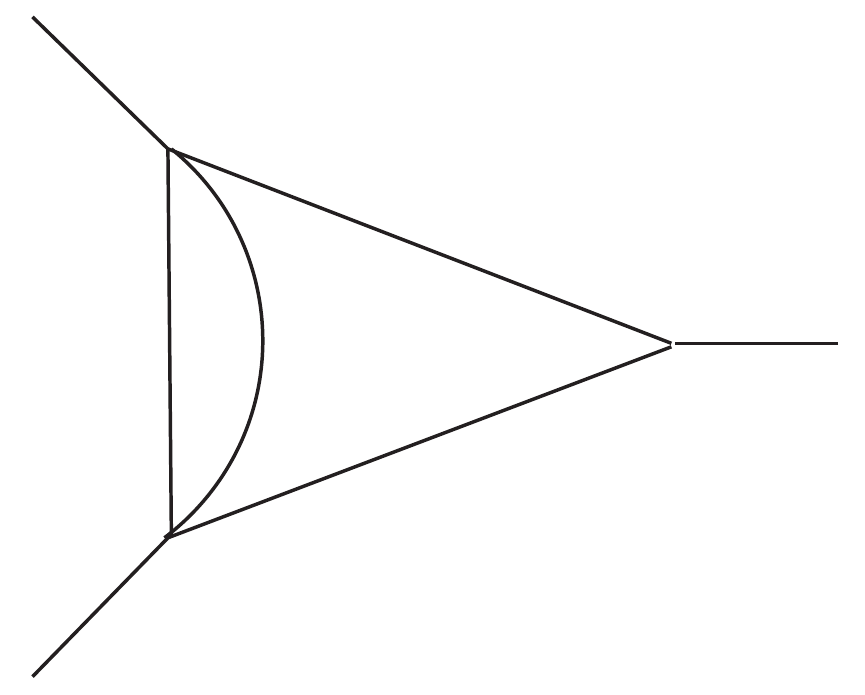}} &$100$&$26$&$16$\\
				&        &$\{1,x_{11},x_{21},x_{31},x_{41},x_{22},x_{32},x_{42}\}$&$\{1,x_{11},x_{21},x_{22},\lambda_{11},\lambda_{22},\lambda_{12}\}$&$\{x_{11},x_{21},x_{22}\}$\\
				\hline
			\end{tabular}
		}
		\caption{\small{Residue parametrization for irreducible three-point two-loop topologies. Denominators depend on the variables $\boldsymbol{\tau}=\{x_{11},x_{21},x_{12},x_{22},\lambda_{11},\lambda_{22},\lambda_{12}\}$. For every step of the reduction algorithm, we list the number of monomials of each residues and the set of variables appearing in it.
						}}
		\label{Tab:3pt}
	\end{table}
	\begin{table}[!ht]
		\centering
		\renewcommand{\arraystretch}{1.2}
		\scalebox{0.75}{
			\begin{tabular}{|c c||c|c|c|}
				\hline
				\multicolumn{2}{|c||}{$\mathcal{I}_{i_1\cdots i_r}$}&$\Delta_{i_1\cdots i_r}$ &$\Delta^{\text{int}}_{i_1\cdots i_r}$& $\Delta'_{i_1\cdots i_r}$ \\
				\hline
				\hline
				\multirow{2}{0.85cm}{\centering $\mathcal{I}_{15610\,11}^{\text{P}}$}&\multirow{2}{1.6cm}{\centering\includegraphics[height=0.3in]{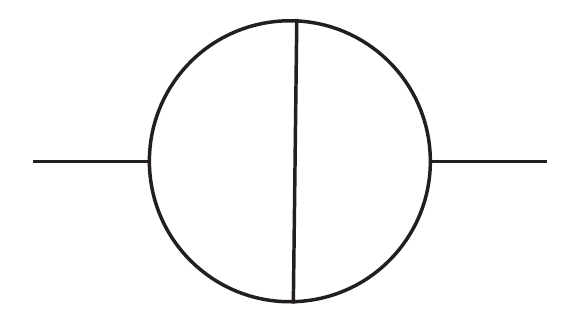}} &$180$&$8$&$1$\\
				&        &$\{1,x_{21},x_{31},x_{41},x_{22},x_{32},x_{42}\}$&$\{1,\lambda_{11},\lambda_{22},\lambda_{12}\}$&$\{1\}$\\
				\hline
				\multirow{2}{0.85cm}{\centering $\mathcal{I}_{1610\,11}^{\text{P}}$}&\multirow{2}{1.6cm}{\centering\includegraphics[height=0.3in]{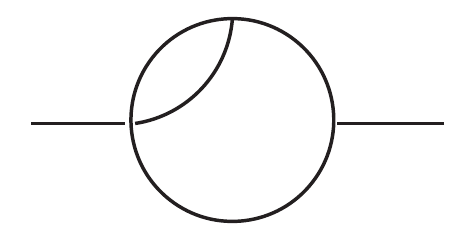}} &$100$&$8$&$3$\\
				&        &$\{1,x_{11},x_{21},x_{31},x_4,x_{22},y_3,x_{42}\}$&$\{1,x_{11},\lambda_{11},\lambda_{22},\lambda_{12}\}$&$\{1,x_{11}\}$\\
				\hline
				\multirow{2}{0.85cm}{\centering $\mathcal{I}_{1310\,11}^{\text{P}}$}&\multirow{2}{1.6cm}{\centering\includegraphics[height=0.3in]{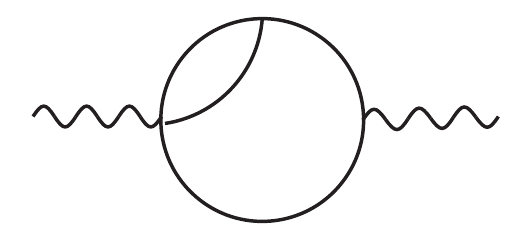}} &$100$&$26$&$16$\\
				&        &$\{1,x_{11},x_{21},x_{31},x_{41},x_{12},x_{32},x_{42}\}$&$\{1,x_{11},x_{21},x_{12},\lambda_{11},\lambda_{22},\lambda_{12}\}$&$\{1,x_{11},x_{21},x_{12}\}$\\
				\hline
				\multirow{2}{0.85cm}{\centering $\mathcal{I}_{210\,11}^{\text{P}}$}&\multirow{2}{1.6cm}{\centering\includegraphics[height=0.3in]{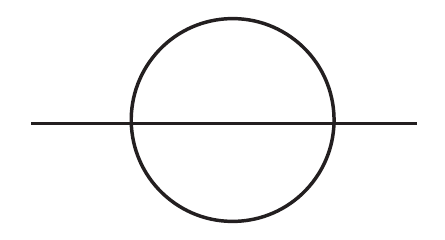}} &$45$&$9$&$6$\\
				&        &$\{1,x_{11},x_{21},x_{31},x_{41},x_{12},x_{22},x_{32},x_{42}\}$&$\{1,x_{11},x_{12},\lambda_{11},\lambda_{22},\lambda_{12}\}$&$\{1,x_{11},x_{12}\}$\\
				\hline
				\multirow{2}{0.85cm}{\centering $\mathcal{I}_{210\,11}^{\text{P}}$}&\multirow{2}{1.6cm}{\centering\includegraphics[height=0.3in]{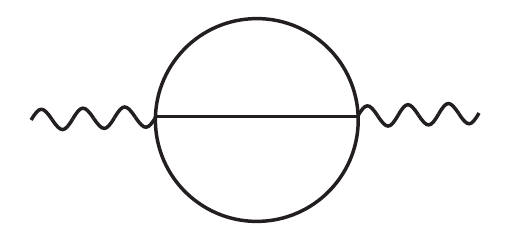}} &$45$&$18$&$15$\\
				&        &$\{1,x_{11},x_{21},x_{31},x_{41},x_{12},x_{22},x_{32},x_{42}\}$&$\{1,x_{11},x_{21},x_{12},x_{22},\lambda_{11},\lambda_{22},\lambda_{12}\}$&$\{1,x_{11},x_{22},x_{21},x_{22}\}$\\
				\hline
			\end{tabular}
		}
		\caption{\small{Residue parametrization for irreducible two-point two-loop topologies. Denominators depend on the variables $\boldsymbol{\tau}=\{x_{11},x_{12},\lambda_{11},\lambda_{22},\lambda_{12}\}$ in the case of massive external momenta and on $\boldsymbol{\tau}=\{x_{11},x_{21},x_{12},x_{22},\lambda_{11},\lambda_{22},\lambda_{12}\}$ in the case of massless one. For every step of the reduction algorithm, we list the number of monomials of each residues and the set of variables appearing in it. In the figures, wavy lines indicate massless particles, whereas solid ones stands for arbitrary masses.
				}}
		\label{Tab:2pt}
	\end{table}
	\begin{table}[!ht]
		\centering
		\renewcommand{\arraystretch}{1.2}
		\scalebox{0.75}{
			\begin{tabular}{|c c||c|c|c|}
				\hline
				\multicolumn{2}{|c||}{$\mathcal{I}_{i_1\cdots i_r}$}&$\Delta_{i_1\cdots i_r}$ &$\Delta^{\text{int}}_{i_1\cdots i_r}$& $\Delta'_{i_1\cdots i_r}$ \\
					\hline
					\hline
					\multirow{2}{0.85cm}{\centering $\mathcal{I}_{110\,11}^{\text{P}}$}&\multirow{2}{1.6cm}{\centering\includegraphics[height=0.3in]{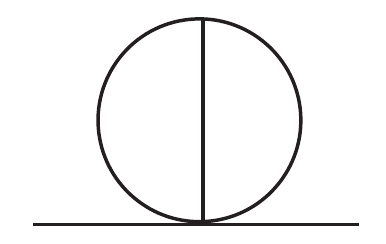}} &$45$&$4$&$1$\\
					&        &$\{1,x_{11},x_{21},x_{31},x_{41},x_{12},x_{22},x_{32},x_{42}\}$&$\{1,\lambda_{11},\lambda_{22},\lambda_{12}\}$&$\{1\}$\\
					\hline
				\end{tabular}
			}
			\caption{\small{Residue parametrization for the irreducible one-point two-loop topology. Denominators depend on the variables $\boldsymbol{\tau}=\{\lambda_{11},\lambda_{22},\lambda_{12}\}$. For every step of the reduction algorithm, we list the number of monomials of the residue and the set of variables appearing in it.
					}}
			\label{Tab:1pt}
	\end{table}
	\begin{table}[h]
		\centering
		\renewcommand{\arraystretch}{1.2}
		\scalebox{0.75}{
			\begin{tabular}{|c c||c|c|c|}
				\hline
				\multicolumn{2}{|c||}{$\mathcal{I}_{i_1\cdots i_r}$}&$\Delta_{i_1\cdots i_r}$ &$\Delta^{\text{int}}_{i_1\cdots i_r}$& $\Delta'_{i_1\cdots i_r}$ \\
				\hline
				\hline
				\multirow{2}{1.3cm}{\centering $\mathcal{I}_{12345678910}^{\text{P}}$}&\multirow{2}{1.6cm}{\centering\includegraphics[height=0.32in]{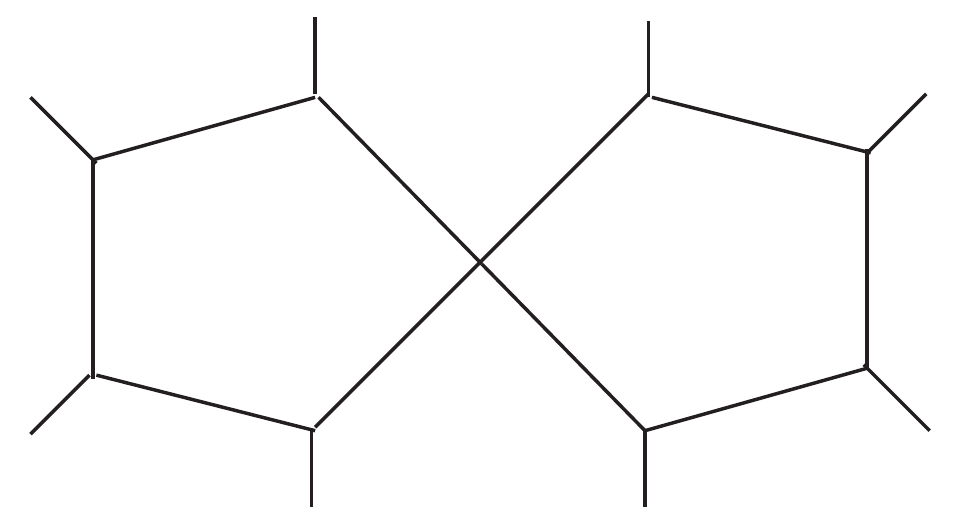}} &$1$&$-$&$-$\\
				&        &$\{1\}$&$-$&$-$\\
				\hline
				\multirow{2}{1.4cm}{\centering $\mathcal{I}_{1245678910}^{\text{P}}$}&\multirow{2}{1.6cm}{\centering\includegraphics[height=0.32in]{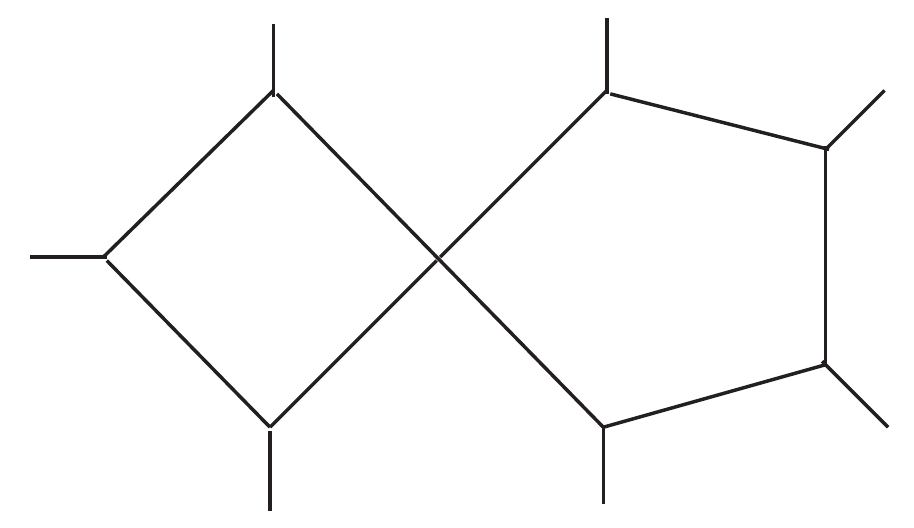}} &$5$&$3$&$1$\\
				&        &$\{1,x_{41}\}$&$\{1,\lambda_{11}\}$&$\{1\}$\\
				\hline
				\multirow{2}{1.4cm}{\centering $\mathcal{I}_{125678910}^{\text{P}}$}&\multirow{2}{1.6cm}{\centering\includegraphics[height=0.4in]{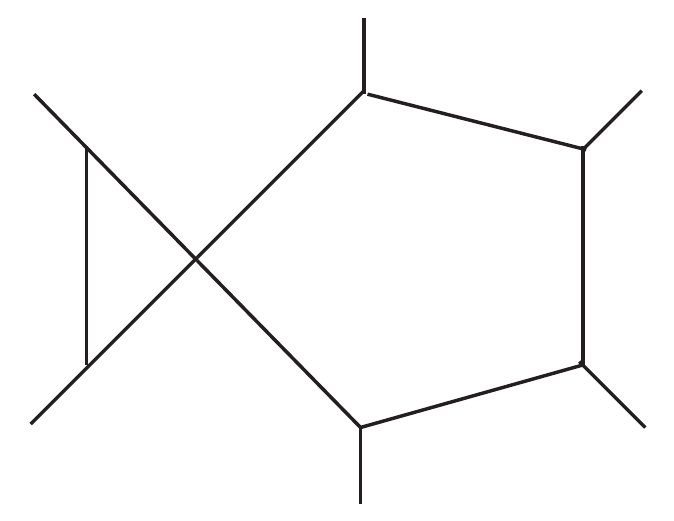}} &$10$&$2$&$1$\\
				&        &$\{1,x_{31},x_{41}\}$&$\{1,\lambda_{11}\}$&$\{1\}$\\
				\hline
				\multirow{2}{1.4cm}{\centering $\mathcal{I}_{15678910}^{\text{P}}$}&\multirow{2}{1.6cm}{\centering\includegraphics[height=0.4in]{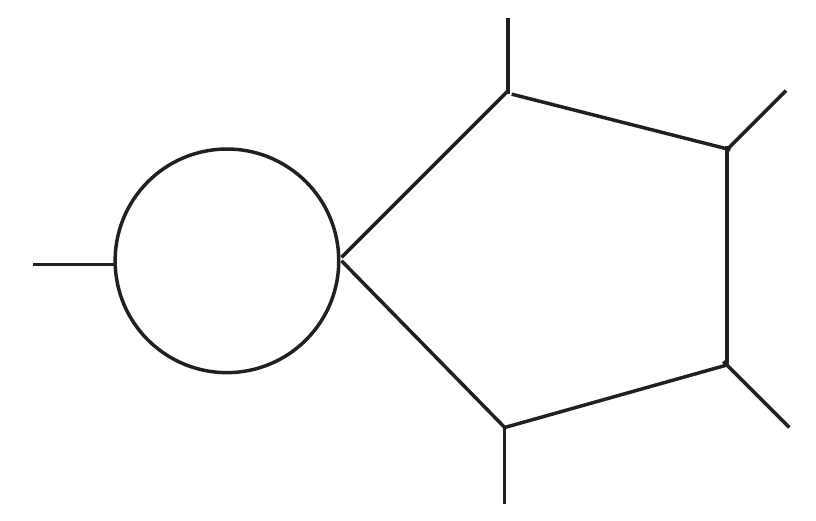}} &$10$&$2$&$1$\\
				&        &$\{1,x_{21},x_{31},x_{41}\}$&$\{1,\lambda_{11}\}$&$\{1\}$\\
				\hline
				\multirow{2}{1.4cm}{\centering $\mathcal{I}_{12678910}^{\text{P}}$}&\multirow{2}{1.6cm}{\centering\includegraphics[height=0.4in]{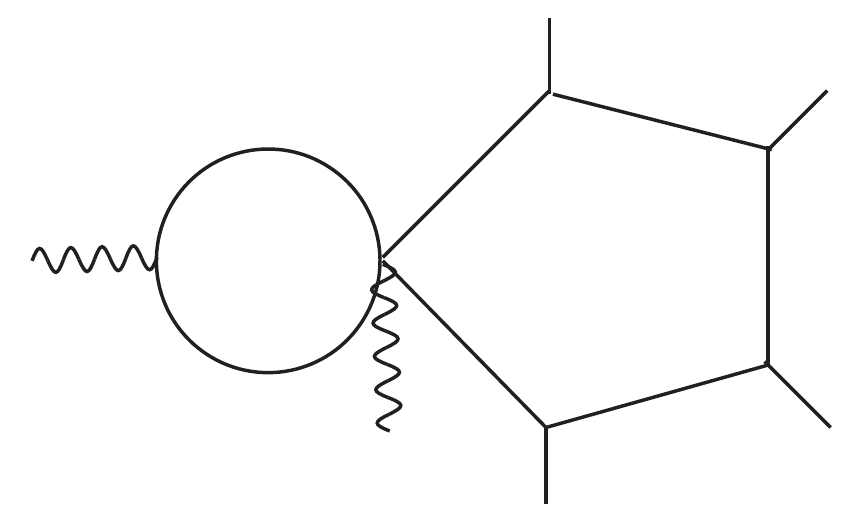}} &$10$&$4$&$3$\\
				&        &$\{1,x_{11},x_{31},x_{41}\}$&$\{1,x_{11},\lambda_{11}\}$&$\{1,x_{11}\}$\\
				\hline
				\multirow{2}{1.4cm}{\centering $\mathcal{I}_{1678910}^{\text{P}}$}&\multirow{2}{1.6cm}{\centering\includegraphics[height=0.4in]{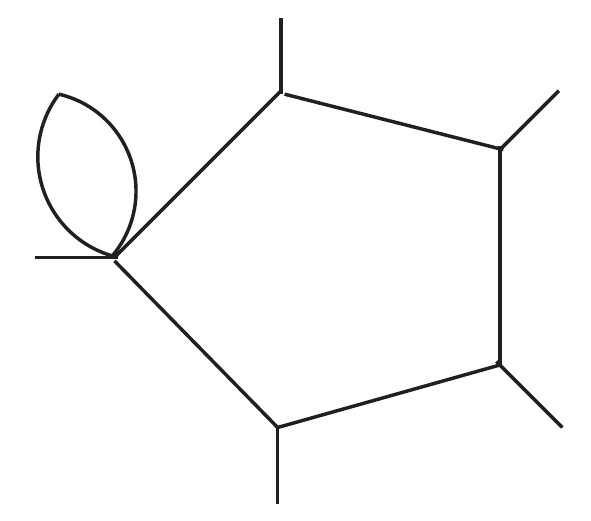}} &$5$&$1$&$-$\\
				&        &$\{1,x_{11},x_{21},x_{31},x_{41}\}$&$\{1\}$&$-$\\
				\hline
			    \multirow{2}{1.4cm}{\centering $\mathcal{I}_{23456789}^{\text{P}}$}&\multirow{2}{1.6cm}{\centering\includegraphics[height=0.4in]{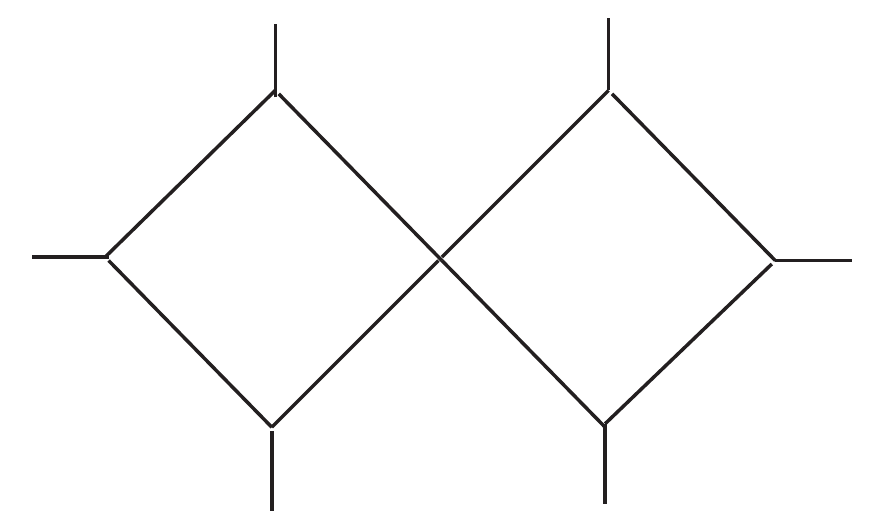}} &$25$&$9$&$1$\\
				&        &$\{1,x_{41},x_{42}\}$&$\{1,\lambda_{11},\lambda_{22}\}$&$\{1\}$\\
				\hline
				\multirow{2}{1.4cm}{\centering $\mathcal{I}_{2356789}^{\text{P}}$}&\multirow{2}{1.6cm}{\centering\includegraphics[height=0.4in]{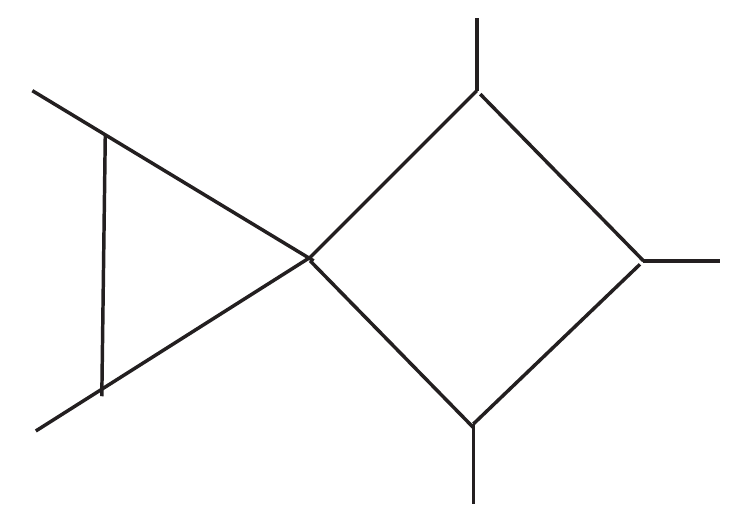}} &$50$&$6$&$1$\\
				&        &$\{1,x_{31},x_{41},x_{42}\}$&$\{1,\lambda_{11},\lambda_{22}\}$&$\{1\}$\\
				\hline
				\multirow{2}{1.4cm}{\centering $\mathcal{I}_{256789}^{\text{P}}$}&\multirow{2}{1.6cm}{\centering\includegraphics[height=0.4in]{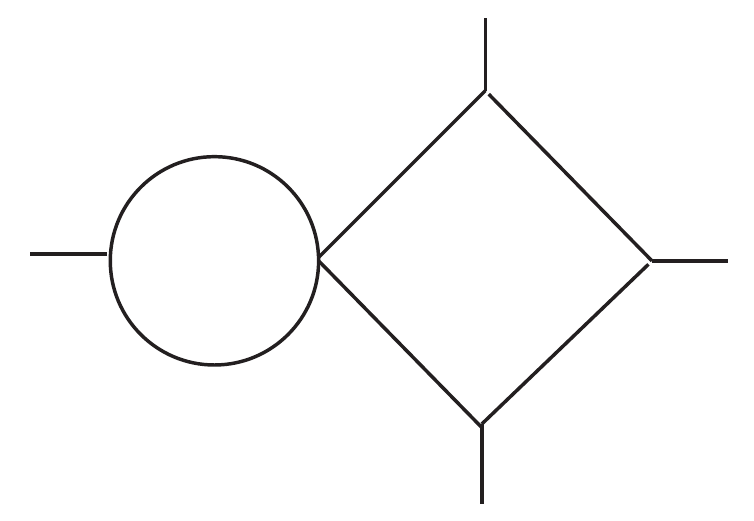}} &$50$&$6$&$1$\\
				&        &$\{1,x_{21},x_{31},x_{41},x_{42}\}$&$\{1,\lambda_{11},\lambda_{22}\}$&$\{1\}$\\
				\hline
				\multirow{2}{1.4cm}{\centering $\mathcal{I}_{236789}^{\text{P}}$}&\multirow{2}{1.6cm}{\centering\includegraphics[height=0.4in]{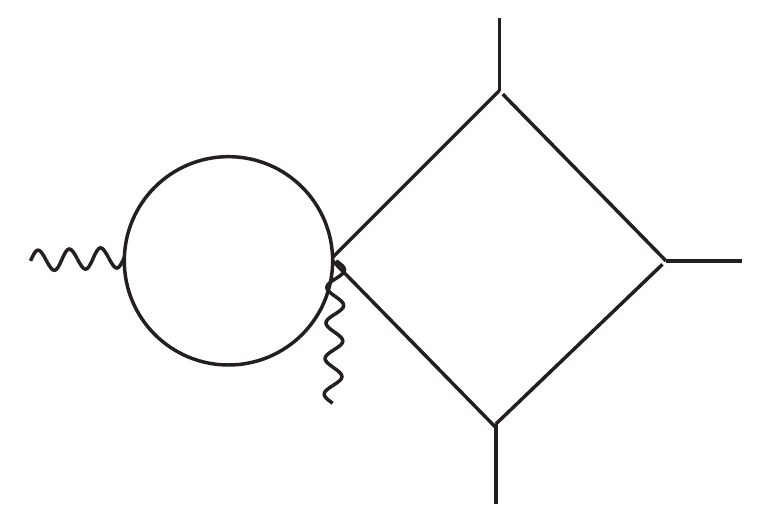}} &$50$&$12$&$3$\\
				&        &$\{1,x_{11},x_{31},x_{41},x_{42}\}$&$\{1,x_{11},\lambda_{11},\lambda_{22}\}$&$\{1,x_{11}\}$\\
				\hline
				\multirow{2}{1.4cm}{\centering $\mathcal{I}_{26789}^{\text{P}}$}&\multirow{2}{1.6cm}{\centering\includegraphics[height=0.45in]{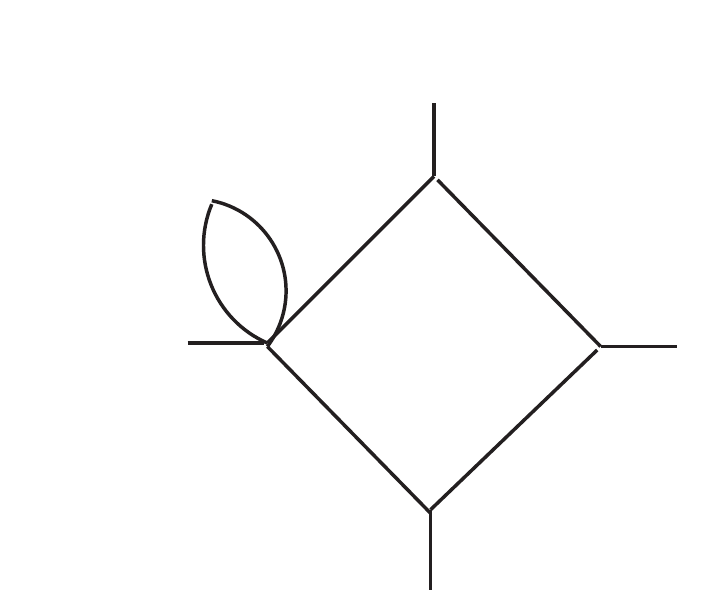}} &$25$&$3$&$1$\\
				&        &$\{1,x_{11},x_{21},x_{31},x_{41},x_{42}\}$&$\{1,\lambda_{22}\}$&$\{1\}$\\
				\hline		
				\multirow{2}{1.4cm}{\centering $\mathcal{I}_{245689}^{\text{P}}$}&\multirow{2}{1.6cm}{\centering\includegraphics[height=0.3in]{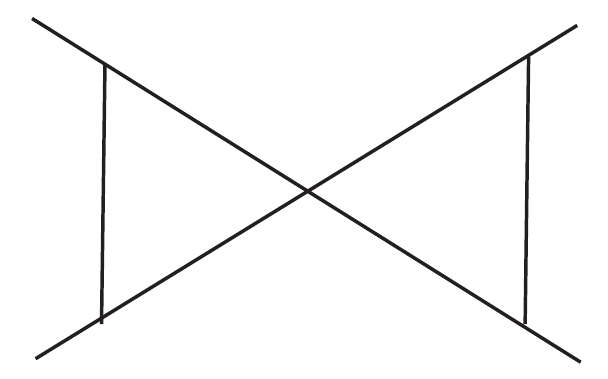}}&$100$&$4$&$1$\\
				&                        &$\{1,x_{31},x_{42},x_{32},x_{42}\}$&$\{1,\lambda_{11},\lambda_{22}\}$&$\{1\}$\\
				\hline
				\multirow{2}{1.4cm}{\centering$\mathcal{I}_{24689}^{\text{P}}$}&\multirow{2}{1.6cm}{\centering\includegraphics[height=0.4in]{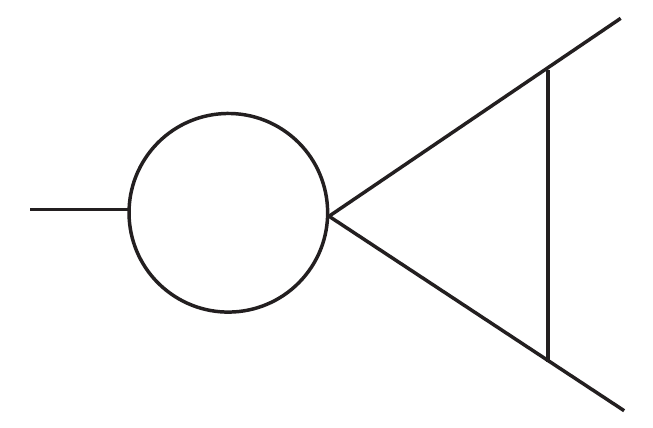}} &$100$&$4$&$1$\\
				&        &$\{1,x_{21},x_{31},x_{41},x_{32},x_{42}\}$&$\{1,\lambda_{11},\lambda_{22}\}$&$\{1\}$\\
				\hline
				\multirow{2}{1.4cm}{\centering $\mathcal{I}_{45689}^{\text{P}}$}&\multirow{2}{1.6cm}{\centering\includegraphics[height=0.4in]{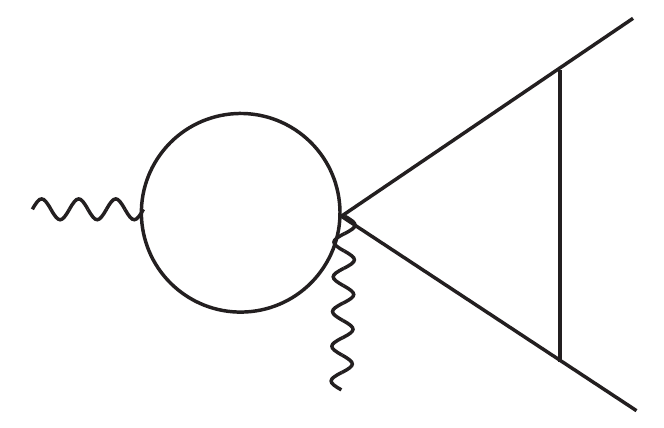}} &$100$&$8$&$3$\\
				&        &$\{1,x_{11},x_{31},x_{41},x_{32},x_{42}\}$&$\{1,x_{11},\lambda_{11},\lambda_{22}\}$&$\{1,x_{11}\}$\\
				\hline
				\multirow{2}{1.4cm}{\centering $\mathcal{I}_{2689}^{\text{P}}$}&\multirow{2}{1.6cm}{\centering\includegraphics[height=0.35in]{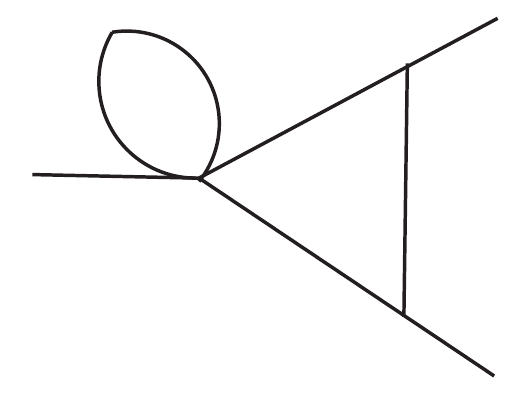}} &$50$&$2$&$1$\\
				&        &$\{1,x_{11},x_{21},x_{31},x_{41},x_{32},x_{42}\}$&$\{1,\lambda_{22}\}$&$\{1\}$\\
				\hline
				\multirow{2}{1.4cm}{\centering $\mathcal{I}_{2569}^{\text{P}}$}&\multirow{2}{1.6cm}{\centering\includegraphics[height=0.22in]{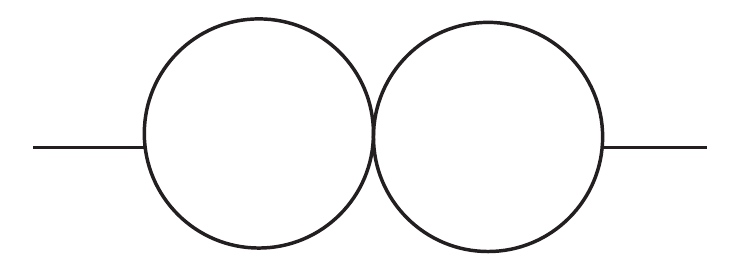}} &$100$&$4$&$1$\\
				&        &$\{1,x_{11},x_{31},x_{41},x_{22},x_{32},x_{42}\}$&$\{1,\lambda_{11},\lambda_{22}\}$&$\{1\}$\\
				\hline
				\multirow{2}{1.4cm}{\centering $\mathcal{I}_{4569}^{\text{P}}$}&\multirow{2}{1.6cm}{\centering\includegraphics[height=0.28in]{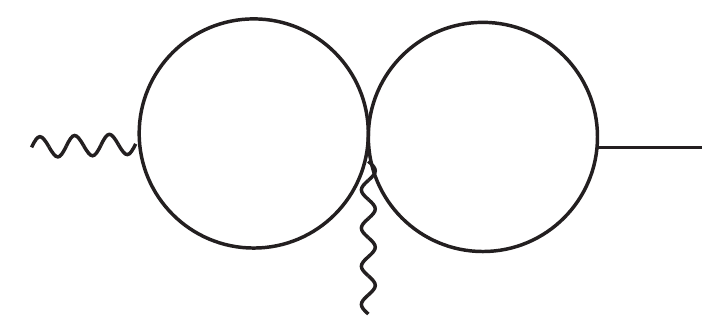}} &$100$&$8$&$3$\\
				&        &$\{1,x_{11},x_{31},x_{41},x_{12},x_{32},x_{42}\}$&$\{1,x_{11},\lambda_{11},\lambda_{22}\}$&$\{1,x_{11}\}$\\
				\hline
				\multirow{2}{1.4cm}{\centering $\mathcal{I}_{4568}^{\text{P}}$}&\multirow{2}{1.6cm}{\centering\includegraphics[height=0.28in]{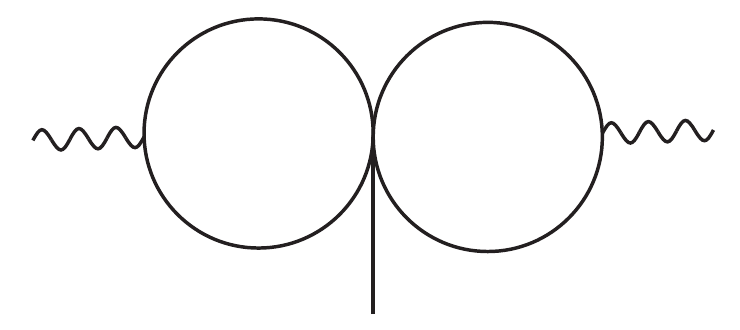}} &$100$&$16$&$9$\\
				&        &$\{1,x_{11},x_{21},x_{31},x_{41},x_{32},x_{42}\}$&$\{1,x_{11},x_{12},\lambda_{11},\lambda_{22}\}$&$\{1,x_{11},x_{12}\}$\\
				\hline
				\multirow{2}{1.4cm}{\centering $\mathcal{I}_{269}^{\text{P}}$}&\multirow{2}{1.6cm}{\centering\includegraphics[height=0.28in]{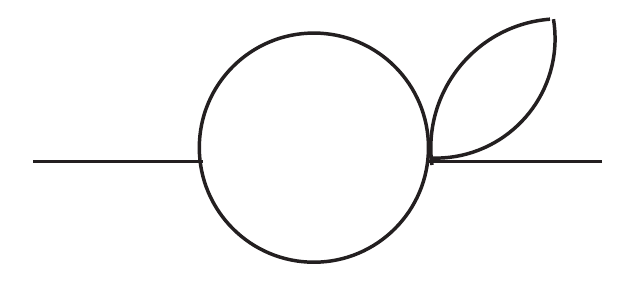}} &$50$&$2$&$1$\\
				&        &$\{1,x_{11},x_{21},x_{31},x_{41},x_{22},x_{32},x_{42}\}$&$\{1,\lambda_{22}\}$&$\{1\}$\\
				\hline
				\multirow{2}{1.4cm}{\centering $\mathcal{I}_{268}^{\text{P}}$}&\multirow{2}{1.6cm}{\centering\includegraphics[height=0.28in]{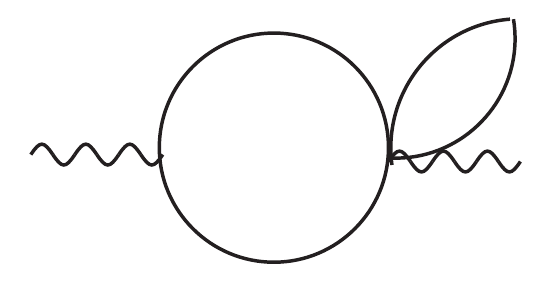}} &$50$&$4$&$3$\\
				&        &$\{1,x_{11},x_{21},x_{31},x_{41},x_{12},x_{32},x_{42}\}$&$\{1,x_{12},\lambda_{22}\}$&$\{x_{12}\}$\\
				\hline
				\multirow{2}{1.4cm}{\centering $\mathcal{I}_{29}^{\text{P}}$}&\multirow{2}{1.6cm}{\centering\includegraphics[height=0.33in]{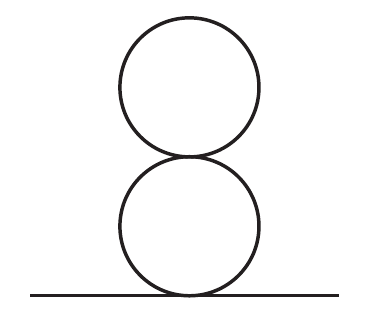}} &$25$&$1$&$-$\\
				&        &$\{1,x_{11},x_{21},x_{31},x_{41},x_{12},x_{22},x_{32},x_{42}\}$&$\{1\}$&$-$\\
				\hline
			\end{tabular}
		}
		\caption{\small{Residue parametrization for factorized two-loop topologies. For every step of the reduction algorithm, we list the number of monomials of each residues and the set of variables appearing in it. In the figures, wavy lines indicate massless particles, whereas solid ones stands for arbitrary masses.
				}}
		\label{Tab:fact}
	\end{table}

%% file: Example.tex
	\clearpage
	\subsubsection{Example: the four-point residue for $A^{2-\text{loop}}(g_1^+,g_2^-,g_3^+,g_4^-)$}
	\label{sec:example}
	As a concrete example of the \textit{adaptive} integrand
    decomposition method, we go back to the helicity amplitude
    $A^{2-\text{loop}}(p_1^+,p_2^-,p_3^+,p_4^-)$ discussed in
    sec.~\ref{sec:esempioAggg} and we compute the residue of the
    double-box topology. We start from the full numerator, which contains {\it 2025} terms up to rank
    four with respect to each loop momentum and rank six in total and we determine the residue three steps:
	\begin{enumerate}[1)]
		\item{\textbf{Divide:}} an easy way to perform the first division step of the procedure consists in observing that, since denominators are independent of $x_{4j}$,
		all coordinates $x_{ij}$ with $i\leq 3$ can be written in terms of differences of
		denominators and irreducible scalar products by solving a linear
		system of equation.  Moreover, the variables $\lambda_{ij}$ can
		also be easily written as combinations of denominators and scalar
		products by solving simple linear relations. After this manipulations, the numerator on the
		cut (\textit{i.e.}\ imposing $D_i=0$) is given by a sum of {\it70} vanishing terms in the
		components of the four dimensional loop momenta $\mathbf{x}=\{\mathbf{x}_{\parallel},\mathbf{x}_{\perp}\}$. The expression of the integrand on the cut found agreement with the results of ~\cite{Badger:2012dp}.
	    \item{\textbf{Integrate:}} after integration over the transverse directions, the numerator acquires again a dependence on
	    $\lambda_{ij}$ and it is expressed a sum of 39 non-vanishing terms in the
	    variables  $\mathbf{x}=\{\mathbf{x}_{\parallel},\lambda_{ij}\}$, whose coefficient now also depend on the dimensional regulator $d$.
	    \item{\textbf{Divide:}} using the same relations as in the first step, the $\lambda_{ij}$ can be expressed in terms of denominators, completing the final division
	    step, after which the numerator of the integrand on the cut is expressed as linear combination of 15 terms depending  on the physical directions $\mathbf{x}_{\parallel}$ left unconstrained by the cut conditions, \textit{i.e.} on the two irreducible scalar products $(q_1 \cdot p_4)$ and $(q_2\cdot p_1)$.
	\end{enumerate}
	Putting everything together, after factoring out a contribution
	proportional to the tree-level result by means of some spinor algebra,
	the gauge invariant decomposition of this cut (\textit{i.e.}\ ignoring contributions
	proportional to denominators) can be written as
	\begin{equation} \label{eq:2ldbcutdec}
	A^{2-\text{loop}}(p_1^+,p_2^-,p_3^+,p_4^-)\Big|_{cut} = i\frac{\la 2\, 4 \ra^4}{\la 1\, 2 \ra \la 2\, 3 \ra \la 3\, 4 \ra \la 4\, 1 \ra}\, \Big(\sum_{\alpha, \beta} c_{\alpha,\beta} \, I_{4}^{d\,(2)}[(q_1 \cdot p_4)^\alpha\, (q_2\cdot p_1)^\beta]  \Big),
	\end{equation}
	where the non-vanishing coefficients $c_{\alpha, \beta}$ only depend
	on the invariants $s_{12}$ and $s_{14}$, as well as on the dimension
	$d$ of the loop integration.  By putting $s_{12}=1$ and $s_{14}=t$ for
	brevity (the dependence on $s_{12}$ is unambiguously determined by
	dimensional analysis) they read
	\begin{align}
	c_{4,0} = {} & -\frac{(d_s-2) (2 t+1)^2}{2 t
		(t+1)^4}-\frac{(d_s-2) \left(2 t^2-2 t-1\right)}{(d-3) t
		(t+1)^4}-\frac{3 (d_s-2)}{2 (d-1) (d-3) t
		(t+1)^4},\nn
	c_{3,1} = {} & -\frac{3 (d_s-2) (2 t+1)}{(d-1) (d-3)
		t (t+1)^4}-\frac{(d_s-2) (2 t+1)}{t (t+1)^4}+\frac{2 (d_s-2)
		\left(4 t^2+2 t+1\right)}{(d-3) t (t+1)^4},\nn
	c_{3,0} = {} & -\frac{(2
		t+1) (d_s-2)}{(t+1)^3}+\frac{2 (d_s-2)}{(d-3)
		(t+1)^2}-\frac{3 (d_s-2)}{(d-1) (d-3) (t+1)^3}, \nn 
	c_{2,2} = {} &
	-\frac{3 (d_s-2) \left(8 t^2+8 t+3\right)}{2 (d-1) (d-3)
		t (t+1)^4}-\frac{32 t^2+32 t+3 (d_s-2)}{2 t (t+1)^4}\nn
	& +\frac{32
		t^3+16 t^2+12 (d_s-2) t-16 t+3 (d_s-2)}{(d-3) t
		(t+1)^4},\nn
	c_{2,1} = {} & -\frac{3 (d_s-2) (4 t+3)}{(d-1) (d-3)
		(t+1)^3}-\frac{(d_s-2)+8 t+4}{(t+1)^3}\nn
	& +\frac{4 \left(8 t^2+2
		(d_s-2) t+2 t+2 (d_s-2)-3\right)}{(d-3)
		(t+1)^3},\nn
	c_{2,0} = {} & -\frac{3 (d_s-2) t (2 t+3)}{2 (d-1)
		(d-3) (t+1)^3}-\frac{(d_s-2) t+8 t+4}{2 (t+1)^2}\nn &+\frac{16
		t^3+7 (d_s-2) t^2+16 t^2+4 (d_s-2) t+4 t+4}{2 (d-3)
		(t+1)^3},\nn
	c_{1,3} = {} & -\frac{3 (d_s-2) (2 t+1)}{(d-1) (d-3)
		t (t+1)^4}-\frac{(d_s-2) (2 t+1)}{t (t+1)^4}+\frac{2 (d_s-2)
		\left(4 t^2+2 t+1\right)}{(d-3) t (t+1)^4},\nn
	c_{1,2} = {} & -\frac{3
		(d_s-2) (4 t+3)}{(d-1) (d-3) (t+1)^3}-\frac{(d_s-2)+8
		t+4}{(t+1)^3}\nn&
	+\frac{4 \left(8 t^2+2 (d_s-2) t+2 t+2
		(d_s-2)-3\right)}{(d-3) (t+1)^3},\nn
	c_{1,1} = {} & -\frac{2 (2
		t+1)}{(t+1)^2}-\frac{3 (d_s-2) t (4 t+3)}{(d-1) (d-3)
		(t+1)^3},\nn
	& +\frac{32 t^3+4 (d_s-2) t^2+32 t^2+7 (d_s-2) t+2
		t+2}{(d-3) (t+1)^3},\nn
	c_{1,0} = {} & -\frac{3 (d_s-2)
		t^2}{(d-1) (d-3) (t+1)^2}+\frac{\left(8 t^2+(d_s-2) t+6
		t+2\right) t}{(d-3) (t+1)^2}-\frac{2 t}{t+1}, \nn
	c_{0,4} = {} &
	-\frac{(d_s-2) (2 t+1)^2}{2 t (t+1)^4}-\frac{(d_s-2) \left(2
		t^2-2 t-1\right)}{(d-3) t (t+1)^4}-\frac{3 (d_s-2)}{2
		(d-1) (d-3) t (t+1)^4},\nn
	c_{0,3} = {} & -\frac{(2 t+1)
		(d_s-2)}{(t+1)^3}+\frac{2 (d_s-2)}{(d-3) (t+1)^2}-\frac{3
		(d_s-2)}{(d-1) (d-3) (t+1)^3},\nn
	c_{0,2} = {} & -\frac{3
		(d_s-2) t (2 t+3)}{2 (d-1) (d-3)
		(t+1)^3}-\frac{(d_s-2) t+8 t+4}{2 (t+1)^2}\nn & +\frac{16 t^3+7
		(d_s-2) t^2+16 t^2+4 (d_s-2) t+4 t+4}{2 (d-3)
		(t+1)^3},\nn
	c_{0,1} = {} & -\frac{3 (d_s-2) t^2}{(d-1) (d-3)
		(t+1)^2}+\frac{\left(8 t^2+(d_s-2) t+6 t+2\right) t}{(d-3)
		(t+1)^2}-\frac{2 t}{t+1},\nn
	c_{0,0} = {} & -\frac{3 (d_s-2) t^3}{4
		(d-1) (d-3) (t+1)^2}+\frac{(2 t+1) t^2}{(d-3)
		(t+1)}-\frac{t}{2},
	\end{align}
	where $d_s$ is the number of dimensions of the internal gluons,
    \textit{i.e.}\ $d_s=d$ in the 't Hooft-Veltman scheme and $d_s=4$
    in the four-dimensional helicity scheme. The integrals appearing
    on the r.h.s.\ of eq.~\eqref{eq:2ldbcutdec}, which only depend on
    the momenta defined by the external kinematic, can then be reduced
    to a minimal set of master integrals by means of traditional
    methods such as integration by parts.  It is worth noticing that,
    while the original integrand had terms up to total rank six in the
    loop momenta, after the reduction the maximum rank is reduced down to four.

%% file: Conclusions.tex
\section{Conclusions}
\label{sec:3}
We presented the integrand reduction of dimensionally regulated integrals in the parallel and orthogonal
space, where the number of space-time dimensions $d = 4 - 2\epsilon$
is decomposed as $d = d_\parallel + d_\perp$. According to the
external topology of each diagram, characterized by the number $n$ of legs, the
parallel space is spanned by the external momenta, $d_\parallel = n-1$,  while the orthogonal
space is spanned by the complementary directions.  For diagrams
with a number of legs $n > 5$, the orthogonal space is generated by the
regulating directions, $d_\perp = -2\epsilon$, while for $n \le 4$, it
embeds also the four-dimensional complement to the parallel space,
namely  $d_\perp = 5 -n -2\epsilon$. 

Owing to this representation of Feynman integrals in
parallel and orthogonal space, numerators and denominators of integrands
with $n\le 4$ appear to
depend on different sets of variables, since the former can depend on
transverse angles which are absent from the latter. 
Therefore, the integration over this subset of transverse variables which do not
appear in the denominators can be carried
out, before any reduction, simply by employing the orthogonality relation of Gegenbauer polynomials.

Because of the reduced number of variables appearing in the
denominators of diagrams with $n\le 4$ legs, the integrand reduction
algorithm, which is based on the multivariate polynomial division, is
simplified. In particular, the Gr\"obner basis generated by the
denominators are linear in the variables reduced by the division
algorithm and the multivariate division is reduced to a mere
substitution of the solutions of a set of linear equations, which is a
consequence of the separation of the physical directions from the
extra-dimensional ones.  Moreover, the residues, namely the remainders
of the polynomial divisions, present a novel simpler structure.  If
the integration over the orthogonal directions is performed before the
reduction, then the residues contain only monomials that correspond to
non-vanishing integrals. On the contrary, if the polynomial division
is applied to the complete numerator, the residues will contain also
spurious monomials. In the latter case, the integration of the
decomposed integrand over the transverse directions by means of
Gegenbauer polynomials automatically detects and annihilates the
spurious terms.

The outcome of the proposed algorithm
is the decomposition of multiloop amplitudes in terms of a
set of integrals which, beside the scalar ones, contains tensor
structures corresponding to irreducible scalar products between loop
momenta and external momenta. These integrals depend on the 
parallel directions and on the lengths of the transverse vectors only.
We have shown that the integration over the transverse angles, which can be
systematically implemented by using Gegenbauer polynomials, plays an
important role in eliminating the superfluous degrees of freedom of 
multiloop integrals any time that a certain subset of integration
variables do not appear in the denominators. We have discussed how, in the
case of factorized diagrams and ladder topologies, such integration can be applied, besides to the transverse angles, to a
larger number of variables. In addition, we have shown that the integration over the transverse directions leads to integrals which
can be subject to additional polynomial divisions, which in some cases
correspond to dimension-shifting
recurrence relations implemented at the integrand level.

We have revisited the one-loop integrand decomposition and we have shown
that it is completely determined by the maximum-cut theorem in
different dimensions.
We have also considered the complete reduction of two-loop planar and
non-planar integrals for arbitrary kinematics, classifying the
corresponding residues and identifying the set of integrals
contributing to the amplitude. 
We have discussed how the whole algorithm can be simply extended to
higher loops, by giving explicit examples of four-point integrals at three loops.

The  dependence of the denominators, hence of the cut-conditions, on a
subset of variables determined by the number of legs suggested us to introduce 
the concept of {\it adaptive cuts}. We
believe that the idea presented in this article of
cutting diagrams in different space-time dimensions, according to the topology under consideration, can be applied, in
general, to any unitarity-based algorithm.

It is known that the number of integrals emerging from the integrand
reduction is not minimal. In fact, because of
the properties of dimensional regularization, the number of integrals appearing in the amplitude decomposition can be further
reduced by applying integration-by-parts identities and ensuing relations.
We believe that novel reduction algorithms, explicitly built for
decomposing integrals that depend on 
parallel directions and on the lengths of the transverse vectors,
may lead to simplified integration-by-parts solving strategies.

As it stands, the proposed variant of a simplified integrand reduction
algorithm can be used in tandem, on the one side, with
automatic diagram generators and, on the other side, with codes dedicated to the automatic
integrals evaluation by means of numerical or semi-analytical
routines.

\section*{Acknowledgments}

We thank Giovanni Ossola for collaboration at earlier stages and fruitful discussions,
Ulrich Schubert and William Torres for useful comparisons
and Andrey Grozin for pointing out the relevant literature on the
orthogonal and transverse decomposition of loop momenta to us.
We wish to acknowledge stimulating discussions with Goran Duplancic,
Hjalte Frellesvig, Ettore Remiddi, Andreas von Manteuffel and Yang
Zhang.

%% file: Appendix.tex
		\section{Spherical coordinates for multiloop integrals}
		\label{Ap:1}
		In this appendix we gibe a derivation the $d=d_{\parallel}+d_{\perp}$ representation \eqref{eq:lambth} of multiloop Feynman integrals,
		\begin{align}
		I_{n}^{d\,(\ell)}[\mathcal{N}]=\int \left(\prod_{1=1}^{\ell}\frac{d^dq_{i}}{\pi^{d/2}}\right)\frac{\mathcal{N}(q_{i})}{\prod_{j}D_{j}(q_{i})}, \qquad n\leq 4,
		\label{eq:lloop}
		\end{align} 
		presented in sec.~\ref{sec:1}. We provide explicit formulae up to three loops and we show how these results can be extended to higher orders.	
		We start by studying the properties of a set of auxiliary integrals that we will later identify with the integrals over the transverse space.
		\begin{itemize}
		\item
		For one-loop calculations, it is useful to consider integrals of the type
		\begin{align}
		I_1=\int\! d^m\!\boldsymbol{\lambda}_1\;\mathcal{I}_1(\boldsymbol{\lambda}_1),
		\label{1lvAp}
		\end{align}
		where  $\boldsymbol{\lambda}_1$ is a vector of an Euclidean space, whose dimension $m$ is first assumed to be an integer and the analytically continued to complex values. We suppose $\boldsymbol{\lambda}_1$ to be decomposed with respect to an orthonormal basis $\{\mathbf{v}_i\}$ as
		\begin{align}
		\label{laAp}
		&\boldsymbol{\lambda}_1=\sum_{i=1}^{m}a_{i1}\mathbf{v}_i.
		\end{align}
		Regardless of the symmetries of the integrand, we can reparametrize $I_1$ in terms of spherical coordinates in $m$ dimensions which, being $\{\mathbf{v}_i\}$ orthonormal, are defined by the well-known change of variables
		\begin{equation}
		\label{cartAp}
		\begin{cases}
		a_{11}&=\sqrt\lambda_{11} \cos\theta_{11},\\
		&\cdots\\
		a_{k1}&=\sqrt\lambda_{11}\cos\theta_{k1}\prod_{i=1}^{k-1} \sin\theta_{i1}\\
		&\cdots\\
		a_{m1}&=\sqrt\lambda_{11}\prod_{i=1}^{m-1} \sin\theta_{i1},
		\end{cases}
		\end{equation}
		where $\sqrt\lambda_{11}\in[0,\infty)$ and all angles range over the interval $[0,\pi]$, except for $\theta_{m-1\,1}\in[0,2\pi]$. Hence, by introducing the differential solid angle in $M$ dimensions
		\begin{align}
		d\Omega_{M-1}=(\sin\theta_1)^{M-3}d\!\cos\theta_1(\sin\theta_2)^{M-4}d\!\cos\theta_2\,\dots\,d\theta_{M-1},
		\end{align}
		such that
		\begin{align}
		\label{solid}
		\Omega_{M-1}=\int\! d\Omega_{M-1}=\frac{2\pi^{\frac{M}{2}}}{\Gamma\left(\frac{M}{2}\right)},
		\end{align}
		we can write \eqref{1lvAp} as
		\begin{align}
		I_1=\frac{1}{2}\int_{0}^{\infty}d\lambda_{11} (\lambda_{11})^{\frac{m-2}{2}}\int d\Omega_{m-1}\mathcal{I}_1(\lambda_{11},\cos\theta_{i1},\sin\theta_{i1}).
		\label{pol1}
		\end{align}
		If the integrand is rotational invariant, \textit{i.e.} it depends on $\lambda_{11}=\boldsymbol{\lambda}_1\cdot\boldsymbol{\lambda}_1$ only, we can integrate over all angular variables in such a way to obtain, by specifying \eqref{solid} for $M=m$
		\begin{align}
		I_1=\frac{\pi^{\frac{m}{2}}}{\Gamma\left(\frac{n}{2}\right)}\int_{0}^{\infty}d\lambda_{11} (\lambda_{11})^{\frac{m-2}{2}}\mathcal{I}_1(\lambda_{11}).
		\label{polrot}
		\end{align}
		However, in general one-loop applications, the integrand can show an explicit dependence on a subset of $\kappa<m-1$ components of $\boldsymbol\lambda_1$ which, with a suitable definition of the reference frame, can always be chosen to correspond to $\{a_{11},\,\dots,\, a_{\kappa1}\}$. In this way, according to \eqref{cartAp}, the integrand will depend only on $\boldsymbol{\Lambda}=\{\lambda_{11}\}$ and $\boldsymbol{\Theta}_{\perp}=\{\theta_{11},\,\dots,\, \theta_{\kappa1}\}$ while all angles $\theta_{i1}$ with $i>\kappa$ can be still integrated out by using \eqref{solid} with $M=m-\kappa$,
		\begin{align}
		I_{1}^{\kappa}=
		\Omega_{(m-\kappa-1)}\int d\boldsymbol{\Lambda}\int d^{\kappa}\boldsymbol{\Theta}_{\perp}\mathcal{I}_{1}(\boldsymbol{\Lambda},\boldsymbol{\Theta}_{\perp}),
		\label{meas1lAp}
		\end{align}
		with
		\begin{align}
		\int d\boldsymbol{\Lambda}=&\int_{0}^{\infty}d\lambda_{11}(\lambda_{11})^{\frac{m-2}{2}},\qquad \int d^{\kappa}\boldsymbol{\Theta}_{\perp}=&\prod_{i=1}^{\kappa}\int_{-1}^{1}d\cos\theta_{i1}(\sin\theta_{i1})^{m-i-2}.
		\label{meas1lpc}
		\end{align}
		\item
		In two-loop computations, we encounter multiple integrals of type
		\begin{align}
		I_2=\int\! d^m\! \boldsymbol{\lambda}_1 d^m \!\boldsymbol{\lambda}_2 \mathcal{I}(\boldsymbol{\lambda}_1,\boldsymbol{\lambda}_2),
		\label{2lvAP}
		\end{align}
		where we suppose the two vectors $\boldsymbol{\lambda}_i$ to be decomposed in terms of the same orthonormal basis $\{\mathbf{v}_i\}$,
		\begin{subequations}
		\begin{align}
		\label{la1Ap}
		&\boldsymbol{\lambda}_{1}=\sum_{i=1}^{m}a_{i1}\mathbf{v}_i,\\
		\label{la2Ap}
		&\boldsymbol{\lambda}_{2}=\sum_{i=1}^{m}a_{i2}\mathbf{v}_i.
		\end{align}
		\end{subequations}
		Analogously to the one-loop case, we would like to map all integrals associated to a subset of $\kappa$ components of each vectors $\boldsymbol{\lambda}_i$ into angular integrals. For $I_1$, due to the choice of an orthonormal basis, this mapping was immediately achieved by parametrizing the integral in terms of spherical coordinates.
		In this case, there is an additional direction, corresponding to $\lambda_{12}=\boldsymbol{\lambda}_1\cdot \boldsymbol{\lambda}_2$, we need to trace back after the change of coordinates is performed, since the integrand will  generally depend on it. The simultaneous factorization of the integral over the relative orientation $\lambda_{12}$ and over all relevant components of the two vectors can be obtained by expressing $\boldsymbol{\lambda}_2$ into a new orthonormal basis $\{\mathbf{e}_i\}$, containing the vector $\mathbf{e}_1\propto\boldsymbol{\lambda}_1$. From \eqref{la1Ap} we see that, indeed, the set of vectors
		\begin{align}
		\{\mathbf{v}'_i\}=\{\boldsymbol{\lambda}_1,\mathbf{v}_1,...,\mathbf{v}_{m-1}\}
		\end{align}
		is a basis, although it is not an orthogonal one. Nevertheless, we can apply the Gram-Schimdt algorithm to pass from the arbitrary basis $\{\mathbf{v}'_i\}$ to an orthonormal one $\{\mathbf{e}_i\}$, given by
		\begin{align}
		\label{basise}
		&\mathbf{e}_1=\frac{\mathbf{u}_1}{|\mathbf{u}_1|},\qquad \mathbf{u}_1=\mathbf{v}'_1,\nn
		&\mathbf{e}_k=\frac{\mathbf{u}_k}{|\mathbf{u}_k|},\qquad \mathbf{u}_k=\mathbf{v}'_k-\sum_{j=1}^{k-1}(\mathbf{v}'_{k}\cdot\mathbf{e}_{j})\mathbf{e}_{j}, \qquad k\neq 1.
		\end{align}
		By construction, the first vector of the new basis exactly corresponds to the direction of $\boldsymbol{\lambda}_1$. Applying the change of basis to \eqref{la2Ap}, we get
		\begin{align}
		&\boldsymbol{\lambda}_{2}=\sum_{i=1}^{m}b_{i2}\mathbf{e}_i,
		\end{align}
		where the coefficients $\{b_{i2}\}$ are related to the components of both $\boldsymbol{\lambda}_1$ and $\boldsymbol{\lambda}_2$ with respect to $\{\mathbf{v}_i\}$ by
		\begin{align}
		\begin{cases}
		b_{12}&=\frac{\lambda_{12}}{\sqrt{\lambda_{11}}}\\
		b_{22}&=\frac{a_{12}\lambda_{11}-a_{11}\lambda_{12}}{\sqrt{\lambda_{11}}\sqrt{\lambda_{11}-a_{11}^2}}\\
		&\cdots\\ b_{k2}&=\frac{a_{k-1\,2}\big(\lambda_{11}-\sum_{i=1}^{k-2}a_{i1}^2\big)-a_{k-1\,1}\big(\lambda_{12}-\sum_{i=1}^{k-2}a_{i1}b_{i2}\big)}{\sqrt{\lambda_{11}-\sum_{i=1}^{k-2}a_{i1}^2}\sqrt{\lambda_{11}-\sum_{i=1}^{k-1}a_{i1}^2}}\\
		&\cdots\\
		b_{m2}=&\frac{a_{m1}a_{m-1\,2}-a_{m-1\,1}a_{m\,2}}{\sqrt{a_{m1}^2+a_{m-1\,1}^2}}.
		\label{newcomp}
		\end{cases}
		\end{align}
		Since both $\boldsymbol{\lambda}_1$ and $\boldsymbol{\lambda}_2$ are now decomposed in two different but still orthonormal basis, we can introduce the change of variables
		\begin{equation}
		\begin{cases}
		a_{11}&=\sqrt{\lambda_{11}} \cos\theta_{11}\\
		&\cdots\\
		a_{k1}&=\sqrt{\lambda_{11}}\cos\theta_{k1}\prod_{i=1}^{k-1}\sin\theta_{i1}\\
		&\cdots\\
		a_{m1}&=\sqrt{\lambda_{11}}\prod_{i=1}^{m-1} \sin\theta_{i1}
		\end{cases} \qquad
		\begin{cases}
		b_{12}&=\sqrt{\lambda_{22}}\cos\theta_{12}\\
		&\cdots\\
		b_{k2}&=\sqrt{\lambda_{22}}\cos\theta_{k2} \prod_{i=1}^{k-1}\sin\theta_{i2}\\
		&\cdots\\
		b_{m2}&=\sqrt{\lambda_{22}}\prod_{i=1}^{m-1} \sin\theta_{i2}
		\end{cases}
		\label{cart2lc}
		\end{equation}
		and express the integral $I_2$ into spherical coordinates as
		\begin{align}
		I_2=&\frac{1}{4}\int_{0}^{\infty}d\lambda_{11}	d\lambda_{22}(\lambda_{11})^{\frac{m-2}{2}}
	(\lambda_{22})^{\frac{m-2}{2}}\int d\Omega_{(m-1)}d\Omega_{(m-1)}\mathcal{I}_2(\lambda_{ij},\cos\theta_{ij},\sin\theta_{ij}).
		\label{pol2}
		\end{align}
		By combining \eqref{newcomp} with the transformation \eqref{cart2lc}, we immediately see that, as expected,
		\begin{align}
		\lambda_{12}=\sqrt{\lambda_{11}\lambda_{22}}\cos\theta_{12}.
		\end{align}
		In addition, with some more algebra, we can express back the components of $\boldsymbol{\lambda}_2$ with respect to $\{\mathbf{v}_i\}$ in terms of the angular variables,
		\begin{equation}
		\resizebox{0.9\hsize}{!}
		{$
		\begin{cases}
		a_{12}&=\sqrt{\lambda_{22}}\big(\cos\theta_{12}\cos\theta_{11}+\cos\theta_{22}\sin\theta_{11}\sin\theta_{12}\big)\\
		a_{i2}&=\sqrt{\lambda_{22}}\big[\cos\theta_{12}\cos\theta_{i1}\prod_{j=1}^{i-1}\sin\theta_{j1}+\cos\theta_{i+1\,2}\sin\theta_{i1}\prod_{j=1}^{i}\sin\theta_{j2}\\
		&-\cos\theta_{i1}\sum_{k=2}^{i}\cos\theta_{k2}\cos\theta_{k-1\,1}\prod_{j=1}^{k-1}\sin\theta_{j2}\big(\delta_{ik}+(1-\delta_{ik})\prod_{l=1}^{i-k}\sin\theta_{k+l-1\,1}\big)\big],\quad i\neq 1.
		\end{cases}
		\label{b}
		$}
		\end{equation}
		In this way, the integral over each component $a_{i1}\notin\{a_{m-1\,1},a_{m\,1}\}$ of $\boldsymbol{\lambda}_1$ is mapped into the integral over the angular variable $\theta_{i1}$ whereas each component $a_{i2}\notin\{a_{m-1\,2},a_{m\,2}\}$ of $\boldsymbol{\lambda}_2$ can be expressed in terms of the angles $\theta_{j1}$ with $j\leq i$ and $\theta_{j2}$ with $j\leq i+1$. Therefore, if we are dealing with and integrand depending on $\kappa<m-1$ components of both vectors, which we can always choose to correspond to $\{a_{11},\,\cdots,\, a_{\kappa1}\}$ and $\{a_{12},\,\cdots,\, a_{\kappa2}\}$, we can integrate out all angular variables $\theta_{i1}$, $j>\kappa$ and $\theta_{i2}$, $j>\kappa+1$. Hence, if we define
		\begin{align}
		\boldsymbol{\Theta}_{\Lambda}=&\{\theta_{12}\},\nn
		\boldsymbol{\Theta}_{\perp}=&\{\theta_{11},\,\dots,\,\theta_{\kappa1},\theta_{22},\,\dots,\,\theta_{\kappa+1\, 2}\},
		\end{align} 
		we can rewrite $I_2$ as
		\begin{align}
		I_2^{\kappa}=&
		\Omega_{(m-\kappa-1)}\Omega_{(m-\kappa-2)}\int d^3\boldsymbol{\Lambda}\int d^{2\kappa}\boldsymbol{\Theta}_{\perp}\mathcal{I}_2(\boldsymbol{\Lambda},\boldsymbol{\Theta}_{\perp}),
		\label{meas2lAp}
		\end{align}
		with
		\begin{align}
		\int d^3\boldsymbol{\Lambda}=&\int_{0}^{\infty}d\lambda_{11}d\lambda_{22}(\lambda_{11})^{\frac{m-2}{2}}(\lambda_{22})^{\frac{m-2}{2}}\int d\boldsymbol{\Theta}_{\Lambda},\nn
		\int d\boldsymbol{\Theta}_{\Lambda}=&\int_{-1}^{1}d\cos\theta_{12}(\sin\theta_{12})^{m-3},\nn
		\int d^{2\kappa}\boldsymbol{\Theta}_{\perp}=&\int_{-1}^{1}\prod_{i=1}^{\kappa}d\!\cos\theta_{i1}d\!\cos\theta_{i+1\,2}(\sin\theta_{i1})^{m-i-2}(\sin\theta_{i+1\,2})^{m-i-3}.
		\label{meas2lAppc}
		\end{align}
		\item
		For three-loop applications, we consider integrals of the type
		\begin{align}
		I_3=\int d^m \boldsymbol{\lambda}_1 d^m \boldsymbol{\lambda}_2 d^m \boldsymbol{\lambda}_3 \mathcal{I}_{3}(\boldsymbol{\lambda}_1,\boldsymbol{\lambda}_2,\boldsymbol{\lambda}_3)
		\label{3lvAP}
		\end{align}
		and, as usual, we assume the vectors $\boldsymbol{\lambda}_i$ to be initially decomposed in terms of the same orthonormal basis $\{\mathbf{v}_i\}$,
			\begin{align}
			\label{la1Ap3}
			\boldsymbol{\lambda}_{1}=\sum_{i=1}^{m}a_{i1}\mathbf{v}_i,\qquad
			\boldsymbol{\lambda}_{2}=\sum_{i=1}^{m}a_{i2}\mathbf{v}_i,\qquad
		    \boldsymbol{\lambda}_{3}=\sum_{i=1}^{m}a_{i3}\mathbf{v}_i.
			\end{align}
		When moving to spherical coordinates, we want to keep trace of the three relative orientations
		\begin{align}
		\lambda_{12}=\boldsymbol{\lambda}_1\cdot \boldsymbol{\lambda}_2,\quad \lambda_{23}=\boldsymbol{\lambda}_2\cdot \boldsymbol{\lambda}_3,\quad\lambda_{31}=\boldsymbol{\lambda}_3\cdot \boldsymbol{\lambda}_1,
		\end{align}
		together with the usual subset of $\kappa$ components of each $\boldsymbol{\lambda}_i$. The proper change of variables can be reach in two steps:
		\begin{enumerate}[1)]
			\item First we express the vectors $\boldsymbol{\lambda}_2$ and $\boldsymbol{\lambda}_3$ in terms of the basis $\{\mathbf{e}_i\}$ defined by eq.\eqref{basise}, which contains the vector $\mathbf{e}_1\propto \boldsymbol{\lambda}_1$,
			\begin{subequations}
			\begin{align}
			\label{labda2e}
			&\boldsymbol{\lambda}_{2}=\sum_{i=1}^{m}b_{i2}\mathbf{e}_i,\\
			&\boldsymbol{\lambda}_{3}=\sum_{i=1}^{m}b_{i3}\mathbf{e}_i,
			\end{align}
		    \end{subequations}
		where, similarly to \eqref{newcomp}, $\{b_{i2}\}$ and $\{b_{i3}\}$ are defined in terms of the components with respect to the basis $\{\mathbf{v}_i\}$ as
		\begin{equation}
		\resizebox{0.9\hsize}{!}
		{$
		\begin{cases}
		b_{12}&=\frac{\lambda_{12}}{\sqrt{\lambda_{11}}}\\
		b_{22}&=\frac{a_{12}\lambda_{11}-a_{11}\lambda_{12}}{\sqrt{\lambda_{11}}\sqrt{\lambda_{11}-a_{11}^2}}\\
		&\cdots\\ b_{k2}&=\frac{a_{k-1\,2}\big(\lambda_{11}-\sum_{i=1}^{k-2}a_{i1}^2\big)-a_{k-1\,1}\big(\lambda_{12}-\sum_{i=1}^{k-2}a_{i1}b_{i2}\big)}{\sqrt{\lambda_{11}-\sum_{i=1}^{k-2}a_{i1}^2}\sqrt{\lambda_{11}-\sum_{i=1}^{k-1}a_{i1}^2}}\\
		&\cdots\\
		b_{m2}&=\frac{a_{m1}a_{m-1\,2}-a_{m-1\,1}a_{m2}}{\sqrt{a_{m1}^2+a_{m-1\,1}^2}},
		\end{cases}
		\quad
		\begin{cases}
		b_{13}&=\frac{\lambda_{13}}{\sqrt{\lambda_{11}}}\\
		b_{23}&=\frac{a_{13}\lambda_{11}-a_{11}\lambda_{13}}{\sqrt{\lambda_{11}}\sqrt{\lambda_{11}-a_{11}^2}}\\
		&\cdots\\ b_{k3}&=\frac{a_{k-1\,3}\big(\lambda_{11}-\sum_{i=1}^{k-2}a_{i1}^2\big)-a_{k-1\,1}\big(\lambda_{13}-\sum_{i=1}^{k-2}a_{i1}a_{i3}\big)}{\sqrt{\lambda_{11}-\sum_{i=1}^{k-2}a_{i1}^2}\sqrt{\lambda_{11}-\sum_{i=1}^{k-1}a_{i1}^2}}\\
		&\cdots\\
		b_{m3}&=\frac{a_{m1}a_{m-1\,3}-a_{m-1\,1}a_{m3}}{\sqrt{a_{m1}^2+a_{m-1\,1}^2}}.
		\label{newcomph}
		\end{cases}
		$}
		\end{equation}
		\item Then we use the fact that the vectors
		\begin{align}
		\mathbf{e}_i'=\{\boldsymbol{\lambda}_2,\mathbf{e}_1,\,\dots,\,\mathbf{e}_{m-1}\}
		\end{align}
		form a (non-orthogonal) basis which can be orthogonalized by applying the Gram-Schmidt algorithm in such a way to obtain an orthonormal basis $\{\mathbf{f}_i\}$,
		\begin{align}
		\label{basisf}
		&\mathbf{f}_1=\frac{\mathbf{w}_1}{|\mathbf{w}_1|},\qquad \mathbf{w}_1=\mathbf{e}'_1,\nn
		&\mathbf{f}_k=\frac{\mathbf{w}_k}{|\mathbf{w}_k|},\qquad \mathbf{w}_k=\mathbf{e}'_k-\sum_{j=1}^{k-1}(\mathbf{e}'_{k}\cdot\mathbf{f}_{j})\mathbf{f}_{j}, \qquad k\neq 1,
		\end{align}
		whose first element is $\mathbf{f}_1\propto\boldsymbol{\lambda}_2$, and decompose $\boldsymbol{\lambda}_3$ as
		\begin{align}
		\label{l3deco}
		\boldsymbol{\lambda}_{3}=\sum_{i=1}^{m}c_{i3}\mathbf{f}_i,
		\end{align}
		with
		\begin{align}
		\begin{cases}
		c_{13}&=\frac{\lambda_{23}}{\sqrt{\lambda_{22}}}\\
		c_{23}&=\frac{b_{13}\lambda_{22}-b_{12}\lambda_{23}}{\sqrt{\lambda_{22}}\sqrt{\lambda_{22}-b_{12}^2}}\\
		&\cdots\\ c_{k3}&=\frac{b_{k-1\,3}\big(\lambda_{22}-\sum_{i=1}^{k-2}b_{i2}^2\big)-b_{k-1\,2}\big(\lambda_{23}-\sum_{i=1}^{k-2}b_{i2}b_{i3}\big)}{\sqrt{\lambda_{22}-\sum_{i=1}^{k-2}b_{i2}^2}\sqrt{\lambda_{22}-\sum_{i=1}^{k-1}b_{i2}^2}}\\
		&\cdots\\
		c_{m3}&=\frac{b_{m2}b_{m-1\,3}-b_{m-1\,2}b_{m\,3}}{\sqrt{b_{m2}^2+b_{m-1\,2}^2}}.
		\label{newcompw}
		\end{cases}
		\end{align}
		\end{enumerate}
		 Eqs.\eqref{la1Ap3}, \eqref{labda2e} and \eqref{l3deco} give us a decomposition of the three vectors $\boldsymbol{\lambda}_i$ in terms of three different but still orthonormal basis. Hence, we can introduce spherical coordinates
		\begin{equation}
		\resizebox{0.95\hsize}{!}
		{$
		\begin{cases}
		a_{11}&=\sqrt{\lambda_{11}} \cos\theta_{11}\\
		&\cdots\\
		a_{k1}&=\sqrt{\lambda_{11}}\cos\theta_{k1}\prod_{i=1}^{k-1}\sin\theta_{i1}\\
		&\cdots\\
		a_{m1}&=\sqrt{\lambda_{11}}\prod_{i=1}^{m-1} \sin\theta_{i1},
		\end{cases}
		\quad
		\begin{cases}
		b_{12}&=\sqrt{\lambda_{22}}\cos\theta_{12}\\
		&\cdots\\
		b_{k2}&=\sqrt{\lambda_{22}}\cos\theta_{k2} \prod_{i=1}^{k-1}\sin\theta_{i2}\\
		&\cdots\\
		b_{m2}&=\sqrt{\lambda_{22}}\prod_{i=1}^{m-1} \sin\theta_{i2}
		\end{cases}
		\quad
		\begin{cases}
		c_{13}&=\sqrt{\lambda_{33}}\cos\theta_{23}\\
		&\cdots\\
		c_{k3}&=\sqrt{\lambda_{33}}\cos\theta_{k3} \prod_{i=1}^{k-1}\sin\theta_{i3}\\
		&\cdots\\
		c_{m3}&=\sqrt{\lambda_{33}}\prod_{i=1}^{m-1} \sin\theta_{i3}
		\end{cases}
		$}
		\end{equation} 
		and rewrite $I_3$ as
		 \begin{align}
		 I_3=&\frac{1}{8}\int_{0}^{\infty}d\lambda_{11}(\lambda_{11})^{\frac{m-2}{2}}\int_{0}^{\infty}d\lambda_{22}(\lambda_{22})^{\frac{m-2}{2}}\int_{0}^{\infty}d\lambda_{33}(\lambda_{33})^{\frac{m-2}{2}}\times\nn
		 &\int d\Omega_{(m-1)}\int d\Omega_{(m-1)}\int d\Omega_{(m-1)}\mathcal{I}_3(\lambda_{ij},\cos\theta_{ij},\sin\theta_{ij}).
		 \label{pol3L}
		 \end{align}
		By construction, the relative orientations of between the vectors $\boldsymbol{\lambda}_i$ are mapped into 
		\begin{align}
		&\lambda_{12}=\sqrt{\lambda_{11}\lambda_{22}}\cos\theta_{12},\nn
		&\lambda_{23}=\sqrt{\lambda_{22}\lambda_{33}}\cos\theta_{13},\nn
		&\lambda_{31}=\sqrt{\lambda_{11}\lambda_{33}}\left(\cos\theta_{12} \cos\theta_{13}+\sin\theta_{12} \sin\theta_{13} \cos\theta_{23}\right),
		\end{align}
		and by inverting \eqref{newcomph} and \eqref{newcompw} one can obtain the expressions of $\{a_{i2}\}$ and $\{a_{i3}\}$ as polynomials in (\textit{sine} and \textit{cosine} of) the angular variables. In particular, one can verify that, as in all previous cases, each integral over $a_{i1}\notin\{a_{m-1\,1},a_{m\,1}\}$ is mapped into the integral over the angular variable $\theta_{i1}$ and, as we have seen for $I_2$, each component $a_{i2}\notin\{a_{m-1\,2},a_{m\,2}\}$ can be expressed in terms of the angles $\theta_{j1}$ with $j\leq i$ and $\theta_{j2}$ with $j\leq i+1$. Moreover, each $a_{i3}\notin\{a_{m-1\,3},a_{m\,3}\}$ turns out to be function of the angles $\theta_{j1}$ with $j\leq i$, $\theta_{j2}$ with $j\leq i+1$ and $\theta_{j3}$ with $j\leq i+2$. Therefore, if we are dealing with and integrand depending on $\kappa<m-1$ components of all $\boldsymbol{\lambda}_i$, which can be always chosen to correspond to the first $\kappa$ ones, we can integrate out all angular variables $\theta_{i1}$, $j>\kappa$, $\theta_{i2}$, $j>\kappa+1$ and $\theta_{i3}$, $j>\kappa+2$ and obtain
		\begin{align}
		I_3^{\kappa}=&
		\prod_{i=1}^{3}\Omega_{(m-\kappa-i)}\int d^6\boldsymbol{\Lambda}\int d^{3\kappa}\boldsymbol{\Theta}_{\perp}
		\mathcal{I}_3(\boldsymbol{\Lambda},\boldsymbol{\Theta}_{\perp}),
		\label{meas3lAp}
		\end{align}
		with
		\begin{align}
		\int d^6\boldsymbol{\Lambda}=&\int_{0}^{\infty}d\lambda_{11}d\lambda_{22}d\lambda_{33}(\lambda_{11})^{\frac{n-2}{2}}(\lambda_{22})^{\frac{n-2}{2}}(\lambda_{33})^{\frac{m-2}{2}}\int d^3\boldsymbol{\Theta}_{\Lambda},\nn
		\int d^3\boldsymbol{\Theta}_{\Lambda}=&\int_{-1}^{1}\!d\!\cos\theta_{12}d\!\cos\theta_{13}d\!\cos\theta_{23}(\sin\theta_{12})^{n-3}(\sin\theta_{12})^{m-3}(\sin\theta_{23})^{n-4},\nn
		\int d^{3\kappa}\boldsymbol{\Theta}_{\perp}=&\int_{-1}^{1}\prod_{i=1}^{\kappa}d\cos\theta_{i1}d\cos\theta_{i+1\,2}d\cos\theta_{i+2\,3}\times\nn
		&(\sin\theta_{i1})^{m-i-2}(\sin\theta_{i+1\,2})^{m-i-3}(\sin\theta_{i+2\,3})^{m-i-4}.
		\label{meas3lAppc}
		\end{align}
		\item
		It is now clear how integrals $I_{\ell}$ involving a number $\ell$ of vectors $\boldsymbol{\lambda}_i$,
		\begin{align}
		I_{\ell}=\int\prod_{i=1}^{\ell}d^{m}\boldsymbol{\lambda}_i\mathcal{I}_{\ell}(\boldsymbol{\lambda}_j),
		\end{align} 
		can be treated in order to define a change of variable which maps a subset of $\kappa$ components of each vector as well as their $\ell(\ell-1)$ relative directions $\lambda_{ij}$ into angular variables. Starting from the decomposition of all vectors in terms of a single orthonormal basis, one can define, by recursively applying the Gram-Schimdt algorithm, $\ell-1$ auxiliary orthonormal basis carrying information both on $\kappa\leq m-1$ directions of the original basis and on the relative orientations $\lambda_{ij}$. After all vectors have been decomposed into the proper orthonormal basis, we can introduce $m$-dimensional polar coordinates and, by inverting the nested chain of transformations, we can obtain the expression of the components of all $\boldsymbol{\lambda}_i$ with respect to $\{\mathbf{v}_i\}$ in terms of the angular variables. The final transformation has the form
		 \begin{align}
		 \begin{cases}
		 \lambda_{ij}\to& P\left[\lambda_{ll},\sin[\boldsymbol{\Theta}_{\Lambda}],\cos[\boldsymbol{\Theta}_{\Lambda}]\right],\qquad\quad i\neq j\\
		 a_{ji}\to& P\left[\lambda_{ll},\sin[\boldsymbol{\Theta}_{\perp,\,\Lambda}],\cos[\boldsymbol{\Theta}_{\perp,\,\Lambda}]\right],\quad  j\leq \kappa,
		 \end{cases}
		 \label{eq:poltrap}
		 \end{align}
		 where $\boldsymbol{\Theta}_{\Lambda}$ and $\boldsymbol{\Theta}_{\perp}$ label the sets of angular variables
		 \begin{align}
		 \boldsymbol{\Theta}_{\Lambda}=&\{\theta_{ij}\},\qquad\quad 1\leq i<j\leq \ell,\nn
		 \boldsymbol{\Theta}_{\perp}=&\{\theta_{ij}\}, \qquad j\leq i\leq \ell+\kappa-1,\:\;1\leq j\leq \ell.
		 \end{align}
		 Therefore, if the integrand $\mathcal I_{\ell}$ only depends on $\kappa$ components of each $\boldsymbol{\lambda}_i$, all angles $\theta_{ij}$, $i\geq j+\kappa$ can be integrated out, producing
		\begin{align}
		I_\ell^{\kappa}=\prod_{i=1}^{\ell}\Omega_{(m-\kappa-i)}\int d^{\frac{\ell(\ell-1)}{2}}\boldsymbol{\Lambda}\int d^{\ell\kappa}\boldsymbol{\Theta}_{\perp}\mathcal{I}_\ell(\boldsymbol{\Lambda},\boldsymbol{\Theta}_\perp),
		\label{eq:Il}
		\end{align}
		where
		\begin{align}
		\int d^{\frac{\ell(\ell+1)}{2}} \boldsymbol{\Lambda}=&\int_{0}^{\infty}\!\prod_{i=1}^{\ell}d\lambda_{ii}(\lambda_{ii})^{\frac{m-2}{2}}
		\int \!d^{\frac{\ell(\ell-1)}{2}}\boldsymbol{\Theta}_{\Lambda},\nn
		\int \!d^{\frac{\ell(\ell-1)}{2}}\boldsymbol{\Theta}_{\Lambda}=&\int_{-1}^{1}\!\prod_{1\leq i<j\leq \ell}^\ell \!d\!\cos\theta_{ij}(\sin\theta_{ij})^{m-2-i},\nn
		\int d^{\ell\kappa}\boldsymbol{\Theta}_{\perp}=&\int_{-1}^{1}\prod_{i=1}^{\kappa}\prod_{j=1}^{\ell}\!d\!\cos\theta_{i+j-1\;j}(\sin\theta_{i+j-1\;j})^{m-i-j-1}.
		\end{align}
		\end{itemize}
		We can now go back to an arbitrary $\ell$ loop integral with $n\leq 4$ external legs and, after introducing the $q_{i}^{\alpha}=q_{\parallel\, i}^{\alpha}+\lambda^{\alpha}_{i}$ parametrization of the loop momenta, we can rewrite eq.~\eqref{eq:lloop} as
		\begin{align}
		I_{n}^{d\,(\ell)}[\mathcal{N}]=\int \prod_{1=1}^{\ell}\frac{d^{n-1}q_{\parallel\,i}}{\pi^{d/2}}\int d^{d-n-1}\lambda_{i}\frac{\mathcal{N}(q_{\parallel\,i}^{\alpha},\lambda_i^{\alpha})}{\prod_{j}D_{j}(q_{\parallel\,i}^{\alpha},\lambda_{ij})},
		\label{eq:par/lamb}
		\end{align}
		where we have explicitly indicated that the denominators depend on the $d_{\parallel}$-dimensional momenta $q_{\parallel\,i}$ and on the scalar products $\lambda_{ij}$ between the transverse vectors living in $d_{\perp}$ dimensions.
		We now observe that the numerator can additionally depend only on the four-dimensional components of each $\lambda_{i}^{\alpha}$,
		\begin{align}
		\mathcal{N}(q_{\parallel\,i}^{\alpha},\lambda_i^{\alpha})\equiv \mathcal{N}(q_{\parallel\,i}^{\alpha},\lambda_{ij},x_{d_{\parallel}+1\,i},\,\dots,\,x_{4i}).
		\end{align}
		Therefore, the integral over the transverse vectors $\lambda^{\alpha}$ corresponds to a $d_{\perp}$-dimensional integral of the type $I_{\ell}^{\kappa}$ with $\kappa=4-d_{\parallel}$ so that, by substituting \eqref{eq:Il} in \eqref{eq:par/lamb}, we obtain
		\begin{align}
		I_{n}^{d\,(\ell)}[\mathcal{N}]=&\Omega^{(\ell)}_d\!\int\prod_{i=1}^{\ell} d^{n-1}q_{\parallel \, i}\int\! d^{\frac{\ell(\ell+1)}{2}} \boldsymbol{\Lambda}\int\! d^{(4-d_\parallel)\ell}\boldsymbol{\Theta}_{\perp}\frac{\mathcal{N}(q_{i \,\parallel},\boldsymbol{\Lambda},\boldsymbol{\Theta}_{\perp})}{\prod_{j}D_{j}(q_{ \parallel\, i},\boldsymbol{\Lambda})},\nn
		\Omega^{(\ell)}_d=&\prod_{i=1}^{\ell}\frac{\Omega_{(d-4-i)}}{2\pi^{\frac{d}{2}}},
		\end{align}
		which reproduces \eqref{eq:lambth}.
		\section{One-loop integrals}
		\label{Ap:1l}
		In this appendix we collect some useful formulae for one-loop integrals in $d=d_{\parallel}+d_{\perp}$. In order to make the notation more intuitive, we hereby indicate as $q_{[d_{\parallel}]}^{\alpha}$ the component of the loop momentum lying in the space spanned by the $d_{\parallel}$ independent external momenta and we denote by $\lambda_{[d_{\perp}]}^{\alpha}$ ($\lambda^2\equiv \lambda_{[d_{\perp}]}\cdot \lambda_{[d_{\perp}]}$) the transverse vector living in $d_{\perp}$ dimensions. The explicit definition of the basis vectors $\{e_i^{\alpha}\}$ can be found in Appendix~\ref{Ap:32}.
		\begin{itemize}
			\item \textbf{Four-point integrals $(\ell=1,\,d_{\parallel}=3)$}
			\begin{align}
			I_4^{d\,(1)}[\,\mathcal{N}\,]=\int \frac{d^dq}{\pi^{d/2}}\frac{\mathcal{N}(q)}{D_0D_1D_2D_3},
			\label{1lint4}
			\end{align}
			- Loop momentum decomposition, $q^{\alpha}=q_{[3]}^{\alpha}+\lambda_{[d-3]}^{\alpha}$:
			\begin{align}
			q_{[3]}^{\alpha}=&\sum_{i=1}^{3}x_{i}e_{i}^{\alpha},\quad\lambda_{[d-3]}^{\alpha}=x_{4}e_4^{\alpha}+\mu^{\alpha},
			\end{align}
			- Denominators:
			\begin{align}
			&D_0=q^2_{[3]}+\lambda^2+m_0^2,
			&\quad
			&D_1=(q^2_{[3]}+p_1)^2+\lambda^2+m_1^2,\nn
			&D_2=(q^2_{[3]}+p_1+p_2)^2+\lambda^2+m_2^2,
			&\quad
			&D_3=(q^2_{[3]}+p_1+p_2+p_3)^2+\lambda^2+m_3^2,
			\end{align}
			- Transverse variable:
			\begin{align}
			x_{4}=\lambda\cos\theta_{1},
			\end{align}
			- $d=d_{\parallel}+d_{\perp}$ parametrization
			 \begin{align}
			 I_4^{d\,(1)}[\,\mathcal{N}\,]=& \frac{1}{\pi^2\Gamma\big(\frac{d-4}{2}\big)}\int\! d^3\!q_{[3]}\int_{0}^{\infty}\!d\lambda^{2} (\lambda^{2})^{\frac{d-5}{2}}\int_{-1}^{1}\!d\!\cos\theta_{1}(\sin\theta_{1})^{d-6}\times\nn
			 &\frac{\mathcal{N}(q_{[3]},\lambda^{2},\cos\theta_1)}{D_0D_1D_2D_3},
			 \label{eq:1Lboxap}
			 \end{align}
			 - Transverse tensor integrals:
			 \begin{align}
			 I_4^{d\,(1)}[\,x_4^{2n_4+1}\,]=&0,\nn
			 I_4^{d\,(1)}[\,x_4^{2n_4}\,]=&\frac{(2n_4-1)!!}{\prod_{i=1}^{n_4}(d-5+2i)}I_4^{d\,(1)}[\,\lambda^{2n_4}\,]=\frac{(2n_4-1)!!}{2^{n4}}I_4^{d+2n_4\,(1)}[\,1\,].
			 \end{align}
			\item\textbf{Three-point integrals $(\ell=1,\,d_{\parallel}=2)$}
			\begin{align}
			I_3^{d\,(1)}[\,\mathcal{N}\,]=\int \frac{d^dq}{\pi^{d/2}}\frac{\mathcal{N}(q)}{D_0D_1D_2},
			\label{1lint3}
			\end{align}
			- Loop momentum decomposition, $q^{\alpha}=q_{[2]}^{\alpha}+\lambda_{[d-2]}^{\alpha}$:
			\begin{align}
			q_{[2]}^{\alpha}=&\sum_{i=1}^{2}x_{i}e_{i}^{\alpha},\quad\lambda_{[d-2]}^{\alpha}=\sum_{i=3}^{4}x_{i}e_i^{\alpha}+\mu^{\alpha},
			\end{align}
			- Denominators:
			\begin{align}
			&D_0=q^2_{[2]}+\lambda^2+m_0^2\nn
			&D_1=(q^2_{[2]}+p_1)^2+\lambda^2+m_1^2,\nn
			&D_2=(q^2_{[2]}+p_1+p_2)^2+\lambda^2+m_2^2,
			\end{align}
			- Transverse variables:
			\begin{align}
			\label{pol3}
			\begin{cases}
			&x_{3}=\lambda\cos\theta_{1}\\
			&x_{4}=\lambda\sin\theta_{1}\cos\theta_{2},
			\end{cases}
			\end{align}
			- $d=d_{\parallel}+d_{\perp}$ parametrization:
			\begin{align}
			I_3^{d\,(1)}[\,\mathcal{N}\,]=& \frac{1}{\pi^2\Gamma\big(\frac{d-4}{2}\big)}\int d^2q_{[2]}\int_{0}^{\infty}d\lambda^{2} (\lambda^{2})^{\frac{d-4}{2}}\int_{-1}^{1}\!d\!\cos\theta_{1}d\!\cos\theta_{2}\times\nn
			&(\sin\theta_{1})^{d-5}
			(\sin\theta_{2})^{d-6}\frac{\mathcal{N}(q_{[2]},\lambda^{2},\{\cos\theta_{1},\sin\theta_{1},\cos\theta_{2}\})}{D_0D_1D_2},
			\label{sub31}
			\end{align}
			- Transverse tensor integrals:
			\begin{align}
			I_3^{d\,(1)}[\,x_3^{m_3}x_4^{m_4}\,]=&0\quad\text{if}\quad m_3\vee m_4 \quad \text{odd},\nn
			I_3^{d\,(1)}[\,x_3^{2n_3}x_4^{2n_4}\,]=&\frac{\prod_{i=3}^4(2n_i-1)!!}{\prod_{i=1}^{n_3+n_4}(d-4+2i)}I_3^{d\,(1)}[\,\lambda^{2(n_3+n_4)}\,]\nn
			=&\prod_{i=3}^4\frac{(2n_i-1)!!}{2^{n_i}}I_3^{d+2(n_3+n_4)\,(1)}[\,1\,].
			\end{align}
			\item\textbf{Two-point integrals with $p^2\neq 0$ $(\ell=1,\,d_{\parallel}=1)$}
			\begin{align}
			I_2^{d\,(1)}[\,\mathcal{N}\,]=\int \frac{d^dq}{\pi^{d/2}}\frac{\mathcal{N}(q)}{D_0D_1},
			\label{1lint2p}
			\end{align}
			- Loop momentum decomposition, $q^{\alpha}=q_{[1]}^{\alpha}+\lambda_{[d-1]}^{\alpha}$:
			\begin{align}
			q_{[1]}^{\alpha}=&x_{1}e_{1}^{\alpha},\quad\lambda_{[d-1]}^{\alpha}=\sum_{i=2}^{4}x_{i}e_i^{\alpha}+\mu^{\alpha},
			\end{align}
			- Denominators:
			\begin{align}
			D_0=q_{[1]}^2+\lambda^2+m_0^2,&\quad&D_1=(q_{[1]}+p)^2+\lambda^2+m_1^2,
			\end{align}
		    - Transverse variables:
			\begin{align}
			\begin{cases}
			&x_{2}=\lambda\cos\theta_{1}\\
			&x_{3}=\lambda\sin\theta_{1}\cos\theta_{2},\\
			&x_{4}=\lambda\sin\theta_{1}\sin\theta_{2}\cos\theta_{3},
			\end{cases}
			\end{align}
		    - $d=d_{\parallel}+d_{\perp}$ parametrization:
			\begin{align}
			I_2^{d\,(1)}[\,\mathcal{N}\,]=&\frac{1}{\pi^2\Gamma\big(\frac{d-4}{2}\big)}\int dq_{[1]}\int_{0}^{\infty}d\lambda^2 (\lambda^{2})^{\frac{d-3}{2}}\int_{-1}^{1}\!d\!\cos\theta_{1}d\!\cos\theta_{2}d\!\cos\theta_{3}\times\nn
			&(\sin\theta_{1})^{d-4}
			(\sin\theta_{2})^{d-5}(\sin\theta_{3})^{d-6}\times\nn
			&\frac{\mathcal{N}(q_{[1]},\lambda_{11},{\cos\theta_{1},\sin\theta_{1},\cos\theta_{2},\sin\theta_{2},\cos\theta_{3}})}{D_0D_1},
			\label{sub21}
			\end{align}
	    	- Transverse tensor integrals:
				\begin{align}
				I_2^{d\,(1)}[\,x_{2}^{m_2}x_3^{m_3}x_4^{m_4}\,]=&0\quad\text{if}\quad m_2\vee m_3\vee m_4 \quad \text{odd},\nn
				I_2^{d\,(1)}[\,x_2^{2n_2}x_3^{2n_3}x_4^{2n_4}\,]=&\frac{\prod_{i=2}^4(2n_i-1)!!}{\prod_{i=1}^{n_2+n_3+n_4}(d-3+2i)}I_2^{d\,(1)}[\,\lambda^{2(n_2+n_3+n_4)}\,]\nn
				=&\prod_{i=2}^4\frac{(2n_i-1)!!}{2^{n_i}}I_2^{d+2(n_2+n_3+n_4)\,(1)}[\,1\,].
				\end{align}
			\item\textbf{Two-point integrals with $p^2= 0$ $(\ell=1,\,d_{\parallel}=2)$}
				\begin{align}
				I_2^{d\,(1)}[\,\mathcal{N}\,]|_{p^{2}=0}=\int \frac{d^dq}{\pi^{d/2}}\frac{\mathcal{N}(q)}{D_0D_1},
				\label{1lint2p0}
				\end{align}
				- Loop momentum decomposition, $q^{\alpha}=q_{[2]}^{\alpha}+\lambda_{[d-2]}^{\alpha}$:
				\begin{align}
				q_{[2]}^{\alpha}=&\sum_{i=1}^{2}x_{i}e_{i}^{\alpha},\quad\lambda_{[d-2]}^{\alpha}=\sum_{i=3}^{4}x_{i}e_i^{\alpha}+\mu^{\alpha},
				\end{align}
				- Denominators:
				\begin{align}
				&D_0=q^2_{[2]}+\lambda^2+m_0^2, &\quad D_1=&(q^2_{[2]}+p)^2+\lambda^2+m_1^2,
				\end{align}
				- Transverse variables:
				\begin{align}
				\label{pol2p0}
				\begin{cases}
				&x_{3}=\lambda\cos\theta_{1}\\
				&x_{4}=\lambda\sin\theta_{1}\cos\theta_{2},
				\end{cases}
				\end{align}
				- $d=d_{\parallel}+d_{\perp}$ parametrization:
				\begin{align}
				I_2^{d\,(1)}[\,\mathcal{N}\,]|_{p^{2}=0}=& \frac{1}{\pi^2\Gamma\big(\frac{d-4}{2}\big)}\int d^2q_{[2]}\int_{0}^{\infty}d\lambda^2 (\lambda^2)^{\frac{d-4}{2}}\int_{-1}^{1}\!d\!\cos\theta_{1}(\sin\theta_{1})^{d-5}\times\nn
				&\int_{-1}^{1}\!d\!\cos\theta_{2}(\sin\theta_{2})^{d-6}\frac{\mathcal{N}(q_{[2]},\lambda^2,\{\cos\theta_{1},\sin\theta_{1},\cos\theta_{2}\})}{D_0D_1D_2},
				\label{sub21p0}
				\end{align}
				- Transverse tensor integrals:
				\begin{align}
				I_2^{d\,(1)}[\,x_3^{m_3}x_4^{m_4}\,]|_{p^{2}=0}=&0\quad\text{if}\quad m_3\vee m_4 \quad \text{odd},\nn
				I_2^{d\,(1)}[\,x_3^{2n_3}x_4^{2n_4}\,]|_{p^{2}=0}=&\frac{(2n_3-1)!!(2n_4-1)!!}{\prod_{i=1}^{n_3+n_4}(d-4+2i)}I_2^{d\,(1)}[\,\lambda^{2(n_3+n_4)}\,]|_{p^{2}=0}\nn
				=&\prod_{i=3}^4\frac{(2n_i-1)!!}{2^{n_i}}I_3^{d+2(n3+n_4)\,(1)}[\,1\,]|_{p^{2}=0}.
				\end{align}
			\item\textbf{One-point integrals $(\ell=1,\,d_{\parallel}=0)$}
			 \begin{align}
			 I_1^{d\,(1)}[\,\mathcal{N}\,]=\int \frac{d^dq}{\pi^{d/2}}\frac{\mathcal{N}(q)}{D_0},
			 \label{1lint1p}
			 \end{align}
			 - Loop momentum decomposition, $q^{\alpha}=\lambda^{\alpha}_{[d]}$: 
			 \begin{align}
			 &q^{\alpha}\equiv\lambda_{[d]}^\alpha=\sum_{i=1}^{4}x_{i}^{\alpha}e_i^{\alpha}+\mu^{\alpha},
			 \label{la1pt}
			 \end{align}
			 - Denominator:
			 \begin{align}
			 D_0=\lambda^2+m_0^2,
			 \end{align}
			 - Transverse variables:
			 \begin{align}
			 \begin{cases}
			 &x_{1}=\lambda\cos\theta_{1},\\
			 &x_{2}=\lambda\sin\theta_{1}\cos\theta_{2},\\
			 &x_{3}=\lambda\sin\theta_{1}\sin\theta_{2}\cos\theta_{3}\\
			 &x_{4}=\lambda\sin\theta_{1}\sin\theta_{2}\sin\theta_{3}\cos\theta_{4},
			 \end{cases}
			 \end{align}  
			 $d=d_{\parallel}+d_{\perp}$ parametrization:
			 \begin{align}
			 I_1^{d\,(1)}[\,\mathcal{N}\,]=& \frac{1}{\pi^2\Gamma\big(\frac{d-4}{2}\big)}\int_{0}^{\infty}d\lambda^2 (\lambda^2)^{\frac{d-2}{2}}\int_{-1}^{1}\!d\!\cos\theta_{1}d\!\cos\theta_{2}d\!\cos\theta_{3}d\cos\theta_{4}\times\nn
			 &(\sin\theta^2)^{d-3}(\sin\theta_{2})^{d-4}(\sin\theta_{3})^{d-5}(\sin\theta_{4})^{d-6}\times\nn
			 &\frac{\mathcal{N}(\lambda^2,\cos\theta_{1},\sin\theta_{1},\cos\theta_{2},\sin\theta_{2},\cos\theta_{3},\sin\theta_{3},\cos\theta_{4})}{D_0},
			 \label{sub11}
			\end{align}	
		    - Transverse tensor integrals:
				\begin{align}
				I_1^{d\,(1)}[\,x_{1}^{m_1}x_{2}^{m_2}x_3^{m_3}x_4^{m_4}\,]=&0\quad\text{if}\quad m_1\vee m_2\vee m_3\vee m_4 \quad \text{odd},\nn
				I_1^{d\,(1)}[\,x_1^{2n_1}x_2^{2n_2}x_3^{2n_3}x_4^{2n_4}\,]=&\frac{\prod_{i=1}^4(2n_i-1)!!}{\prod_{i=1}^{n_1+n_2+n_3+n_4}(d-3+2i)}I_2^{d\,(1)}[\,\lambda^{2(n_1+n_2+n_3+n_4)}\,]\nn
				=&\prod_{i=1}^3\frac{(2n_i-1)!!}{2^{n_i}}I_2^{d+2(n_1+n_2+n_3+n_4)\,(1)}[\,1\,].
			   \end{align}
		\end{itemize}
		\section{Two-loop integrals}
		\label{Ap:2l}
		In this appendix we collect some useful formulae for two-loop integrals in $d=d_{\parallel}+d_{\perp}$. As for Appendix~\ref{Ap:1l}, we indicate as $q_{[d_{\parallel}]\,i}^{\alpha}$ the component of the loop momenta lying in the space spanned by the $d_{\parallel}$ independent external momenta and we denote by $\lambda_{[d_{\perp}]\, i}^{\alpha}$ ($\lambda_{ij}\equiv \lambda_{[d_{\perp}]\,i}\cdot \lambda_{[d_{\perp}]\, j}$) the transverse vectors living in $d_{\perp}$ dimensions. The explicit definition of the basis vectors $\{e_i^{\alpha}\}$ can be found in Appendix~\ref{Ap:32}. In all cases, the relative orientation of the transverse vectors is defined as
		\begin{align}
			\lambda_{12}=&\sqrt{\lambda_{11}\lambda_{22}}\cos\theta_{12}.
		\end{align}
		\begin{itemize}
			\item \textbf{Four-point integrals $(\ell=2,\,d_{\parallel}=3)$}
			\begin{align}
			I_4^{d\,(2)}[\,\mathcal{N}\,]=\int \frac{d^dq_1d^dq_2}{\pi^{d}}\frac{\mathcal{N}(q_1,q_2)}{D_1\dots D_n},
			\label{2lint4}
			\end{align}
			- Loop momenta decomposition, $q^{\alpha}=q_{[3]\,i}^{\alpha}+\lambda_{[d-3]\,i}^{\alpha}$:
			\begin{align}
			q_{[3]\,i}^{\alpha}=&\sum_{j=1}^{3}x_{ji}e_{j}^{\alpha},\quad\lambda_{[d-3]\,i}^{\alpha}=x_{4i}e_4^{\alpha}+\mu_{i}^{\alpha},
			\end{align}
			- Transverse variables:
			\begin{align}
			\begin{cases}
			x_{41}=&\sqrt{\lambda_{11}}\cos\theta_{11}\\
			x_{42}=&\sqrt{\lambda_{22}}\big(\cos\theta_{11}\cos\theta_{12}+\sin\theta_{11}\sin\theta_{12}\cos\theta_{22}\big),
			\end{cases}
			\end{align}
			- $d=d_{\parallel}+d_{\perp}$ parametrization:
			\begin{align}
				I_4^{d\,(2)}[\,\mathcal{N}\,]=&\frac{2^{d-6}}{\pi^5\Gamma(d-5)}\int\! d^3q_{[3]\,1}d^3q_{[3]\, 2}
				\int_{0}^{\infty}\!d\lambda_{11}d\lambda_{22} (\lambda_{11})^{\frac{d-5}{2}}(\lambda_{22})^{\frac{d-5}{2}}\times\nn
				&\int_{-1}^{1}\!d\!\cos\theta_{12}d\!\cos\theta_{22}d\!\cos\theta_{11}\left(\sin\theta_{12}\right)^{d-6}(\sin\theta_{11})^{d-6}
				(\sin\theta_{22})^{d-7}\times\nn
				&\frac{\mathcal{N}(q_1,q_2)}{{D_1\dots D_n}},
			\label{eq:2Lboxap}
			\end{align}
			- Transverse tensor integrals (unless otherwise stated, we assume $i\neq j$):
				\begin{align}
			    I_4^{d\,(2)}[\,x_{4i}x_{4j}\,]=&\frac{1}{(d-3)}I_4^{d\,(2)}[\,\lambda_{ij}]\quad \forall i,j,\nn
				I_4^{d\,(2)}[\,x_{4i}^4\,]=&\frac{3}{(d-3)(d-1)}I_4^{d\,(2)}[\,\lambda_{ii}^2]\quad \forall i,j,\nn
				I_4^{d\,(2)}[\,x_{4i}^3x_{4j}\,]=&\frac{3}{(d-3)(d-1)}I_4^{d\,(2)}[\,\lambda_{12}\lambda_{ii}],\nn
				I_4^{d\,(2)}[\,x_{41}^2x_{42}^2\,]=&\frac{3}{(d-3)(d-1)}I_4^{d\,(2)}[\,2\lambda_{12}^2+\lambda_{11}\lambda_{22}],\nn
				I_4^{d\,(2)}[\,x_{4i}^6\,]=&\frac{15}{(d-3)(d-1)(d+1)}I_4^{d\,(2)}[\,\lambda_{ii}^3],\nn
				I_4^{d\,(2)}[\,x_{4i}^5x_{4j}\,]=&\frac{1}{(d-3)(d-1)(d+1)}I_4^{d\,(2)}[\,\lambda_{12}\lambda_{ii}^2],\nn
				I_4^{d\,(2)}[\,x_{4i}^4x_{4j}^2\,]=&\frac{3}{(d-3)(d-1)(d+1)}I_4^{d\,(2)}[\lambda_{ii}(4\lambda_{12}^2+\lambda_{11}\lambda_{22})],\nn
				I_4^{d\,(2)}[\,x_{42}^3x_{41}^3\,]=&\frac{3}{(d-3)(d-1)(d+1)}I_4^{d\,(2)}[\lambda_{12}(2\lambda_{12}^2+3\lambda_{11}\lambda_{22})].
				\end{align}
				 Moreover, in general we have
				 \begin{align}
				 	I_4^{d\,(2)}[\,x_{41}^{\alpha_4}x_{42}^{\beta_4}\,]=&0,\quad \text{if}\quad \alpha_4+\beta_4=2n+1.
				 \end{align}
			\item \textbf{Three-point integrals $(\ell=2,\,d_{\parallel}=2)$}
			\begin{align}
			I_3^{d\,(2)}[\,\mathcal{N}\,]=\int \frac{d^dq_1d^dq_2}{\pi^{d}}\frac{\mathcal{N}(q_1,q_2)}{D_1\dots D_n},
			\label{2lint3}
			\end{align}
			- Loop momenta decomposition, $q^{\alpha}=q_{[2]\,i}^{\alpha}+\lambda_{[d-2]\,i}^{\alpha}$:
			\begin{align}
			q_{[2]\,i}^{\alpha}=&\sum_{j=1}^{2}x_{ji}e_{j}^{\alpha},\quad\lambda_{[d-2]\,i}^{\alpha}=\sum_{j=3}^{4}x_{ji}e_{i}^{\alpha}+\mu_{i}^{\alpha},
			\end{align}
			- Transverse variables:
		   \begin{align}
		   \begin{cases}
		   x_{31}=&\sqrt{\lambda_{11}}\cos\theta_{11}\\
		   x_{41}=&\sqrt{\lambda_{11}}\sin\theta_{11}\cos\theta_{21}\\
		   x_{32}=&\sqrt{\lambda_{22}}\big(\cos\theta_{12}\cos\theta_{11}+\sin\theta_{12}\cos\theta_{22}\sin\theta_{11}\big)\\
		   x_{42}=&\sqrt{\lambda_{22}}\big[\cos\theta_{12}\cos\theta_{21}\sin\theta_{11}+\sin\theta_{12}\big(\cos\theta_{32}\sin\theta_{21}\sin\theta_{22}\\
		   &-\cos\theta_{11}\cos\theta_{21}\cos\theta_{22}\big)\big],
		   \end{cases}
		   \label{coor3pt}
		\end{align}
			- $d=d_{\parallel}+d_{\perp}$ parametrization:
			\begin{align}
			I_3^{d}[\,\mathcal{N}\,]=&\frac{2^{d-6}}{\pi^5\Gamma(d-5)}\int\! d^2q_{[2]\,1}d^2q_{[2]\,2}
			\int_{0}^{\infty}\!d\lambda_{11}d\lambda_{22}(\lambda_{11})^{\frac{d-4}{2}}(\lambda_{22})^{\frac{d-4}{2}}\times\nn
			&\int_{-1}^{1}\!d\!\cos\theta_{12}d\!\cos\theta_{11}d\!\cos\theta_{21}d\!\cos\theta_{22}d\!\cos\theta_{32}\left(\sin\theta_{12}\right)^{d-5}(\sin\theta_{11})^{d-5}\times\nn
			&(\sin\theta_{21})^{d-6}(\sin\theta_{22})^{d-6}(\sin\theta_{32})^{d-7}\frac{\mathcal{N}(q_1,q_2)}{{D_1\dots D_n}},
			\label{eq:2Ltriap}
			\end{align}
			- Transverse tensor integrals (unless otherwise stated, we assume $i\neq j$):
				\begin{align}
				I_3^{d\,(2)}[\,x_{3i}x_{3j}\,]=&I_3^{d\,(2)}[\,x_{4i}x_{4j}\,]=\frac{1}{(d-2)}I_3^{d\,(2)}[\,\lambda_{ij}]\quad \forall i,j,\nn
				I_3^{d\,(2)}[\,x_{3i}^4\,]=&I_3^{d\,(2)}[\,x_{4i}^4\,]=\frac{3}{(d-2)d}I_3^{d\,(2)}[\,\lambda_{ii}^2],\nn
				I_3^{d\,(2)}[\,x_{3i}^3x_{3j}\,]=&I_3^{d\,(2)}[\,x_{4i}^3x_{4j}\,]=\frac{3}{(d-2)d}I_3^{d\,(2)}[\,\lambda_{ii}\lambda_{ij}],\nn
				I_3^{d\,(2)}[\,x_{31}^2x_{32}^2\,]=& I_3^{d\,(2)}[\,x_{41}^2x_{42}^2\,]=\frac{1}{(d-2)d}I_3^{d\,(2)}[\,2\lambda_{12}^2+\lambda_{11}\lambda_{22}],\nn
			    I_3^{d\,(2)}[\,x_{3i}^2x_{4j}^2\,]=&\frac{1}{(d-3)(d-2)d}I_3^{d\,(2)}[\,-2\lambda_{12}^2+(d-1)\lambda_{11}\lambda_{22}],\nn
			    I_3^{d\,(2)}[\,x_{3i}^2x_{4i}x_{4j}\,]=&I_3^{d\,(2)}[\,x_{4i}^2x_{3i}x_{3j}\,]=\frac{1}{(d-2)d}I_3^{d\,(2)}[\,\lambda_{12}\lambda_{ii}],\nn
			    I_3^{d\,(2)}[\,x_{31}x_{41}x_{32}x_{32}\,]=&\frac{1}{(d-3)(d-2)d}I_3^{d\,(2)}[\,(d-2)\lambda_{12}^2-\lambda_{11}\lambda_{22}].
				\end{align}
			  Moreover, in general we have
			  \begin{align}
			  	I_3^{d\,(2)}[\,x_{31}^{\alpha_3}x_{41}^{\alpha_4}x_{32}^{\beta_3}x_{42}^{\beta_4}\,]=&0,\quad \text{if}\quad \alpha_i+\beta_i=2n+1.
			  \end{align}
			\item \textbf{Two-point integrals with $p^2\neq 0$ $(\ell=2,\,d_{\parallel}=1)$}
			\begin{align}
			I_2^{d\,(2)}[\,\mathcal{N}\,]=\int \frac{d^dq_1d^dq_2}{\pi^{d}}\frac{\mathcal{N}(q_1,q_2)}{D_1\dots D_n},
			\label{2lint2}
			\end{align}
			- Loop momenta decomposition, $q^{\alpha}=q_{[1]\,i}^{\alpha}+\lambda_{[d-1]\,i}^{\alpha}$:
			\begin{align}
			q_{[1]\,i}^{\alpha}=&x_{1i}e_{1}^{\alpha},\quad\lambda_{[d-1]\,i}^{\alpha}=\sum_{j=2}^{4}x_{ji}e_{i}^{\alpha}+\mu_{i}^{\alpha},
			\end{align}
			- Transverse variables:
			\begin{align}
			\begin{cases}
			x_{21}=&\sqrt{\lambda_{11}}\cos\theta_{11},\\
			x_{31}=&\sqrt{\lambda_{11}}\sin\theta_{11}\cos\theta_{21},\\
			x_{41}=&\sqrt{\lambda_{11}}\sin\theta_{11}\sin\theta_{21}\cos\theta_{31},\\
			x_{22}=&\sqrt{\lambda_{22}}\big(\cos\theta_{12}\cos\theta_{11}+\sin\theta_{12}\cos\theta_{22}\sin\theta_{11}\big)\\
			x_{32}=&\sqrt{\lambda_{22}}\left[\cos\theta_{12}\cos\theta_{21}\sin\theta_{11}+\sin\theta_{12}\right(\cos\theta_{32}\sin\theta_{21}\sin\theta_{22}\\
			&-\cos\theta_{11}\cos\theta_{21}\cos\theta_{22}\left)\right]\\
			x_{42}=&\sqrt{\lambda_{22}}[\cos\theta_{12}\cos\theta_{31}\sin\theta_{11}\sin\theta_{21}+\sin\theta_{12}(\cos\theta_{42}\sin\theta_{31}\sin\theta_{22}\sin\theta_{32}\\
			&\qquad\quad-\cos\theta_{11}\cos\theta_{31}\cos\theta_{22}\sin\theta_{21}-\cos\theta_{21}\cos\theta_{31}\cos\theta_{32}\sin\theta_{22})],
			\end{cases}
			\label{coor2pt}
			\end{align} 
			- $d=d_{\parallel}+d_{\perp}$ parametrization:
			\begin{align}
			I_2^{d}[\,\mathcal{N}\,]=&\frac{2^{d-6}}{\pi^5\Gamma(d-5)}\int\! dq_{[1]\,1}dq_{[1]\,2}
			\int_{0}^{\infty}\!d\lambda_{11}d\lambda_{22}(\lambda_{11})^{\frac{d-3}{2}}(\lambda_{22})^{\frac{d-3}{2}}\times\nn
			&\int_{-1}^{1}\!d\!\cos\theta_{12}d\!\cos\theta_{11}d\!\cos\theta_{21}d\!\cos\theta_{31}d\!\cos\theta_{22}d\!\cos\theta_{32}d\!\cos\theta_{42}\times\nn
			&\left(\sin\theta_{12}\right)^{d-4}(\sin\theta_{11})^{d-4}(\sin\theta_{21})^{d-5}(\sin\theta_{31})^{d-6}(\sin\theta_{22})^{d-5}\times\nn
			&(\sin\theta_{32})^{d-6}(\sin\theta_{42})^{d-7}\frac{\mathcal{N}(q_1,q_2)}{{D_1\dots D_n}},
			\label{eq:2Lbubap}
			\end{align}
			- Transverse tensor integrals (unless otherwise stated, we assume $i\neq j$):
				\begin{align}
				I_2^{d\,(2)}[\,x_{2i}x_{2j}\,]=&I_2^{d\,(2)}[\,x_{3i}x_{3j}\,]=I_2^{d\,(2)}[\,x_{4i}x_{4j}\,]=\frac{1}{(d-1)}I_2^{d\,(2)}[\,\lambda_{ij}],\quad \forall i,j,\nn
				I_2^{d\,(2)}[\,x_{2i}^4\,]=&I_2^{d\,(2)}[\,x_{3i}^4\,]=I_2^{d\,(2)}[\,x_{3i}^4\,]=\frac{3}{(d-1)(d+1)}I_2^{d\,(2)}[\,\lambda_{ii}^2],\nn
				I_2^{d\,(2)}[\,x_{2i}^3x_{2j}\,]=&I_2^{d\,(2)}[\,x_{3i}^3x_{3j}\,]=I_2^{d\,(2)}[\,x_{4i}^3x_{4j}\,]=\frac{3}{(d-1)(d+1)}I_3^{d\,(2)}[\,\lambda_{ii}\lambda_{12}],\nn
				I_2^{d\,(2)}[\,x_{2i}^2x_{2i}^2\,]=&I_2^{d\,(2)}[\,x_{3i}^2x_{3i}^2\,]=I_2^{d\,(2)}[\,x_{4i}^2x_{4i}^2\,]\nn
				&=\frac{1}{(d-1)(d+1)}I_2^{d\,(2)}[\,2\lambda_{12}^2+\lambda_{11}\lambda_{22}],\nn
				I_2^{d\,(2)}[\,x_{2i}^2x_{3i}^2\,]=&I_2^{d\,(2)}[\,x_{2i}^2x_{4i}^2\,]=I_2^{d\,(2)}[\,x_{3i}^2x_{4i}^2\,]=\frac{1}{(d-1)(d+1)}I_2^{d\,(2)}[\,\lambda_{ii}^2],\nn
				I_2^{d\,(2)}[\,x_{2i}^2x_{3j}^2\,]=&I_2^{d\,(2)}[\,x_{2i}^2x_{4j}^2\,]=I_2^{d\,(2)}[\,x_{3i}^2x_{4j}^2\,]\nn
				&=\frac{1}{(d-2)(d-1)(d+1)}I_2^{d\,(2)}[\,-2\lambda_{12}^2+d\lambda_{11}\lambda_{22}],\nn
				I_2^{d\,(2)}[\,x_{2i}^2x_{3i}x_{3j}\,]=&I_2^{d\,(2)}[\,x_{2i}^2x_{4i}x_{4j}\,]=\frac{1}{(d-1)(d+1)}I_2^{d\,(2)}[\,\lambda_{12}\lambda_{ii}],\nn
				I_2^{d\,(2)}[\,x_{3i}^2x_{2i}x_{2j}\,]=&I_2^{d\,(2)}[\,x_{3i}^2x_{4i}x_{4j}\,]=\frac{1}{(d-1)(d+1)}I_2^{d\,(2)}[\,\lambda_{12}\lambda_{ii}],\nn
				I_2^{d\,(2)}[\,x_{4i}^2x_{2i}x_{2j}\,]=&I_2^{d\,(2)}[\,x_{4i}^2x_{3i}x_{3j}\,]=\frac{1}{(d-1)(d+1)}I_2^{d\,(2)}[\,\lambda_{12}\lambda_{ii}],\nn
				I_2^{d\,(2)}[\,x_{21}x_{31}x_{22}x_{32}\,]=&\frac{1}{(d-2)(d-1)(d+1)}I_2^{d\,(2)}[\,(d-1)\lambda_{12}^2-\lambda_{11}\lambda_{22}],\nn
				I_2^{d\,(2)}[\,x_{21}x_{41}x_{22}x_{42}\,]=&\frac{1}{(d-2)(d-1)(d+1)}I_2^{d\,(2)}[\,(d-1)\lambda_{12}^2-\lambda_{11}\lambda_{22}],\nn
				I_2^{d\,(2)}[\,x_{31}x_{41}x_{32}x_{42}\,]=&\frac{1}{(d-2)(d-1)(d+1)}I_2^{d\,(2)}[\,(d-1)\lambda_{12}^2-\lambda_{11}\lambda_{22}].
				\end{align}
	           Moreover, in general we have
	           \begin{align}
	           	I_2^{d\,(2)}[\,x_{21}^{\alpha_2}x_{31}^{\alpha_3}x_{41}^{\alpha_4}x_{22}^{\beta_2}x_{32}^{\beta_3}x_{42}^{\beta_4}\,]=&0,\quad \text{if}\quad \alpha_i+\beta_i=2n+1.
	           \end{align}
			\item \textbf{Two-point integrals with $p^2=0$ $(\ell=2,\,d_{\parallel}=2)$}
			\begin{align}
			I_2^{d\,(2)}[\,\mathcal{N}\,]|_{p^2=0}=\int \frac{d^dq_1d^dq_2}{\pi^{d}}\frac{\mathcal{N}(q_1,q_2)}{D_1\dots D_n},
			\label{2lint2p0}
			\end{align}
			- Loop momenta decomposition, $q^{\alpha}=q_{[2]\,i}^{\alpha}+\lambda_{[d-2]\,i}^{\alpha}$:
			\begin{align}
			q_{[2]\,i}^{\alpha}=&\sum_{j=1}^{2}x_{ji}e_{j}^{\alpha},\quad\lambda_{[d-2]\,i}^{\alpha}=\sum_{j=3}^{4}x_{ji}e_{i}^{\alpha}+\mu_{i}^{\alpha},
			\end{align}
			- Transverse variables:
			\begin{align}
			\begin{cases}
			x_{31}=&\sqrt{\lambda_{11}}\cos\theta_{11}\\
			x_{41}=&\sqrt{\lambda_{11}}\sin\theta_{11}\cos\theta_{21}\\
			x_{32}=&\sqrt{\lambda_{22}}\big(\cos\theta_{12}\cos\theta_{11}+\sin\theta_{12}\cos\theta_{22}\sin\theta_{11}\big)\\
			x_{42}=&\sqrt{\lambda_{22}}\big[\cos\theta_{12}\cos\theta_{21}\sin\theta_{11}+\sin\theta_{12}\big(\cos\theta_{32}\sin\theta_{21}\sin\theta_{22}\\
			&-\cos\theta_{11}\cos\theta_{21}\cos\theta_{22}\big)\big],
			\end{cases}
			\label{coor2ptp0}
			\end{align}
			- $d=d_{\parallel}+d_{\perp}$ parametrization:
			\begin{align}
			I_3^{d}[\,\mathcal{N}\,]=&\frac{2^{d-6}}{\pi^5\Gamma(d-5)}\int\! d^2q_{[2]\,1}d^2q_{[2]\,2}
			\int_{0}^{\infty}\!d\lambda_{11}d\lambda_{22}(\lambda_{11})^{\frac{d-4}{2}}(\lambda_{22})^{\frac{d-4}{2}}\times\nn
			&\int_{-1}^{1}\!d\!\cos\theta_{12}d\!\cos\theta_{11}d\!\cos\theta_{21}d\!\cos\theta_{22}d\!\cos\theta_{32}\left(\sin\theta_{12}\right)^{d-5}(\sin\theta_{11})^{d-5}\times\nn
			&(\sin\theta_{21})^{d-6}(\sin\theta_{22})^{d-6}(\sin\theta_{32})^{d-7}\frac{\mathcal{N}(q_1,q_2)}{{D_1\dots D_n}},
			\label{eq:2Lbubp0}
			\end{align}
			- Transverse tensor integrals(unless specified we assume $i\neq j$):
				\begin{align}
					I_2^{d\,(2)}[\,x_{3i}x_{3j}\,]|_{p^2=0}=&I_2^{d\,(2)}[\,x_{4i}x_{4j}\,]|_{p^2=0}=\frac{1}{(d-2)}I_2^{d\,(2)}[\,\lambda_{ij}]|_{p^2=0}\quad \forall i,j,\nn
					I_2^{d\,(2)}[\,x_{3i}^4\,]|_{p^2=0}=&I_2^{d\,(2)}[\,x_{4i}^4\,]|_{p^2=0}=\frac{3}{(d-2)d}I_2^{d\,(2)}[\,\lambda_{ii}^2]|_{p^2=0},\nn
					I_2^{d\,(2)}[\,x_{3i}^3x_{3j}\,]|_{p^2=0}=&I_2^{d\,(2)}[\,x_{4i}^3x_{4j}\,]|_{p^2=0}=\frac{3}{(d-2)d}I_2^{d\,(2)}[\,\lambda_{ii}\lambda_{ij}]|_{p^2=0},\nn
					I_2^{d\,(2)}[\,x_{31}^2x_{32}^2\,]|_{p^2=0}=& I_2^{d\,(2)}[\,x_{41}^2x_{42}^2\,]|_{p^2=0}=\frac{1}{(d-2)d}I_2^{d\,(2)}[\,2\lambda_{12}^2+\lambda_{11}\lambda_{22}]|_{p^2=0},\nn
					I_2^{d\,(2)}[\,x_{3i}^2x_{4j}^2\,]|_{p^2=0}=&\frac{1}{(d-3)(d-2)d}I_2^{d\,(2)}[\,-2\lambda_{12}^2+(d-1)\lambda_{11}\lambda_{22}]|_{p^2=0},\nn
					I_2^{d\,(2)}[\,x_{3i}^2x_{4i}x_{4j}\,]|_{p^2=0}=&I_2^{d\,(2)}[\,x_{4i}^2x_{3i}x_{3j}\,]=\frac{1}{(d-2)d}I_2^{d\,(2)}[\,\lambda_{12}\lambda_{ii}]|_{p^2=0},\nn
					I_2^{d\,(2)}[\,x_{31}x_{41}x_{32}x_{32}\,]|_{p^2=0}=&\frac{1}{(d-3)(d-2)d}I_2^{d\,(2)}[\,(d-2)\lambda_{12}^2-\lambda_{11}\lambda_{22}]|_{p^2=0}.
				\end{align}
				Moreover, in general we have
				\begin{align}
				I_2^{d\,(2)}[\,x_{31}^{\alpha_3}x_{41}^{\alpha_4}x_{32}^{\beta_3}x_{42}^{\beta_4}\,]|_{p^2=0}=&0,\quad \text{if}\quad \alpha_i+\beta_i=2n+1.
				\end{align}
			\item \textbf{One-point integrals $(\ell=2,\,d_{\parallel}=0)$}
			\begin{align}
			I_1^{d\,(2)}[\,\mathcal{N}\,]=\int \frac{d^dq_1d^dq_2}{\pi^{d}}\frac{\mathcal{N}(q_1,q_2)}{D_1\dots D_n},
			\label{2lint1}
			\end{align}
			- Loop momenta decomposition, $q^{\alpha}=\lambda_{[d]\,i}^{\alpha}$:
			\begin{align}
			\lambda_{[d]\,i}^{\alpha}=\sum_{j=1}^{4}x_{ji}e_{i}^{\alpha}+\mu_{i}^{\alpha},
			\end{align}
			- Transverse variables:
			\begin{align}
			\begin{cases}
			x_{11}=&\sqrt{\lambda_{11}}\cos\theta_{11}\\
			x_{21}=&\sqrt{\lambda_{11}}\sin\theta_{11}\cos\theta_{21}\\
			x_{31}=&\sqrt{\lambda_{11}}\sin\theta_{11}\sin\theta_{21}\cos\theta_{31}\\
			x_{41}=&\sqrt{\lambda_{11}}\sin\theta_{11}\sin\theta_{21}\sin\theta_{31}\cos\theta_{41}\\
			x_{12}=&\sqrt{\lambda_{22}}\big(\cos\theta_{12}\cos\theta_{11}+\sin\theta_{12}\cos\theta_{22}\sin\theta_{11}\big)\\
			x_{22}=&\sqrt{\lambda_{22}}\big[\cos\theta_{12}\cos\theta_{21}\sin\theta_{11}+\sin\theta_{12}\big(\cos\theta_{32}\sin\theta_{21}\sin\theta_{22}\\
			&-\cos\theta_{11}\cos\theta_{21}\cos\theta_{22}\big)\big]\\
			x_{32}=&\sqrt{\lambda_{22}}[\cos\theta_{12}\cos\theta_{31}\sin\theta_{11}\sin\theta_{21}+\sin\theta_{12}(\cos\theta_{42}\sin\theta_{31}\sin\theta_{22}\sin\theta_{32}\\
			&-\cos\theta_{11}\cos\theta_{31}\cos\theta_{22}\sin\theta_{21}-\cos\theta_{21}\cos\theta_{31}\cos\theta_{32}\sin\theta_{22})]\nn
			x_{42}=&\sqrt{\lambda_{22}}[\cos\theta_{12}\cos\theta_{41}\sin\theta_{11}\sin\theta_{21}\sin\theta_{31}\\
			&+\sin\theta_{12}(\cos\theta_{52}\sin\theta_{41}\sin\theta_{22}\sin\theta_{32}\sin\theta_{42}\\
			&-\cos\theta_{11}\cos\theta_{41}\cos\theta_{22}\sin\theta_{21}\sin\theta_{31}\\
			&-\cos\theta_{21}\cos\theta_{41}\cos\theta_{32}\sin\theta_{22}\sin\theta_{31}\\
			&-\cos\theta_{31}\cos\theta_{41}\cos\theta_{42}\sin\theta_{22}\sin\theta_{32})],
			\end{cases}
			\label{coo12pt}
			\end{align}
			- $d=d_{\parallel}+d_{\perp}$ parametrization:
			\begin{align}
			I_1^{d}[\,\mathcal{N}\,]=&\frac{2^{d-6}}{\pi^5\Gamma(d-5)}
			\int_{0}^{\infty}\!d\lambda_{11}d\lambda_{22}(\lambda_{11})^{\frac{d-2}{2}}(\lambda_{11})^{\frac{d-2}{2}}\times\nn
			&\int_{-1}^{1}\!d\!\cos\theta_{12}d\!\cos\theta_{11}d\!\cos\theta_{21}d\!\cos\theta_{31}d\!\cos\theta_{41}d\!\cos\theta_{22}d\!\cos\theta_{32}d\!\cos\theta_{52}\times\nn
			&\left(\sin\theta_{12}\right)^{d-3}(\sin\theta_{11})^{d-3}(\sin\theta_{21})^{d-3}(\sin\theta_{31})^{d-5}(\sin\theta_{41})^{d-6}\times\nn
			&(\sin\theta_{22})^{d-4}(\sin\theta_{32})^{d-5}d\cos\theta_{42}(\sin\theta_{42})^{d-6}(\sin\theta_{52})^{d-7}\frac{\mathcal{N}(q_1,q_2)}{{D_1\dots D_n}},
			\end{align}
			- Transverse tensor integrals:
				\begin{align}
				I_1^{d\,(2)}[\,x_{1i}x_{1j}\,]=&I_1^{d\,(2)}[\,x_{2i}x_{2j}\,]=I_1^{d\,(2)}[\,x_{3i}x_{3j}\,]=I_1^{d\,(2)}[\,x_{4i}x_{4j}\,]=\frac{1}{d}I_1^{d\,(2)}[\,\lambda_{ij}],\quad \forall i,j. 
				\end{align}
				 Moreover, in general we have
				 \begin{align}
				 I_1^{d\,(2)}[\,x_{11}^{\alpha_1}x_{21}^{\alpha_2}x_{31}^{\alpha_3}x_{41}^{\alpha_4}x_{12}^{\beta_1}x_{22}^{\beta_2}x_{32}^{\beta_3}x_{42}^{\beta_4}\,]=&0,\quad \text{if}\quad \alpha_i+\beta_i=2n+1.
				 \end{align}
		\end{itemize}
		\section{Gegenbauer polynomials}
		\label{Ap:2}
		In this appendix we recall the most relevant properties of Gegenbauer polynomials.
		\textit{Gegenbauer polynomials} $C^{(\alpha)}_{n}(x)$ are  orthogonal polynomials over the interval $[-1,1]$ with respect to the weight function 
		\begin{align}
		\omega_\alpha(x)=(1-x^2)^{\alpha-\frac{1}{2}}
		\end{align}
		and they can be defined through the generating function
		\begin{align}
		\frac{1}{(1-2xt+t^2)^{\alpha}}=\sum_{n=0}^{\infty}C^{(\alpha)}_{n}(x)t^n.
		\end{align}
		These polynomials obey the orthogonality relation
		\begin{align}
		\int_{-1}^{1}dx \;\omega_\alpha(x)C^{(\alpha)}_{n}(x)C^{(\alpha)}_{m}(x)=\delta_{mn}\frac{2^{1-2\alpha}\pi\Gamma(n+2\alpha)}{n!(n+\alpha)\Gamma^2(\alpha)}.
		\label{ort1}
		\end{align}
		The explicit expression of the first Gegenbauer polynomials is given by
		\begin{align}
		C^{(\alpha)}_{0}(x)&=1,\nn
		C^{(\alpha)}_{1}(x)&=2\alpha x,\nn
		C^{(\alpha)}_{2}(x)&=-\alpha+2\alpha(1+\alpha)x^2,\nn
		&\cdots
		\end{align}
		and it can inverted in order to express arbitrary powers of the variable $x$ in terms of products of Gegenbauer polynomials,
		\begin{align}
		x&=\frac{1}{2\alpha}C_0^{(\alpha)}(x)C_1^{(\alpha)}(x),\nn
		x^2&=\frac{1}{4\alpha^2}[C_1^{(\alpha)}(x)]^2,\nn
		x^3&=\frac{1}{4\alpha^2(1+\alpha)}C_1^{(\alpha)}(x)[\alpha C^{(\alpha)}_0(x)+C_2^{(\alpha)}(x)],\nn
		x^4&=\frac{1}{4\alpha^2(1+\alpha)^2}[\alpha C^{(\alpha)}_0(x)+C_2^{(\alpha)}(x)]^2,\nn
		&\cdots
		\label{xs}
		\end{align}
		These identities can be used in order to evaluate the integral of any polynomial in $x$, convoluted with the weight function $\omega_\alpha(x)$, by means of the orthogonality relation \eqref{ort1}.
		\section{Four-dimensional basis}
		\label{Ap:3}
		In this appendix we provide the explicit definitions of the four-dimensional basis $\{e_i^{\alpha}\}$ used throughout the text to decompose the four-dimensional part of the loop momenta $q_{[4]\,i}$,
		 \begin{align}
		q_{[4]\,i}^{\alpha}=p_{0\,i}^{\alpha}+x_{1i}e_1^{\alpha}+x_{2i}e_2^{\alpha}+x_{3i}e_3^{\alpha}+x_{4i}e_4^{\alpha}.
		\end{align}
		In the following, for any pair of massless vectors $q_1^{\alpha}$ and $q_{2}^{\alpha}$, we denote by $\varepsilon^{\alpha}_{q_1,q_2}$ the spinor product
		\begin{align}
		\varepsilon^{\alpha}_{q_1,q_2}=\frac{1}{2}\langle q_1\gamma^{\alpha}q_2].
		\end{align}
		\subsection{$d=4-2\epsilon$ basis}
		\label{Ap:31}
		In the $d=4-2\epsilon$ parametrization of Feynman integrals we choose, independently from the number of external legs, a basis of massless vectors $\{e_i^{\alpha}\}$ defined in terms of two adjacent external momenta $p_1$ and $p_2$ by
		\begin{align}
		e_{1}^{\alpha}=\frac{1}{1-r_1r_2}(p_1^{\alpha}-r_1p_2^{\alpha}),\quad e_{2}^{\alpha}=\frac{1}{1-r_1r_2}(p_2^{\alpha}-r_2p_1^{\alpha}),\quad e_{3}^{\alpha}=\varepsilon^{\alpha}_{e_1,e_2},\quad e_{4}^{\alpha}=\varepsilon^{\alpha}_{e_2,e_1},
		\end{align}
		where
		\begin{align}
		r_i=\frac{p_i^2}{\gamma}\qquad\text{with}\qquad \gamma=(p_1\cdot p_2)\left(1+\sqrt{1-\frac{p_1^2p_2^2}{(p_1\cdot p_2)^2}}\right).
		\label{eq:r12}
		\end{align}
		In the case of two-point integrals, $p_1$ corresponds to the external momentum and $p_2$ is an arbitrary massless vector. In the case of one-point integrals, both $p_1$ and $p_2$ are chosen to be arbitrary massless vectors.
		\subsection{$d=d_{\parallel}+d_{\perp}$ basis}
		\label{Ap:32}
		In the $d=d_{\parallel}+d_{\perp}$ parametrization of Feynman integrals with $n\leq 4$ external legs, the four-dimensional basis $\{e_i^{\alpha}\}$ is chosen in such a way to satisfy the requirements
		 \begin{subequations}
		 	\begin{align}
		 	&e_i\cdot p_j=0,\:\qquad i> n-1,\quad \forall j=1,\dots n-1,\\
		 	&e_i\cdot e_j =\delta_{ij},\quad \;i,j> n-1,
		 	\end{align}
		  \end{subequations}
		  where $\{p_1,p_2,\dots,p_{n-1}\}$ is the set of independent external momenta.
		\begin{itemize}
		\item{\textbf{Four-point integrals}}\\
		In case of four-point integrals $\{e_i^{\alpha}\}$ is defined as
		\begin{align}
		e_{1}^{\alpha}=&\frac{1}{1-r_1r_2}(p_1^{\alpha}-r_1p_2^{\alpha}),\nn
		e_{2}^{\alpha}=&\frac{1}{1-r_1r_2}(p_2^{\alpha}-r_2p_1^{\alpha}),\nn
		e_{3}^{\alpha}=&\frac{1}{i\sqrt{\beta}}\left[\left(\varepsilon_{e_2,e_1}\cdot p_3\right) \varepsilon^{\alpha}_{e_1,e_2}+\left(\varepsilon_{e_1,e_2}\cdot p_3\right) \varepsilon^{\alpha}_{e_2,e_1}\right],\nn
		e_{4}^{\alpha}=&\frac{1}{\sqrt{\beta}}\left[\left(\varepsilon_{e_2,e_1}\cdot p_3\right) \varepsilon^{\alpha}{e_1,e_2}-\left(\varepsilon_{e_1,e_2}\cdot p_3\right) \varepsilon^{\alpha}_{e_2,e_1}\right].
		\end{align}
		with $r_{1,2}$ given by \eqref{eq:r12} and $\beta=2 e_1\cdot e_2\left(\varepsilon_{e_1,e_2}\cdot p_3\right)\left(\varepsilon_{e_1,e_2}\cdot p_3\right)$.
		\item{\textbf{Three-point integrals}}\\
		For three-point integrals $\{e_i^{\alpha}\}$ is defined as
		\begin{align}
		e_{1}^{\alpha}=&\frac{1}{1-r_1r_2}(p_1^{\alpha}-r_1p_2^{\alpha}),&\qquad e_{2}^{\alpha}=&\frac{1}{1-r_1r_2}(p_2^{\alpha}-r_2p_1^{\alpha}),\nn
		e_{3}^{\alpha}=&\frac{1}{i\sqrt{2 e_1\cdot e_2}}\left(\varepsilon^{\alpha}_{e_1,e_2}+\varepsilon^{\alpha}_{e_2,e_1}\right),&\qquad e_{4}^{\alpha}=&\frac{1}{\sqrt{2 e_1\cdot e_2}}\left(\varepsilon^{\alpha}_{e_1,e_2}-\varepsilon^{\alpha}_{e_2,e_1}\right),
		\end{align}
		with $r_{1,2}$ given by \eqref{eq:r12}.
		\item{\textbf{Two-point integrals with $p^2\neq 0$}}\\
		For a two-point integral with massive external momentum $p$, we introduce two massless vectors $q_1$ and $q_2$ satisfying
		\begin{align}
		p^\alpha=&q_1^{\alpha}+\frac{p^2}{2q_1\cdot q_2}q_{2}^\alpha
		\end{align}
		and we define the massive auxiliary momentum $q$
		\begin{align}
		q^\alpha=&q_1^{\alpha}-\frac{p^2}{2q_1\cdot q_2}q_{2}^\alpha.
		\end{align}
		The basis $\{e_i^{\alpha}\}$ is therewith defined as
		\begin{align}
		e_{1}^{\alpha}=&\frac{1}{\sqrt{p^2}}p^\alpha,&\qquad e_{2}^{\alpha}=&\frac{1}{i\sqrt{p^2}}q^\alpha,\nn
		e_{3}^{\alpha}=&\frac{1}{i\sqrt{2 q_1\cdot q_2}}(\varepsilon_{q_1,q_2}^{\alpha}+\varepsilon_{q_2,q_1}^{\alpha}),&\qquad
		e_{4}^{\alpha}=&\frac{1}{\sqrt{2 q_1\cdot q_2}}(\varepsilon_{q_1,q_2}^{\alpha}-\varepsilon_{q_2,q_1}^{\alpha}).
		\end{align}
		\item{\textbf{Two-point integrals with $p^2=0$}}\\
		In the case of two-point integrals with massless external momentum $p$, we introduce a massless auxiliary vector $q_1$ and we define the basis $\{e_i^{\alpha}\}$ as
		\begin{align}
		e_{1}^{\alpha}=&p^\alpha,&\qquad 
		e_{2}^{\alpha}=&q_1^\alpha,\nn
		e_{3}^{\alpha}=&\frac{1}{i\sqrt{2 p\cdot q_1}}(\varepsilon_{p,q_1}^{\alpha}+\varepsilon_{q_1,p}^{\alpha}),&\qquad
		e_{4}^{\alpha}=&\frac{1}{\sqrt{2 p\cdot q_1}}(\varepsilon_{p,q_1}^{\alpha}-\varepsilon_{q_1,p}^{\alpha}).
		\end{align}
		\item{\textbf{One-point integrals}}\\
		For one-point integrals we introduce two arbitrary independent massless vectors $q_1$ and $q_2$ and we build a completely orthonormal basis $\{e_i^{\alpha}\}$,
		\begin{align}
		e_{1}^{\alpha}=&\frac{1}{\sqrt{2q_1\cdot q_2}}(q_1^\alpha+q_2^{\alpha}),&\qquad
		e_{2}^{\alpha}=&\frac{1}{i\sqrt{2q_1\cdot q_2}}(q_1^\alpha-q_2^{\alpha}),\nn
		e_{3}^{\alpha}=&\frac{1}{i\sqrt{2 q_1\cdot q_2}}(\varepsilon_{q_1,q_2}^{\alpha}+\varepsilon_{q_2,q_1}^{\alpha}),&\qquad
		e_{4}^{\alpha}=&\frac{1}{\sqrt{2 q_1\cdot q_2}}(\varepsilon_{q_1,q_2}^{\alpha}-\varepsilon_{q_2,q_1}^{\alpha}).
		\end{align}
		\end{itemize}